\documentclass[11pt]{article}
 \newcommand{\blind}{0}
 
    \makeatletter
    \renewcommand\section{\@startsection {section}{1}{\z@}%
                                       {-3.5ex \@plus -1ex \@minus -.2ex}%
                                       {2.3ex \@plus.2ex}%
                                       {\normalfont\fontfamily{phv}\fontsize{16}{19}\bfseries}}
    \renewcommand\subsection{\@startsection{subsection}{2}{\z@}%
                                         {-3.25ex\@plus -1ex \@minus -.2ex}%
                                         {1.5ex \@plus .2ex}%
                                         {\normalfont\fontfamily{phv}\fontsize{14}{17}\bfseries}}
    \renewcommand\subsubsection{\@startsection{subsubsection}{3}{\z@}%
                                        {-3.25ex\@plus -1ex \@minus -.2ex}%
                                         {1.5ex \@plus .2ex}%
                                         {\normalfont\normaslsize\fontfamily{phv}\fontsize{14}{17}\selectfont}}
    \makeatother
\usepackage{epsfig}
\usepackage{graphics,graphicx,color}
\usepackage{amssymb,amsfonts,amsthm,amsmath}
\usepackage{multirow}
\usepackage{mathrsfs}
\usepackage{enumerate}
\usepackage{natbib}
\usepackage{float}
\usepackage{algorithm}
\usepackage{algpseudocode}
\usepackage{subcaption} 
\usepackage{multirow}
\usepackage{url}
\usepackage[colorlinks=true, linkcolor=blue, citecolor=blue]{hyperref}
\usepackage{bm}
\usepackage{verbatim}
\usepackage{booktabs}
\usepackage{makecell}
\usepackage{threeparttable}
\usepackage{authblk}
\usepackage{xcolor}

\usepackage[normalem]{ulem}

\newtheorem{proposition}{Proposition}%[section]
\newcommand{\diag}{\mbox{diag}}
\newcommand{\dd}{\mbox{d}}
\newcommand{\cov}{\mbox{cov}}

\graphicspath{ {./Figures/} }

\begin{document}
%%%%%%%%%%%%%%%%%%%%%%%%%%%%%%%%%%%%%%%%%%%%%%%%%%%%%%%%%%%%%%%%%%%%%%%%%%%%%%
\def\spacingset#1{\renewcommand{\baselinestretch}%
{#1}\small\normalsize} \spacingset{1}
%%%%%%%%%%%%%%%%%%%%%%%%%%%%%%%%%%%%%%%%%%%%%%%%%%%%%%%%%%%%%%%%%%%%%%%%%%%%%%

\if0\blind
{
\title{\bf Bayesian Bridge Gaussian Process Regression}
\author{Minshen Xu $^a$, Shiwei Lan $^b$, Lulu Kang $^a$ \\
$^a$ Department of Mathematics and Statistics, \\
University of Massachusetts Amherst, Amherst, MA, United States \\
$^b$ School of Mathematical and Statistical Sciences, \\
Arizona State University, Tempe, AZ, United States}
\date{}
\maketitle
}\fi
  
\if1\blind
{
\title{\bf Bayesian Bridge Gaussian Process Regression}
\author{Author information is purposely removed for double-blind review}
  
\bigskip
\bigskip
\bigskip
\begin{center}
{\LARGE\bf Bayesian Bridge Gaussian Process Regression}
\end{center}
\medskip
} \fi
\bigskip

\date{}

\begin{abstract}

The performance of Gaussian Process (GP) regression is often hampered by the curse of dimensionality, which inflates computational cost and reduces predictive power in high-dimensional problems. 
Variable selection is thus crucial for building efficient and accurate GP models. 
Inspired by Bayesian bridge regression, we propose the Bayesian Bridge Gaussian Process Regression (B\textsuperscript{2}GPR) model. 
This framework places $\ell_q$-norm constraints on key GP parameters to automatically induce sparsity and identify active variables. 
We formulate two distinct versions: one for $q=2$ using conjugate Gaussian priors, and another for $0<q<2$ that employs constrained flat priors, leading to non-standard, norm-constrained posterior distributions. 
To enable posterior inference, we design a Gibbs sampling algorithm that integrates Spherical Hamiltonian Monte Carlo (SphHMC) to efficiently sample from the constrained posteriors when $0<q<2$. 
Simulations and a real-data application confirm that B\textsuperscript{2}GPR offers superior variable selection and prediction compared to alternative approaches.


\end{abstract}

\spacingset{1.5} 

\section{Introduction}

Gaussian Process (GP) regression is a widely used statistical method for modeling complex nonlinear relationships, particularly in computer experiments and simulations. 
However, it faces the curse of dimensionality due to a high-dimensional input space, resulting in high computational demands and diminished predictive performance.
In practice, many simulation models primarily depend on a small subset of influential inputs. 
Identifying these key variables is critical for enhancing the efficiency and accuracy of GP regression.

\subsection{Related Literature}

Regularization methods are essential tools for preventing overfitting and identifying significant predictors in linear and generalized linear models. 
For the linear regression model $Y(\bm x)=\bm x^\top \bm \beta+\epsilon$, frequentist regularization approaches introduce penalty terms to constrain the coefficients $\bm \beta$. 
Ridge regression \citep{hoerl1970ridge} employs an $l_2$-norm penalty, which shrinks coefficients uniformly without performing variable selection. 
In contrast, LASSO by \cite{tibshirani1996regression} uses an $l_1$-norm penalty to induce sparsity by driving some coefficients to zero, thereby enabling variable selection.
The Elastic Net \citep{zou2005regularization} combines $l_1$ and $l_2$ penalties, offering a balance between shrinkage and selection, which is particularly useful in high-dimensional settings where the number of predictors exceeds the number of observations. 
Bridge regression \citep{frank1993statistical} generalizes these methods with a flexible $\ell_q$-norm penalty ($q\geq 0$), allowing adaptive control over both shrinkage intensity and selection behavior. 

While closed-form solutions for arbitrary $q$ were not derived, the authors highlighted the importance of optimizing $q$ based on the data.

Bayesian regularization methods consider these penalties as prior distributions, updating them with observed data to obtain posterior estimates. 
Bayesian ridge regression by \cite{bishop2006pattern} assigns Gaussian priors to coefficients, yielding posteriors whose mode aligns with the $l_2$-penalized estimates from the frequentist ridge regression. 
Similarly,  \cite{park2008bayesian} proposed the Bayesian LASSO that adopts the Laplace prior distribution, replicating the sparsity-inducing properties of the $l_1$ penalty. 
For more flexible regularization, Bayesian bridge regression \citep{polson2014bayesian} employs priors analogous to the $\ell_q$ penalty, enabling adaptive shrinkage and selection. 
While Bayesian methods often involve computationally intensive Markov Chain Monte Carlo (MCMC) sampling, they provide a natural framework for uncertainty quantification, unlike their frequentist counterparts, which rely on convex optimization.

In GP modeling, regularization techniques have been adapted to address the challenges of the curse of dimensionality. 
Frequentist approaches include penalized blind kriging \citep{hung2011penalized} that introduces penalties on the regression coefficients of the mean function of the GP, and methods like \cite{li2005analysis} that impose penalties directly on covariance hyperparameters. 
Bayesian GP methods, on the other hand, often employ threshold or indicator-based variable selection. 
For instance, \cite{hu2024generalized} identified active variables by thresholding on the samples from the posterior of the inverse length-scale hyperparameters of the correlation function, while \cite{zhang2023indicator} proposed to use an indicator to decide if an input variable is significant or not. 
In their framework, an input variable is deemed to be active if its corresponding parameters are significant in either the mean function or the covariance function. 
We broadly divide these existing GP regularization methods into three categories:
(i) penalization methods acting on regression coefficients \citep{hung2011penalized} or covariance hyperparameters \citep{li2005analysis}; (ii) indicator-based Bayesian variable selection approaches \citep{zhang2023indicator}; and (iii) thresholding procedures based on posterior summaries of kernel parameters \citep{hu2024generalized}.
While these methods successfully promote sparsity, most focus on either the mean function or covariance structure separately, and many rely on discrete selection mechanisms or fixed $\ell_1$ or $\ell_2$ regularization schemes. 
In contrast, our proposed framework introduces a unified bridge-type regularization mechanism for both components simultaneously, allowing adaptive sparsity through general $\ell_q$ constraints and enabling full posterior inference under constrained Bayesian formulations.

\subsection{Our Contribution}

The aforementioned regularization methods for GP models reveal the trade-off between computational tractability and modeling flexibility, and they motivate our proposed \emph{Bayesian Bridge Gaussian Process Regression (B\textsuperscript{2}GPR)} framework. 
It integrates the adaptive shrinkage property of the Bridge regression with the probabilistic foundations of Gaussian Processes.
Unlike \cite{hung2011penalized} and \cite{li2005analysis}, our method simultaneously regularizes both the mean function's regression coefficients and the covariance function's hyperparameters through the flexible $\ell_q$ penalty. 
While initially conceptualized within a frequentist penalized optimization framework, we identified two critical challenges: (1) the non-convexity of the optimization landscape, and (2) the computational burden of gradient evaluations for covariance hyperparameters under the $\ell_q$-norm constraints.

To resolve these limitations, we reformulate the problem within the Bayesian framework, extending the Bayesian bridge regression to the GP context. 
This transition introduces a challenge—particularly in constructing appropriate prior distributions that enforce $\ell_q$-norm constraints on GP parameters. 
The traditional exponential-power prior (including the Laplace prior for $q=1$) relies on a scale-mixture representation to enable easy Gibbs sampling. 
While this works for the linear fixed-effect coefficients $\bm \beta$, it fails for the covariance hyperparameters $\bm \omega$ because they enter the likelihood nonlinearly through the kernel matrix $\bm K_n$. 
As a result, the conditional posterior of $\bm \omega$ does not permit a conjugate mixture representation, rendering standard unconstrained scale-mixture samplers inapplicable.
We address this challenge by adopting truncated prior distributions that impose explicit $\ell_q$-norm constraints on both $\bm \beta$ and $\bm \omega$.
This construction ensures the mode of the posterior distributions of the parameters exactly corresponds to the estimates of the regularized loss function in the frequentist approach. 
The bridge regularization framework is particularly attractive for GP models because covariance hyperparameters often exhibit heterogeneous importance across input dimensions. 
Standard Gaussian or Laplace-type shrinkage may either overshrink important variables or insufficiently suppress inactive ones. 
The $\ell_q$ bridge formulation provides a continuum of shrinkage behaviors between subset-selection-like sparsity and ridge-type regularization, allowing the model to adapt more flexibly to varying sparsity structures in high-dimensional GP settings.

However, sampling from the constrained posterior distribution requires special techniques. 
The Spherical Hamiltonian Monte Carlo (SphHMC) method \citep{lan2014spherical}, originally designed for spherical constraint spaces, provides a suitable solution to our problem. 
We use the SphHMC to sample the parameters of the mean and the correlation function and integrate it into the overall Gibbs sampling framework for all the parameters when $\ell_q$-norm with $0<q<2$ is used. 
This constrained MCMC procedure enables efficient exploration of the posterior distribution. 
It shows three main advantages: (1) flexibility of the choice in prior distributions; (2) variable selection through the $\ell_q$ penalty; (3) accurate prediction. 
For $q=2$, Gaussian prior distributions on the parameters naturally correspond to the $l_2$-norm penalty, and thus we only need to use a conventional sampling approach. 
Here, we use Hamiltonian Monte Carlo for the $q=2$ case. 

Compared to the frequentist regularization approach, Bayesian framework has the advantage of fully exploring the multimodal posterior distribution in the high-dimensional space, which returns more accurate estimates and predictions than the single point estimates as in the frequentist methods. 
More importantly, statistical inferences are possible based on the posterior samples for all parameters, including hyperparameters. 
Unfortunately, frequentist methods only return single point estimates without any inferences. 
More importantly, the frequentist methods choose tuning parameter for the regularization term using cross-validation. 
This is far more computational than the Bayesian framework in which we combine the sampling of all the tuning parameters with other parameters in full Bayes fashion and proves to be more efficient. 
Table \ref{tab:gp_regularization_comparison} summaries the differences between the the proposed method and existing GP regularization methods. 
The proposed B$^2$GPR checks all the boxes and some boxes are desirable features for GP models.

\linespread{1}
\begin{table}[ht]
\centering
\caption{Comparison of existing GP regularization approaches and the proposed B$^2$GPR framework.
\label{tab:gp_regularization_comparison}}
\renewcommand{\arraystretch}{1.2}
\begin{tabular}{lcccccc}
\hline
\multirow{2}{*}{Method}  & Mean & Covariance & \multirow{2}{*}{Bayesian} & \multirow{2}{*}{General $\ell_q$} & Joint & Full Posterior\\
                         & Reg. & Reg. &              &     &  Selection   & Inference  \\\hline
Penalized Likelihood &  &  &  &  &  &   \\
GP \cite{li2005analysis} & $\times$ & \checkmark & $\times$  & Limited  & $\times$ & $\times$ \\\hline
Penalized Blind &  &  &  &  &  &   \\
Kriging \cite{hung2011penalized} & \checkmark & $\times$  & $\times$  & Limited  & $\times$  & $\times$ \\\hline

Indicator-based GP &  &  &  &  &  &   \\
\cite{zhang2023indicator} & Partial  & \checkmark  & \checkmark  & $\times$  & Partial  & \checkmark \\\hline

Thresholding-based &  &  &  &  &  &   \\
GP \cite{hu2024generalized} & $\times$  & \checkmark  & \checkmark  & $\times$  & $\times$  & \checkmark \\\hline
B$^2$GPR & \checkmark  & \checkmark  & \checkmark  & \checkmark  & \checkmark  & \checkmark \\\hline
\end{tabular}
\end{table}
\linespread{1.5}

The paper proceeds as follows. 
Section \ref{sec:background} establishes the theoretical basis, reviewing GP models and bridge regularization in both frequentist and Bayesian settings. 
Section \ref{sec:B2GPR} introduces the B\textsuperscript{2}GPR model, formalizing the generalized bridge prior and its frequentist-Bayesian duality. 
Section \ref{sec:MCMC} details our computational methods, presenting the modified SphHMC algorithm for constrained posterior sampling. 
Sections \ref{sec:simu} and \ref{sec:real} validate the method through simulated experiments and a real-case study, demonstrating superior performance in variable selection and prediction accuracy. 
We conclude with discussion and future directions in Section \ref{sec:end}.

\section{Preliminaries}\label{sec:background}

\subsection{Gaussian Process model}\label{subsec:GPreview}

A Gaussian Process (GP) is a stochastic process characterized by the property that any finite collection of the function values, evaluated at different input locations, follows a multivariate normal distribution.
We can fully specify a GP by the mean and covariance function. 
Given the input domain $\Omega \subset \mathbb{R}^d$, we focus on the univariate GP.
Specially, we assume the random process $Y(\bm x)\in \mathbb{R}^{1}$ at any input point $\bm x \in \Omega$ is 
\begin{equation}\label{eq:model}
Y(\bm x)=\underbrace{\bm g(\bm x)^\top \bm \beta}_\text{fixed effect}+ \underbrace{Z(\bm x)}_\text{random effect}+\underbrace{\epsilon}_\text{noise}.
\end{equation}
In the fixed effect term, $\bm g(\bm x)$ is a $p$-dimensional vector of user-specified basis functions.
Often we use polynomial functions of $\bm x$ up to a chosen degree.
For instance, a two-dimensional quadratic polynomial basis functions are $\bm g(\bm x)=[1, x_1, x_2, x_1^2, x_1x_2, x_2^2]^\top$.
Correspondingly, $\bm \beta$ is a $p$-dimensional vector of linear coefficients.
The random noise term, $\epsilon\sim \mathcal{N}(0,\sigma^2)$, is independent of all the other model components.
The random effect term $Z(\bm x)$ follows a zero-mean stationary Gaussian Process, denoted as $Z(\bm x) \sim \mathcal{GP}(0,\tau^2 K(\cdot,\cdot;\bm \omega))$, with variance parameter $\tau^2$ and correlation function $K(\cdot, \cdot;\bm \omega)$ with hyperparameters $\bm \omega$. 
The correlation function $K(\cdot, \cdot;\bm \omega): \Omega \times \Omega \mapsto \mathbb{R}_{+}$ must be a symmetric, positive definite kernel function.
A common choice for the correlation kernel is the separable anisotropic Gaussian kernel, 
\[
    K(\bm x_i, \bm x_j; \bm \omega)=\exp\left\{-\sum_{k=1}^d \omega_k^2(x_{ik}-x_{jk})^2\right\},
\]
with the correlation hyperparameters $\bm \omega \in \mathbb{R}^d$ which control the shape of the kernel functions.
They have been referred to as \emph{correlation strength} or \emph{inverse length-scale} hyperparameters in the literature. 

Under the model assumption, given the parameters $\bm \theta=(\bm \beta, \bm \omega, \tau^2, \eta)$, where $\eta=\sigma^2/\tau^2$, $Y(\bm x)$ is a GP with mean and covariance as follows:
\begin{align*}
    \mathbb{E}[Y(\bm x)]&=\bm g(\bm x)^\top \bm \beta, \quad \forall \bm x \in \Omega\\
    \cov[Y(\bm x_i), Y(\bm x_j)]&=\tau^2 K(\bm x_i, \bm x_j;\bm \omega) + \sigma^2\delta(\bm x_i, \bm x_j), \quad \forall \bm x_i, \bm x_j\in \Omega,\\
    &=\tau^2\left[ K(\bm x_i, \bm x_j;\bm \omega) + \eta\delta(\bm x_i, \bm x_j)\right], 
\end{align*}
where $\delta(\bm x_i,\bm x_j)=1$ if $\bm x_i=\bm x_j$ and $0$ otherwise.
Denote the observed data from a certain computer or physical experiment by $\{\bm x_i,y_i\}_{i=1}^n$, the set of $n$ design points by $\mathcal{X}_n=\{\bm x_1,\ldots, \bm x_n\}$, and the response data by $\bm y=[y_1,\dots,y_n]^\top$. 
Given the parameter $\bm \theta$, the posterior predictive distribution of $Y(\bm x)$ at any query point $\bm x$ is the following normal distribution
\[
Y(\bm x)|\bm y, \bm \theta \sim \mathcal{N}(\mu_n(\bm x), \sigma^2_{\bm x|n}), 
\]
where 
\begin{align*}
\mu_n(\bm x)&=\mathbb{E}(Y(\bm x)|\bm y, \bm \theta)=\bm g(\bm x)^\top \bm \beta + \bm k(\bm x)^\top(\bm K+\eta\bm I_n)^{-1}(\bm y-\bm G \bm \beta),\\
\sigma^2_{\bm x|n}(\bm x)&=\tau^2 \left\{ 1-\bm k(\bm x)^\top(\bm K+\eta\bm I_n)^{-1}\bm k(\bm x)\right\},\\
\bm k(\bm x) &=K(\bm x,\mathcal{X}_n)=K(\mathcal{X}_n, \bm x)=[K(\bm x, \bm x_1),\ldots, K(\bm x, \bm x_n)]^\top 
\end{align*}
The derivations can be found in \cite{santner2003design} and \cite{williams2006gaussian}. 

For physical or stochastic computer experiments, $\sigma^2$ is typically estimated by the pooled variance from replications if there are replicated observations at some design points. 
For deterministic computer experiments, since there is no measurement noise involved, $\sigma^2$ should be specified as $0$, and thus $\eta=0$. 
In this case, the posterior predictive mean $\mu_n(\bm x)$ interpolates all the training data.
However, it is usually advantageous to specify a small positive value for $\eta$ known as the \emph{nugget} value \citep{peng2014choice} to overcome the possible ill-conditioning of the kernel matrix $\bm K$.
In this paper, the estimation and inference of the unknown parameters $\bm \theta$ and prediction inference are conducted under the Bayesian framework. 

\subsection{Bayesian Bridge Regression}\label{subsec:BBR}

Bayesian Bridge regression is a Bayesian approach that achieves bridge regularization for regression models.
Consider a linear regression model defined by  $\bm y = \bm X \bm \beta + \bm \epsilon$ for data $\{\bm x_i, y_i\}_{i=1}^n$ with the error term $\bm \epsilon \sim \mathcal{MVN}_n(\bm 0, \sigma^2 \bm I_n)$.
The bridge regression, introduced by \cite{frank1993statistical}, estimates coefficients $\bm \beta$ by solving the following minimization problem:
\begin{equation} \label{eq:bridge1}
 \min_{\bm \beta} \frac{1}{2}\|\bm y-\bm X\bm \beta\|^2_2+\lambda \sum_{j=1}^p |\beta_j|^q, 
\end{equation}
where $\bm y$ is the $n \times 1$ response vector, $\bm X$ is the $n \times p$ matrix of regressors, and $\bm \beta$ is the $p \times 1$ vector of regression coefficients to be estimated.
The tuning parameter $\lambda > 0$ controls the strength of penalization, and the parameter $q \geq 0$ adjusts the shape of the penalty \citep{mallick2018bayesian}.
Notably, when $q=0$ (with the convention $0^0=0$), the penalty corresponds to the number of nonzero coefficients, resembling the best subset selection.
Other special cases include LASSO regression \citep{tibshirani1996regression} with $q=1$ and ridge regression \citep{hoerl1970ridge} with $q=2$. 
Thus, bridge regression generalizes these methods into a unified framework.
It bridges a class of shrinkage and selection operators, allowing for flexible regularization and variable selection.

For the penalty term, we can also write it as $\lambda \| \bm \beta\|_q$, where $\| \bm \cdot\|_q$ is the $\ell_q$-norm of a vector.
Based on \cite{hastie2009elements}, bridge regression problem \eqref{eq:bridge1} can be equivalently expressed as a constrained minimization problem:
\begin{equation} \label{eq:bridge2}
    \min_{\bm \beta} \frac{1}{2}\|\bm y-\bm X\bm \beta\|^2_2 \quad \text{subject to} \quad \|\bm \beta\|_q =\left(\sum_{j=1}^p \beta_j^q\right)^{1/q} \leq t.
\end{equation}
This explicitly imposes a size constraint on the coefficients $\bm \beta$, establishing a direct one-to-one correspondence between the penalty parameter $\lambda$ and the constraint $t$.
The constraint is concave when $0<q < 1$ and convex when $q\geq 1$. 
%In this paper, we require $0<q\leq 2$, which is also the range of bridge regression. 

The above is the frequentist formulation.
From the Bayesian perspective, bridge regression naturally corresponds to specifying a particular prior distribution for the regression coefficients.
Following the penalization form in \eqref{eq:bridge1}, \cite{park2008bayesian} and \cite{polson2014bayesian} introduced the exponential-power prior distribution for $\bm \beta$ given $\lambda$
\begin{equation}
\label{eq:exp-power}
    p(\bm \beta|\lambda)\propto \prod_{j=1}^p \exp\left(-\lambda|\beta_j|^q \right). 
\end{equation}
This prior is the Laplace distribution when $q=1$ and the Gaussian distribution when $q=2$.
Combining with the prior distribution of $\lambda$, the joint prior of all parameters has a mixture representation which enables efficient Gibbs sampling of the posterior distribution. 
It is one of the key advantages of the Bayesian bridge approach. 

Alternative to the exponential-power prior, the constrained formulation \eqref{eq:bridge2} leads to a constrained prior distribution for $\bm \beta$ (omitting the normalizing constant):
\begin{equation*}
    p(\bm \beta|t) \propto p(\bm\beta) \mathbb{I}(\|\bm\beta\|_q\leq t),
\end{equation*}
and it leads to the posterior distribution for $\bm \beta$ 
\begin{equation*}
    p(\bm \beta|\bm y, t) \propto p(\bm y|\bm \beta) p(\bm \beta) \mathbb{I}(\|\bm\beta\|_q\leq t). 
\end{equation*}
Here $\mathbb{I}$ is an indicator function, and it is equal to one if the input statement is true and zero otherwise. 
The posterior mode can be obtained by minimizing $-\log p(\bm \beta|\bm y, t)$, which is equal to minimizing $- \log p(\bm y|\bm \beta) - \log p(\bm \beta)$ subject to $\|\bm\beta\|_q\leq t$. 
Hence, if using the non-informative prior for $\bm \beta$, finding the posterior mode is equivalent to solving the frequentist constrained optimization problem \eqref{eq:bridge2}. 
However, this constrained prior is not restrictive to non-informative priors.
It offers greater flexibility by allowing a wide variety of alternative distributions that encode additional information or preferences about the predictors.

\section{Bayesian Bridge Gaussian Proces Regression}\label{sec:B2GPR}

Extending the Bridge regression concept to the GP model described in \eqref{eq:model}, we introduce a regularization strategy that penalizes both the linear coefficients $\bm \beta$ and the correlation parameters $\bm \omega$.
Under the frequentist framework, to estimate the unknown parameters of the penalized GP regression is to solve the minimization problem
\begin{equation} \label{eq:bridge-GP1}
\mathop{\min}_{\bm \beta, \bm \omega} \left \{-\log p(\bm y | \bm \beta, \bm \omega, \tau^2, \eta)+\lambda_1 \sum_{j=1}^p |\beta_j|^q+\lambda_2 \sum_{j=1}^d |\omega_j|^q \right \},
\end{equation}
where $-\log p(\bm y | \bm \beta, \bm \omega, \tau^2, \eta)$ is the negative log-likelihood function, and the two regularization terms represent the $\ell_q$-norm penalties on the parameters $(\bm \beta,\bm \omega)$, controlling shrinkage intensity through tuning parameters $\lambda_1$ and $\lambda_2$.
Given $(\bm \beta, \bm \omega, \eta)$, the parameter $\tau^2$ can be estimated via maximum likelihood estimation. 
%\[\hat{\tau}^2=\frac{1}{n}(\bm y-\bm G\bm \beta)^\top (\bm K+\eta\bm I_n)^{-1}\bm (\bm y-\bm G\bm \beta).\]
The nugget effect $\eta$ can be computed as a tuning parameter via cross-validation or through optimization. 
The overall estimation of all $(\bm \beta, \bm \omega, \tau^2, \eta)$ can take iterative steps by separately computing $(\bm \beta, \bm \omega)$, $\tau^2$, and $\eta$. 
We call it \emph{Bridge Gaussian Process Regression} with $0<q\leq 2$. 

Just like the equivalency between \eqref{eq:bridge1} and \eqref{eq:bridge2}, \eqref{eq:bridge-GP1} is equivalent to the constrained minimization problem:
\begin{equation} \label{eq:bridge-GP2}
\mathop{\min}_{\bm\beta, \bm\omega} \left \{-\log p(\bm y | \bm \beta, \bm \omega, \tau^2, \eta) \right \} \quad \text{subject to} \quad \|\bm \beta\|_q \leq r_{\beta}, \quad \|\bm \omega\|_q \leq r_{\omega}. 
\end{equation}
This formulation explicitly introduces constraints on $\ell_q$-norm of $\bm \beta$ and $\bm \omega$. 
There is also an one-to-one correspondence between penalty parameters $(\lambda_1, \lambda_2)$ and constraints $(r_{\beta}, r_{\omega})$.

Given that $\bm y$ still follows a multivariate normal distribution, the exponential-power prior for $\bm \beta$ can still lead to a mixture representation of the conditional posterior for $\bm \beta$. 
For example, when $q=1$, this corresponds to the Laplace prior used in the Bayesian LASSO, which can be sampled using standard scale-mixture Gibbs steps. 
However, this tractable mixture representation does not extend to the correlation parameters $\bm \omega$. 
Because $\bm \omega$ enters the likelihood nonlinearly through the covariance function, its conditional posterior remains non-conjugate regardless of the prior choice.
To regularize $\bm \omega$ under a unified framework, we utilize the constrained formulation \eqref{eq:bridge-GP2} via truncated priors. 
When $0<q<2$, the boundary of the constraint $\|\bm \omega\|_q$ is non-smooth (especially at $q=1$, where it has sharp vertices) or non-convex (when $q<1$). 
Standard HMC algorithms struggle to negotiate these sharp boundaries, frequently rejecting proposals or getting trapped. 
Hence, SphHMC is essential for the entire range of $0<q<2$ to handle these boundary constraints efficiently by mapping the parameter space onto a smooth hypersphere.
In contrast, when $q=2$, the constraint represents a smooth $\ell_2$ ball. 
Rather than enforcing a hard boundary $\|\bm \omega\|_2\leq r_{\omega}$, we can reformulate this via smooth, unconstrained Gaussian priors. 
Because the resulting posterior surface is smooth and lacks sharp boundary interfaces, standard HMC is highly efficient and SphHMC is no longer required. 
This motivates the development of our two distinct algorithms: sph-B2GPR (for $0<q<2$) and hmc-B2GPR (for $q=2$).

The bridge constraints in \eqref{eq:bridge-GP2} can be implemented through truncated prior distributions, i.e., 
\begin{equation*}
\bm \beta \sim p(\bm\beta) \mathbb{I}(\|\bm\beta\|_q\leq r_{\beta}) \quad \text{and} \quad \bm \omega \sim p(\bm\omega) \mathbb{I}(\|\bm\omega\|_q\leq r_{\omega}),
\end{equation*}
omitting the normalizing constants. 
Based on the truncated prior distributions, the conditional posterior distributions for $\bm \beta$ and $\bm \omega$ are 
\begin{equation*}
\begin{aligned}
p(\bm \beta | \bm y, \bm \omega, \tau^2, \eta, r_{\beta}) &\propto p(\bm y | \bm \beta, \bm \omega, \tau^2, \eta) p(\bm \beta) \mathbb{I}(\|\bm\beta\|_q\leq r_{\beta}),\\
p(\bm \omega | \bm y, \bm \beta, \tau^2, \eta, r_{\omega}) &\propto p(\bm y | \bm \beta, \bm \omega, \tau^2, \eta) p(\bm \omega) \mathbb{I}(\|\bm\omega\|_q\leq r_{\omega}).
\end{aligned}
\end{equation*}
To find the posterior modes of $\bm \beta$ and $\bm \omega$, we just need to solve the following two minimization problems.
\begin{equation}\label{eq:bridge-GP-minus-log-posterior}
\begin{aligned}
& \min_{\bm \beta} -\log p(\bm y | \bm \beta, \bm \omega, \tau^2, \eta) -\log p(\bm \beta) & \text{s.t. } \|\bm\beta\|_q\leq r_{\beta}, \\
& \min_{\bm \omega} -\log p(\bm y | \bm \beta, \bm \omega, \tau^2, \eta) -\log p(\bm \omega) & \text{s.t. } \|\bm\omega\|_q\leq r_{\omega}.
\end{aligned}
\end{equation}
As discussed on Bayesian bridge regression, if non-informative priors are used, i.e., $p(\bm \beta)\propto 1$ and $p(\bm \omega)\propto 1$, \eqref{eq:bridge-GP-minus-log-posterior} and \eqref{eq:bridge-GP2} are the same, with the minor difference that \eqref{eq:bridge-GP2} jointly computes $\bm \beta$ and $\bm \omega$ whereas \eqref{eq:bridge-GP-minus-log-posterior} computes one given the other. 
Other than the non-informative priors, we can use any other suitable alternatives, such as normal distributions, for $\bm \beta$ and $\bm \omega$ for more flexibility. 
However, in that case, there is a redundancy in regularization. 
The normal prior distributions have applied the $l_2$-norm regularization on $(\bm \beta, \bm \omega)$, but the constraints would apply $\ell_q$-norm regularization again. 

In the next section, we introduce two versions of the B\textsuperscript{2}GPR. 
\begin{enumerate}
\item For $0<q<2$, we adopt the non-informative prior distributions for $(\bm \beta, \bm \omega)$ and impose the $\ell_q$-norm constraints. 
To overcome the challenge of sampling from the constrained posterior distributions, we incorporate the Spherical Hamiltonian Monte Carlo within the Gibbs sampling framework. 
For short, we name this version sph-B\textsuperscript{2}GPR. 
\item For $q=2$, we adopt informative Gaussian distributions for $(\bm \beta, \bm \omega)$. 
But to avoid the redundancy of $\ell_2$ regularization and $\ell_q$-norm constraint, we remove the $\ell_q$-norm constraints in this case. 
The full Bayes framework is developed, $\bm \beta$ is sampled directly from its analytically available conditional posterior distribution, while $\bm \omega$ is sampled using Hamiltonian Monte Carlo without constraints.
For short, we name the second version hmc-B\textsuperscript{2}GPR. 
\end{enumerate}

As pointed out by \cite{polson2014bayesian}, both the frequentist and Bayesian approaches for the GP regression face the same practical challenge: exploring and summarizing the multimodal likelihood or posterior distribution in the high-dimensional Euclidean space. 
\cite{polson2014bayesian} made several compelling arguments in favor of the fully Bayesian approach, many of which apply to our setting.
First, it is misleading to summarize a multimodal surface via a single point estimate, and the MCMC sampling approach is more effective at exploring the entire likelihood surface. 
Second, statistical inference is possible for all the hyperparameters based on the posterior sampling. 
Third, the frequentist approach chooses the tuning parameter $(r_{\beta}, r_{\omega})$ via cross-validation, which is computational demanding and lacks uncertainty quantification. 
In the Bayesian approach, sph-B\textsuperscript{2}GPR, we assume non-informative priors on $(r_{\beta}, r_{\omega})$ and sample their values from their respective conditional posterior distribution. 
Compared to cross-validation, which has to estimate all parameters for each pair of possible $(r_{\beta}, r_{\omega})$, the full Bayes approach combines the sampling of $(r_{\beta}, r_{\omega})$ with other parameters and proves to be more efficient. 
Similarly, for hmc-B\textsuperscript{2}GPR, the regularization is applied through the variance parameters of the prior distribution, which are also incorporated in the MCMC sampling of the posterior distributions.

Compared to Bayesian bridge regression, the sampling procedure for B\textsuperscript{2}GPR is more difficult since the computation for the correlation parameters and nugget effect is more time-consuming due to the inverse of the $n\times n$ covariance matrix in each iteration. 
Therefore, we develop the MCMC sampling procedure based on the Hamiltonian Monte Carlo, which is more efficient compared to conventional Metropolis-Hastings type of algorithms. 

\section{Posterior Sampling}\label{sec:MCMC}

\subsection{HMC and Spherical HMC}

Hamiltonian Monte Carlo (HMC) \citep{neal2011mcmc} is an MCMC method that leverages Hamiltonian dynamics to efficiently explore complex target distributions. The sampler operates on an augmented state space of position parameters $\bm{\theta}$ and momentum variables $\bm{\phi} \sim \mathcal{N}(\mathbf{0}, \bm{M})$. The system's evolution is governed by the Hamiltonian $H(\bm{\theta}, \bm{\phi}) = U(\bm{\theta}) + K(\bm{\phi})$, where $U(\bm{\theta}) = -\log p(\bm{\theta})$ is the potential energy and $K(\bm{\phi}) = \frac{1}{2} \bm{\phi}^\top \bm{M}^{-1} \bm{\phi}$ is the kinetic energy. Numerical simulation via the leapfrog integrator (Algorithm \ref{alg:HMC}) generates distant, high-acceptance proposals, enabling efficient sampling from high-dimensional posteriors.

While HMC is powerful, it does not inherently handle parameter constraints. Spherical HMC (SphHMC) \citep{lan2014spherical, Lan2016b} addresses this for norm-constrained domains. 
The core idea is \emph{spherical augmentation}. 
The domain defined by the $\ell_q$-norm constraint on the parameter is $\bm \beta\in \mathbb{R}^d$ as $\mathcal{Q}^d:=\{\bm \beta \in \mathbb{R}^d| \|\bm \beta\|_q \leq 1\}$, without loss of generality. 
It can be bijectively mapped to the $d$-dimensional unit ball, $\mathcal{B}_0^{d}(1)=\{\bm \theta\in \mathbb{R}^{d}|\|\bm \theta\|_2\leq 1\}$, by either $\beta_i \mapsto \theta_i=\text{sgn}(\beta_i)|\beta_i|^{q/2}$ or $\bm \beta\mapsto \bm \theta =\bm \beta \frac{\|\bm \beta\|_{\infty}}{\|\bm \beta\|_2}$. 
Then it is embedded onto a sphere $\mathcal{S}^d$ in $\mathbb{R}^{d+1}$ by adding $\theta_{d+1} = \pm\sqrt{1 - |\bm{\theta}|_2^2}$. 
This transforms the boundary into the sphere's equator, allowing the sampler (Algorithm \ref{alg:SphHMC}) to move freely on $\mathcal{S}^d$ while implicitly respecting the original constraints. 
We invite readers to go to the Supplement for more in depth review and the Algorithms \ref{alg:HMC} and \ref{alg:SphHMC}. 

In this article, we impose the $\ell_q$-norm constraints on both $\bm \omega$ and $\bm \beta$ as defined in \eqref{eq:bridge-GP-minus-log-posterior} for $0<q \leq 2$. 
As explained in the previous section, hmc-B\textsuperscript{2}GPR (for $q=2$) only needs to use HMC in Algorithm \ref{alg:HMC}, whereas sph-B\textsuperscript{2}GPR (for $0<q<2$) calls the spherical HMC in Algorithm \ref{alg:SphHMC} to deal with the norm constraint. 
In sph-B\textsuperscript{2}GPR, the bijectively map from the $\ell_q$ norm constraints $\|\bm \beta\|_q \leq r_{\beta}$ and $\|\bm \omega\|_q \leq r_{\omega}$ are
\begin{equation*}
\beta_i \mapsto \theta_{\beta_i} = \text{sgn}(\beta_i) \left|\frac{\beta_i}{r_{\beta}} \right|^{q/2}, \quad \omega_i \mapsto \theta_{\omega_i} = \text{sgn}(\omega_i) \left|\frac{\omega_i}{r_{\omega}} \right|^{q/2}. 
\end{equation*}
We sample the auxiliary variables $\bm \theta_{\beta}$ and $\bm \theta_{\omega}$ from the unit ball using the SphHMC method, and then transform the samples $\bm \theta_{\beta}$ and $\bm \theta_{\omega}$ back to the original constrained parameter spaces $\bm \beta$ and $\bm \omega$ using the inverse mappings:
\begin{equation*}
\theta_{\beta_i} \mapsto \beta_i = r_{\beta} \cdot \text{sgn}(\theta_{\beta_i}) |\theta_{\beta_i}|^{2/q}, \quad
\theta_{\omega_i} \mapsto \omega_i = r_{\omega} \cdot \text{sgn}(\theta_{\omega_i}) |\theta_{\omega_i}|^{2/q}.
\end{equation*}

\subsection{Sampling for sph-B\textsuperscript{2}GPR}

For the sph-B\textsuperscript{2}GPR approach, we use the non-informative prior for $\bm \beta$ and $\bm \omega$, but still need to specify the prior distributions for $(\tau^2,\eta, r_{\beta}, r_{\omega})$. 
Below is a summary of the prior distributions for all the parameters. 
\begin{equation} \label{eq:prior}
\begin{aligned}
& p(\bm \beta|r_{\beta}) \propto \mathbb{I}(\|\bm \beta\|_q \leq r_{\beta}), &&\text{constrained non-informative prior},\\
& p(r_{\beta}) \propto 1, &&\text{non-informative prior},\\
& p(\bm \omega|r_{\omega}) \propto \mathbb{I}(\|\bm \omega\|_q \leq r_{\omega}), &&\text{constrained non-informative prior},\\
& p(r_{\omega}) \propto 1, &&\text{non-informative prior},\\
& \tau^2 \sim \text{Inverse-}\chi^2(df_{\tau^2}),&& \eta \sim \Gamma(a_{\eta}, b_{\eta}).
\end{aligned}
\end{equation}
The hyperparameters $df_{\tau^2}, a_{\eta}, b_{\eta}$ are user-specified. 

Given data, the sampling distribution is:
\begin{equation*}
\bm y_n | \bm \theta \sim \mathcal{MVN}_n(\bm G\bm \beta, \tau^2 (\bm K_n  + \eta \bm I_n))
\end{equation*}
where $\bm y_n$ is the response vector of $y_i$'s, $\bm G$ is the basis function matrix and each row is $\bm g(\bm x_i)^\top$. 
The matrix $\bm K_n$ is the $n\times n$ kernel matrix with entries $K_{ij}=K(\bm x_i,\bm x_j;\bm \omega)$, which is symmetric and positive definite.
The likelihood function of $\bm y_n|\bm \theta$ is 
\begin{equation*} \label{eq:pdfy2}
p(\bm y_n|\bm \theta) \propto (\tau^2)^{-\frac{n}{2}}\det(\bm K_n+\eta\bm I_n)^{-1/2}\exp\left(-\frac{1}{2\tau^2}(\bm y_n-\bm G\bm \beta)^\top (\bm K_n+\eta \bm I_n)^{-1} (\bm y_n-\bm G\bm \beta)\right).
\end{equation*}
The joint posterior density is: 
\begin{equation*} 
p(\bm \theta|\bm y_n)\propto p(\bm \beta|r_{\beta})p(r_{\beta})p(\bm \omega|r_{\omega})p(r_{\omega}) p(\tau^2) p(\eta) p(\bm y_n|\bm \theta).
\end{equation*}
We summarize the conditional posterior distributions of all the parameters in Proposition \ref{prop:sph}. 
The detailed derivation can be found in the Supplement. 
Algorithm \ref{alg:bridge-GP-MCMC} summarizes the sampling procedure of sph-B\textsuperscript{2}GPR. 
\begin{proposition}\label{prop:sph}
Under the prior assumptions in \eqref{eq:prior}, the conditional posterior distributions for $(\bm \beta, \bm \omega, \tau^2, \eta, r_{\beta}, r_{\omega})$ are listed below. 
\begin{enumerate}
\item The linear coefficients $\bm \beta$ has the following full conditional distribution
\begin{align*}
& p(\bm \beta|\bm y_n, \bm \omega, \tau^2, \eta) \propto \\
& \exp\left(-\frac{1}{2\tau^2} \left(\bm \beta^\top \bm G^\top (\bm K_n+\eta \bm I_n)^{-1} \bm G\bm \beta - 2\bm \beta^\top \bm G^\top (\bm K_n+\eta \bm I_n)^{-1} \bm y_n \right) \right) \mathbb{I}(\|\bm \beta\|_q \leq r_{\beta}).
\end{align*}
Therefore, $\bm \beta$ follows a truncated multivariate normal distribution specified below
\begin{equation*}
\bm \beta | \bm y_n, \bm \omega, \tau^2,\eta \sim \mathcal{MVN}_p(\hat{\bm \beta}_n, \bm \Sigma_{\bm \beta|n})
\end{equation*}
with
\begin{equation*}
\bm \Sigma_{\bm \beta|n}=\left[\frac{1}{\tau^2}\bm G^\top (\bm K_n+\eta \bm I_n)^{-1}\bm G\right]^{-1}, \quad
\hat{\bm \beta}_n =\bm \Sigma_{\bm \beta|n}\frac{\bm G^\top (\bm K_n+\eta \bm I_n)^{-1}\bm y_n}{\tau^2}.
\end{equation*}
\item The radius parameter $r_{\beta}$ for $\bm \beta$ has the following full conditional distribution
{\small
\begin{align*}
& p(r_{\beta} | \bm y_n, \bm \theta_{\beta}, \bm \omega, \tau^2, \eta) \\
& \propto \exp\left(-\frac{1}{2\tau^2}(\bm y_n-\bm G(\text{sgn}(\bm \theta_{\beta})\circ|\bm \theta_{\beta}|^{\frac{2}{q}} r_{\beta}))^\top (\bm K_n+\eta \bm I_n)^{-1} (\bm y_n-\bm G(\text{sgn}(\bm \theta_{\beta})\circ|\bm \theta_{\beta}|^{\frac{2}{q}} r_{\beta}))\right)
\end{align*}
}
where $\bm \theta_{\beta}=(\theta_{\beta_1}, \ldots, \theta_{\beta_p})^\top$ with $\theta_{\beta_i}$ defined at the end of Section 4.2 and ``$\circ$'' is the Hadamard product (element-wise product).
\item The correlation parameters $\bm \omega$ have the following full conditional distribution
\begin{align*}
p(\bm \omega | \bm y_n, \bm \beta, \tau^2,\eta) & \propto  \det(\bm K_n+\eta\bm I_n)^{-\frac{1}{2}}\\
& \times \exp\left(-\frac{1}{2\tau^2}(\bm y_n-\bm G\bm \beta)^\top (\bm K_n+\eta \bm I_n)^{-1} (\bm y_n-\bm G\bm \beta)\right) \mathbb{I}(\|\bm \omega\|_q \leq r_{\omega})
\end{align*}
where the kernel matrix $\bm K_n$ has entries $K_{ij}=K(\bm x_i, \bm x_j, \bm \omega)$.
\item The radius parameter $r_{\omega}$ for $\bm \omega$ has the following full conditional distribution
\begin{align*}
p(r_{\omega} | \bm y_n, \bm \beta, \bm \theta_{\omega}, \tau^2, \eta) & \propto  \det(\bm K_n+\eta\bm I_n)^{-\frac{1}{2}}\\
& \times \exp\left(-\frac{1}{2\tau^2}(\bm y_n-\bm G\bm \beta)^\top (\bm K_n+\eta \bm I_n)^{-1} (\bm y_n-\bm G\bm \beta)\right).
\end{align*}
The kernel matrix $\bm K_n$ depends on $r_{\omega}$ and $\bm \theta_{\omega}$
\begin{equation*}
K_{ij}=K(\bm x_i, \bm x_j; \bm \omega)=\exp\left\{-r_{\omega}^2\sum_{k=1}^d |\theta_{\omega_k}|^{\frac{4}{q}} (x_{ik}-x_{jk})^2\right\}.
\end{equation*}
where $\bm \theta_{\omega}=(\theta_{\omega_1}, \ldots, \theta_{\omega_d})^\top$ with $\theta_{\omega_i}$ defined at the end of Section 4.2.
\item The variance parameter $\tau^2$ has the following full conditional distribution
\begin{equation*}
\tau^2 | \bm y_n, \bm \beta,  \bm \omega, \eta   \sim \text{Scaled Inverse-}\chi^2(df_{\tau^2}+n,\hat{\tau}^2),
\end{equation*}
with
\begin{equation*}
\hat{\tau}^2 = \frac{1+(\bm y_n-\bm G\bm \beta)^\top (\bm K_n+\eta \bm I_n)^{-1} (\bm y_n-\bm G\bm \beta)}{df_{\tau^2}+n}.
\end{equation*}
\item The nugget parameter $\eta$ has the following full conditional distribution
\begin{align*}
p(\eta | \bm y_n, \bm \beta, \bm \omega, \tau^2) \propto \eta^{a_{\eta}-1} e^{-b_{\eta} \eta} & \det(\bm K_n+\eta\bm I_n)^{-\frac{1}{2}}\\
& \times \exp\left(-\frac{1}{2\tau^2}(\bm y_n-\bm G\bm \beta)^\top (\bm K_n+\eta \bm I_n)^{-1} (\bm y_n-\bm G\bm \beta)\right).
\end{align*}
\end{enumerate}
\end{proposition}

\spacingset{1} 
\begin{algorithm}
\caption{Sampling procedure for sph-B\textsuperscript{2}GPR.}\label{alg:bridge-GP-MCMC}
\begin{algorithmic}[1]
\State Set the hyperparameters $a_{\eta}, b_{\eta}, df_{\tau^2}$, the number of samples in the burn-in period (denoted by $B$), and the total number of samples of the MCMC chain (denoted by $T$). 
\State Initialize $\bm \beta^{(0)}, r_{\beta}^{(0)}, \bm \omega^{(0)}, r_{\omega}^{(0)}, \tau^{2^{(0)}}, \eta^{(0)}$.
\For{$t=1$ to $T$}
\State Sample $\bm \beta^{(t)}$ from $p(\bm \beta|\bm y_n, \bm \omega^{(t-1)}, \tau^{2^{(t-1)}}, \eta^{(t-1)})$ through Algorithm \ref{alg:SphHMC} with constraint $\| \bm \beta^{(t)} \|_q \leq r_{\beta}^{(t-1)}$.
\State Sample $r_{\beta}^{(t)}$ from $p(r_{\beta}|\bm y_n, \bm \theta_\beta^{(t)}, \bm \omega^{(t-1)}, \tau^{2^{(t-1)}}, \eta^{(t-1)})$ through Metropolis-Hastings (MH) method.
\State Sample $\bm \omega^{(t)}$ from $p(\bm \omega|\bm y_n, \bm \beta^{(t)}, \tau^{2^{(t-1)}}, \eta^{(t-1)})$ through Algorithm \ref{alg:SphHMC} with constraint $\| \bm \omega^{(t)} \|_q \leq r_{\omega}^{(t-1)}$.
\State Sample $r_{\omega}^{(t)}$ from $p(r_{\omega}|\bm y_n, \bm \beta^{(t)}, \bm \theta_\omega^{(t)}, \tau^{2^{(t-1)}}, \eta^{(t-1)})$ through MH.
\State Sample $\tau^{2^{(t)}}$ from $p(\tau^2|\bm y_n, \bm \beta^{(t)}, \bm \omega^{(t)}, \eta^{(t-1)})$ through Inverse-$\chi^2$ posterior distribution.
\State Sample $\eta^{(t)}$ from $p(\eta|\bm y_n, \bm \beta^{(t)}, \bm \omega^{(t)}, \tau^{2^{(t)}}, \eta^{(t-1)})$ through MH. 
\EndFor
\State Remove the initial $B$ samples from the burn-in period. Use the remaining samples to perform prediction and inference tasks. 
\end{algorithmic}
\end{algorithm}
\spacingset{1.5} 

\subsection{Sampling for hmc-B\textsuperscript{2}GPR}
For the hmc-B\textsuperscript{2}GPR method, we use the informative prior distributions on all the parameters $\bm \theta=(\bm \beta, \nu_{\beta}^2, \bm \omega, \nu_{\omega}^2, \tau^2, \eta)$, which are specified as follows. 
\begin{equation}\label{eq:prior2}
\begin{aligned}
& \bm \beta | \nu_{\beta}^2 \sim \mathcal{MVN}_p({\bf 0}, \nu_{\beta}^2 \bm R), \quad  \nu_{\beta}^2 \sim \text{Inverse-} \Gamma (a_{\beta},b_{\beta}) \\
& \omega_i | \nu_{\omega}^2 \overset{\text{i.i.d.}}{\sim} \mathcal{N}(0, \nu_{\omega}^2), \text{ for } i=1,\ldots, d, \quad  \nu_{\omega}^2 \sim \text{Inverse-} \Gamma (a_{\omega},b_{\omega}) \\
& \tau^2 \sim \text{Inverse-}\chi^2(df_{\tau^2}), \quad \eta \sim \Gamma(a_{\eta}, b_{\eta}).
\end{aligned}
\end{equation}
The correlation matrix $\bm R$ is taken to be diagonal, $\bm R=\diag\{1, \rho, \ldots,\rho, \rho^2,\ldots, \rho^2,\ldots\}$. 
Here we let $\rho=0.5$. 
The power of $\rho$ is the same as the order of the corresponding polynomial term. 
For example, for the model $\bm g(\bm x)=[1, x_1, x_2,x_1^2, x_1x_2, x_2^2]$, the correlation matrix should be $\bm R=\diag\{1, \rho, \rho, \rho^2, \rho^2, \rho^2\}$. 
In this way the prior variance decrease exponentially as the order of the polynomial terms, following the hierarchy ordering principle \citep{wu2011experiments}, which reduces the size of the model and avoid including higher order and less significant model terms.
Such prior distribution was firstly proposed by \cite{joseph2006bayesian}, and later used by \cite{kang2009bayesian, ai2009bayesian, kang2018bayesian}.

The sampling distribution of $\bm y_n | \bm \theta$ is the same as in sph-B\textsuperscript{2}GPR and the joint posterior density is
\begin{equation*} 
\begin{aligned} 
p(\bm \theta|\bm y_n)&\propto p(\bm \theta) p(\bm y_n|\bm \theta)\\
&\propto p(\bm \beta|\nu_{\beta}^2)p(\nu_{\beta}^2)p(\bm \omega| \nu_{\omega}^2)p(\nu_{\omega}^2) p(\tau^2) p(\eta) p(\bm y_n|\bm \theta).
\end{aligned}
\end{equation*}
Due to conditional conjugacy, some posterior distributions are directly available, as summarized in the following proposition.
\begin{proposition} \label{prop:hmc}
Under the prior distributions in \eqref{eq:prior2}, the conditional posterior distributions for all the parameters are derived as follows. 
\begin{enumerate}
\item The full conditional distribution for $\bm \beta$ is
\begin{equation*}
\bm \beta | \bm y_n, \nu_{\beta}^2,\bm \omega,\tau^2,\eta \sim \mathcal{MVN}_p(\hat{\bm \beta}_n, \bm \Sigma_{\bm \beta|n}),
\end{equation*}
where
\begin{equation*}
\bm \Sigma_{\bm \beta|n} =\left[\frac{1}{\tau^2}\bm G^\top (\bm K_n+\eta \bm I_n)^{-1}\bm G+\frac{1}{\nu_{\beta}^2} \bm R^{-1}\right]^{-1}, \quad 
\hat{\bm \beta}_n =\bm \Sigma_{\bm \beta|n}\frac{\bm G^\top (\bm K_n+\eta \bm I_n)^{-1}\bm y_n}{\tau^2}.
\end{equation*}
\item The full conditional distribution for $\nu_{\beta}^2$ is
\begin{equation*}
\nu_{\beta}^2 | \bm \beta, \bm \omega,\tau^2, \eta, \bm y_n \sim \text{Inverse-} \Gamma \left(a_{\beta}+\frac{p}{2},\frac{1}{2} \bm \beta^\top \bm R^{-1} \bm \beta+b_{\beta}\right).
\end{equation*}
\item The full conditional distribution for the correlation parameters $\bm \omega$ is
\begin{equation*}
\begin{aligned}
p(\bm \omega | \bm \beta,\nu_{\omega}^2,\tau^2, \eta, \bm y_n) & \propto \det(\bm K_n+\eta\bm I_n)^{-\frac{1}{2}}\\
& \times \exp\left(-\frac{1}{2\tau^2}(\bm y_n-\bm G\bm \beta)^\top (\bm K_n+\eta \bm I_n)^{-1} (\bm y_n-\bm G\bm \beta)-\frac{\bm \omega^\top \bm \omega}{2\nu_{\omega}^2}\right).\\
\end{aligned}
\end{equation*}
\item The full conditional distribution for $\nu_{\omega}^2$ is
\begin{equation*}
\nu_{\omega}^2 | \bm \beta,\bm \omega, \tau^2, \eta, \bm y_n \sim \text{Inverse-} \Gamma \left(a_{\omega}+\frac{d}{2},\frac{1}{2} \sum_{i=1}^{d} \omega_i^2+b_{\omega}\right).
\end{equation*}
\item The full conditional distribution for the variance parameter $\tau^2$ and $\eta$ are the same as in Proposition \ref{prop:sph}. 
\iffalse
\item The full conditional distribution for the variance parameter $\tau^2$ is 
\begin{equation*}
\tau^2 | \bm y_n, \bm \beta,  \bm \omega, \eta   \sim \text{Scaled Inverse-}\chi^2(df_{\tau^2}+n,\hat{\tau}^2),
\end{equation*}
with
\begin{equation*}
\hat{\tau}^2 = \frac{1+(\bm y_n-\bm G\bm \beta)^\top (\bm K_n+\eta \bm I_n)^{-1} (\bm y_n-\bm G\bm \beta)}{df_{\tau^2}+n}.
\end{equation*}
\item The full conditional for the nugget parameter $\eta$ is
\begin{align*}
p(\eta | \bm y_n, \bm \beta, \bm \omega, \tau^2) \propto \eta^{a_{\eta}-1} e^{-b_{\eta} \eta} &\det(\bm K_n+\eta\bm I_n)^{-\frac{1}{2}}\\
&\times \exp\left(-\frac{1}{2\tau^2}(\bm y_n-\bm G\bm \beta)^\top (\bm K_n+\eta \bm I_n)^{-1} (\bm y_n-\bm G\bm \beta)\right).
\end{align*}
\fi
\end{enumerate}
\end{proposition}

Unlike sph-B\textsuperscript{2}GPR, which explicitly imposes hard $\ell_q$-norm constraints on parameters to directly achieve sparsity via regularization, the hmc-B\textsuperscript{2}GPR method uses hierarchical shrinkage priors to implicitly push sparsity through posterior concentration. 
Consequently, the standard HMC method relies solely on the data-driven shrinkage induced by the priors to identify and suppress irrelevant parameters. 
The complete MCMC sampling algorithm for hmc-B\textsuperscript{2}GPR is summarized in Algorithm \ref{alg:bridge-GP-MCMC2}.

\spacingset{1} 
\begin{algorithm}
\caption{Sampling procedure for hmc-B\textsuperscript{2}GPR.}\label{alg:bridge-GP-MCMC2}
\begin{algorithmic}[1]
\State Set the hyperparameters $a_{\eta}, b_{\eta}, df_{\tau^2}, a_{\beta}, b_{\beta}, a_{\omega}, b_{\omega}$, the number of samples in the burn-in period (denoted by $B$), and the total number of samples of the MCMC chain (denoted by $T$). 
\State Initialize $\bm \beta^{(0)}, \nu_{\beta}^{2^{(0)}}, \bm \omega^{(0)}, \nu_{\omega}^{2^{(0)}}, \tau^{2^{(0)}}, \eta^{(0)}$
\For{$t=1$ to $T$}
\State Sample $\bm \beta^{(t)}$ from $p(\bm \beta|\bm y_n, \nu_{\beta}^{2^{(t-1)}}, \bm \omega^{(t-1)}, \tau^{2^{(t-1)}}, \eta^{(t-1)})$ using multivariate normal distribution. 
\State Sample $\nu_{\beta}^{2^{(t)}}$ from $p(\nu_{\beta}^{2^{(t)}} | \bm y_n, \bm \beta^{(t)}, \bm \omega^{(t-1)},\tau^{2^{(t-1)}}, \eta^{(t-1)})$ through the inverse-$\Gamma$ distribution.
\State Sample $\bm \omega^{(t)}$ from $p(\bm \omega|\bm y_n, \nu_{\omega}^{2^{(t-1)}}, \bm \beta^{(t)}, \tau^{2^{(t-1)}}, \eta^{(t-1)})$ through Algorithm \ref{alg:HMC}.
\State Sample $\nu_{\omega}^{2^{(t)}}$ from $p(\nu_{\omega}^{2^{(t)}} | \bm y_n, \bm \beta^{(t)}, \bm \omega^{(t)},\tau^{2^{(t-1)}}, \eta^{(t-1)})$ through inverse-$\Gamma$ distribution. 
\State Sample $\tau^{2^{(t)}}$ from $p(\tau^{2^{(t)}}|\bm y_n, \bm \beta^{(t)}, \bm \omega^{(t)}, \eta^{(t-1)})$ through inverse-$\chi^2$  distribution.
\State Sample $\eta^{(t)}$ from $p(\eta^{(t)}|\bm y_n, \bm \beta^{(t)}, \bm \omega^{(t)}, \tau^{2^{(t)}}, \eta^{(t-1)})$ through MH method. 
\EndFor
\State Remove the initial $B$ samples from the burn-in period. Use the remaining samples to perform prediction and inference tasks. 
\end{algorithmic}
\end{algorithm}
\spacingset{1.5} 

\section{Simulation Examples}\label{sec:simu}

In this section, we evaluate the performance of the proposed sph- and hmc-B\textsuperscript{2}GPR methods through a series of simulation studies and compare them against six alternative GP or linear regression approaches implemented via different \texttt{R} packages and \texttt{Matlab} codes.
\begin{itemize}
\item Maximum likelihood estimation (MLE)-based Gaussian Process regression.
\item Traditional kriging method with constant, linear, and quadratic mean functions \citep{santner2003design}.
\item Sparse Bayesian linear regression with the horseshoe prior based on \citep{carvalho2010horseshoe}
\item Sparse Bayesian linear regression with the spike-and-slab prior based on \citep{ishwaran2005spike}
\item Bayesian Bridge linear regression by \citep{polson2014bayesian}
\item Indicator-based Bayesian variable selection for Gaussian process models by \citep{zhang2023indicator}
\end{itemize}
\iffalse
Each of these methods has distinct characteristics and is suitable for different scenarios.
Specifically, MLE-based GP and kriging methods are popular approaches widely employed in surrogate modeling.
The two are essentially the same except that they are implemented in different packages. 
In MLE-based GP, the mean is a linear function of the input. 
In the kriging methods, we use three mean functions of the input: constant, linear, and quadratic functions.
These methods perform well when $n$ is not large and $d$ is small. 
However, their predictive accuracy may degrade for increasing $d$ due to the absence of built-in sparsity mechanisms.
The laGP, on the other hand, is tailored for larger data sets. 
It partitions data and fits a local GP model around each prediction point, offering computational efficiency and scalability.
While laGP often provides accurate predictions with larger $n$, it may not adequately capture global trends or complex interactions as effectively as the full GP model.
All numerical experiments in this study were performed in \texttt{R} \citep{R}. 
Specifically, the MLE-based GP, kriging, and local GP approaches were implemented using the R packages \texttt{mlegp} \citep{dancik2008mlegp}, \texttt{DiceKriging} \citep{roustant2012dicekriging}, and \texttt{laGP} \citep{gramacy2016lagp}, respectively.
Table \ref{tab:abbreviation} summarizes the abbreviated names of the methods employed throughout this study, along with corresponding descriptions for reference in subsequent discussions.
\fi

Each comparison method has distinct characteristics and is included to assess both prediction accuracy and variable-selection behavior.
The MLE-based GP and kriging methods are standard GP-based surrogate modeling benchmarks; they are closely related approaches implemented in different packages.
For the MLE-based GP implemented by \texttt{mlegp} R package , the mean function is linear in the input, whereas for kriging implemented by \texttt{DiceKriging} package we can consider different kinds of basis functions such as linear, and quadratic mean functions.
These full GP methods use ARD-type kernels in their covariance functions, so the relevance of each input variable can be assessed heuristically through the estimated correlation or length-scale parameters. 
Although such parameters can provide variable-importance information, these methods do not directly impose sparsity on the input variables.
Therefore, they are effective when the sample size is moderate and the input dimension is small, but their performance can degrade as the dimension increases due to the lack of explicit sparsity mechanisms.
In contrast, Bayesian Bridge, Horseshoe, and Spike-and-Slab regression are sparse Bayesian regression methods developed specifically to shrink or select regression coefficients, and therefore provide natural variable-selection baselines.
Because widely used and mature GP implementations of these sparse Bayesian regression priors are not readily available, we implement them in the linear regression setting to provide reference comparisons for the variable-selection performance of B\textsuperscript{2}GPR.
The indicator-based GP method is also a variable-selection method, but it performs selection within a GP framework by introducing latent binary indicators for the revlevance of input variables.
Together, these methods allow us to compare B\textsuperscript{2}GPR with both prediction-oriented GP competitors and sparsity-oriented variable-selection competitors.

All numerical experiments in this study were performed in \texttt{R} \citep{R}, except that the indicator-based GP method was implemented in \texttt{MATLAB} \citep{MATLAB} using the code provided by the authors of \cite{zhang2023indicator}.
Specifically, the MLE-based GP and kriging approaches were implemented using the R packages \texttt{mlegp} \citep{dancik2008mlegp} and \texttt{DiceKriging} \citep{roustant2012dicekriging}, respectively.
Bayesian linear regression with horseshoe and spike-and-slab priors was implemented using the \texttt{bayesreg} package \citep{bayesreg}, and Bayesian bridge linear regression was implemented using the \texttt{bridge.reg} function from the archived \texttt{BayesBridge} package, version 0.6, based on \citep{BayesBridge}.
Table \ref{tab:abbreviation_modified} summarizes the abbreviated names of the methods employed throughout this study, along with corresponding descriptions for reference in subsequent discussions.

\iffalse
\spacingset{1} 
\begin{table}[ht]
\centering
\caption{Abbreviated names of the methods}
\label{tab:abbreviation}
\begin{tabular}{ll}
\toprule
\textbf{Abbreviation} & \textbf{Full Description} \\
\midrule
\textbf{mlegp} & Maximum likelihood estimated GP with linear mean \\
\textbf{laGP} & Local approximate GP \\
\textbf{krig-const} & Kriging with constant mean \\
\textbf{krig-linear} & Kriging with linear mean \\
\textbf{krig-quad} & Kriging with quadratic mean \\
\textbf{hmc-B\textsuperscript{2}GPR} & B\textsuperscript{2}GPR using HMC with normal shrinkage priors\\
\textbf{sph-B\textsuperscript{2}GPR-$q$} & B\textsuperscript{2}GPR using the SphHMC with $\ell_q$-norm constraint ($0<q<2$)\\
\bottomrule
\end{tabular}
\end{table} 
\spacingset{1.5} 
\fi

\spacingset{1} 
\begin{table}[ht]
\centering
\caption{Abbreviated names of the methods}
\label{tab:abbreviation_modified}
\begin{tabular}{ll}
\toprule
\textbf{Abbreviation} & \textbf{Full Description} \\
\midrule
\textbf{mlegp} & Maximum likelihood estimated GP with linear mean \\
\textbf{mlegp const} & Maximum likelihood estimated GP with constant mean \\
\textbf{Bayesian Bridge} & Bayesian Bridge regression \\
\textbf{Horseshoe} & Bayesian regression with Horseshoe priors \\
\textbf{Spike Slab} & Bayesian regression with Spike-and-Slab priors \\
\textbf{DiceKriging const} & Kriging with constant mean \\
\textbf{DiceKriging linear} & Kriging with linear mean \\
\textbf{DiceKriging quadratic} & Kriging with quadratic mean \\
\textbf{Indicator-based GP} & Indicator-based GP \\
\textbf{hmc-B\textsuperscript{2}GPR} & B\textsuperscript{2}GPR using HMC with normal shrinkage priors\\
\textbf{B\textsuperscript{2}GPR-$q$} & B\textsuperscript{2}GPR using the SphHMC with $\ell_q$-norm constraint ($0<q<2$)\\
\bottomrule
\end{tabular}
\end{table} 
\spacingset{1.5}

\paragraph{Algorithm Settings}
We summarize the algorithmic settings for all methods.
For the sph-B\textsuperscript{2}GPR method described in Algorithm~\ref{alg:bridge-GP-MCMC}, we set the hyperparameters in the prior distributions as follows: $a_{\eta}=0.5, b_{\eta}=0.5, df_{\tau^2}=4$.
For the hmc-B\textsuperscript{2}GPR approach with shrinkage priors described in Algorithm \ref{alg:bridge-GP-MCMC2}, we use the following hyperparameters in prior distributions: $a_{\eta}=0.5, b_{\eta}=0.5, df_{\tau^2}=4, a_{\beta}=1, b_{\beta}=1, a_{\omega}=1, b_{\omega}=1$.
For both the HMC and the Spherical HMC (Algorithm~\ref{alg:HMC} and \ref{alg:SphHMC}), we need to specify the step size $\epsilon$ and the number of leapfrog integration steps $L$ (the for-loop part).
Following the recommendations in \cite{lan2019flexible} and \cite{hoffman2014no}, we adopt the dual averaging scheme \citep{nesterov2009primal} to adaptively tune the step size $\epsilon$ during the burn-in phase.
We employ the ``No-U-Turn'' criterion as the stopping rule for the leapfrog integration, terminating when the trajectory exits the initial state, which is defined by the condition $\langle \tilde{\bm \theta}^{(0)}, \tilde{\bm \theta}^{(l)} \rangle <0$.
Additionally, we set the maximum leapfrog integration step number $L_{max}$ randomly from a uniform integer distribution between $1$ and $10$.
The initial step size is heuristically chosen following the approach in \cite{hoffman2014no}.

For both Algorithm \ref{alg:bridge-GP-MCMC} and \ref{alg:bridge-GP-MCMC2}, we run two parallel chains.
For each chain, we set the first $B=1{,}600$ samples as burn-in.
The total length of the chain (including burn-in) ranges from $T=3{,}000$ to $T=6{,}000$. 
The actual chain length may differ across runs because we employ a simple convergence diagnostic, i.e., if the two chains become sufficiently close at some point, the simulation is stopped early.

For the convergence diagnostic, we use the \texttt{gelman.diag} function in the \texttt{coda} package to compute the potential scale reduction factor (PSRF) for scalar parameters and the multivariate PSRF (MPSRF) for vector-valued parameters.
After the burn-in period, the diagnositic is evaluated every $100$ iterations.
Specifically, we monitor the MPSRFs for the vector-valued parameters $\bm \beta$ and $\bm \omega$, and the PSRFs for the scalar parameters $\tau^2$ and $\eta$.
If all monitored PSRF and MPSRF values are below $1.1$, we regard the chains as reached the prescribed convergence criterion and stop the simulation early.

\paragraph{Performance Metrics} 
We evaluate the performance of different methods using the standardized root-mean-square error (SRMSE) on the test data sets, defined as the RMSE normalized by the standard deviation of the true outputs:
\begin{equation*}
\text{SRMSE}=\frac{\sqrt{\frac{1}{n}\sum_{i=1}^n(y_i-\hat{y}_i)^2}}{\sqrt{\frac{1}{n}\sum_{i=1}^n(y_i-\bar{y})^2}}
\end{equation*}
where $n$ is the test sample size, $y_i$ denotes the true output, $\hat{y}_i$ is the predicted output from the model, and $\bar{y}$ is the mean of the true outputs.
Additionally, we illustrate the box plots of the point estimates (sample medians) of the parameters $\bm \beta$ and correlation parameters $\bm \omega$ from all repeated simulations.
Note that they are not the posterior samples from one simulation, but just the point estimates, which is why all the box plots show little variation. 
In each simulation, we check whether the 95\% credible intervals (obtained from sample quantiles) of the parameters $\bm \beta$ and $\bm \omega$ include zero. 
If zero is included in the confidence intervals, it means that the corresponding $\beta_i$ or $\omega_i$ are not statistical significant, otherwise significant. 
The color shade of the box plots indicates the frequencies when $\beta_i$ or $\omega_i$ are identified to be significant. 
The darker the color is, the more significant the corresponding parameter is. 
Specifically, for B\textsuperscript{2}GPR, an input variable is retained as active if either the corresponding correlation parameter $\omega_i$ is significant or at least one mean-function coefficient involving that input is significant.
If both the relevant mean-function coefficient(s) and $\omega_i$ have credible intervals containing zero, then the variable is discarded.
Therefore, when the mean and correlation components disagree, we use a union rule: the variable is selected as long as at least one component provides evidence of importance.
Then we also evaluate the variable-selection performance of different methods by calculating the number of false positives (FP), defined as inactive variables incorrectly selected and false negatives (FN), defined as active variables incorrectly not selected.
For the MLE-based GP and kriging methods, we use the estimated correlation parameters $\bm \omega$ and the estimated regression coefficients $\bm \beta$ to determine variable importance.
Under our $\bm \omega$ parameterization, a larger $\omega_i$ indicates a more important variable, so we use the threshold of $\omega_i > 0.01$ to select a variable as active.
This threshold is based on the scaled input designs in $[0,1]^d$.
For each regression coefficient $\beta_i$, significance is assessed using a Wald-type 95\% confidence interval based on the standard error of the MLE estimates.
A coefficient $\beta_i$ is declared significant if the interval does not contain zero, and the same union rule is used to determine variable selection for the MLE-based GP and kriging methods.
For the Bayesian Bridge, Horseshoe, and Spike-and-Slab regression methods, we use the posterior samples of $\bm \beta$ to determine variable importance: a variable is selected as active if the 95\% credible interval of its corresponding $\beta_i$ does not contain zero.
For the indicator-based GP method, active dimensions are determined from posterior samples of the latent binary indicators, with a variable selected as active if the posterior mean of its corresponding indicator exceeds $0.5$.

\subsection{Pre-specified Gaussian Process}

\begin{figure}[tbp]
\centering
\begin{subfigure}[htb]{.49\linewidth}
\includegraphics[width=\linewidth]{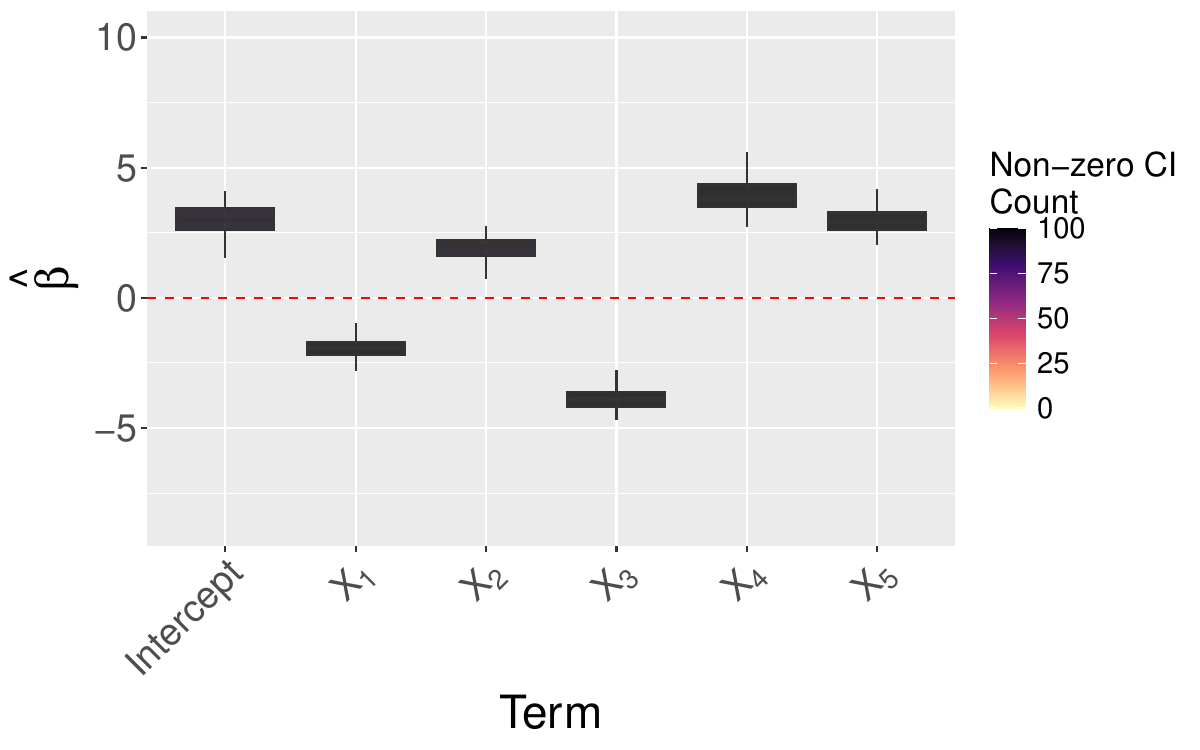}
\caption{Point estimates for $\bm \beta$ and significance.}
\end{subfigure}
\begin{subfigure}[htb]{.49\linewidth}
\includegraphics[width=\linewidth]{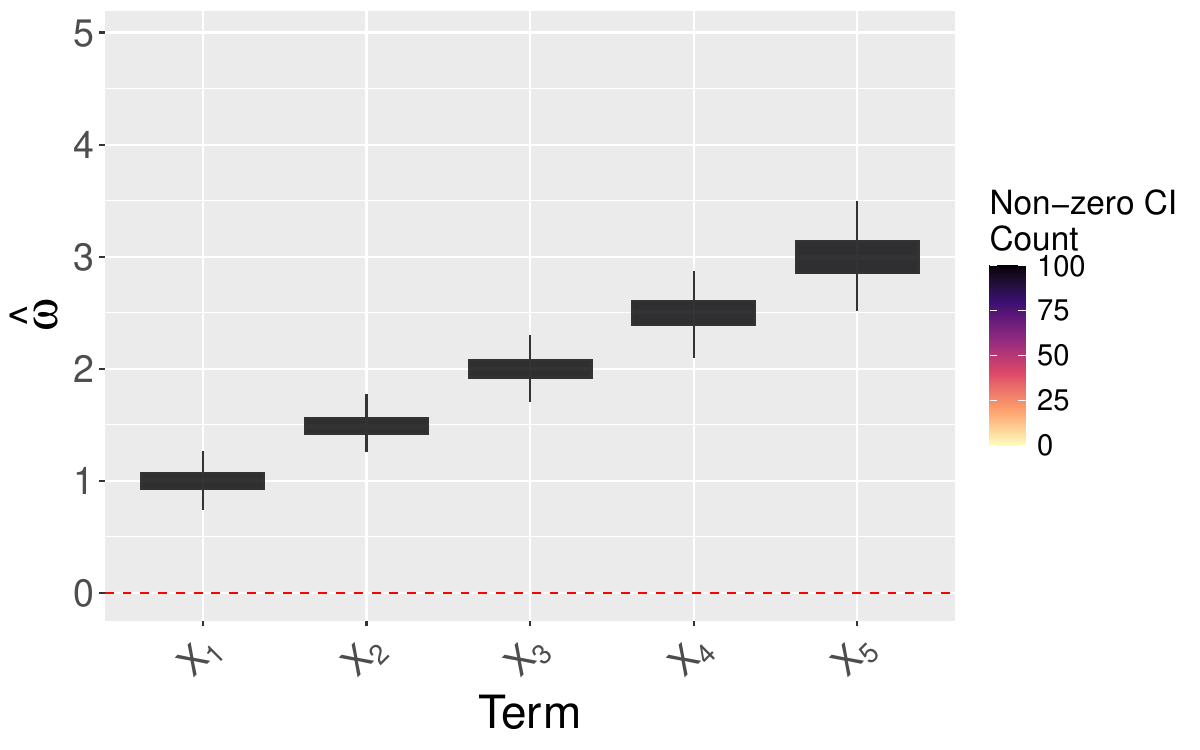}
\caption{Point estimates for $\bm \omega$ and significance.}
\end{subfigure}
\caption{Box plots of the point estimates of the parameters for the pre-specified GP.}\label{fig:boxplot_sim_para}
\end{figure}

\begin{figure}[tbp]
\centering
\includegraphics[width=0.9\linewidth]{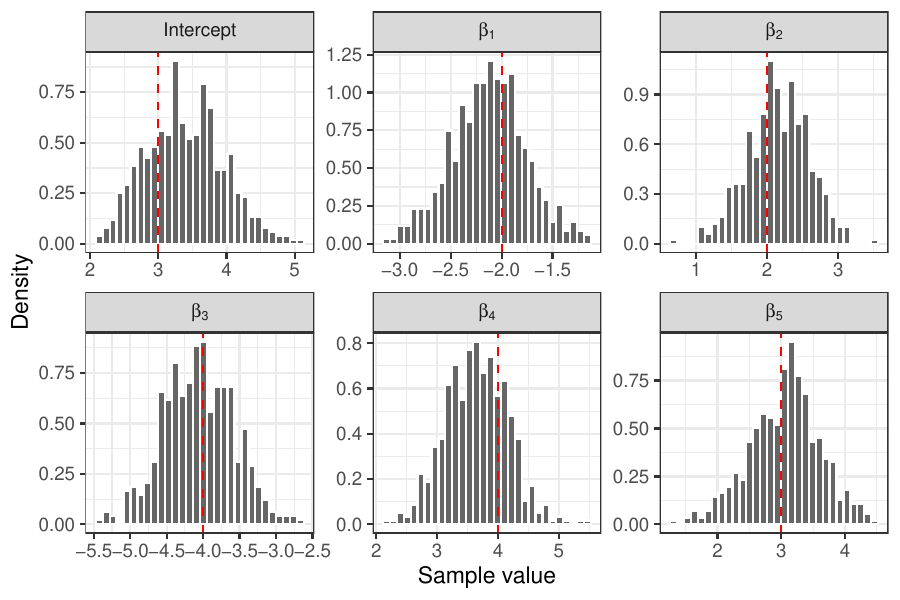}
\caption{Posterior histograms for $\bm \beta$ in one simulation by sph-B\textsuperscript{2}GPR with $q=1.8$ for the pre-specified GP.}\label{fig:histogram_sim_beta}
\end{figure}

\begin{figure}[tbp]
\centering
\includegraphics[width=0.9\linewidth]{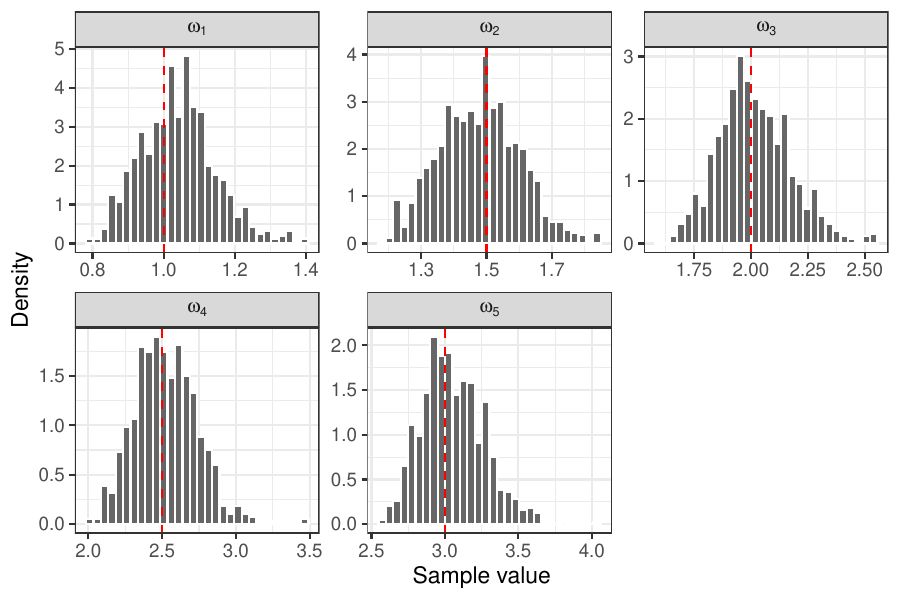}
\caption{Posterior histograms for $\bm \omega$ in one simulation by sph-B\textsuperscript{2}GPR with $q=1.8$ for the pre-specified GP.}\label{fig:histogram_sim_omega}
\end{figure}

In this example, we simulate data from a pre-specified Gaussian process model~\eqref{eq:model} in Section~\ref{sec:background} with input dimension $d=5$. 
The true parameters of the GP are set as follows: the intercept $\beta_0=3$, the coefficient of the linear terms $\bm \beta = (-2, 2, -4, 4, 3)$, the correlation parameters $\bm \omega = (1,1.5,2,2.5,3)$, and the variance parameter $\tau^2=1$.
For each simulation, we generate $n=200$ training data. 
The input data $\bm x_i$'s are from a $5$-dimensional maximin Latin hypercube design \citep{lhs} within the domain $[0,1]^5$ and the output $y(\bm x_i)$ is simulated from the GP without noise $\epsilon$. 
An additional design of $1000$ input points is generated using a random Latin hypercube design within the same domain, and the responses are simulated from the GP based on these inputs to form the test data set for evaluating prediction accuracy.
Table \ref{tab:rmse_fn_fp_simulation2_n200} shows the average standardized RMSE, false negative (FN), and false positive (FP) rates over $100$ simulations.
Figure \ref{fig:boxplot_sim_para} shows the box plots of the point estimates of the parameters for $\bm \beta$ and $\bm \omega$ from the method sph-B\textsuperscript{2}GPR with $q=1.8$. 
The dark shade of the box plots indicates that all $\bm \beta$ and $\bm \omega$ are significant. 
Figure \ref{fig:histogram_sim_beta} and \ref{fig:histogram_sim_omega} present the posterior histograms for each $\bm \beta$ and $\bm \omega$ in one simulation produced by sph-B\textsuperscript{2}GPR with $q=1.8$.
The red vertical lines are the true parameter values of the underlying GP model. 

\spacingset{1} 
\begin{table}[htb]
\centering
\caption{Standardized RMSE ($100$ replicates for $n=200$) for pre-specified Gaussian Process}
\label{tab:rmse_sim}
\begin{tabular}{lcccccccc}
\toprule
 & \makecell{mlegp}
 & \makecell{laGP}
 & \makecell{krig-\\const}
 & \makecell{krig-\\linear}
 & \makecell{hmc-\\B\textsuperscript{2}GPR}
 & \makecell{sph-\\B\textsuperscript{2}GPR-$0.8$}
 & \makecell{sph-\\B\textsuperscript{2}GPR-$1$}
 & \makecell{sph-\\B\textsuperscript{2}GPR-$1.8$}\\
\midrule
Mean   & $0.0260$ & $0.1091$ & $0.0799$ & $0.0270$ & $0.0303$ & \bm{$0.0267$} & \bm{$0.0264$} & \bm{$0.0260$}\\
SD     & $0.0082$ & $0.0427$ & $0.0139$ & $0.0087$ & $0.0139$ & \bm{$0.0086$} & \bm{$0.0082$} & \bm{$0.0083$}\\
\bottomrule
\end{tabular}
\end{table}

\spacingset{1}

\begin{table}[htbp]
\scriptsize
\setlength{\tabcolsep}{3.0pt}
\centering
\begin{threeparttable}
\caption{Prediction and variable-selection performance for the pre-specified Gaussian process example with $d=5$ and $n=200$.}
\label{tab:rmse_fn_fp_simulation2_n200}
\begin{tabular}{ccccccccccc}
\toprule
\makecell{\textbf{Metric}} &
\makecell{\textbf{mlegp}} &
\makecell{\textbf{Bayesian}\\\textbf{Bridge}} &
\makecell{\textbf{Horse-}\\\textbf{shoe}} &
\makecell{\textbf{Spike}\\\textbf{Slab}} &
\makecell{\textbf{DiceKriging}\\\textbf{linear}} &
\makecell{\textbf{Indicator-}\\\textbf{based GP}} &
\makecell{\textbf{hmc-}\\\textbf{B\textsuperscript{2}GPR}} &
\makecell{\textbf{B\textsuperscript{2}GPR}\\\textbf{$q=0.8$}} &
\makecell{\textbf{B\textsuperscript{2}GPR}\\\textbf{$q=1$}} &
\makecell{\textbf{B\textsuperscript{2}GPR}\\\textbf{$q=1.8$}}\\
\midrule
SRMSE &
\makecell{$0.0266$\\$(0.0090)$} &
\makecell{$0.2961$\\$(0.0397)$} &
\makecell{$0.2960$\\$(0.0398)$} &
\makecell{$0.3185$\\$(0.0435)$} &
\makecell{$0.0274$\\$(0.0090)$} &
\makecell{$0.0548$\\$(0.0107)$} &
\makecell{$0.0303$\\$(0.0368)$} &
\makecell{$0.0267$\\$(0.0086)$} &
\makecell{$0.0264$\\$(0.0082)$} &
\makecell{$\bm{0.0260}$\\$\bm{(0.0083)}$}\\
FN &
\makecell{$\bm{0.00}$\\$\bm{(0.00)}$} &
\makecell{$0.38$\\$(0.60)$} &
\makecell{$0.45$\\$(0.63)$} &
\makecell{$0.03$\\$(0.17)$} &
\makecell{$\bm{0.00}$\\$\bm{(0.00)}$} &
\makecell{$\bm{0.00}$\\$\bm{(0.00)}$} &
\makecell{$\bm{0.00}$\\$\bm{(0.00)}$} &
\makecell{$\bm{0.00}$\\$\bm{(0.00)}$} &
\makecell{$\bm{0.00}$\\$\bm{(0.00)}$} &
\makecell{$\bm{0.00}$\\$\bm{(0.00)}$}\\
FP &
\makecell{$\bm{0.00}$\\$\bm{(0.00)}$} &
\makecell{$\bm{0.00}$\\$\bm{(0.00)}$} &
\makecell{$\bm{0.00}$\\$\bm{(0.00)}$} &
\makecell{$\bm{0.00}$\\$\bm{(0.00)}$} &
\makecell{$\bm{0.00}$\\$\bm{(0.00)}$} &
\makecell{$\bm{0.00}$\\$\bm{(0.00)}$} &
\makecell{$\bm{0.00}$\\$\bm{(0.00)}$} &
\makecell{$\bm{0.00}$\\$\bm{(0.00)}$} &
\makecell{$\bm{0.00}$\\$\bm{(0.00)}$} &
\makecell{$\bm{0.00}$\\$\bm{(0.00)}$}\\
\bottomrule
\end{tabular}
\begin{tablenotes}\footnotesize
\item \textit{Note:} Each cell reports the mean on the first line and the standard deviation in parentheses on the second line. SRMSE denotes standardized root-mean-square error. FN and FP denote the numbers of false negatives and false positives in variable selection, respectively. Lower values are better. Bold indicates the best mean in each row; ties are all bold. Results are computed across $100$ replicates.
\end{tablenotes}
\end{threeparttable}
\end{table}

\spacingset{1.5}

\spacingset{1.5} 

\iffalse
{\bf Performance Summary}
The method sph-B\textsuperscript{2}GPR with $q=1.8$ and mlegp both have the best performance in terms of average standardized RMSE ($\approx 2.6 \times 10^{-2}$) and stability (small standard deviation).
As $q$ becomes smaller, sph-B\textsuperscript{2}GPR performs slightly worse. 
But compared to sph-B\textsuperscript{2}GPR, hmc-B\textsuperscript{2}GPR is only better than laGP and krig-const. 
For the parameter $\bm \beta$ and $\bm \omega$, the posterior medians closely match the true values, and the 95\% confidence intervals rarely include zero, correctly flagging all as significant. 
Each posterior histogram for $\bm \beta$ and $\bm \omega$ is unimodal and centered around the true parameter values.
This aligns with the low standardized RMSE observed, indicating accurate parameter estimation by the proposed method.
\fi

In this controlled example, the data are generated from a $d=5$ pre-specified Gaussian process with all five non-zero fixed-effect coefficients $\bm\beta$ and all five non-zero correlation parameters $\bm\omega$.
Thus, the ideal variable-selection result has zero false negatives and zero false positives.
Table \ref{tab:rmse_fn_fp_simulation2_n200} shows that B\textsuperscript{2}GPR with $q=1.8$ attains the smallest average SRMSE ($0.0260$), with the $q=1$ and $q=0.8$ variants close behind.
The predictive performance of these three B\textsuperscript{2}GPR variants is comparable to \texttt{mlegp} and \texttt{DiceKriging} with a linear mean, while Bayesian Bridge, Horseshoe, and Spike Slab have substantially larger standardized RMSEs in this nonlinear GP setting.
The supplementary box plots in Figure \ref{fig:ggplot_simulation2_200} add a distributional view: the B\textsuperscript{2}GPR SRMSE distributions are concentrated near the lowest values, whereas hmc-B\textsuperscript{2}GPR has a similar median but noticeably larger variability.
For variable selection, all three B\textsuperscript{2}GPR variants achieve zero false negatives and zero false positives across the $100$ replicates, matching the performance of \texttt{mlegp}, \texttt{DiceKriging}, and the indicator-based GP.
Because all inputs are truly active in this simulation, false positives are zero for all methods; the meaningful selection difference is in false negatives, where Bayesian Bridge and Horseshoe tend to miss active variables, and Spike Slab reduces false negatives but at the cost of larger prediction error.
The parameter-estimation box plots in Figure \ref{fig:boxplot_sim_para} and the posterior histograms in Figures \ref{fig:histogram_sim_beta} and \ref{fig:histogram_sim_omega} further show that the B\textsuperscript{2}GPR posterior medians are centered near the true values and that the active $\bm\beta$ and $\bm\omega$ components are correctly identified as significant.
Together, these results show that B\textsuperscript{2}GPR achieves accurate prediction, stable posterior inference, and reliable recovery of active variables in the controlled pre-specified GP example.

\subsection{Benchmark functions: Borehole, OTL Circuit, and Piston}

In this example, we use three benchmark test functions from literature: the Borehole function to model water flow through a borehole, the OTL Circuit function to simulate the behavior of an electronic circuit, and the Piston function to model the circular motion of a piston within a cylinder \citep{simulationlib}.
Their definitions are given in the Supplement. 
Due to limited space, we moved all the figures for this subsection to the Supplement. 

For each benchmark, we generate training data sets of size $n\in\{200,500\}$ with input from a maximin Latin hypercube design \citep{lhs} in $[0,1]^d$.
Then we scale the inputs to their respective physical ranges defined by the test functions.
Outputs are generated from the benchmark functions with added small noise.
A separate test data set of size $1000$ with inputs generated from a random Latin hypercube design and outputs generated from the benchmark functions with added noise of the same distribution as the training data. 
We report the joint prediction and variable-selection results over $100$ simulations in Table \ref{tab:rmse_fn_fp_combined_baseline_n200} for $n=200$ and in Table \ref{tab:rmse_fn_fp_combined_baseline_n500} for $n=500$ for the original input dimensions.
\ref{fig:ggplot_borehole_200}--\ref{fig:ggplot_piston_500} further display the replicate-level distributions of standardized RMSE, false positives, and false negatives for all methods and benchmarks with the original input dimensions.
These particular box plots are especially useful for assessing stability.
Figures \ref{fig:boxplot_borehole_para_200}-\ref{fig:boxplot_piston_para_500} show the box plots of the point estimates for the parameters $\bm \beta$ and $\bm \omega$ with $q=0.8$ for each benchmark function with $n=200$ and $n=500$.
The mean functions of the GP for both sph- and hmc-B\textsuperscript{2}GPR are quadratic polynomials including intercept, linear, quadratic, and interaction terms. 

We then augment the input with additional irrelevant input variables to test the variable selection capability of the proposed B\textsuperscript{2}GPR methods in a high-dimensional setting.
Specifically, we increase the input dimension to $d=20$ for each of the three functions.
Only the original input variables should be significant.
The training and testing data sets are generated in the same way as earlier.
Like the original input dimension cases, we also report the joint prediction and variable-selection results in Table \ref{tab:rmse_fn_fp_combined_hd_n200} for $n=200$ and in Table \ref{tab:rmse_fn_fp_combined_hd_n500} for $n=500$ for the high-dimensional cases.
For the $n=500$ and $d=20$ setting, the indicator-based GP method is omitted because its computation is prohibitively time-consuming in this high dimensional setting.
The corresponding performance box plots are shown in Figures \ref{fig:ggplot_borehole_200_20}--\ref{fig:ggplot_piston_500_20}.
Figures \ref{fig:boxplot_borehole_20_para_200}--\ref{fig:boxplot_piston_20_para_500} show the box plots of the point estimates for the parameters $\bm \beta$ and $\bm \omega$ with $q=0.8$ for all test functions of $n=200$ and 500.
The mean functions of the GP for both sph- and hmc-B\textsuperscript{2}GPR are linear polynomials for $d=20$.

{\bf Borehole Function}
For the Borehole function with the original input dimension $d=8$, the proposed sph-B\textsuperscript{2}GPR methods deliver the strongest predictive performance. At $n=200$, sph-B\textsuperscript{2}GPR with $q=0.8$ achieves the lowest average SRMSE ($0.0182$) and standard deviation ($0.0019$), closely followed by the $q=1$ and $q=1.8$ variants. At $n=500$, all methods show improvement, but sph-B\textsuperscript{2}GPR remains the most accurate and stable. Parameter-estimation box plots (Figures \ref{fig:boxplot_borehole_para_200} and \ref{fig:boxplot_borehole_para_500}) show that most fixed-effect coefficients in $\bm \beta$ are shrunk toward zero, while the correlation parameters $\bm \omega$ correctly identify the active inputs ($r_w$, $H_u$, $H_l$, $L$, and $K_w$), in line with \cite{harper1983sensitivity}.

Because we conservatively treat all eight original inputs as active for evaluation, there are no inactive variables in the $d=8$ setting, resulting in zero false positives for all methods. Consequently, selection differences are primarily reflected in false negatives (FN). Sensitivity analysis by \cite{harper1983sensitivity} indicates that five of these eight inputs are dominant. An ideal sparse selection would exclude the remaining three non-dominant inputs, corresponding to three false negatives and zero false positives. Tables \ref{tab:rmse_fn_fp_combined_baseline_n200} and \ref{tab:rmse_fn_fp_combined_baseline_n500} show that all three B\textsuperscript{2}GPR variants successfully recover this sparse selection pattern while achieving the lowest or near-lowest SRMSEs. This is further supported by Figures \ref{fig:ggplot_borehole_200} and \ref{fig:ggplot_borehole_500}, where B\textsuperscript{2}GPR SRMSEs are tightly concentrated at low values, and FN box plots confirm the consistent selection of the five dominant inputs.

After augmenting the input to $d=20$ by adding 12 irrelevant variables, the sph-B\textsuperscript{2}GPR variants remain robust, maintaining low SRMSEs and stable variability. In contrast, hmc-B\textsuperscript{2}GPR exhibits higher variability, and kriging/mlegp degrade notably. Figures \ref{fig:ggplot_borehole_200_20} and \ref{fig:ggplot_borehole_500_20} confirm that sph-B\textsuperscript{2}GPR errors remain concentrated, whereas hmc-B\textsuperscript{2}GPR displays a long upper tail, indicating occasional unstable runs. The FN box plots also show that the B\textsuperscript{2}GPR variants maintain a stable selection pattern across both sample sizes, with FN counts concentrated around three, representing the consistent selection of the five dominant inputs. Finally, Tables \ref{tab:rmse_fn_fp_combined_hd_n200} and \ref{tab:rmse_fn_fp_combined_hd_n500}, along with the false-positive box plots, confirm that B\textsuperscript{2}GPR variants effectively suppress false positives from the irrelevant variables, whereas mlegp, DiceKriging, and hmc-B\textsuperscript{2}GPR select many inactive inputs.

{\bf OTL Circuit Function}
For the OTL Circuit function with the original input dimension $d=6$, sph-B\textsuperscript{2}GPR provides superior prediction accuracy. At $n=200$, the $q=0.8$ variant achieves the lowest average SRMSE ($0.0188$) and standard deviation ($0.0019$), with the $q=1$, $q=1.8$, and hmc-B\textsuperscript{2}GPR variants close behind. At $n=500$, all three B\textsuperscript{2}GPR variants remain highly stable and competitive, with $q=1.8$ yielding the lowest average SRMSE (Table \ref{tab:rmse_fn_fp_combined_baseline_n500}). Parameter estimation indicates that the lower-order fixed effects primarily involve $X_3$ and $X_4$ (with weaker effects for $X_1$ and $X_2$ that clarify at $n=500$), while $X_3X_4$ and $X_4^2$ dominate the second-order terms. For the correlation parameters, $X_1$ and $X_2$ are consistently identified as important. Together, these estimates correctly identify $X_1$, $X_2$, $X_3$, and $X_4$ as the active inputs, in agreement with \cite{moon2010design}.

Treating all six original inputs as active for evaluation leads to zero false positives for all methods when $d=6$. The ideal sparse selection identifies the four dominant inputs ($X_1$--$X_4$) and excludes the two non-dominant ones, which corresponds to two false negatives and zero false positives. Tables \ref{tab:rmse_fn_fp_combined_baseline_n200} and \ref{tab:rmse_fn_fp_combined_baseline_n500} show that the B\textsuperscript{2}GPR variants, particularly $q=0.8$, successfully recover or closely approximate this sparse selection pattern while achieving the lowest or near-lowest SRMSEs. This concentration of low SRMSEs is illustrated in Figures \ref{fig:ggplot_otlcircuit_200} and \ref{fig:ggplot_otlcircuit_500}, where the $q=0.8$ variant centers near the target two false negatives, while the $q=1$ and $q=1.8$ variants remain slightly more conservative.

After increasing the input dimension to $d=20$, the sph-B\textsuperscript{2}GPR variants remain the most accurate methods (Tables \ref{tab:rmse_fn_fp_combined_hd_n200} and \ref{tab:rmse_fn_fp_combined_hd_n500}). Figures \ref{fig:ggplot_otlcircuit_200_20} and \ref{fig:ggplot_otlcircuit_500_20} show that their SRMSE concentration persists, whereas hmc-B\textsuperscript{2}GPR exhibits a wider error distribution with high-SRMSE outliers. The B\textsuperscript{2}GPR variants maintain stable selection across both sample sizes, with FN counts concentrated around two, and high-dimensional parameter plots identify only the first four correlation parameters ($\omega_1,\ldots,\omega_4$) as significant. Furthermore, the B\textsuperscript{2}GPR variants successfully control false positives from the added irrelevant variables, whereas mlegp, DiceKriging, and hmc-B\textsuperscript{2}GPR select numerous inactive inputs.

{\bf Piston Function}
For the Piston function with original dimension $d=7$, sph-B\textsuperscript{2}GPR achieves the lowest average SRMSE and smallest variability across both sample sizes. Tables \ref{tab:rmse_fn_fp_combined_baseline_n200} and \ref{tab:rmse_fn_fp_combined_baseline_n500} indicate that the three B\textsuperscript{2}GPR variants are nearly tied, with the $q=1.8$ variant delivering the lowest average SRMSE for both $n=200$ and $n=500$. Parameter estimation shows that none of the fixed-effect coefficients in $\bm \beta$ are significant. For the correlation parameters $\bm \omega$, inputs $X_1$ through $X_4$ are highly significant, $X_5$ shows weak significance, and $X_6$ and $X_7$ are close to zero, which matches the findings in \citep{moon2010design}.

Treating all seven original inputs as active results in zero false positives for all methods when $d=7$. The desired sparse selection identifies the four dominant inputs ($X_1$--$X_4$) and the weakly active $X_5$ while excluding $X_6$ and $X_7$, corresponding to two false negatives and zero false positives. Tables \ref{tab:rmse_fn_fp_combined_baseline_n200} and \ref{tab:rmse_fn_fp_combined_baseline_n500} show that the B\textsuperscript{2}GPR variants closely approximate this target pattern while yielding the lowest SRMSEs. Figures \ref{fig:ggplot_piston_200} and \ref{fig:ggplot_piston_500} confirm that B\textsuperscript{2}GPR SRMSE distributions are the lowest and most stable, with FN counts tightly concentrated around two.

When the input dimension is increased to $d=20$, the sph-B\textsuperscript{2}GPR variants continue to outperform competing methods, identifying the same active variables as in the $d=7$ case. Figures \ref{fig:ggplot_piston_200_20} and \ref{fig:ggplot_piston_500_20} show that this predictive advantage persists; hmc-B\textsuperscript{2}GPR sometimes achieves a low median but displays a much wider spread and high-SRMSE outliers. The B\textsuperscript{2}GPR variants maintain stable selection patterns after dimension augmentation, with FN counts concentrated near two across both sample sizes. Finally, false-positive box plots and Tables \ref{tab:rmse_fn_fp_combined_hd_n200}--\ref{tab:rmse_fn_fp_combined_hd_n500} confirm that B\textsuperscript{2}GPR variants maintain zero or near-zero false positives, whereas mlegp, DiceKriging, and hmc-B\textsuperscript{2}GPR select many irrelevant inputs.

Across all three benchmarks, the proposed B\textsuperscript{2}GPR framework consistently delivers strong predictive accuracy and robust variable selection in both original and high-dimensional settings. In the original-dimensional cases, the B\textsuperscript{2}GPR variants concentrate error distributions near the lowest values and successfully recover the desired sparse selection patterns. Upon augmenting the input space to $d=20$, B\textsuperscript{2}GPR preserves its predictive advantages and maintains stable selection patterns. Most importantly, while alternative methods like mlegp, DiceKriging, and hmc-B\textsuperscript{2}GPR suffer from notable performance degradation and select numerous irrelevant inputs, the B\textsuperscript{2}GPR variants robustly control false positives at or near zero. Overall, these results demonstrate that B\textsuperscript{2}GPR offers a highly favorable balance of prediction accuracy, sparse input recovery, and strong false-positive control in high-dimensional settings.

\spacingset{1} 
\begin{table}[htbp]
\scriptsize
\setlength{\tabcolsep}{3.5pt}
\centering
\begin{threeparttable}
\caption{Standardized RMSE for the three benchmark test functions: Borehole, OTL Circuit, Piston.}
\label{tab:rmse_combined_baseline}
\begin{tabular}{ccccccccccc}
\toprule
\makecell{\textbf{Function}} & \makecell{\textbf{Size}} &
\makecell{\textbf{mlegp}} & \makecell{\textbf{laGP}} &
\makecell{\textbf{krig-}\\\textbf{const}} &
\makecell{\textbf{krig-}\\\textbf{linear}} &
\makecell{\textbf{krig-}\\\textbf{quad}} &
\makecell{\textbf{hmc-}\\\textbf{B\textsuperscript{2}GPR}} &
\makecell{\textbf{sph-}\\\textbf{B\textsuperscript{2}GPR-$0.8$}} &
\makecell{\textbf{sph-}\\\textbf{B\textsuperscript{2}GPR-$1$}} &
\makecell{\textbf{sph-}\\\textbf{B\textsuperscript{2}GPR-$1.8$}}\\
\midrule
\multirow{2}{*}{\makecell{Borehole \\ ($8$-dim)}}
& $n=200$ &
\makecell{$0.0608$\\$(0.0043)$} &
\makecell{$0.0389$\\$(0.0029)$} &
\makecell{$0.0342$\\$(0.0027)$} &
\makecell{$0.0306$\\$(0.0024)$} &
\makecell{$0.0355$\\$(0.0069)$} &
\makecell{$0.0398$\\$(0.0077)$} &
\makecell{$\bm{0.0182}$\\$\bm{(0.0019)}$} &
\makecell{$0.0187$\\$(0.0024)$} &
\makecell{$0.0207$\\$(0.0034)$}\\
& $n=500$ &
\makecell{$0.0539$\\$(0.0026)$} &
\makecell{$0.0212$\\$(0.0013)$} &
\makecell{$0.0216$\\$(0.0014)$} &
\makecell{$0.0196$\\$(0.0013)$} &
\makecell{$0.0171$\\$(0.0014)$} &
\makecell{$0.0151$\\$(0.0036)$} &
\makecell{$\bm{0.0121}$\\$\bm{(0.0011)}$} &
\makecell{$0.0124$\\$(0.0011)$} &
\makecell{$0.0125$\\$(0.0015)$}\\
\midrule
\multirow{2}{*}{\makecell{OTL Circuit \\ ($6$-dim)}}
& $n=200$ &
\makecell{$0.0539$\\$(0.0040)$} &
\makecell{$0.0284$\\$(0.0034)$} &
\makecell{$0.0272$\\$(0.0019)$} &
\makecell{$0.0250$\\$(0.0019)$} &
\makecell{$0.0271$\\$(0.0045)$} &
\makecell{$0.0216$\\$(0.0032)$} &
\makecell{$\bm{0.0188}$\\$\bm{(0.0019)}$} &
\makecell{$0.0209$\\$(0.0060)$} &
\makecell{$0.0198$\\$(0.0038)$}\\
& $n=500$ &
\makecell{$0.0484$\\$(0.0026)$} &
\makecell{$0.0172$\\$(0.0011)$} &
\makecell{$0.0173$\\$(0.0012)$} &
\makecell{$0.0164$\\$(0.0012)$} &
\makecell{$0.0193$\\$(0.0034)$} &
\makecell{$0.0143$\\$(0.0018)$} &
\makecell{$0.0142$\\$(0.0016)$} &
\makecell{$0.0141$\\$(0.0019)$} &
\makecell{$\bm{0.0138}$\\$\bm{(0.0025)}$}\\
\midrule
\multirow{2}{*}{\makecell{Piston \\ ($7$-dim)}}
& $n=200$ &
\makecell{$0.0749$\\$(0.0075)$} &
\makecell{$0.0721$\\$(0.0111)$} &
\makecell{$0.0637$\\$(0.0052)$} &
\makecell{$0.0585$\\$(0.0051)$} &
\makecell{$0.0645$\\$(0.0262)$} &
\makecell{$0.0522$\\$(0.0248)$} &
\makecell{$0.0438$\\$(0.0047)$} &
\makecell{$0.0436$\\$(0.0049)$} &
\makecell{$\bm{0.0435}$\\$\bm{(0.0046)}$}\\
& $n=500$ &
\makecell{$0.0597$\\$(0.0042)$} &
\makecell{$0.0422$\\$(0.0081)$} &
\makecell{$0.0383$\\$(0.0029)$} &
\makecell{$0.0360$\\$(0.0030)$} &
\makecell{$0.0337$\\$(0.0026)$} &
\makecell{$0.0313$\\$(0.0210)$} &
\makecell{$0.0272$\\$(0.0024)$} &
\makecell{$0.0272$\\$(0.0023)$} &
\makecell{$\bm{0.0271}$\\$\bm{(0.0024)}$}\\
\bottomrule
\end{tabular}
\begin{tablenotes}\footnotesize
\item \textit{Note:} Each cell reports the mean (first line) and standard deviation (second line, in parentheses) of standardized RMSE across $100$ replicates; lower is better. Bold indicates the best (lowest) mean in each row. All ties are in bold.
\end{tablenotes}
\end{threeparttable}
\end{table}

\begin{table}[htbp]
\scriptsize
\setlength{\tabcolsep}{2.5pt}
\centering
\begin{threeparttable}
\caption{Prediction and variable-selection performance for the three benchmark test functions with $n=200$: Borehole, OTL Circuit, and Piston.}
\label{tab:rmse_fn_fp_combined_baseline_n200}
\begin{tabular}{cccccccccccc}
\toprule
\makecell{\textbf{Function}} & \makecell{\textbf{Metric}} &
\makecell{\textbf{mlegp}} &
\makecell{\textbf{Bayesian}\\\textbf{Bridge}} &
\makecell{\textbf{Horse-}\\\textbf{shoe}} &
\makecell{\textbf{Spike}\\\textbf{Slab}} &
\makecell{\textbf{DiceKriging}\\\textbf{quadratic}} &
\makecell{\textbf{Indicator-}\\\textbf{based GP}} &
\makecell{\textbf{hmc-}\\\textbf{B\textsuperscript{2}GPR}} &
\makecell{\textbf{B\textsuperscript{2}GPR}\\\textbf{$q=0.8$}} &
\makecell{\textbf{B\textsuperscript{2}GPR}\\\textbf{$q=1$}} &
\makecell{\textbf{B\textsuperscript{2}GPR}\\\textbf{$q=1.8$}}\\
\midrule
\multirow{3}{*}{\makecell{Borehole \\ ($8$-dim)}}
& SRMSE &
\makecell{$0.0608$\\$(0.0052)$} &
\makecell{$0.0396$\\$(0.0022)$} &
\makecell{$0.0393$\\$(0.0023)$} &
\makecell{$0.0454$\\$(0.0053)$} &
\makecell{$0.0364$\\$(0.0079)$} &
\makecell{$0.1353$\\$(0.0168)$} &
\makecell{$0.0398$\\$(0.0077)$} &
\makecell{$\bm{0.0182}$\\$\bm{(0.0019)}$} &
\makecell{$0.0187$\\$(0.0024)$} &
\makecell{$0.0207$\\$(0.0034)$}\\
& FN &
\makecell{$0.43$\\$(0.61)$} &
\makecell{$2.68$\\$(0.62)$} &
\makecell{$2.86$\\$(0.38)$} &
\makecell{$3.00$\\$(0.00)$} &
\makecell{$\bm{0.00}$\\$\bm{(0.00)}$} &
\makecell{$\bm{0.00}$\\$\bm{(0.00)}$} &
\makecell{$\bm{0.00}$\\$\bm{(0.00)}$} &
\makecell{$3.06$\\$(0.28)$} &
\makecell{$3.21$\\$(0.48)$} &
\makecell{$3.53$\\$(0.63)$}\\
& FP &
\makecell{$\bm{0.00}$\\$\bm{(0.00)}$} &
\makecell{$\bm{0.00}$\\$\bm{(0.00)}$} &
\makecell{$\bm{0.00}$\\$\bm{(0.00)}$} &
\makecell{$\bm{0.00}$\\$\bm{(0.00)}$} &
\makecell{$\bm{0.00}$\\$\bm{(0.00)}$} &
\makecell{$\bm{0.00}$\\$\bm{(0.00)}$} &
\makecell{$\bm{0.00}$\\$\bm{(0.00)}$} &
\makecell{$\bm{0.00}$\\$\bm{(0.00)}$} &
\makecell{$\bm{0.00}$\\$\bm{(0.00)}$} &
\makecell{$\bm{0.00}$\\$\bm{(0.00)}$}\\
\midrule
\multirow{3}{*}{\makecell{OTL Circuit \\ ($6$-dim)}}
& SRMSE &
\makecell{$0.0542$\\$(0.0043)$} &
\makecell{$0.0346$\\$(0.0014)$} &
\makecell{$0.0339$\\$(0.0013)$} &
\makecell{$0.0327$\\$(0.0010)$} &
\makecell{$0.0270$\\$(0.0042)$} &
\makecell{$0.0867$\\$(0.0102)$} &
\makecell{$0.0216$\\$(0.0032)$} &
\makecell{$\bm{0.0188}$\\$\bm{(0.0019)}$} &
\makecell{$0.0209$\\$(0.0060)$} &
\makecell{$0.0198$\\$(0.0038)$}\\
& FN &
\makecell{$0.07$\\$(0.26)$} &
\makecell{$1.71$\\$(0.50)$} &
\makecell{$1.89$\\$(0.31)$} &
\makecell{$2.00$\\$(0.00)$} &
\makecell{$\bm{0.00}$\\$\bm{(0.00)}$} &
\makecell{$0.09$\\$(0.29)$} &
\makecell{$\bm{0.00}$\\$\bm{(0.00)}$} &
\makecell{$2.22$\\$(0.50)$} &
\makecell{$2.55$\\$(0.74)$} &
\makecell{$3.10$\\$(0.87)$}\\
& FP &
\makecell{$\bm{0.00}$\\$\bm{(0.00)}$} &
\makecell{$\bm{0.00}$\\$\bm{(0.00)}$} &
\makecell{$\bm{0.00}$\\$\bm{(0.00)}$} &
\makecell{$\bm{0.00}$\\$\bm{(0.00)}$} &
\makecell{$\bm{0.00}$\\$\bm{(0.00)}$} &
\makecell{$\bm{0.00}$\\$\bm{(0.00)}$} &
\makecell{$\bm{0.00}$\\$\bm{(0.00)}$} &
\makecell{$\bm{0.00}$\\$\bm{(0.00)}$} &
\makecell{$\bm{0.00}$\\$\bm{(0.00)}$} &
\makecell{$\bm{0.00}$\\$\bm{(0.00)}$}\\
\midrule
\multirow{3}{*}{\makecell{Piston \\ ($7$-dim)}}
& SRMSE &
\makecell{$0.0724$\\$(0.0060)$} &
\makecell{$0.1541$\\$(0.0054)$} &
\makecell{$0.1536$\\$(0.0055)$} &
\makecell{$0.1621$\\$(0.0076)$} &
\makecell{$0.0611$\\$(0.0200)$} &
\makecell{$0.1122$\\$(0.0135)$} &
\makecell{$0.0522$\\$(0.0248)$} &
\makecell{$0.0438$\\$(0.0047)$} &
\makecell{$0.0436$\\$(0.0049)$} &
\makecell{$\bm{0.0435}$\\$\bm{(0.0046)}$}\\
& FN &
\makecell{$0.27$\\$(0.45)$} &
\makecell{$2.55$\\$(0.70)$} &
\makecell{$2.84$\\$(0.47)$} &
\makecell{$2.99$\\$(0.10)$} &
\makecell{$\bm{0.00}$\\$\bm{(0.00)}$} &
\makecell{$0.78$\\$(0.48)$} &
\makecell{$\bm{0.00}$\\$\bm{(0.00)}$} &
\makecell{$2.44$\\$(0.50)$} &
\makecell{$2.18$\\$(0.39)$} &
\makecell{$2.11$\\$(0.37)$}\\
& FP &
\makecell{$\bm{0.00}$\\$\bm{(0.00)}$} &
\makecell{$\bm{0.00}$\\$\bm{(0.00)}$} &
\makecell{$\bm{0.00}$\\$\bm{(0.00)}$} &
\makecell{$\bm{0.00}$\\$\bm{(0.00)}$} &
\makecell{$\bm{0.00}$\\$\bm{(0.00)}$} &
\makecell{$\bm{0.00}$\\$\bm{(0.00)}$} &
\makecell{$\bm{0.00}$\\$\bm{(0.00)}$} &
\makecell{$\bm{0.00}$\\$\bm{(0.00)}$} &
\makecell{$\bm{0.00}$\\$\bm{(0.00)}$} &
\makecell{$\bm{0.00}$\\$\bm{(0.00)}$}\\
\bottomrule
\end{tabular}
\begin{tablenotes}\footnotesize
\item \textit{Note:} Each cell reports the mean on the first line and the standard deviation in parentheses on the second line, computed across $100$ replicates. SRMSE denotes standardized root-mean-square error. FN and FP denote the numbers of false negatives and false positives in variable selection, respectively. Lower values are better. Bold indicates the best mean in each row; ties are all bold.
\end{tablenotes}
\end{threeparttable}
\end{table}
\spacingset{1.5} 

\spacingset{1}

\begin{table}[htbp]
\scriptsize
\setlength{\tabcolsep}{2.5pt}
\centering
\begin{threeparttable}
\caption{Prediction and variable-selection performance for the three benchmark test functions with $n=500$: Borehole, OTL Circuit, and Piston.}
\label{tab:rmse_fn_fp_combined_baseline_n500}
\begin{tabular}{cccccccccccc}
\toprule
\makecell{\textbf{Function}} & \makecell{\textbf{Metric}} &
\makecell{\textbf{mlegp}} &
\makecell{\textbf{Bayesian}\\\textbf{Bridge}} &
\makecell{\textbf{Horse-}\\\textbf{shoe}} &
\makecell{\textbf{Spike}\\\textbf{Slab}} &
\makecell{\textbf{DiceKriging}\\\textbf{quadratic}} &
\makecell{\textbf{Indicator-}\\\textbf{based GP}} &
\makecell{\textbf{hmc-}\\\textbf{B\textsuperscript{2}GPR}} &
\makecell{\textbf{B\textsuperscript{2}GPR}\\\textbf{$q=0.8$}} &
\makecell{\textbf{B\textsuperscript{2}GPR}\\\textbf{$q=1$}} &
\makecell{\textbf{B\textsuperscript{2}GPR}\\\textbf{$q=1.8$}}\\
\midrule
\multirow{3}{*}{\makecell{Borehole \\ ($8$-dim)}}
& SRMSE &
\makecell{$0.0544$\\$(0.0033)$} &
\makecell{$0.0361$\\$(0.0017)$} &
\makecell{$0.0360$\\$(0.0017)$} &
\makecell{$0.0367$\\$(0.0018)$} &
\makecell{$0.0177$\\$(0.0031)$} &
\makecell{$0.1068$\\$(0.0103)$} &
\makecell{$0.0151$\\$(0.0035)$} &
\makecell{$\bm{0.0121}$\\$\bm{(0.0011)}$} &
\makecell{$0.0124$\\$(0.0011)$} &
\makecell{$0.0125$\\$(0.0015)$}\\
& FN &
\makecell{$0.05$\\$(0.22)$} &
\makecell{$2.72$\\$(0.59)$} &
\makecell{$2.87$\\$(0.44)$} &
\makecell{$3.00$\\$(0.00)$} &
\makecell{$\bm{0.00}$\\$\bm{(0.00)}$} &
\makecell{$\bm{0.00}$\\$\bm{(0.00)}$} &
\makecell{$\bm{0.00}$\\$\bm{(0.00)}$} &
\makecell{$3.00$\\$(0.00)$} &
\makecell{$2.98$\\$(0.14)$} &
\makecell{$2.97$\\$(0.17)$}\\
& FP &
\makecell{$\bm{0.00}$\\$\bm{(0.00)}$} &
\makecell{$\bm{0.00}$\\$\bm{(0.00)}$} &
\makecell{$\bm{0.00}$\\$\bm{(0.00)}$} &
\makecell{$\bm{0.00}$\\$\bm{(0.00)}$} &
\makecell{$\bm{0.00}$\\$\bm{(0.00)}$} &
\makecell{$\bm{0.00}$\\$\bm{(0.00)}$} &
\makecell{$\bm{0.00}$\\$\bm{(0.00)}$} &
\makecell{$\bm{0.00}$\\$\bm{(0.00)}$} &
\makecell{$\bm{0.00}$\\$\bm{(0.00)}$} &
\makecell{$\bm{0.00}$\\$\bm{(0.00)}$}\\
\midrule
\multirow{3}{*}{\makecell{OTL Circuit \\ ($6$-dim)}}
& SRMSE &
\makecell{$0.0485$\\$(0.0022)$} &
\makecell{$0.0322$\\$(0.0010)$} &
\makecell{$0.0320$\\$(0.0010)$} &
\makecell{$0.0318$\\$(0.0009)$} &
\makecell{$0.0194$\\$(0.0036)$} &
\makecell{$0.0899$\\$(0.0060)$} &
\makecell{$0.0143$\\$(0.0018)$} &
\makecell{$0.0142$\\$(0.0016)$} &
\makecell{$0.0141$\\$(0.0019)$} &
\makecell{$\bm{0.0138}$\\$\bm{(0.0025)}$}\\
& FN &
\makecell{$\bm{0.00}$\\$\bm{(0.00)}$} &
\makecell{$1.80$\\$(0.40)$} &
\makecell{$1.95$\\$(0.22)$} &
\makecell{$2.00$\\$(0.00)$} &
\makecell{$\bm{0.00}$\\$\bm{(0.00)}$} &
\makecell{$0.01$\\$(0.10)$} &
\makecell{$\bm{0.00}$\\$\bm{(0.00)}$} &
\makecell{$2.05$\\$(0.39)$} &
\makecell{$2.11$\\$(0.35)$} &
\makecell{$2.50$\\$(0.73)$}\\
& FP &
\makecell{$\bm{0.00}$\\$\bm{(0.00)}$} &
\makecell{$\bm{0.00}$\\$\bm{(0.00)}$} &
\makecell{$\bm{0.00}$\\$\bm{(0.00)}$} &
\makecell{$\bm{0.00}$\\$\bm{(0.00)}$} &
\makecell{$\bm{0.00}$\\$\bm{(0.00)}$} &
\makecell{$\bm{0.00}$\\$\bm{(0.00)}$} &
\makecell{$\bm{0.00}$\\$\bm{(0.00)}$} &
\makecell{$\bm{0.00}$\\$\bm{(0.00)}$} &
\makecell{$\bm{0.00}$\\$\bm{(0.00)}$} &
\makecell{$\bm{0.00}$\\$\bm{(0.00)}$}\\
\midrule
\multirow{3}{*}{\makecell{Piston \\ ($7$-dim)}}
& SRMSE &
\makecell{$0.0601$\\$(0.0050)$} &
\makecell{$0.1469$\\$(0.0053)$} &
\makecell{$0.1467$\\$(0.0053)$} &
\makecell{$0.1489$\\$(0.0057)$} &
\makecell{$0.0339$\\$(0.0026)$} &
\makecell{$0.1184$\\$(0.0106)$} &
\makecell{$0.0313$\\$(0.0210)$} &
\makecell{$0.0272$\\$(0.0024)$} &
\makecell{$0.0272$\\$(0.0023)$} &
\makecell{$\bm{0.0271}$\\$\bm{(0.0024)}$}\\
& FN &
\makecell{$0.02$\\$(0.14)$} &
\makecell{$2.77$\\$(0.51)$} &
\makecell{$2.89$\\$(0.40)$} &
\makecell{$2.34$\\$(0.48)$} &
\makecell{$\bm{0.00}$\\$\bm{(0.00)}$} &
\makecell{$0.02$\\$(0.14)$} &
\makecell{$\bm{0.00}$\\$\bm{(0.00)}$} &
\makecell{$1.94$\\$(0.42)$} &
\makecell{$1.95$\\$(0.44)$} &
\makecell{$1.95$\\$(0.30)$}\\
& FP &
\makecell{$\bm{0.00}$\\$\bm{(0.00)}$} &
\makecell{$\bm{0.00}$\\$\bm{(0.00)}$} &
\makecell{$\bm{0.00}$\\$\bm{(0.00)}$} &
\makecell{$\bm{0.00}$\\$\bm{(0.00)}$} &
\makecell{$\bm{0.00}$\\$\bm{(0.00)}$} &
\makecell{$\bm{0.00}$\\$\bm{(0.00)}$} &
\makecell{$\bm{0.00}$\\$\bm{(0.00)}$} &
\makecell{$\bm{0.00}$\\$\bm{(0.00)}$} &
\makecell{$\bm{0.00}$\\$\bm{(0.00)}$} &
\makecell{$\bm{0.00}$\\$\bm{(0.00)}$}\\
\bottomrule
\end{tabular}
\begin{tablenotes}\footnotesize
\item \textit{Note:} Each cell reports the mean on the first line and the standard deviation in parentheses on the second line. SRMSE denotes standardized root-mean-square error. FN and FP denote the numbers of false negatives and false positives in variable selection, respectively. Lower values are better. Bold indicates the best mean in each row; ties are all bold. Results are computed across $100$ replicates, except for the indicator-based GP method for Piston, which is based on $99$ successful runs.
\end{tablenotes}
\end{threeparttable}
\end{table}

\spacingset{1.5}

\spacingset{1} 
\begin{table}[htbp]
\footnotesize
\setlength{\tabcolsep}{4pt}
\centering
\begin{threeparttable}
\caption{Standardized RMSE with increased input dimensions $d=20$ for the benchmark test functions.}
\label{tab:rmse_combined_hd}
\begin{tabular}{clcccccccc}
\toprule
\makecell{\textbf{Function}} & \makecell{\textbf{Size}} &
\makecell{\textbf{mlegp}} & \makecell{\textbf{laGP}} &
\makecell{\textbf{krig-}\\\textbf{const}} &
\makecell{\textbf{krig-}\\\textbf{linear}} &
\makecell{\textbf{hmc-}\\\textbf{B\textsuperscript{2}GPR}} &
\makecell{\textbf{sph-}\\\textbf{B\textsuperscript{2}GPR-$0.8$}} &
\makecell{\textbf{sph-}\\\textbf{B\textsuperscript{2}GPR-$1$}} &
\makecell{\textbf{sph-}\\\textbf{B\textsuperscript{2}GPR-$1.8$}}\\
\midrule
\multirow{2}{*}{\makecell{Borehole \\ ($20$-dim)}}
& $n=200$ &
\makecell{$0.0629$\\$(0.0107)$} &
\makecell{$0.3869$\\$(0.6684)$} &
\makecell{$0.0994$\\$(0.0086)$} &
\makecell{$0.0577$\\$(0.0045)$} &
\makecell{$0.0665$\\$(0.0855)$} &
\makecell{$\bm{0.0179}$\\$\bm{(0.0018)}$} &
\makecell{$0.0181$\\$(0.0019)$} &
\makecell{$0.0185$\\$(0.0021)$}\\
& $n=500$ &
\makecell{$0.0540$\\$(0.0153)$} &
\makecell{$0.0666$\\$(0.0037)$} &
\makecell{$0.0664$\\$(0.0037)$} &
\makecell{$0.0385$\\$(0.0024)$} &
\makecell{$0.0520$\\$(0.0823)$} &
\makecell{$\bm{0.0121}$\\$\bm{(0.0010)}$} &
\makecell{$0.0122$\\$(0.0010)$} &
\makecell{$0.0123$\\$(0.0010)$}\\
\midrule
\multirow{2}{*}{\makecell{OTL \\ ($20$-dim)}}
& $n=200$ &
\makecell{$0.0631$\\$(0.0116)$} &
\makecell{$0.1571$\\$(0.0122)$} &
\makecell{$0.1094$\\$(0.0069)$} &
\makecell{$0.0494$\\$(0.0033)$} &
\makecell{$0.0635$\\$(0.0615)$} &
\makecell{$\bm{0.0189}$\\$\bm{(0.0015)}$} &
\makecell{$0.0190$\\$(0.0015)$} &
\makecell{$0.0201$\\$(0.0041)$}\\
& $n=500$ &
\makecell{$0.0481$\\$(0.0082)$} &
\makecell{$0.0866$\\$(0.0065)$} &
\makecell{$0.0722$\\$(0.0036)$} &
\makecell{$0.0334$\\$(0.0018)$} &
\makecell{$0.0358$\\$(0.0512)$} &
\makecell{$\bm{0.0121}$\\$\bm{(0.0011)}$} &
\makecell{$0.0122$\\$(0.0011)$} &
\makecell{$0.0126$\\$(0.0011)$}\\
\midrule
\multirow{2}{*}{\makecell{Piston \\ ($20$-dim)}}
& $n=200$ &
\makecell{$0.0827$\\$(0.0188)$} &
\makecell{$0.2179$\\$(0.0316)$} &
\makecell{$0.1369$\\$(0.0109)$} &
\makecell{$0.0949$\\$(0.0089)$} &
\makecell{$0.1395$\\$(0.1251)$} &
\makecell{$\bm{0.0438}$\\$\bm{(0.0041)}$} &
\makecell{$0.0445$\\$(0.0067)$} &
\makecell{$0.0451$\\$(0.0046)$}\\
& $n=500$ &
\makecell{$0.0608$\\$(0.0141)$} &
\makecell{$0.1147$\\$(0.0073)$} &
\makecell{$0.0936$\\$(0.0055)$} &
\makecell{$0.0645$\\$(0.0045)$} &
\makecell{$0.0761$\\$(0.1024)$} &
\makecell{$\bm{0.0276}$\\$\bm{(0.0027)}$} &
\makecell{$\bm{0.0276}$\\$\bm{(0.0026)}$} &
\makecell{$0.0278$\\$(0.0026)$}\\
\bottomrule
\end{tabular}
\begin{tablenotes}\footnotesize
\item \textit{Note:} Each cell reports the mean (first line) and standard deviation (second line, in parentheses) of standardized RMSE across $100$ replicates; lower is better. Bold indicates the best (lowest) mean in each row. All ties are in bold.
\end{tablenotes}
\end{threeparttable}
\end{table}
\spacingset{1.5}

\spacingset{1} 
\begin{table}[htbp]
\scriptsize
\setlength{\tabcolsep}{2.5pt}
\centering
\begin{threeparttable}
\caption{Prediction and variable-selection performance for the three benchmark test functions with increased input dimension $d=20$ and $n=200$.}
\label{tab:rmse_fn_fp_combined_hd_n200}
\begin{tabular}{cccccccccccc}
\toprule
\makecell{\textbf{Function}} & \makecell{\textbf{Metric}} &
\makecell{\textbf{mlegp}} &
\makecell{\textbf{Bayesian}\\\textbf{Bridge}} &
\makecell{\textbf{Horse-}\\\textbf{shoe}} &
\makecell{\textbf{Spike}\\\textbf{Slab}} &
\makecell{\textbf{DiceKriging}\\\textbf{linear}} &
\makecell{\textbf{Indicator-}\\\textbf{based GP}} &
\makecell{\textbf{hmc-}\\\textbf{B\textsuperscript{2}GPR}} &
\makecell{\textbf{B\textsuperscript{2}GPR}\\\textbf{$q=0.8$}} &
\makecell{\textbf{B\textsuperscript{2}GPR}\\\textbf{$q=1$}} &
\makecell{\textbf{B\textsuperscript{2}GPR}\\\textbf{$q=1.8$}}\\
\midrule
\multirow{3}{*}{\makecell{Borehole \\ ($20$-dim)}}
& SRMSE &
\makecell{$0.0620$\\$(0.0132)$} &
\makecell{$0.2361$\\$(0.0069)$} &
\makecell{$0.2347$\\$(0.0068)$} &
\makecell{$0.2322$\\$(0.0066)$} &
\makecell{$0.0579$\\$(0.0048)$} &
\makecell{$0.1371$\\$(0.0133)$} &
\makecell{$0.0665$\\$(0.0855)$} &
\makecell{$\bm{0.0179}$\\$\bm{(0.0018)}$} &
\makecell{$0.0181$\\$(0.0019)$} &
\makecell{$0.0185$\\$(0.0021)$}\\
& FN &
\makecell{$1.54$\\$(0.88)$} &
\makecell{$3.00$\\$(0.00)$} &
\makecell{$3.00$\\$(0.00)$} &
\makecell{$3.11$\\$(0.31)$} &
\makecell{$\bm{0.00}$\\$\bm{(0.00)}$} &
\makecell{$1.07$\\$(0.67)$} &
\makecell{$\bm{0.00}$\\$\bm{(0.00)}$} &
\makecell{$3.00$\\$(0.00)$} &
\makecell{$3.00$\\$(0.00)$} &
\makecell{$3.00$\\$(0.00)$}\\
& FP &
\makecell{$7.23$\\$(2.82)$} &
\makecell{$0.25$\\$(0.50)$} &
\makecell{$0.08$\\$(0.31)$} &
\makecell{$\bm{0.00}$\\$\bm{(0.00)}$} &
\makecell{$12.00$\\$(0.00)$} &
\makecell{$10.59$\\$(0.99)$} &
\makecell{$12.00$\\$(0.00)$} &
\makecell{$\bm{0.00}$\\$\bm{(0.00)}$} &
\makecell{$\bm{0.00}$\\$\bm{(0.00)}$} &
\makecell{$\bm{0.00}$\\$\bm{(0.00)}$}\\
\midrule
\multirow{3}{*}{\makecell{OTL Circuit \\ ($20$-dim)}}
& SRMSE &
\makecell{$0.0647$\\$(0.0139)$} &
\makecell{$0.1699$\\$(0.0041)$} &
\makecell{$0.1678$\\$(0.0038)$} &
\makecell{$0.1659$\\$(0.0036)$} &
\makecell{$0.0490$\\$(0.0051)$} &
\makecell{$0.0665$\\$(0.0057)$} &
\makecell{$0.0635$\\$(0.0615)$} &
\makecell{$\bm{0.0189}$\\$\bm{(0.0015)}$} &
\makecell{$0.0190$\\$(0.0015)$} &
\makecell{$0.0201$\\$(0.0041)$}\\
& FN &
\makecell{$0.43$\\$(0.67)$} &
\makecell{$2.00$\\$(0.00)$} &
\makecell{$2.00$\\$(0.00)$} &
\makecell{$2.00$\\$(0.00)$} &
\makecell{$\bm{0.00}$\\$\bm{(0.00)}$} &
\makecell{$1.43$\\$(0.66)$} &
\makecell{$\bm{0.00}$\\$\bm{(0.00)}$} &
\makecell{$2.00$\\$(0.00)$} &
\makecell{$2.00$\\$(0.00)$} &
\makecell{$2.00$\\$(0.00)$}\\
& FP &
\makecell{$11.50$\\$(2.78)$} &
\makecell{$0.35$\\$(0.59)$} &
\makecell{$0.09$\\$(0.32)$} &
\makecell{$\bm{0.00}$\\$\bm{(0.00)}$} &
\makecell{$14.00$\\$(0.00)$} &
\makecell{$5.15$\\$(0.89)$} &
\makecell{$14.00$\\$(0.00)$} &
\makecell{$\bm{0.00}$\\$\bm{(0.00)}$} &
\makecell{$\bm{0.00}$\\$\bm{(0.00)}$} &
\makecell{$0.02$\\$(0.20)$}\\
\midrule
\multirow{3}{*}{\makecell{Piston \\ ($20$-dim)}}
& SRMSE &
\makecell{$0.0880$\\$(0.0271)$} &
\makecell{$0.3106$\\$(0.0081)$} &
\makecell{$0.3071$\\$(0.0078)$} &
\makecell{$0.3045$\\$(0.0074)$} &
\makecell{$0.0975$\\$(0.0217)$} &
\makecell{$0.1221$\\$(0.0135)$} &
\makecell{$0.1395$\\$(0.1251)$} &
\makecell{$\bm{0.0438}$\\$\bm{(0.0041)}$} &
\makecell{$0.0445$\\$(0.0067)$} &
\makecell{$0.0451$\\$(0.0046)$}\\
& FN &
\makecell{$1.40$\\$(0.67)$} &
\makecell{$2.86$\\$(0.35)$} &
\makecell{$2.97$\\$(0.17)$} &
\makecell{$3.00$\\$(0.00)$} &
\makecell{$\bm{0.00}$\\$\bm{(0.00)}$} &
\makecell{$1.77$\\$(0.53)$} &
\makecell{$\bm{0.00}$\\$\bm{(0.00)}$} &
\makecell{$2.17$\\$(0.38)$} &
\makecell{$2.03$\\$(0.17)$} &
\makecell{$2.00$\\$(0.00)$}\\
& FP &
\makecell{$5.35$\\$(2.69)$} &
\makecell{$0.52$\\$(0.67)$} &
\makecell{$0.15$\\$(0.39)$} &
\makecell{$\bm{0.00}$\\$\bm{(0.00)}$} &
\makecell{$13.00$\\$(0.00)$} &
\makecell{$1.70$\\$(0.70)$} &
\makecell{$13.00$\\$(0.00)$} &
\makecell{$0.01$\\$(0.10)$} &
\makecell{$0.05$\\$(0.33)$} &
\makecell{$0.03$\\$(0.17)$}\\
\bottomrule
\end{tabular}
\begin{tablenotes}\footnotesize
\item \textit{Note:} Each cell reports the mean on the first line and the standard deviation in parentheses on the second line. SRMSE denotes standardized root-mean-square error. FN and FP denote the numbers of false negatives and false positives in variable selection, respectively. Lower values are better. Bold indicates the best mean in each row; ties are all bold. Results are computed across $100$ replicates, except for the indicator-based GP method for Borehole, which is based on $69$ successful runs.
\end{tablenotes}
\end{threeparttable}
\end{table}
\spacingset{1.5}

\spacingset{1}

\begin{table}[htbp]
\scriptsize
\setlength{\tabcolsep}{2.5pt}
\centering
\begin{threeparttable}
\caption{Prediction and variable-selection performance for the three benchmark test functions with increased input dimension $d=20$ and $n=500$.}
\label{tab:rmse_fn_fp_combined_hd_n500}
\begin{tabular}{ccccccccccc}
\toprule
\makecell{\textbf{Function}} & \makecell{\textbf{Metric}} &
\makecell{\textbf{mlegp}} &
\makecell{\textbf{Bayesian}\\\textbf{Bridge}} &
\makecell{\textbf{Horse-}\\\textbf{shoe}} &
\makecell{\textbf{Spike}\\\textbf{Slab}} &
\makecell{\textbf{DiceKriging}\\\textbf{linear}} &
\makecell{\textbf{hmc-}\\\textbf{B\textsuperscript{2}GPR}} &
\makecell{\textbf{B\textsuperscript{2}GPR}\\\textbf{$q=0.8$}} &
\makecell{\textbf{B\textsuperscript{2}GPR}\\\textbf{$q=1$}} &
\makecell{\textbf{B\textsuperscript{2}GPR}\\\textbf{$q=1.8$}}\\
\midrule
\multirow{3}{*}{\makecell{Borehole \\ ($20$-dim)}}
& SRMSE &
\makecell{$0.0585$\\$(0.0412)$} &
\makecell{$0.2332$\\$(0.0057)$} &
\makecell{$0.2325$\\$(0.0056)$} &
\makecell{$0.2312$\\$(0.0055)$} &
\makecell{$0.0389$\\$(0.0024)$} &
\makecell{$0.0520$\\$(0.0823)$} &
\makecell{$\bm{0.0121}$\\$\bm{(0.0010)}$} &
\makecell{$0.0122$\\$(0.0010)$} &
\makecell{$0.0123$\\$(0.0010)$}\\
& FN &
\makecell{$0.87$\\$(1.35)$} &
\makecell{$3.00$\\$(0.00)$} &
\makecell{$3.00$\\$(0.00)$} &
\makecell{$3.00$\\$(0.00)$} &
\makecell{$\bm{0.00}$\\$\bm{(0.00)}$} &
\makecell{$\bm{0.00}$\\$\bm{(0.00)}$} &
\makecell{$3.00$\\$(0.00)$} &
\makecell{$2.99$\\$(0.10)$} &
\makecell{$2.99$\\$(0.10)$}\\
& FP &
\makecell{$9.55$\\$(2.38)$} &
\makecell{$0.39$\\$(0.62)$} &
\makecell{$0.16$\\$(0.37)$} &
\makecell{$\bm{0.00}$\\$\bm{(0.00)}$} &
\makecell{$12.00$\\$(0.00)$} &
\makecell{$12.00$\\$(0.00)$} &
\makecell{$\bm{0.00}$\\$\bm{(0.00)}$} &
\makecell{$\bm{0.00}$\\$\bm{(0.00)}$} &
\makecell{$\bm{0.00}$\\$\bm{(0.00)}$}\\
\midrule
\multirow{3}{*}{\makecell{OTL Circuit \\ ($20$-dim)}}
& SRMSE &
\makecell{$0.0498$\\$(0.0103)$} &
\makecell{$0.1672$\\$(0.0038)$} &
\makecell{$0.1663$\\$(0.0038)$} &
\makecell{$0.1657$\\$(0.0038)$} &
\makecell{$0.0331$\\$(0.0017)$} &
\makecell{$0.0358$\\$(0.0512)$} &
\makecell{$\bm{0.0121}$\\$\bm{(0.0011)}$} &
\makecell{$0.0122$\\$(0.0011)$} &
\makecell{$0.0126$\\$(0.0011)$}\\
& FN &
\makecell{$0.30$\\$(0.50)$} &
\makecell{$2.00$\\$(0.00)$} &
\makecell{$2.00$\\$(0.00)$} &
\makecell{$2.00$\\$(0.00)$} &
\makecell{$\bm{0.00}$\\$\bm{(0.00)}$} &
\makecell{$\bm{0.00}$\\$\bm{(0.00)}$} &
\makecell{$2.00$\\$(0.00)$} &
\makecell{$2.00$\\$(0.00)$} &
\makecell{$2.00$\\$(0.00)$}\\
& FP &
\makecell{$12.68$\\$(1.81)$} &
\makecell{$0.50$\\$(0.66)$} &
\makecell{$0.08$\\$(0.27)$} &
\makecell{$\bm{0.00}$\\$\bm{(0.00)}$} &
\makecell{$14.00$\\$(0.00)$} &
\makecell{$14.00$\\$(0.00)$} &
\makecell{$\bm{0.00}$\\$\bm{(0.00)}$} &
\makecell{$\bm{0.00}$\\$\bm{(0.00)}$} &
\makecell{$\bm{0.00}$\\$\bm{(0.00)}$}\\
\midrule
\multirow{3}{*}{\makecell{Piston \\ ($20$-dim)}}
& SRMSE &
\makecell{$0.0628$\\$(0.0261)$} &
\makecell{$0.3034$\\$(0.0077)$} &
\makecell{$0.3020$\\$(0.0077)$} &
\makecell{$0.3021$\\$(0.0077)$} &
\makecell{$0.0647$\\$(0.0045)$} &
\makecell{$0.0761$\\$(0.1024)$} &
\makecell{$\bm{0.0276}$\\$\bm{(0.0027)}$} &
\makecell{$0.0276$\\$(0.0026)$} &
\makecell{$0.0278$\\$(0.0026)$}\\
& FN &
\makecell{$0.71$\\$(0.69)$} &
\makecell{$2.40$\\$(0.49)$} &
\makecell{$2.74$\\$(0.44)$} &
\makecell{$3.00$\\$(0.00)$} &
\makecell{$\bm{0.00}$\\$\bm{(0.00)}$} &
\makecell{$\bm{0.00}$\\$\bm{(0.00)}$} &
\makecell{$1.75$\\$(0.46)$} &
\makecell{$1.80$\\$(0.40)$} &
\makecell{$1.75$\\$(0.44)$}\\
& FP &
\makecell{$8.07$\\$(2.31)$} &
\makecell{$0.44$\\$(0.67)$} &
\makecell{$0.12$\\$(0.33)$} &
\makecell{$\bm{0.00}$\\$\bm{(0.00)}$} &
\makecell{$13.00$\\$(0.00)$} &
\makecell{$13.00$\\$(0.00)$} &
\makecell{$\bm{0.00}$\\$\bm{(0.00)}$} &
\makecell{$0.01$\\$(0.10)$} &
\makecell{$\bm{0.00}$\\$\bm{(0.00)}$}\\
\bottomrule
\end{tabular}
\begin{tablenotes}\footnotesize
\item \textit{Note:} Each cell reports the mean on the first line and the standard deviation in parentheses on the second line. SRMSE denotes standardized root-mean-square error. FN and FP denote the numbers of false negatives and false positives in variable selection, respectively. Lower values are better. Bold indicates the best mean in each row; ties are all bold. Results are computed across $100$ replicates, except for the mlegp SRMSE result for Borehole, which is based on $98$ successful runs.
\end{tablenotes}
\end{threeparttable}
\end{table}

\spacingset{1.5}

\section{Case Study: H\textsubscript{2}O Potential Energy}\label{sec:real}

We demonstrate the proposed B\textsuperscript{2}GPR method on a real molecular data set.
Following the ``H$_2$O Potential Energy: A Hands-On Example'' in \cite{deringer2021gaussian}, we consider the problem of learning the potential energy surface for a water molecule using Gaussian process regression.
The dataset comprises $121$ configurations arranged on an $11 \times 11$ grid, uniquely determined by two O-H bond lengths and the H-O-H angle.

Following the preprocessing pipeline in \cite{deringer2021gaussian}, we construct descriptors using the Smooth Overlap of Atomic Positions (SOAP) method, incorporating O-H bond lengths, H-O-H angle, molecular dipole vector, and the so-called atomic lines (element, position, and force for each atom).
The SOAP hyperparameters are set as in \cite{deringer2021gaussian}.
This results in an initial feature matrix with $952$ channels per configuration.
Because this is a small three-atom system sampled on a structured grid, several SOAP channels are identical across configurations; we remove zero-variance columns, leaving $316$ features.

We then split the data set into a training set of size $n=100$ and a test set of size $21$.
However, with $316$ features and only $100$ training samples, this remains a challenging problem, and directly applying any regression model is likely to lead to severe overfitting and numerical instability.
To address this, we perform principal component analysis (PCA) on the feature matrix for the training set only and retain the first $50$ principal components as the input variables.
Test configurations are projected using the same loadings.
This yields a more manageable $50$-dimensional input space for regression.
We evaluate out-of-sample prediction accuracy on the $21$ test configurations and repeat the random train-test split $100$ times to assess variability.

Although PCA substantially reduces the original feature dimension, the resulting input dimension is still large relative to the training sample size, so we use a constant fixed-effect basis for this case study.
Because the true active molecular descriptors are not known in this real-data application, the comparison focuses on SRMSE for test configurations rather than FN/FP variable-selection errors.
Table \ref{tab:rmse_water_pca} reports the mean and standard deviation of SRMSE over $100$ replicates for the PCA-feature setting.
It shows that methods without an effective sparsity or dimension-reduction mechanism  perform poorly in this high-dimensional PCA-feature setting: \texttt{mlegp} and \texttt{DiceKriging} have large SRMSEs, and hmc-B\textsuperscript{2}GPR also has larger error and variability.
The sparse Bayesian regression competitors perform better, with Bayesian Bridge and Horseshoe achieving small errors and Spike-and-Slab improving substantially over the non-sparse or less strongly regularized GP competitors.
Nevertheless, the proposed B\textsuperscript{2}GPR variants provide the best overall prediction performance, with mean SRMSEs around $2\times 10^{-4}$ to $3\times 10^{-4}$ and small variability across replicates.
These results indicate that the spherical $\ell_q$-constrained B\textsuperscript{2}GPR posterior is especially effective for stabilizing prediction in this small-sample, high-dimensional PCA-feature setting.
Overall, the case study demonstrates the practical utility of B\textsuperscript{2}GPR for learning complex functions from limited data in high-dimensional settings, which is common in molecular modeling and other scientific applications.

\iffalse
\spacingset{1} 
\begin{table}[tbp]
\footnotesize
\centering
\caption{Standardized RMSE ($100$ replicates for $n=100$) for the water-molecule dataset}
\label{tab:rmse_water}
\begin{tabular}{cccccccc}
\toprule
 & \makecell{\textbf{mlegp}}
 & \makecell{\textbf{laGP}}
 & \makecell{\textbf{krig-}\\\textbf{const}}
 & \makecell{\textbf{hmc-}\\\textbf{B\textsuperscript{2}GPR}}
 & \makecell{\textbf{sph-}\\\textbf{B\textsuperscript{2}GPR-$0.8$}}
 & \makecell{\textbf{sph-}\\\textbf{B\textsuperscript{2}GPR-$1$}}
 & \makecell{\textbf{sph-}\\\textbf{B\textsuperscript{2}GPR-$1.8$}}\\
\midrule
Mean & $0.9780$ & $0.0342$ & $0.3537$ & $0.1877$ & \bm{$0.0002$} & \bm{$0.0003$} & \bm{$0.0002$}\\
\midrule
SD & $0.0508$ & $0.0827$ & $0.1551$ & $0.3921$ & \bm{$0.0003$} & \bm{$0.0014$} & \bm{$0.0005$}\\
\bottomrule
\end{tabular}
\end{table}
\spacingset{1.5} 
\spacingset{1} 
\fi

\begin{table}[tbp]
\scriptsize
\setlength{\tabcolsep}{3.2pt}
\centering
\begin{threeparttable}
\caption{Standardized RMSE for the water-molecule dataset with PCA features.}
\label{tab:rmse_water_pca}
\begin{tabular}{cccccccccc}
\toprule
& \makecell{\textbf{mlegp}\\\textbf{const}}
& \makecell{\textbf{Bayesian}\\\textbf{Bridge}}
& \makecell{\textbf{Horse-}\\\textbf{shoe}}
& \makecell{\textbf{Spike}\\\textbf{Slab}}
& \makecell{\textbf{DiceKriging}\\\textbf{const}} 
& \makecell{\textbf{hmc-}\\\textbf{B\textsuperscript{2}GPR}}
& \makecell{\textbf{B\textsuperscript{2}GPR}\\\textbf{$q=0.8$}} 
& \makecell{\textbf{B\textsuperscript{2}GPR}\\\textbf{$q=1$}} 
& \makecell{\textbf{B\textsuperscript{2}GPR}\\\textbf{$q=1.8$}} \\
\midrule
Mean 
& $0.9739$
& $0.0007$
& $\bm{0.0006}$
& $0.0069$
& $0.3682$
& $0.1877$
& $\bm{0.0002}$
& $\bm{0.0003}$
& $\bm{0.0002}$\\
\midrule
SD 
& $0.0591$
& $0.0008$
& $\bm{0.0007}$
& $0.0021$
& $0.1569$
& $0.3921$
& $\bm{0.0003}$
& $0.0014$
& $\bm{0.0005}$\\
\bottomrule
\end{tabular}
\begin{tablenotes}\footnotesize
\item \textit{Note:} Each cell reports the mean standardized RMSE and its standard deviation across $100$ replicates. Lower values indicate better prediction performance. Bold indicates the best-performing methods with standardized RMSE close to zero.
\end{tablenotes}
\end{threeparttable}
\end{table}
\spacingset{1.5}

\section{Conclusion}\label{sec:end}

In this paper, we proposed Bayesian Bridge Gaussian Process Regression (B\textsuperscript{2}GPR), a Gaussian process regression framework that regularizes both the linear fixed-effect coefficients $\bm \beta$ and correlation parameters $\bm \omega$ via $\ell_q$-norm constraints.
We suggested two different versions of B\textsuperscript{2}GPR, depending on the regularization norm $q$. 
When $0<q<2$, we use non-informative bounded priors for $\bm \beta$ and $\bm \omega$, and the modes of the conditional posterior distributions of $\bm \beta$ and $\bm \omega$ are equivalent to the solutions of MLE with $\ell_q$-norm regularization. 
To sample efficiently from the resulting constrained posterior, we adopted the Spherical Hamiltonian Monte Carlo scheme that enforces the constraints exactly by mapping the parameter space to a sphere and running Hamiltonian dynamics on that manifold.
This preserves the advantages of standard HMC (geometry-aware proposals and effective global moves) while providing direct control over sparsity and regularization strength. 
This version is named sph-B\textsuperscript{2}GPR for short. 
When $q=2$, we use the informative Gaussian priors for $\bm \beta$ and $\bm \omega$, and the posterior modes are equivalent to the solution of MLE with $\ell_2$-norm regularization. 
Regular sampling methods can be used here. 
We choose the standard HMC for $\bm \omega$ and the Metropolis-Hastings sampler for other parameters if their conditional posteriors are not any known distributions. 
Compared with other GP regression methods, the proposed B\textsuperscript{2}GPR has the same computation complexity of $O(n^3)$ for evaluating the GP likelihood in each iteration. 
For larger-scale problems, further computational improvements, such as sparse GP approximations, low-rank covariance approximations, would be needed. 
When the input dimension is high, our simulation and example show that choosing $q$ between 1 and 2 is likely to lead to the most accurate prediction and variable selection.

Through the controlled simulations (a pre-specified GP), three standard benchmarks (Borehole, OTL circuit, and Piston functions), and a real molecular dataset (the water potential energy surface), the proposed B\textsuperscript{2}GPR, particularly sph-B\textsuperscript{2}GPR,  consistently delivers 
(i) \emph{high predictive accuracy}, achieving the lowest or near-lowest prediction error, often by a clear margin over competing approaches;
(ii) \emph{greater robustness}, with much smaller variability across replicates; and
(iii) \emph{enhanced interpretability}, with posterior credible intervals for $\bm\beta$ and $\bm\omega$ that clearly recover known active inputs in the benchmark problems.
These advantages are persistent in high-dimensional settings with many irrelevant inputs, where classical methods degrade noticeably and hierarchical-prior HMC exhibits high variability.
The proposed sph-B\textsuperscript{2}GPR sampler therefore offers a practical way to simultaneously perform variable selection and inference in GP regression.

\section*{Supplementary materials}
The supplement includes the in-depth reviews of HMC and SphHMC, definitions of the three benchmark functions, figures for the estimated parameter values for all simulation settings for the three benchmark examples, and the derivations for Proposition \ref{prop:sph} and \ref{prop:hmc}. 

\section*{Data Availability Statement}
The case study is based on published data, which are cited and included in the codes and data.
All the existing methods for comparison are published \texttt{R} packages and are cited.  
Codes and data are available for review from this GitHub link \url{https://github.com/tonyxms/B-2GPR}. 

\section*{Author Biographies}

{\bf Minshen Xu} is a Ph.D. candidate in the Department of Mathematics and Statistics at the University of Massachusetts
Amherst. His research area is in statistical design of experiments and uncertainty quantification. Minshen is expected to graduate and obtain a Ph.D. in Statistics degree in the summer of 2027. 

{\bf Shiwei Lan} is an Associate Professor in the School of Mathematical and Statistical Sciences at Arizona State University. He obtained his Ph.D. in the Department of Statistics, University of California, Irvine. His research focuses on Bayesian statistics, uncertainty quantification and machine learning.

{\bf Lulu Kang} is a Professor in the Department of Mathematics and Statistics at the University of Massachusetts
Amherst. Her research interests include statistical learning/machine learning, statistical design of experiments, uncertainty quantification, and Bayesian statistics. 
She serves as an Associate Editor for Technometrics and SIAM/ASA Journal on Uncertainty Quantification. 
Her website is \url{https://sites.google.com/umass.edu/lulukang/home}.

\bibliographystyle{ECA_jasa}
\bibliography{Ref}

\spacingset{1} 
\newpage 
\phantomsection\label{supplementary-material}
\bigskip

\setcounter{page}{1}
\setcounter{figure}{0}
\setcounter{table}{0}
\setcounter{lemma}{0}
\setcounter{theorem}{0}
\setcounter{proposition}{0}
\setcounter{algorithm}{0}

\makeatletter 
\renewcommand{\thefigure}{S\@arabic\c@figure}
\renewcommand{\thetable}{S\@arabic\c@table}
\renewcommand{\theproposition}{S\@arabic\c@proposition}
\renewcommand{\thetheorem}{S\@arabic\c@theorem}
\renewcommand{\thealgorithm}{S\@arabic\c@algorithm}
\makeatother

\begin{center}

{\Large\bf SUPPLEMENTARY MATERIAL}

\end{center}

\section*{Review of Hamiltonian Monte Carlo and Spherical Hamiltonian Monte Carlo }
Hamiltonian Monte Carlo (HMC) \citep{neal2011mcmc} is an MCMC method that uses Hamiltonian dynamics to efficiently sample from complex target distributions. 
By leveraging gradient information, HMC reduces random-walk behavior and yields lower autocorrelation than standard Metropolis-Hastings algorithms, enabling faster convergence.

In HMC, the sampling state consists of a position vector $\bm{\theta}$ (the parameters of interest) and a momentum vector $\bm{\phi}$ (auxiliary variables). 
The total energy of the system is given by the Hamiltonian:
\begin{equation*}
    H(\bm \theta,\bm \phi) = U(\bm \theta) + K(\bm \phi),
\end{equation*}
where $U(\bm{\theta}) = -\log p(\bm{\theta})$ is the potential energy (negative log-posterior) and $K(\bm{\phi}) = \frac{1}{2} \bm{\phi}^\top \bm{M}^{-1} \bm{\phi}$ is the kinetic energy. 
The momentum is typically distributed as $\bm{\phi} \sim \mathcal{N}(\mathbf{0}, \bm{M})$,  where $\bm M$ is a symmetric, positive-definite matrix known as the mass matrix. 
Often, it is chosen for simplicity as the identity matrix $\bm M = \bm I$.

The system evolves via Hamilton’s equations:
\begin{equation*}
    \begin{aligned}
        \frac{\dd\bm \theta}{\dd t} &= \frac{\partial H(\bm \theta, \bm \phi)}{\partial \bm \phi}= \nabla_{\bm \phi} K(\bm \phi)= M^{-1} \bm \phi\\
        \frac{\dd \bm \phi}{\dd t} &= -\frac{\partial H(\bm \theta, \bm \phi)}{\partial \bm \theta} = -\nabla_{\bm \theta} U(\bm \theta)
    \end{aligned}
\end{equation*}
In practice, these equations are discretized using the leapfrog integrator. 
The resulting HMC algorithm (summarized in Algorithm \ref{alg:HMC}) proposes new states by simulating Hamiltonian dynamics, which are then accepted or rejected via a Metropolis–Hastings step to correct for integration error. 
This approach enables efficient exploration of high-dimensional and correlated posterior distributions.

\begin{algorithm}[htb]
\caption{Hamiltonian Monte Carlo}\label{alg:HMC}
\begin{algorithmic}[1]
\State Given the step size $\epsilon$ and current $\bm \theta_{t-1}$, initialize $\bm \theta^{(0)}$ as $\bm \theta_{t-1}$ and randomly sample $\bm \phi^{(0)} \sim \mathcal{N}({\bf 0}, \bm M)$.
\For {$l=0$ to $L-1$} 
\State Update $\bm \phi$ by a half-step of $\epsilon$: $\bm \phi^{(l+\frac{1}{2})} = \bm \phi^{(l)} - \frac{\epsilon}{2} \nabla_{\bm \theta} U(\bm \theta^{(l)})$.
\State Update $\bm \theta$ by a full-step of $\epsilon$: $\bm \theta^{(l+1)} = \bm \theta^{(l)} + \epsilon \bm M^{-1} \bm \phi^{(l+\frac{1}{2})}$.
\State Update $\bm \phi$ by another half-step of $\epsilon$: $\bm \phi^{(l+1)} = \bm \phi^{(l+\frac{1}{2})} - \frac{\epsilon}{2} \nabla_{\bm \theta} U(\bm \theta^{(l+1)})$.
\EndFor
\State Compute the acceptance probability $\alpha = \min\left(1, \frac{p(\bm \theta^{(L)})N(\bm \phi^{(L)} | {\bf 0},\bm M)}{p(\bm \theta_{t-1})N(\bm \phi_{t-1} |{\bf 0},{\bm M})}\right)$.
\State Accept the new state $\bm \theta_t = \bm \theta^{(L)}$ with probability $\alpha$, otherwise $\bm \theta_t = \bm \theta_{t-1}$.
\end{algorithmic}
\end{algorithm}

While HMC efficiently explores complex posterior distributions, it does not inherently handle the geometric constraints often present in statistical models. 
To address this, \cite{lan2014spherical} introduced the Spherical Hamiltonian Monte Carlo (SphHMC) as a general framework for sampling from probability distributions confined to domains bounded by arbitrary norm constraints.

The core idea of SphHMC is to use a \emph{spherical augmentation} technique. 
We denote domain defined by the $\ell_q$-norm constraint on the parameter $\bm \beta\in \mathbb{R}^d$ as $\mathcal{Q}^d:=\{\bm \beta \in \mathbb{R}^d| \|\bm \beta\|_q \leq 1\}$, without loss of generality. 
It can be bijectively mapped to the $d$-dimensional unit ball, $\mathcal{B}_0^{d}(1)=\{\bm \theta\in \mathbb{R}^{d}|\|\bm \theta\|_2\leq 1\}$, by either $\beta_i \mapsto \theta_i=\text{sgn}(\beta_i)|\beta_i|^{q/2}$ or $\bm \beta\mapsto \bm \theta =\bm \beta \frac{\|\bm \beta\|_{\infty}}{\|\bm \beta\|_2}$. 
\cite{Lan2016b} used the spherical augmentation to further map the unit ball $\mathcal{B}_0^d(1)$ to the hyper-sphere $\mathcal{S}^d$ by adding the auxiliary variable $\theta_{d+1}$ to $\bm \theta$. 
Define the extended parameter as $\tilde{\bm \theta}=(\bm \theta, \theta_{d+1})\in \mathcal{S}^d$. 
The lower hemisphere of $\mathcal{S}^{d}$ corresponds to $\theta_{d+1}=-\sqrt{1-\|\bm \theta\|_2^2}$ and the upper hemisphere $\theta_{d+1}=\sqrt{1-\|\bm \theta\|_2^2}$. 
As a result, the hyper-sphere is $\mathcal{S}^d=\{\tilde{\bm \theta}\in \mathbb{R}^{d+1}|\|\tilde{\bm \theta}\|_2=1\}$ and the domain of the unit ball $\mathcal{B}_0^d(1)$ is changed to $\mathcal{S}^d$ by 
\[
T_{\mathcal{B}\mapsto\mathcal{S}}: \mathcal{B}\mapsto\mathcal{S}^d_{\pm}, \quad \bm \theta\mapsto \tilde{\bm \theta}=\left(\bm \theta, \pm \sqrt{1-\|\bm \theta\|_2^2}\right).
\]
The map $\mathcal{T}_{\mathcal{B}\to\mathcal{S}}$ changes the geometry of the problem. 
The original boundary ($\|\bm \theta\|_2 = 1$) becomes the equator of $\mathcal{S}^d$. 
A sampler moving freely on $\mathcal{S}^d$ will implicitly ``bounce off'' this equator when mapped back to $\mathcal{B}_0^d(1)$, thus automatically respecting the constraint without explicit boundary checks. 

The Hamiltonian dynamics are defined on the Riemannian manifold $\mathcal{S}^d$ \citep{lan2014spherical}, endowed with the canonical spherical metric $\bm G_{\mathcal{S}} = \bm I_d + \bm \theta\bm \theta^\top / (1 - \|\bm \theta\|_2^2)$. 
The Hamiltonian on the sphere is given by:
\begin{equation*}
H^*(\tilde{\bm \theta}, \tilde{\bm v}) = U^*(\tilde{\bm \theta}) + K(\tilde{\bm v})=U(\tilde{\bm \theta})+\frac{1}{2}\tilde{\bm v}^{\top} \bm I_{d+1} \tilde{\bm v}=U(\bm \theta)+\frac{1}{2}\bm v^{\top} \bm G_{\mathcal{S}} \bm v 
\end{equation*}
where $\bm v$ is sampled from the tangent space of $\mathcal{B}_0^d(1)$, and $\tilde{\bm v}=(\bm v, v_{d+1})$ is sampled from the tangent space of the unit sphere $\mathcal{S}^d$.
The potential energy $U(\tilde{\bm \theta})=U(\bm \theta)$ and the kinetic energy $K(\tilde{\bm v})=\frac{1}{2}\bm v^\top \bm G_{\mathcal{S}}\bm v=\frac{1}{2}\|\tilde{\bm v}^2\|_2^2$. 
Given this Hamiltonian function defined on the sphere, the associated Hamiltonian dynamics are described by the following system of differential equations:
\begin{equation*}
\begin{aligned}
\dot{\tilde{\bm \theta}} &= \tilde{\bm v} \\
\dot{\tilde{\bm v}} &= -\|\tilde{\bm v}\|_2^2 \tilde{\bm \theta} - \mathcal{P}(\tilde{\bm \theta})\nabla_{\tilde{\bm \theta}} U(\bm \theta)
\end{aligned}
\end{equation*}
where the projection matrix $\mathcal{P}(\tilde{\bm \theta})=\bm I_{d+1}-\tilde{\bm \theta}\tilde{\bm \theta}^\top$ maps the directional derivative $\nabla_{\tilde{\bm \theta}}(\bm \theta)$ onto the tangent space at $\tilde{\bm \theta}$.
The velocity random variable $\tilde{\bm v}\sim \mathcal{N}({\bf 0}, \mathcal{P}(\tilde{\bm \theta})$. 

After obtaining the samples $\{\tilde{\bm \theta}\}$ using the above HMC, we discard the last component $\theta_{d+1}$ and obtain the samples $\{\bm \theta\}$ that automatically satisfy the constraint $\|\bm \theta\|_2 \leq 1$. 
The sign of $\theta_{d+1}$ does not affect the Monte Carlo estimates, but we need to adjust our estimates according to the change of variables theorem. 
We omit these details and summarize the SphHMC into the following Algorithm \ref{alg:SphHMC}. 
\cite{lan2023sampling} provided a comprehensive review of HMC and all constrained HMC, including SphHMC. 

\begin{algorithm}[htb]
\caption{Spherical Hamiltonian Monte Carlo}\label{alg:SphHMC}
\begin{algorithmic}[1]
\State Given the step size $\epsilon$ and current $\bm \theta_{t-1}$, initialize $\tilde{\bm \theta}^{(0)}$ with $\tilde{\bm \theta}$ after transformation.
\State Sample a new velocity value $\tilde{\bm v}^{(0)} \sim \mathcal{N}({\bf 0}, \bm I_{d+1})$, and set $\tilde{\bm v}^{(0)}\leftarrow \tilde{\bm v}^{(0)}-\tilde{\bm \theta}^{(0)}(\tilde{\bm \theta}^{(0)})^\top \tilde{\bm v}^{(0)}$.
\State Calculate $H^*(\tilde{\bm \theta}^{(0)}, \tilde{\bm v}^{(0)})=U(\bm \theta^{(0)})+ K(\tilde{\bm v}^{(0)})$.
\For {$l=0$ to $L-1$} 
\State $\tilde{\bm v}^{(l+\frac{1}{2})} = \tilde{\bm v}^{(l)} - \frac{\epsilon}{2}\mathcal{P}(\tilde{\bm \theta}^{(l)})\nabla_{\tilde{\bm \theta}} U(\bm \theta^{(l)})$.
%
%\left(
%\begin{bmatrix} \bm I_d \\ {\bf 0}^\top \end{bmatrix} - \tilde{\bm \theta}^{(l)}(\bm \theta^{(l)})^{\top}
%\right)\nabla_{\bm \theta}U(\bm \theta^{(l)})$
\State $\tilde{\bm \theta}^{(l+1)} = \tilde{\bm \theta}^{(l)} \cos(\| \tilde{\bm v}^{(l+\frac{1}{2})} \| \epsilon) + \frac{\tilde{\bm v}^{(l+\frac{1}{2})}}{\| \tilde{\bm v}^{(l+\frac{1}{2})} \|} \sin(\| \tilde{\bm v}^{(l+\frac{1}{2})} \| \epsilon)$
\State $\tilde{\bm v}^{(l+\frac{1}{2})} \leftarrow - \tilde{\bm \theta}^{(l)} \| \tilde{\bm v}^{(l+\frac{1}{2})} \| \sin(\| \tilde{\bm v}^{(l+\frac{1}{2})} \| \epsilon) + \tilde{\bm v}^{(l+\frac{1}{2})} \cos(\| \tilde{\bm v}^{(l+\frac{1}{2})} \| \epsilon)$
\State $\tilde{\bm v}^{(l+1)} = \tilde{\bm v}^{(l+\frac{1}{2})} - \frac{\epsilon}{2} \mathcal{P}(\tilde{\bm \theta}^{(l+1)}) \nabla_{\bm \theta} U(\bm \theta^{(l+1)})$
\EndFor
\State Calculate $H^*(\tilde{\bm \theta}^{(L)}, \tilde{\bm v}^{(L)})=U(\bm \theta^{(L)})+ K(\tilde{\bm v}^{(L)})$.
\State Compute the acceptance probability $\alpha = \min\left\{1, \exp(-H^*(\tilde{\bm \theta}^{(L)},\tilde{\bm v}^{(L)})+H^*(\tilde{\bm \theta}^{(0)},\tilde{\bm v}^{(0)}))\right\}$
\State Accept or reject the proposal according to $\alpha$ for the next state $\tilde{\bm \theta}'$. 
\State Apply the importance weights $|\frac{\dd \bm \theta_{\mathcal{B}}}{\dd \tilde{\bm \theta}}|$ to the sample $\bm \theta$ based on $\tilde{\bm \theta}'$. 
\end{algorithmic}
\end{algorithm}

\section*{Three Benchmark Test Functions}
\paragraph{Borehole function}
The function depends on eight inputs as follows \cite{simulationlib}:
\begin{equation}
    f(\bm x)=\frac{2 \pi T_u(H_u-H_l)}{\ln(r/r_w)\left(1+\frac{2LT_u}{\ln(r/r_w)r_w^2 K_w}+\frac{T_u}{T_l}\right)}
\end{equation}
The following table~\ref{table-4} lists each input variable along with its corresponding parameter range and unit.
\footnote{Our mapping $X_1,\dots,X_8$ corresponds to $(r_w,r,T_u,H_u,T_l,H_l,L,K_w)$.}
\begin{table}[H]
    \centering
    \caption{Borehole inputs and ranges}\label{table-4}
    \begin{tabular}{cccc}
        \toprule
        Symbol & Parameter & Range & Unit \\ \midrule
        $r_w$ & radius of borehole & $[0.05,0.15]$ & $m$ \\
        $r$ & radius of influence & $[100,50000]$ & $m$ \\
        $T_u$ & transmissivity of upper aquifer & $[63070,115600]$ & $m^2/yr$ \\
        $H_u$ & potentiometric head of upper aquifer & $[990,1110]$ & $m$ \\
        $T_l$ & transmissivity of lower aquifer & $[63.1,116]$ & $m^2/yr$ \\
        $H_l$ & potentiometric head of lower aquifer & $[700,820]$ & $m$ \\
        $L$ & length of borehole & $[1120,1680]$ & $m$ \\
        $K_w$ & hydraulic conductivity of borehole & $[9855,12045]$ & $m/yr$ \\
        \bottomrule
    \end{tabular}
\end{table}

\paragraph{OTL Circuit function.}
The function depends on six inputs as follows \cite{simulationlib}:
\begin{equation}
    V_m(\mathbf{x}) =\frac{(V_{b1} + 0.74) \beta (R_{c2} + 9)}{\beta (R_{c2} + 9) + R_f} + \frac{11.35 R_f}{\beta (R_{c2} + 9) + R_f} + \frac{0.74 R_f \beta (R_{c2} + 9)}{(\beta (R_{c2} + 9) + R_f) R_{c1}}
\end{equation}
where 
\[
V_{b1} = \frac{12 R_{b2}}{R_{b1} + R_{b2}}.
\]
The following table~\ref{table-5} lists each input variable along with its corresponding parameter range and unit.
\footnote{Our mapping $X_1,\dots,X_6$ corresponds to $(R_{b1},R_{b2},R_f,R_{c1},R_{c2},\beta)$.}

\begin{table}[H]
    \centering
    \caption{OTL inputs and ranges}\label{table-5}
    \begin{tabular}{cccc}
        \toprule
        Symbol & Parameter & Range & Unit \\ \midrule
        $R_{b1}$ & resistance $b1$ & $[50,150]$ & $\text{K-Ohms}$ \\
        $R_{b2}$ & resistance $b2$ & $[25,70]$ & $\text{K-Ohms}$ \\
        $R_f$ & resistance $f$ & $[0.5,3]$ & $\text{K-Ohms}$ \\
        $R_{c1}$ & resistance $c1$ & $[1.2,2.5]$ & $\text{K-Ohms}$ \\
        $R_{c2}$ & resistance $c2$ & $[0.25,1.2]$ & $\text{K-Ohms}$ \\
        $\beta$ & current gain & $[50,300]$ & $\text{Amperes}$ \\
        \bottomrule
    \end{tabular}
\end{table}

\paragraph{Piston function}
The function depends on seven inputs as follows \cite{simulationlib}:
\begin{equation}
    \begin{aligned}
        C(\mathbf{x}) &= 2\pi \sqrt{\frac{M}{k + S^2 \frac{P_0 V_0}{T_0} \frac{T_a}{V^2}}}, \quad \text{where} \\[2ex]
        V &= \frac{S}{2k} \left( \sqrt{A^2 + 4k \frac{P_0 V_0}{T_0} T_a} - A \right) \\[2ex]
        A &= P_0 S + 19.62 M - \frac{k V_0}{S}
    \end{aligned}
\end{equation}

The following table~\ref {table-6} lists each input variable along with its corresponding parameter range and unit.
\footnote{Our mapping $X_1,\dots,X_7$ corresponds to $(M,S,V_0,k,P_0,T_a,T_0)$.}

\begin{table}[H]
    \centering
    \caption{Piston inputs and ranges}\label{table-6}
    \begin{tabular}{cccc}
        \toprule
        Symbol & Parameter & Range & Unit \\ \midrule
        $M$ & piston weight & $[30,60]$ & $kg$ \\
        $S$ & piston surface area & $[0.005,0.020]$ & $m^2$ \\
        $V_0$ & initial gas volume & $[0.002,0.010]$ & $m^3$ \\
        $k$ & spring coefficient & $[1000,5000]$ & $N/m$ \\
        $P_0$ & atmospheric pressure & $[90000,110000]$ & $N/m^2$ \\
        $T_a$ & ambient temperature & $[290,296]$ & $K$ \\
        $T_0$ & filling gas temperature & $[340,360]$ & $K$ \\
        \bottomrule
    \end{tabular}
\end{table}

\section*{Figures for Three Benchmark Examples}

\begin{figure}[tbp]
    \centering
    \begin{subfigure}[t]{.46\linewidth}
    \includegraphics[width=\linewidth]{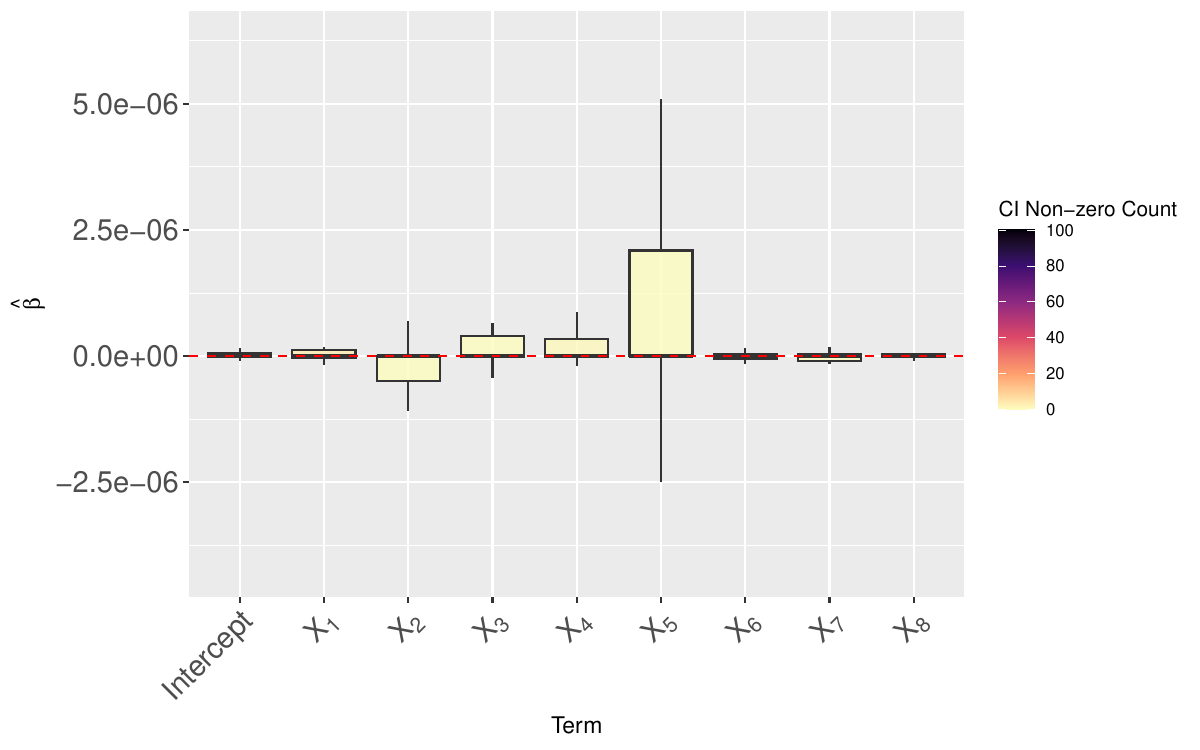}
    \caption{Box plots of point estimates for $\bm \beta$ of linear terms.}
    \end{subfigure}
    \begin{subfigure}[t]{.46\linewidth}
    \includegraphics[width=\linewidth]{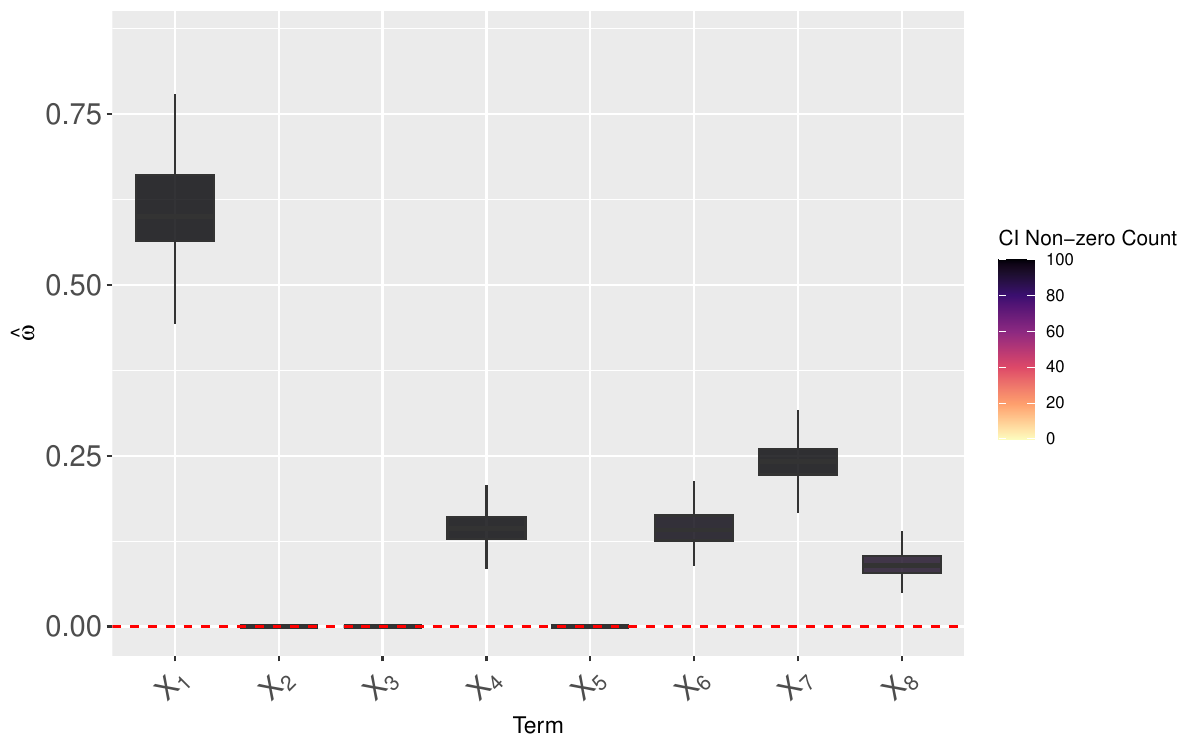}
    \caption{Box plots of point estimates for $\bm \omega$.}
    \end{subfigure}
    \begin{subfigure}[t]{0.8\linewidth}
    \includegraphics[width=1\linewidth]{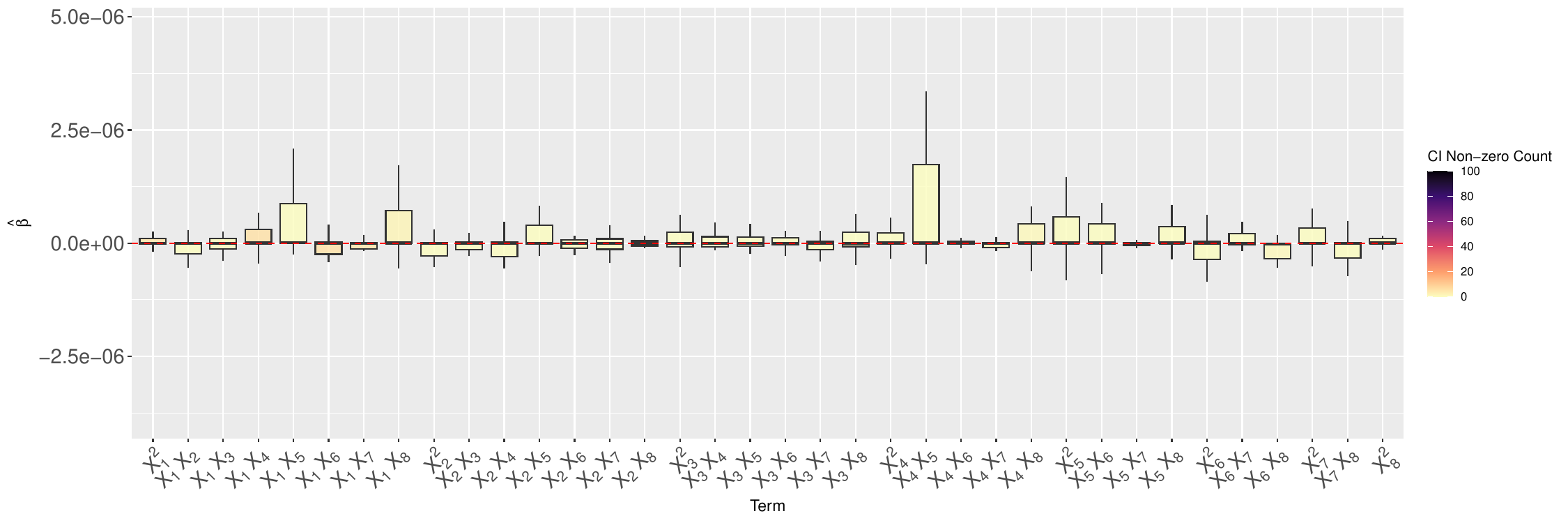}
    \caption{Box plots of point estimates for  $\boldsymbol{\beta}$ of the 2nd order terms.}
    \end{subfigure}
    \caption{Box plot of point estimates for parameters for $n=200$ for Borehole function.}\label{fig:boxplot_borehole_para_200}
\end{figure}

\begin{figure}[tbp]
    \centering
    \begin{subfigure}[t]{.46\linewidth}
    \includegraphics[width=\linewidth]{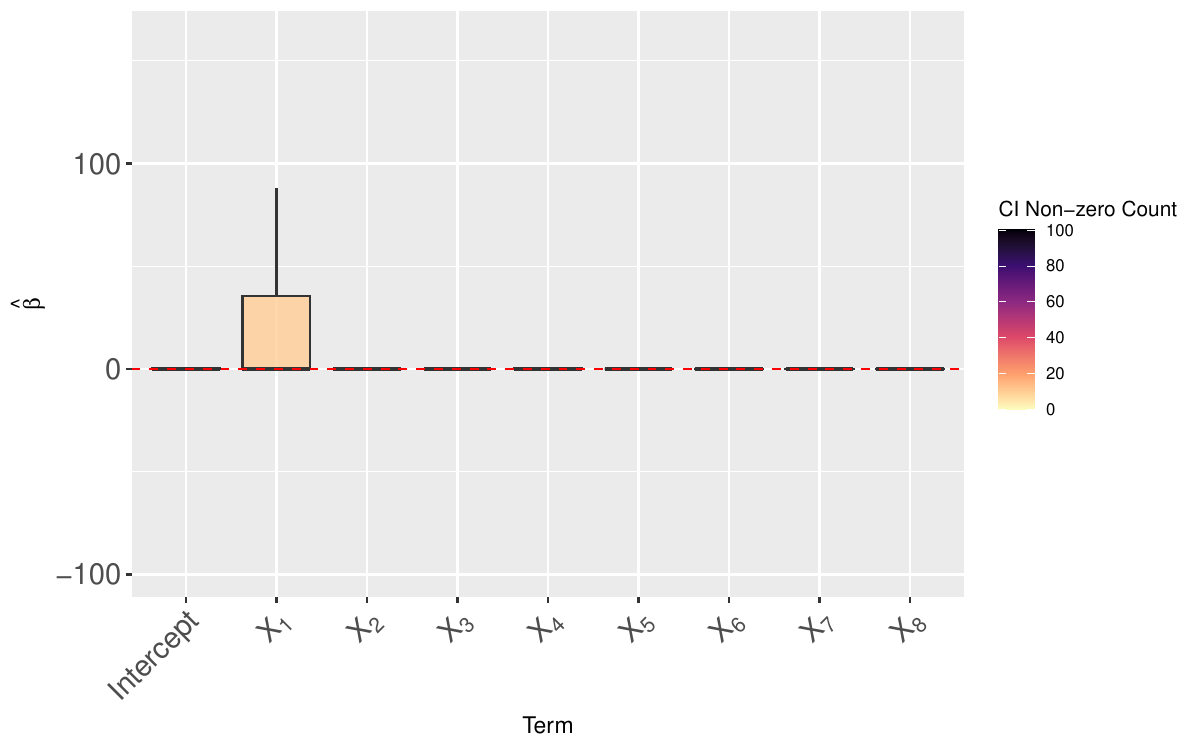}
    \caption{Box plots of point estimates $\bm \beta$ of lower-order terms. }
    \end{subfigure}
    \begin{subfigure}[t]{.46\linewidth}
    \includegraphics[width=\linewidth]{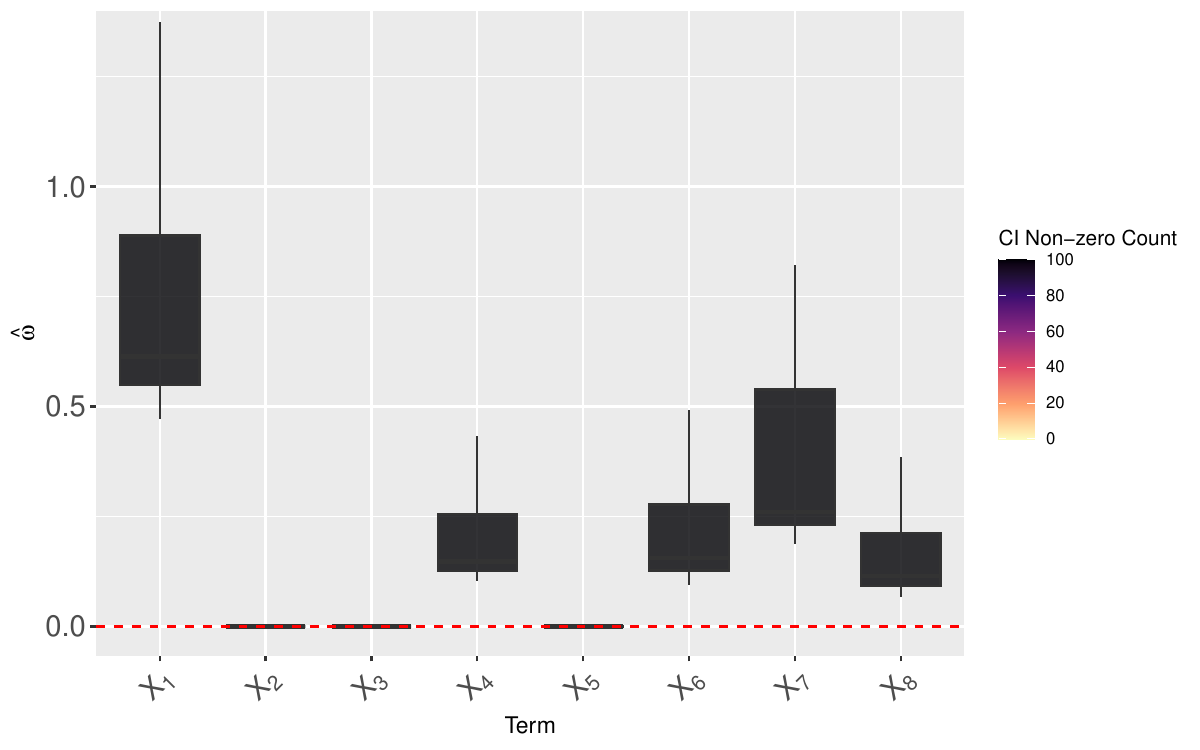}
    \caption{Box plots of point estimates for $\bm \omega$.}
    \end{subfigure}
    \begin{subfigure}[t]{0.8\linewidth}
    \includegraphics[width=\linewidth]{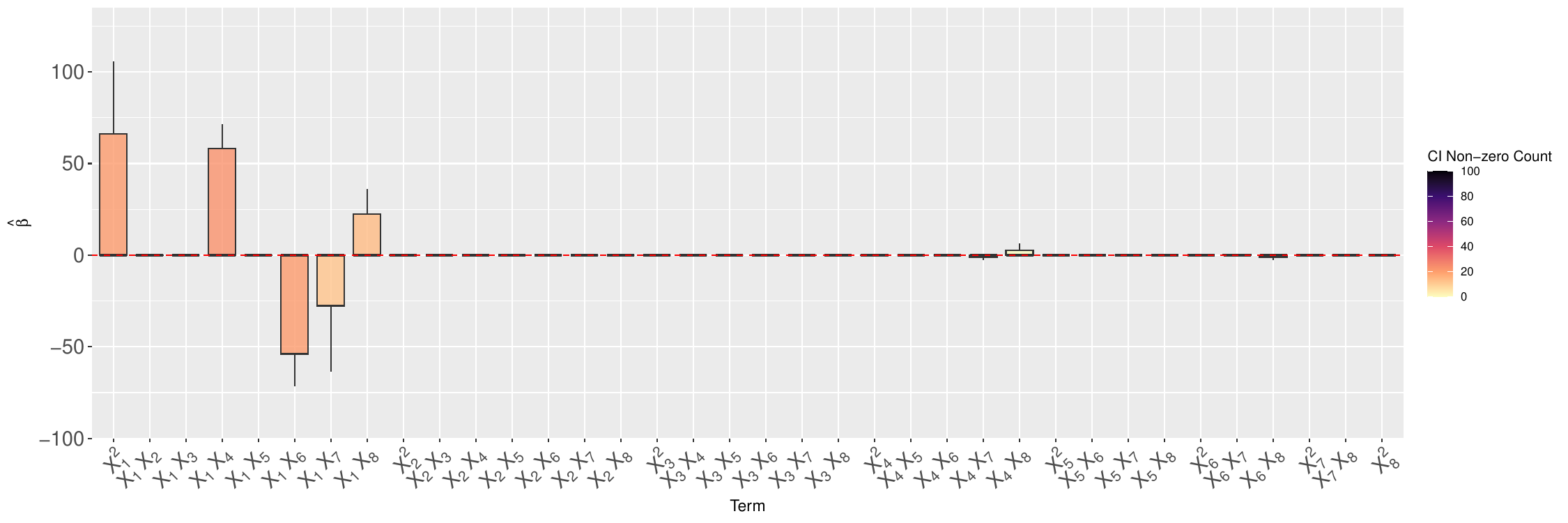}
    \caption{Box plots of point estimates for  $\boldsymbol{\beta}$ of the 2nd order terms.}
    \end{subfigure}
    \caption{Box plot of point estimates for parameters for $n=500$ for Borehole function.}\label{fig:boxplot_borehole_para_500}
\end{figure}

\begin{figure}[tbp]
    \centering
    \begin{subfigure}[t]{.4\linewidth}
    \includegraphics[width=\linewidth]{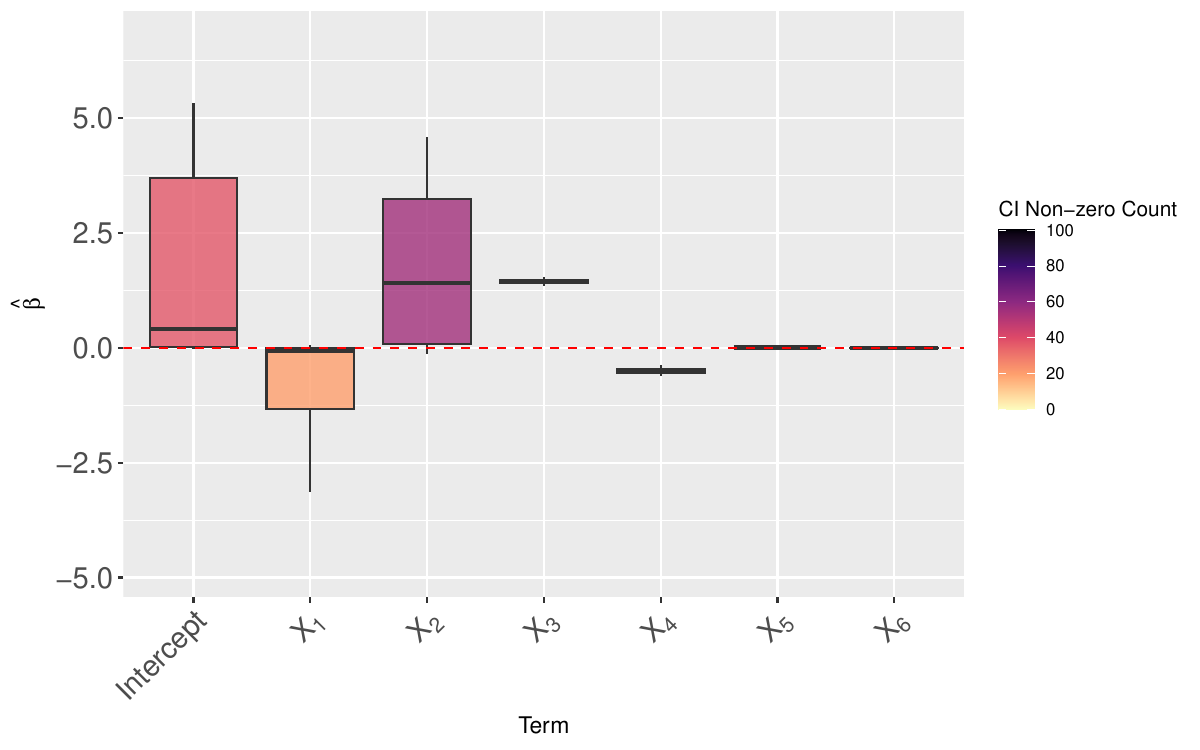}
    \caption{Box plots of point estimates for $\bm \beta$ of lower-order terms.}
    \end{subfigure}
    \begin{subfigure}[t]{.44\linewidth}
    \includegraphics[width=\linewidth]{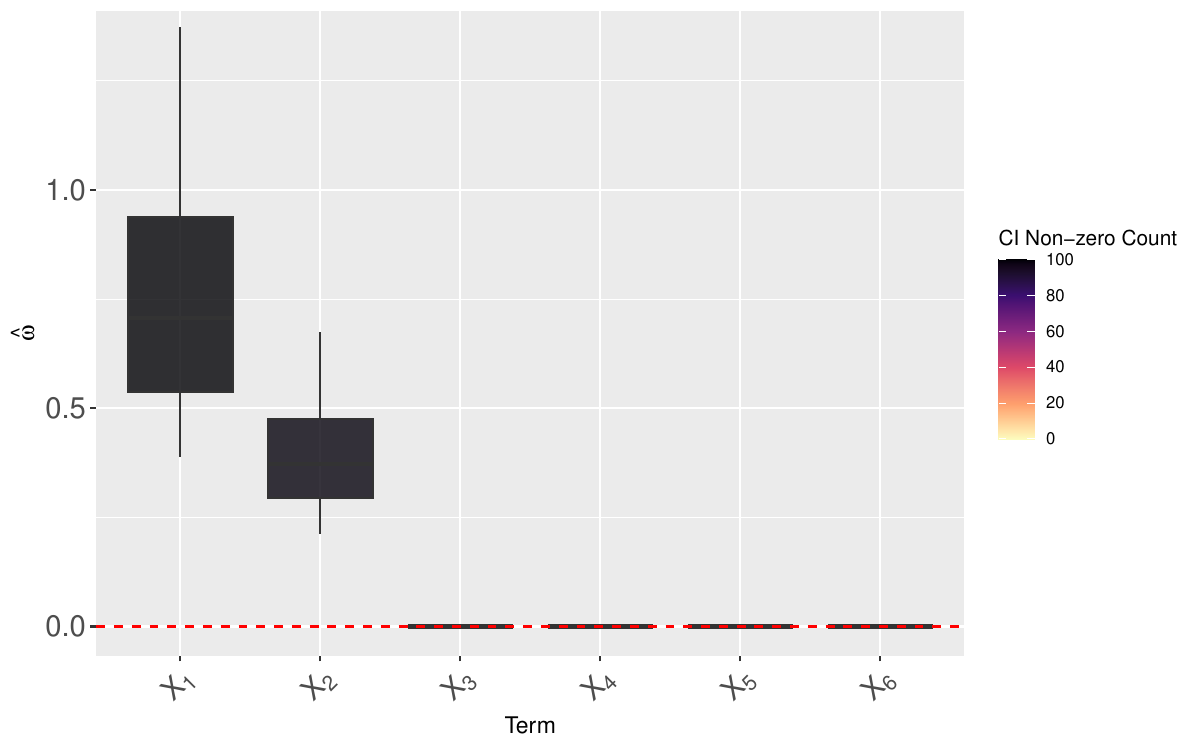}
    \caption{Box plots of point estimates for $\bm \omega$.}
    \end{subfigure}
    \begin{subfigure}[t]{0.6\linewidth}
    \includegraphics[width=\linewidth]{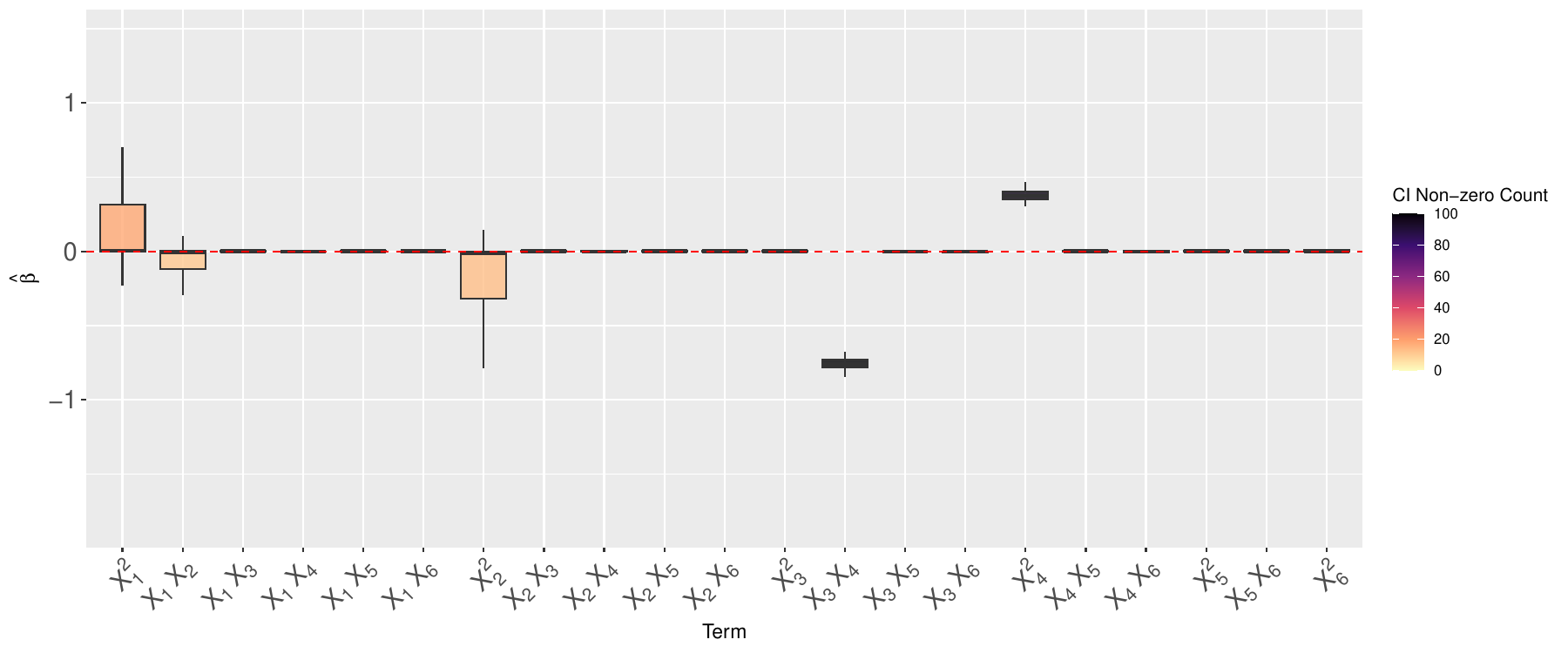}
    \caption{Box plots of point estimates for  $\boldsymbol{\beta}$ of the 2nd order terms.}
    \end{subfigure}
    \caption{Box plot of point estimates for parameters for $n=200$ for OTL Circuit function.}\label{fig:boxplot_otlcircuit_para_200}
\end{figure}

\begin{figure}[tbp]
    \centering
    \begin{subfigure}[t]{.4\linewidth}
    \includegraphics[width=\linewidth]{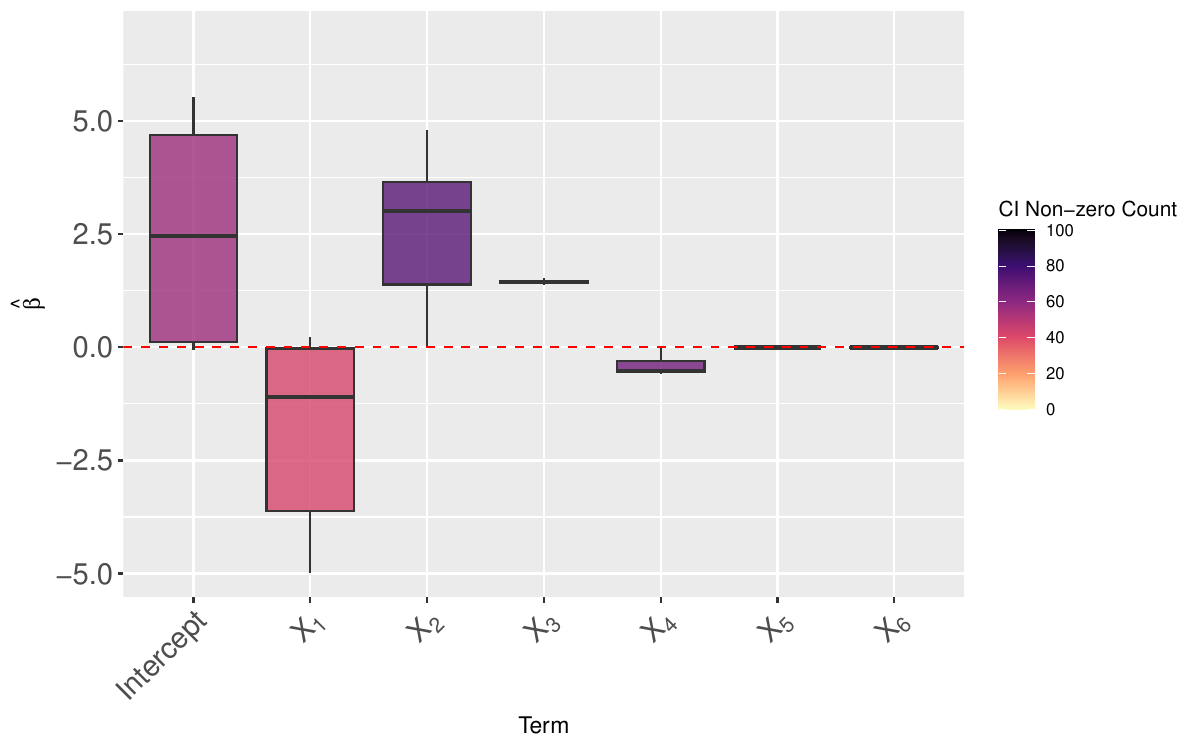}
    \caption{Box plots of point estimates for $\bm \beta$ of lower-order terms.}
    \end{subfigure}
    \begin{subfigure}[t]{.4\linewidth}
    \includegraphics[width=\linewidth]{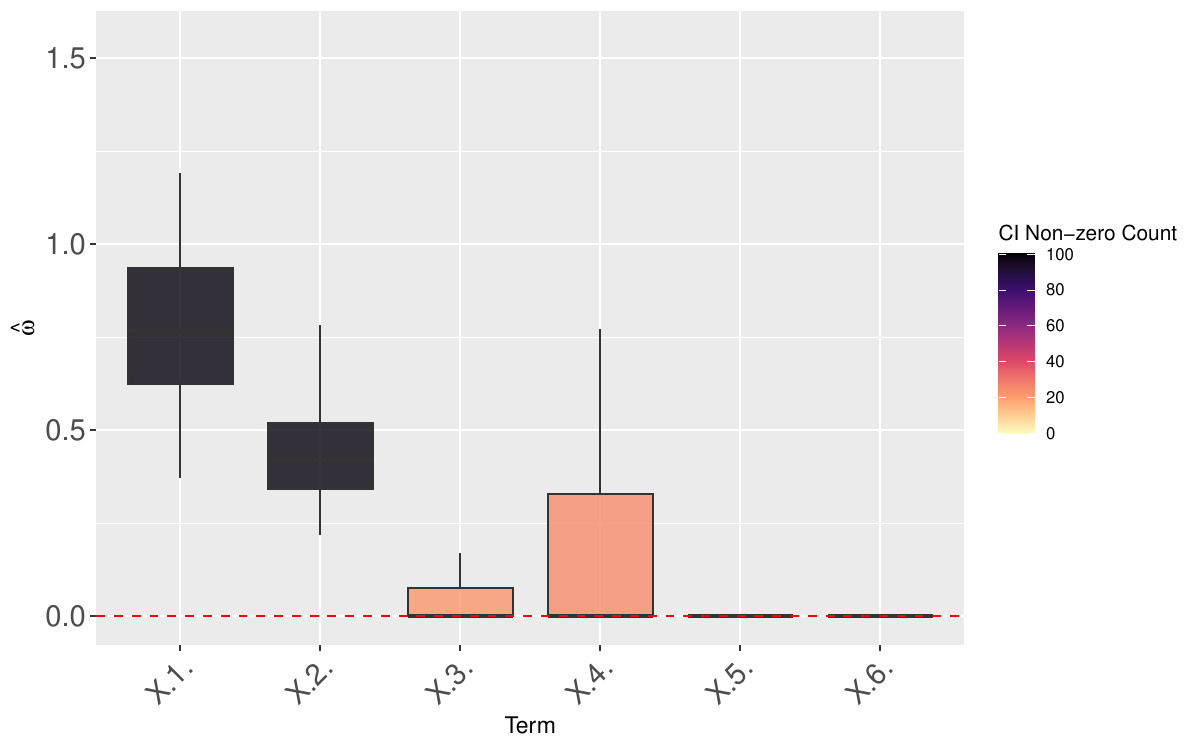}
    \caption{Box plots of point estimates for $\bm \omega$.}
    \end{subfigure}
    \begin{subfigure}[t]{0.6\linewidth}
    \includegraphics[width=\linewidth]{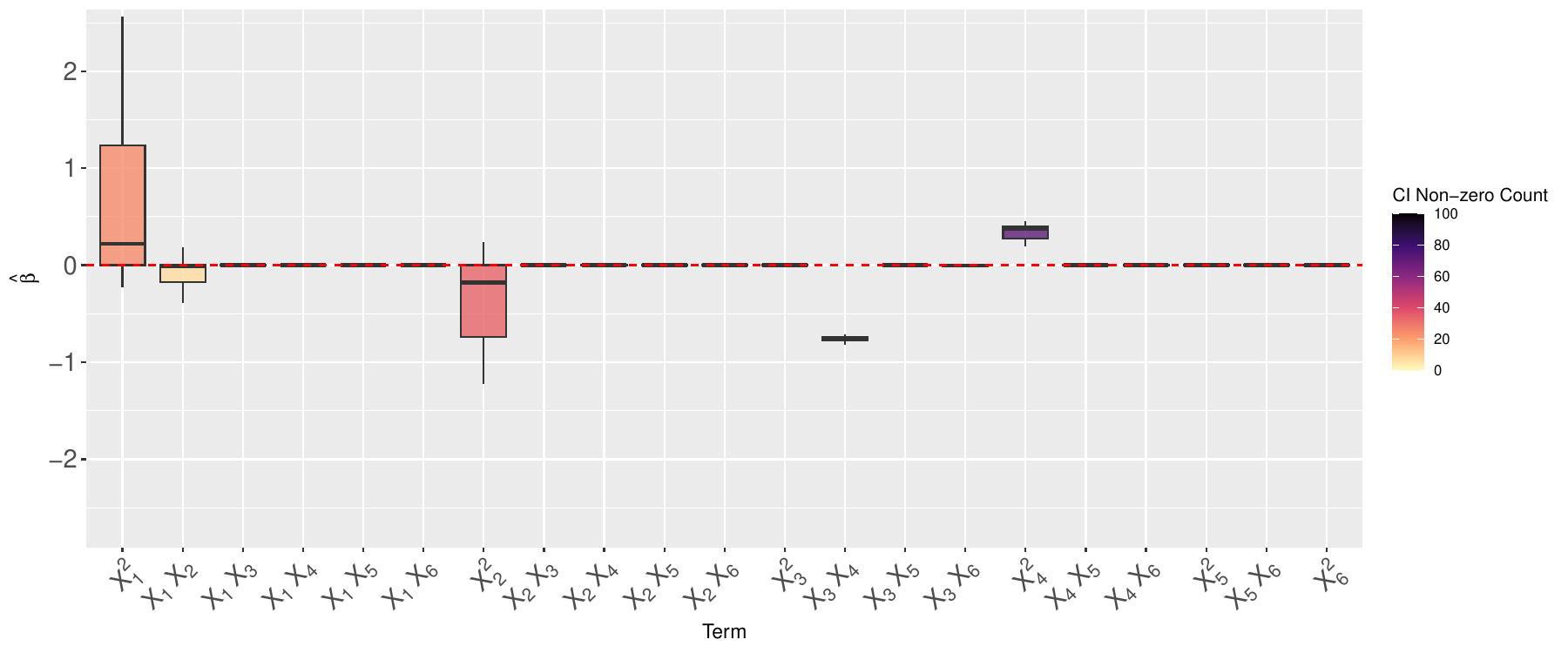}
    \caption{Box plots of point estimates for  $\boldsymbol{\beta}$ of the 2nd order terms.}
    \end{subfigure}
    \caption{Box plot of point estimates for parameters for $n=500$ for OTL Circuit function.}\label{fig:boxplot_otlcircuit_para_500}
\end{figure}

\begin{figure}[tbp]
    \centering
    \begin{subfigure}[t]{.4\linewidth}
    \includegraphics[width=\linewidth]{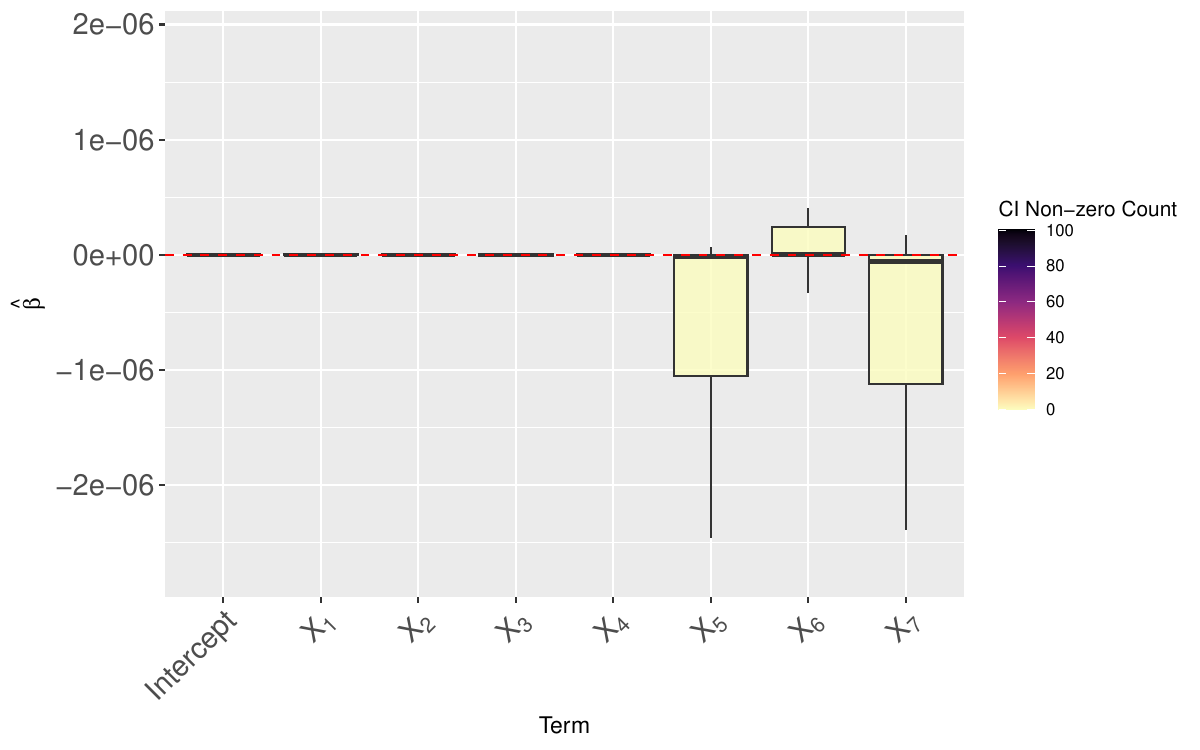}
    \caption{Box plots of point estimates for $\bm \beta$ of lower-order terms.}
    \end{subfigure}
    \begin{subfigure}[t]{.4\linewidth}
    \includegraphics[width=\linewidth]{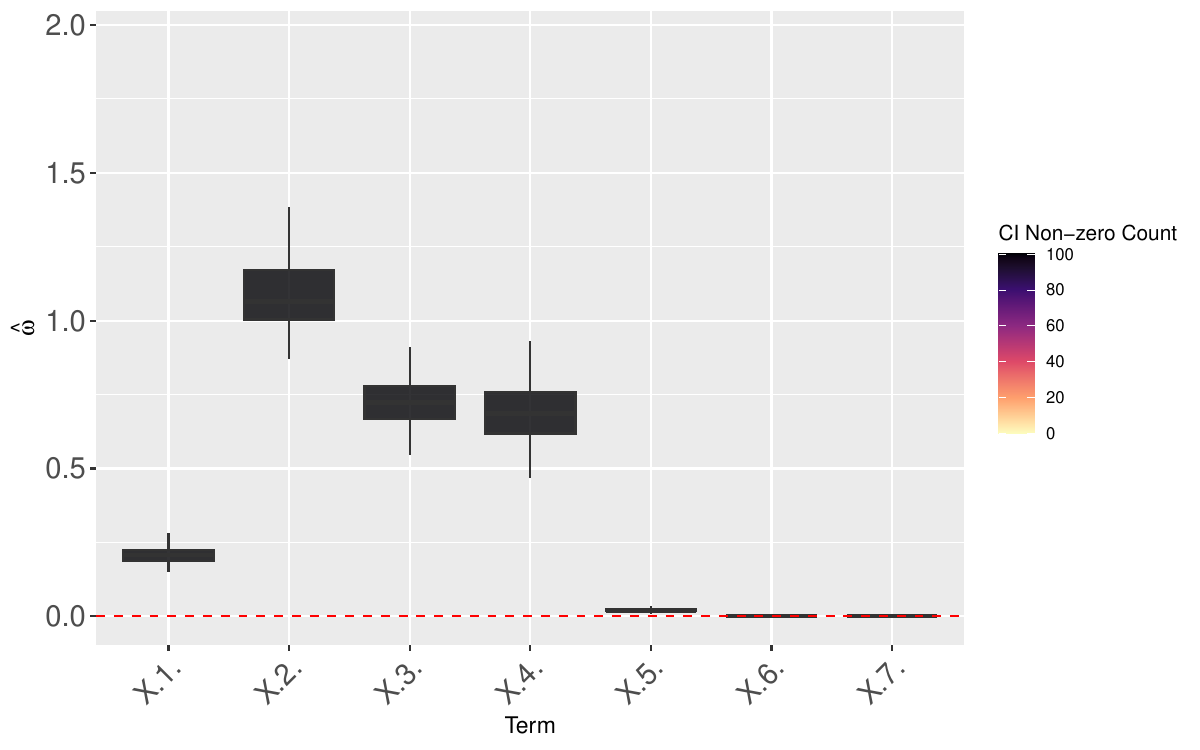}
    \caption{Box plots of point estimates for $\bm \omega$.}
    \end{subfigure}
    \begin{subfigure}[t]{0.6\linewidth}
    \includegraphics[width=\linewidth]{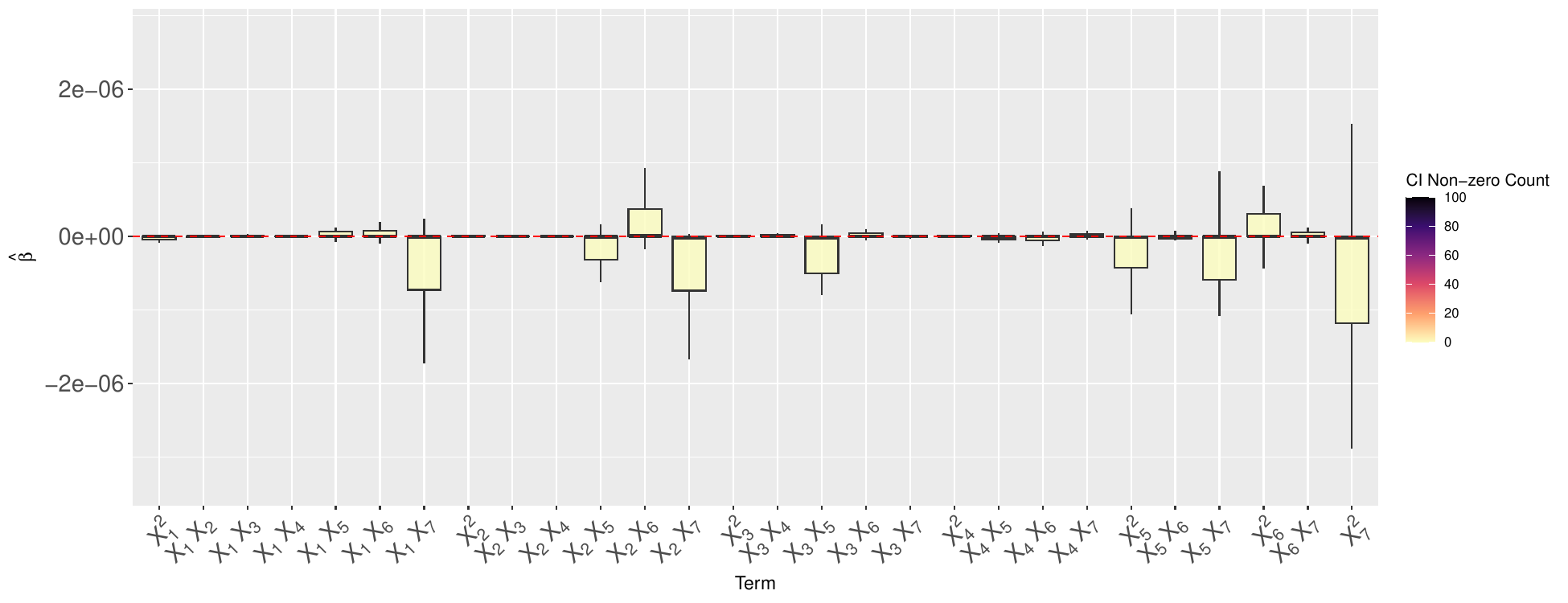}
    \caption{Box plots of point estimates for  $\boldsymbol{\beta}$ of the 2nd order terms.}
    \end{subfigure}
    \caption{Box plot of point estimates for parameters for $n=200$ for Piston function.}\label{fig:boxplot_piston_para_200}
\end{figure}

\begin{figure}[tbp]
    \centering
    \begin{subfigure}[t]{.4\linewidth}
    \includegraphics[width=\linewidth]{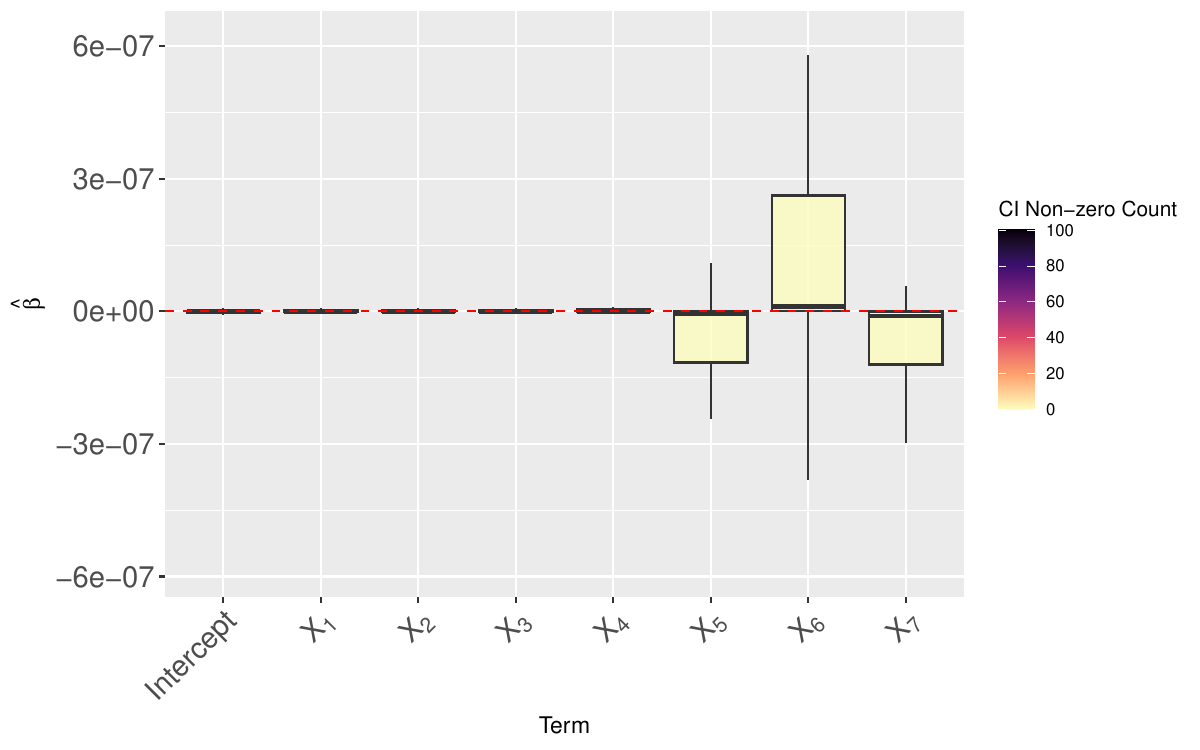}
    \caption{Box plots of point estimates for $\bm \beta$ of lower-order terms.}
    \end{subfigure}
    \begin{subfigure}[t]{.4\linewidth}
    \includegraphics[width=\linewidth]{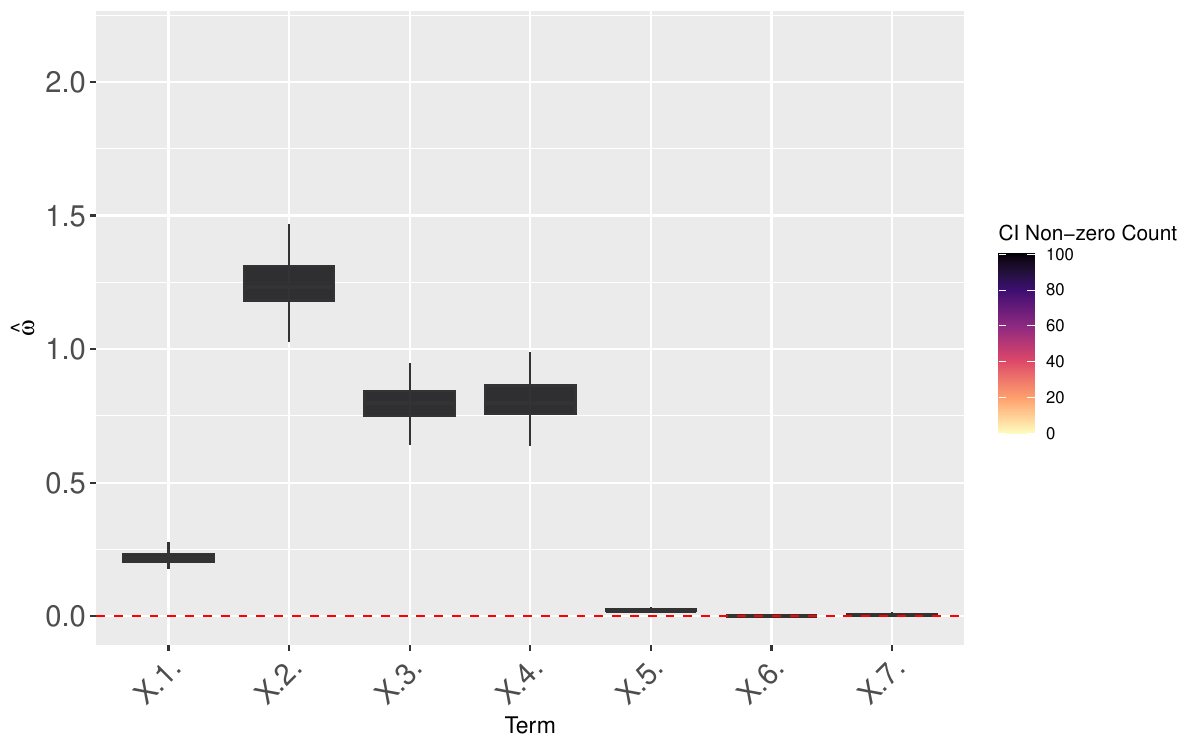}
    \caption{Box plots of point estimates for $\bm \omega$.}
    \end{subfigure}
    \begin{subfigure}[t]{0.6\linewidth}
    \includegraphics[width=\linewidth]{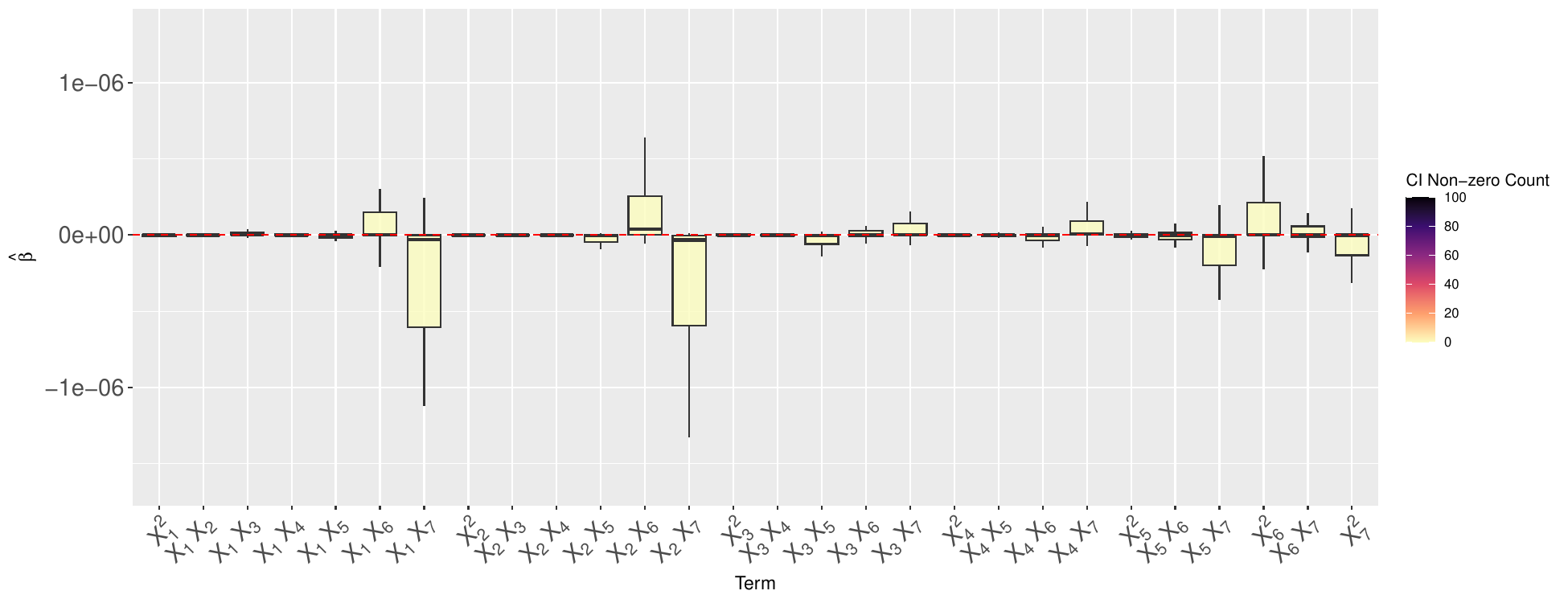}
    \caption{Box plots of point estimates for  $\boldsymbol{\beta}$ of the 2nd order terms.}
    \end{subfigure}
    \caption{Box plot of point estimates for parameters for $n=500$ for Piston function.}\label{fig:boxplot_piston_para_500}
\end{figure}

\begin{figure}[tbp]
    \centering
    \begin{subfigure}[t]{.9\linewidth}
    \includegraphics[width=\linewidth,height=4cm]{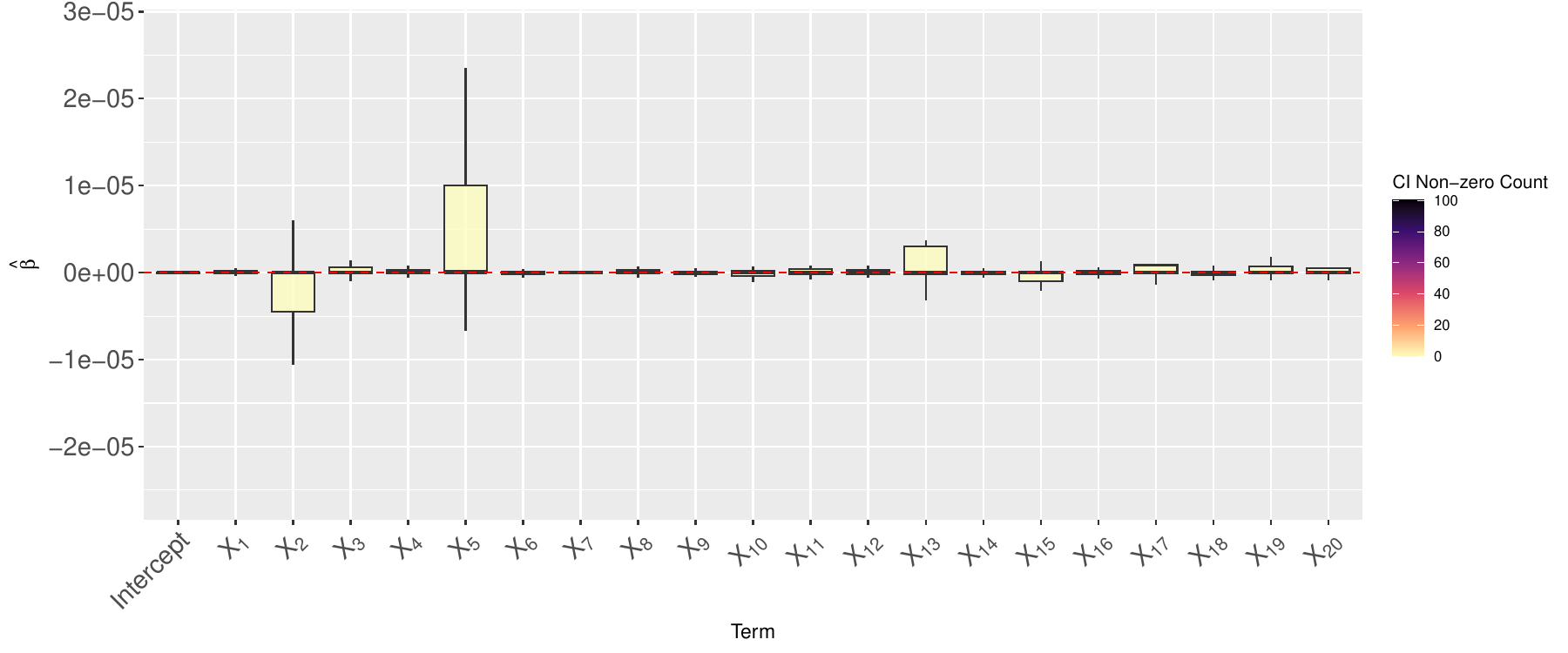}
    \caption{Box plots of point estimates for $\bm \beta$.}
    \end{subfigure}
    \begin{subfigure}[t]{.9\linewidth}
    \includegraphics[width=\linewidth,height=4cm]{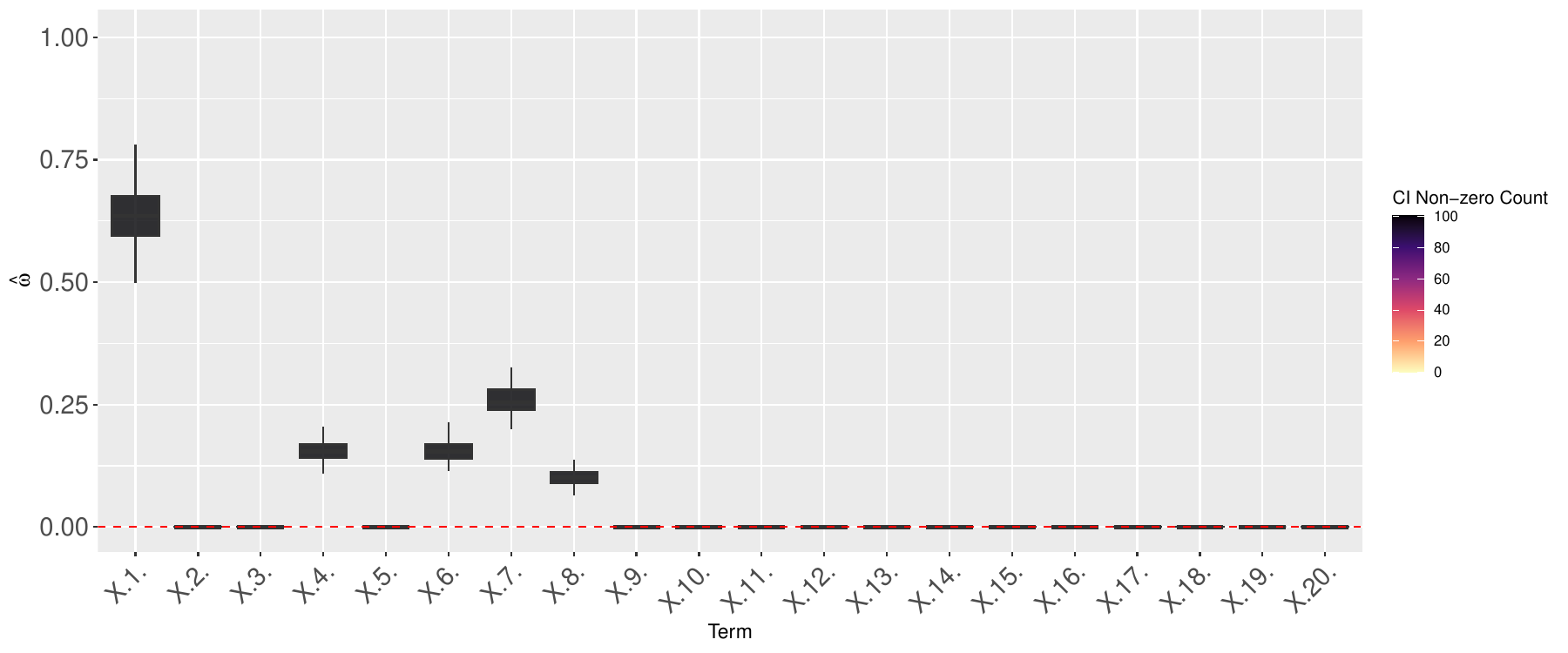}
    \caption{Box plots of point estimates for $\bm \omega$.}
    \end{subfigure}
    \caption{Box plot of point estimates for parameters for the Borehole function with $n=200$ and $d=20$. }\label{fig:boxplot_borehole_20_para_200}
\end{figure}

\begin{figure}[tbp]
    \centering
    \begin{subfigure}[t]{.9\linewidth}
    \includegraphics[width=\linewidth,height=4cm]{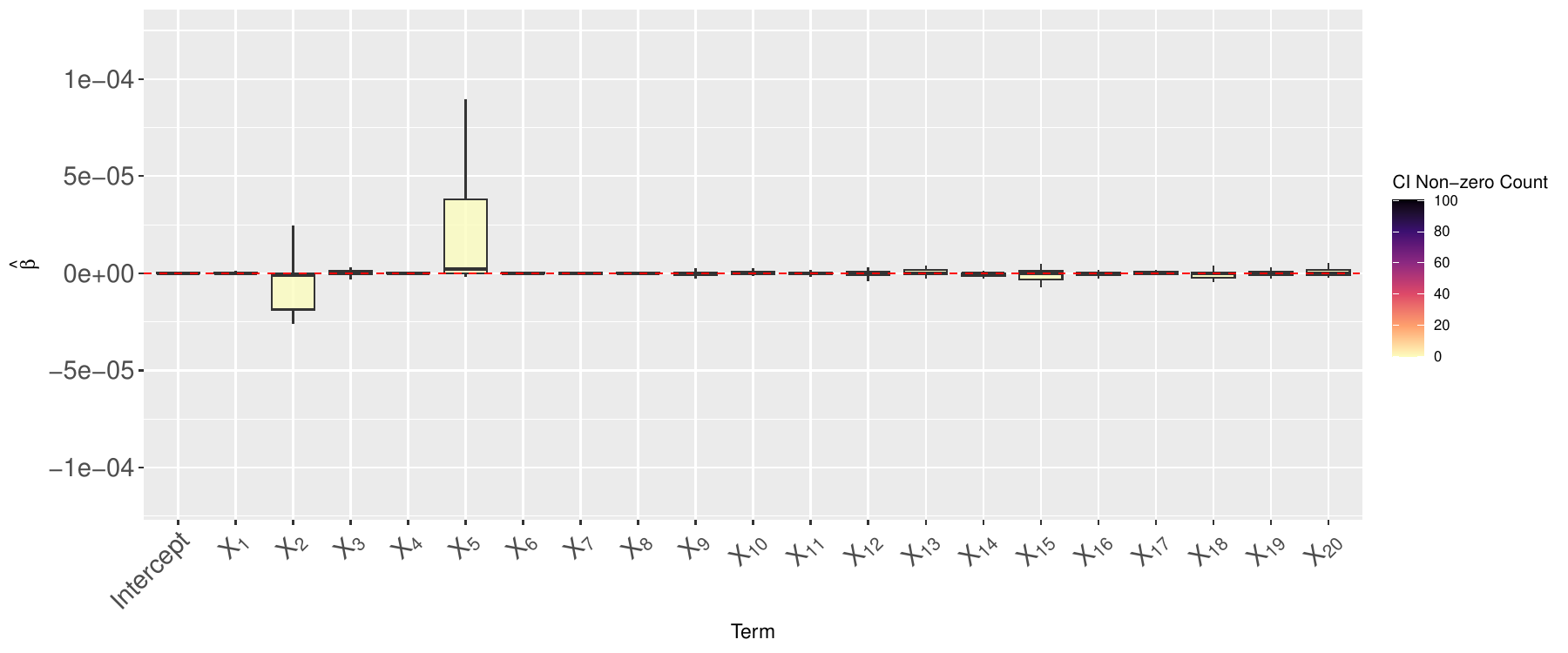}
    \caption{Box plots of point estimates for $\bm \beta$.}
    \end{subfigure}
    \begin{subfigure}[t]{.9\linewidth}
    \includegraphics[width=\linewidth,height=4cm]{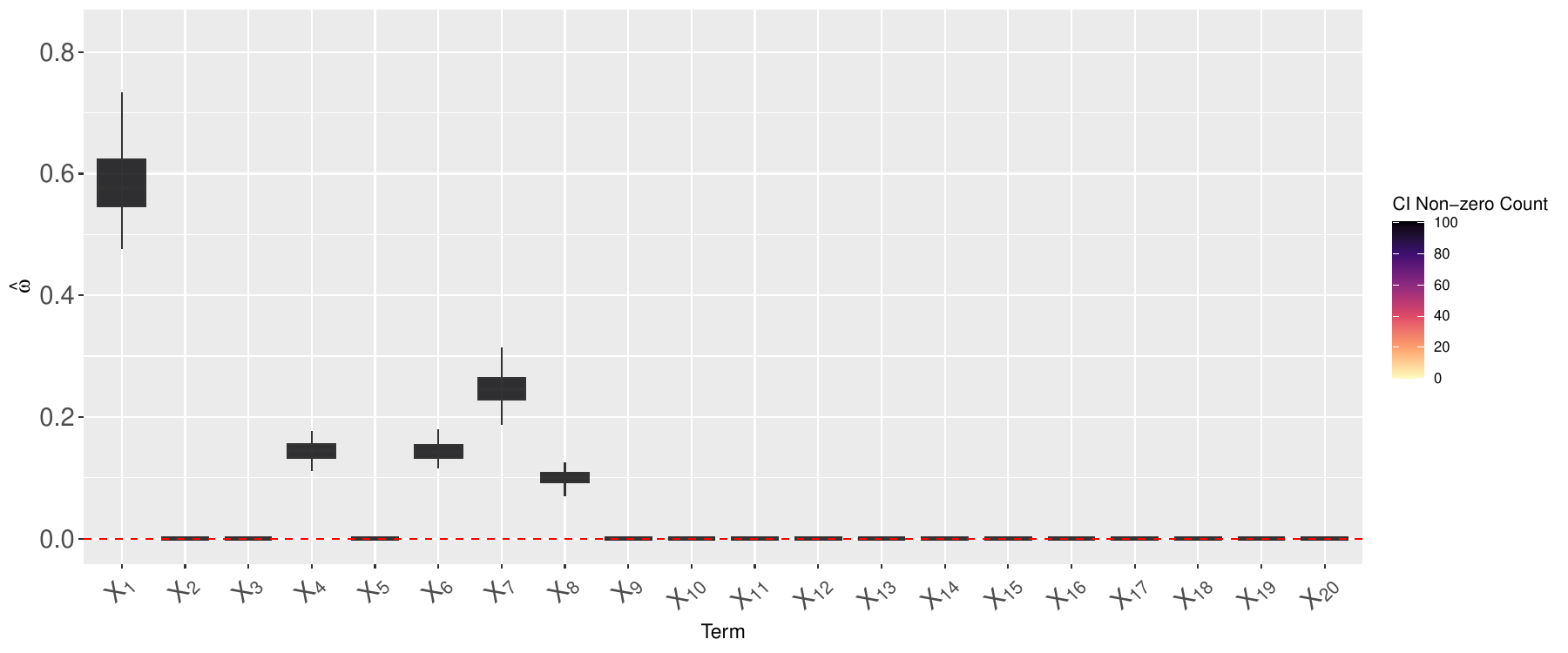}
    \caption{Box plots of point estimates for $\bm \omega$.}
    \end{subfigure}
    \caption{Box plot of point estimates for parameters for the Borehole function with $n=500$ and $d=20$.}\label{fig:boxplot_borehole_20_para_500}
\end{figure}

\begin{figure}[tbp]
    \centering
    \begin{subfigure}[t]{.9\linewidth}
    \includegraphics[width=\linewidth,height=4cm]{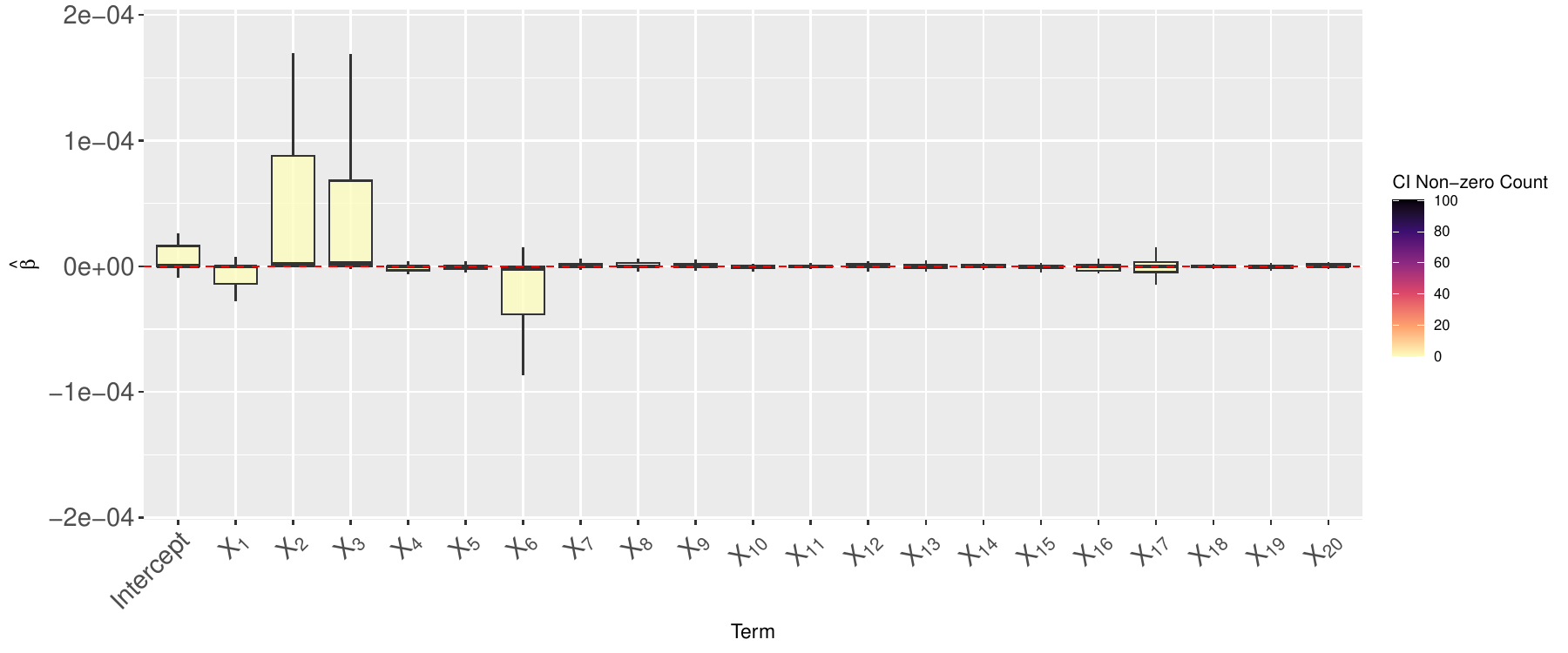}
    \caption{Box plots of point estimates for $\bm \beta$.}
    \end{subfigure}
    \begin{subfigure}[t]{.9\linewidth}
    \includegraphics[width=\linewidth,height=4cm]{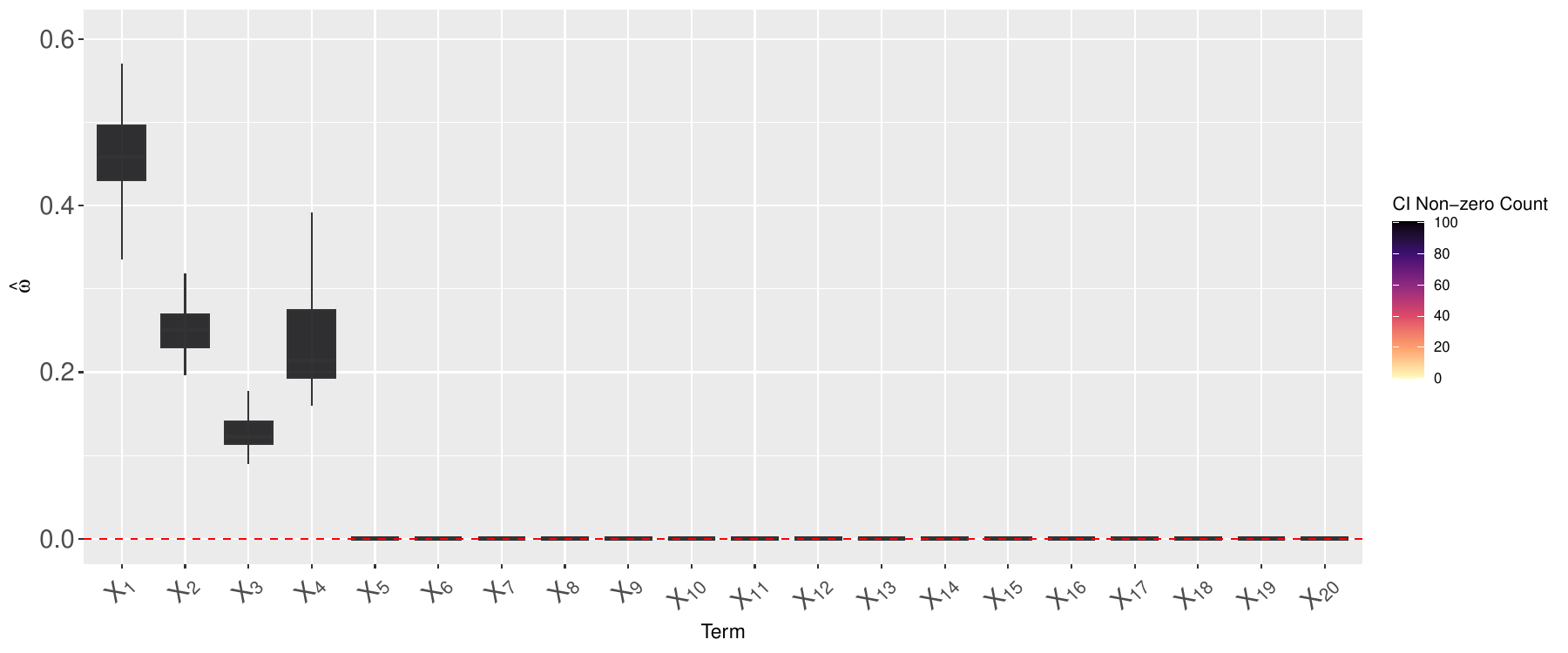}
    \caption{Box plots of point estimates for $\bm \omega$.}
    \end{subfigure}
    \caption{Box plot of point estimates for parameters for the OTL Circuit function with $n=200$ and $d=20$.}\label{fig:boxplot_otlcircuit_20_para_200}
\end{figure}

\begin{figure}[tbp]
    \centering
    \begin{subfigure}[t]{.9\linewidth}
    \includegraphics[width=\linewidth,height=4cm]{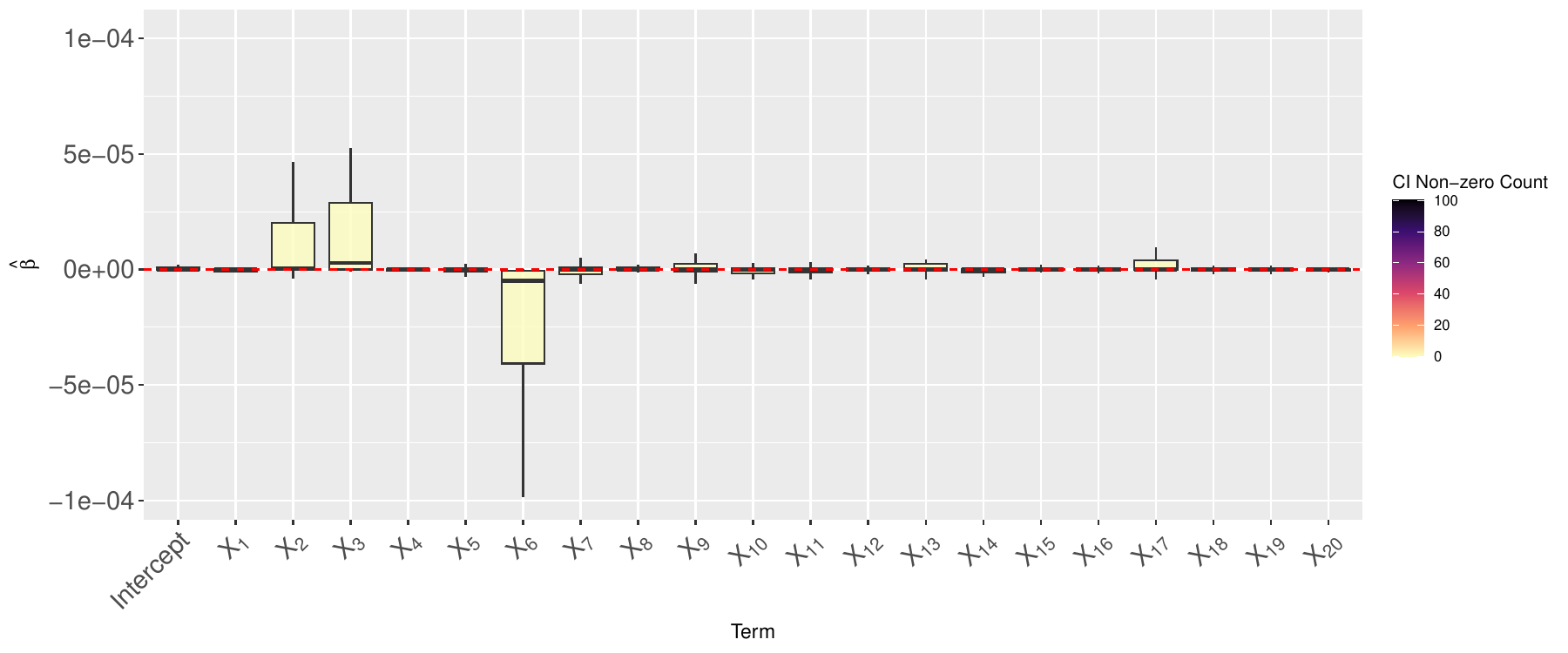}
    \caption{Box plots of point estimates for $\bm \beta$.}
    \end{subfigure}
    \begin{subfigure}[t]{.9\linewidth}
    \includegraphics[width=\linewidth,height=4cm]{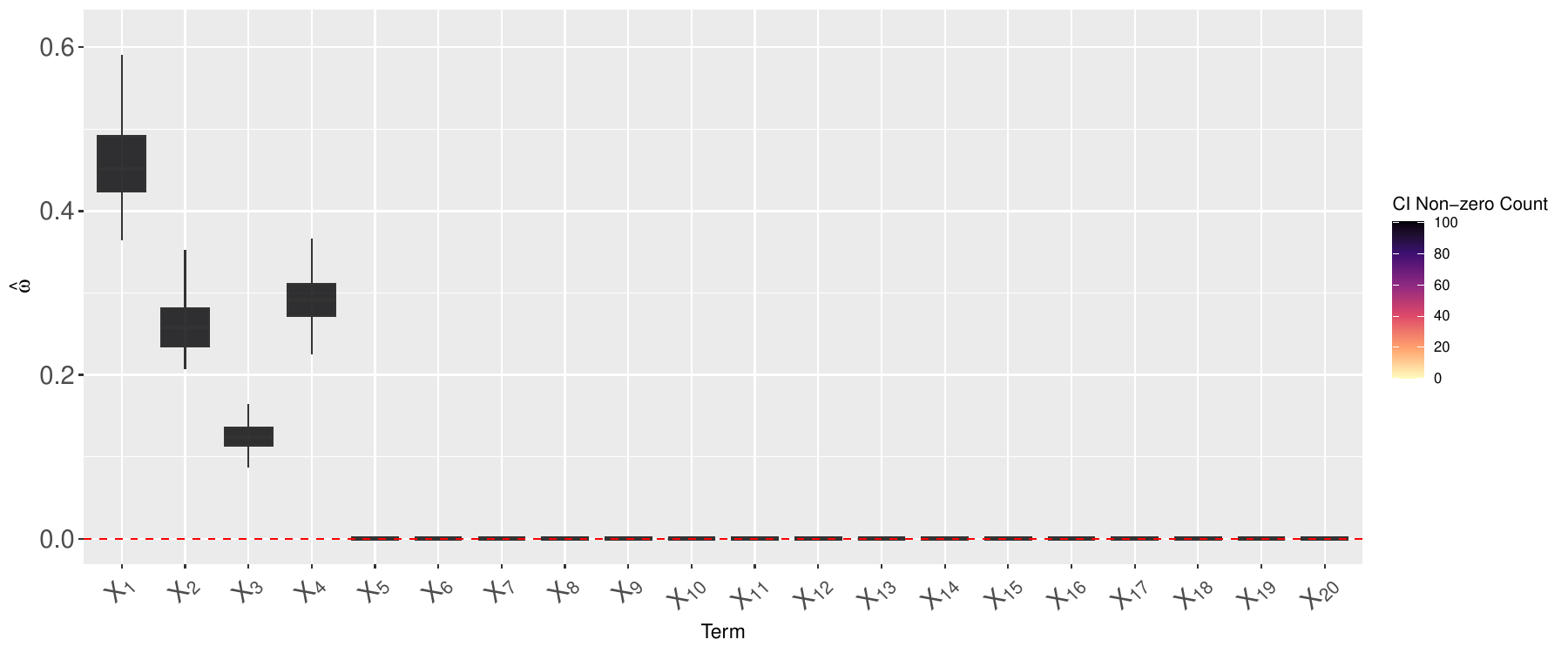}
    \caption{Box plots of point estimates for $\bm \omega$.}
    \end{subfigure}
    \caption{Box plot of point estimates for parameters for the OTL Circuit function with $n=500$ and $d=20$.}\label{fig:boxplot_otlcircuit_20_para_500}
\end{figure}

\begin{figure}[tbp]
    \centering
    \begin{subfigure}[t]{.9\linewidth}
    \includegraphics[width=\linewidth,height=4cm]{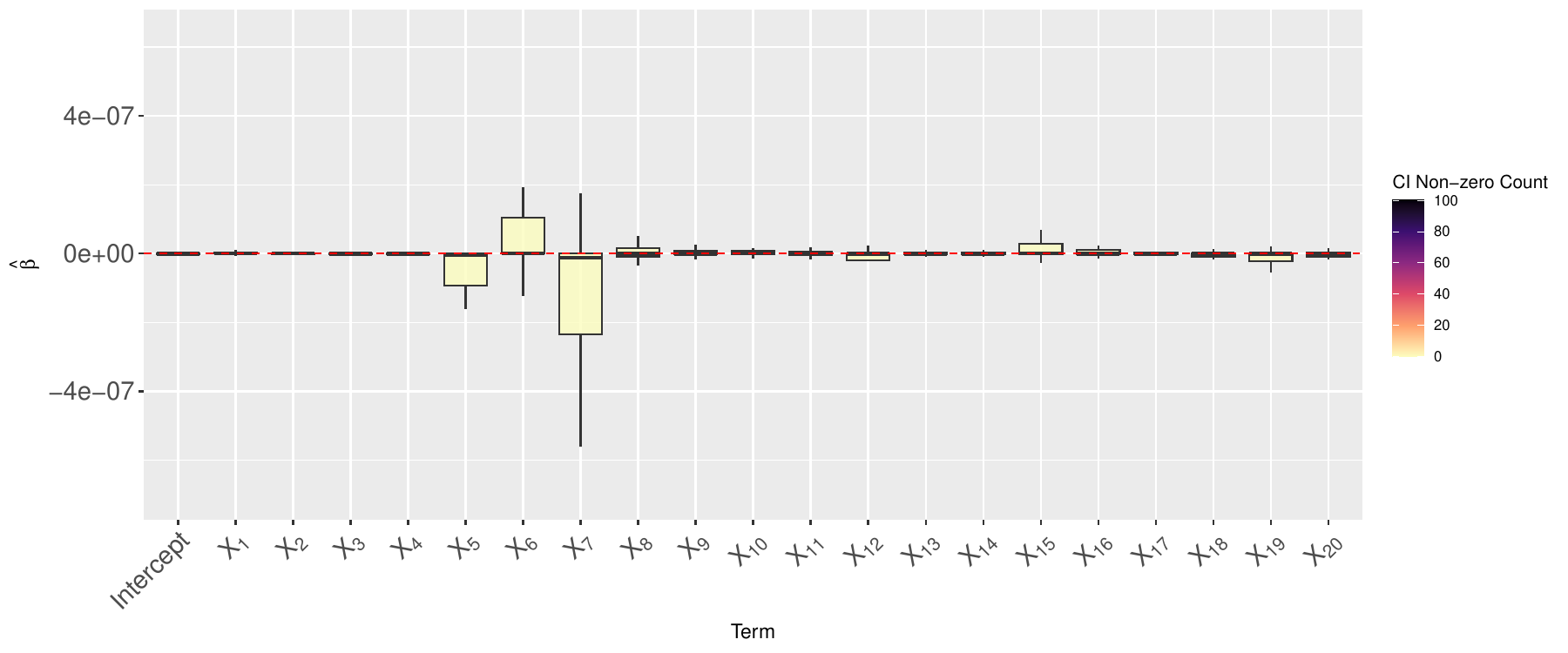}
    \caption{Box plots of point estimates for $\bm \beta$.}
    \end{subfigure}
    \begin{subfigure}[t]{.9\linewidth}
    \includegraphics[width=\linewidth,height=4cm]{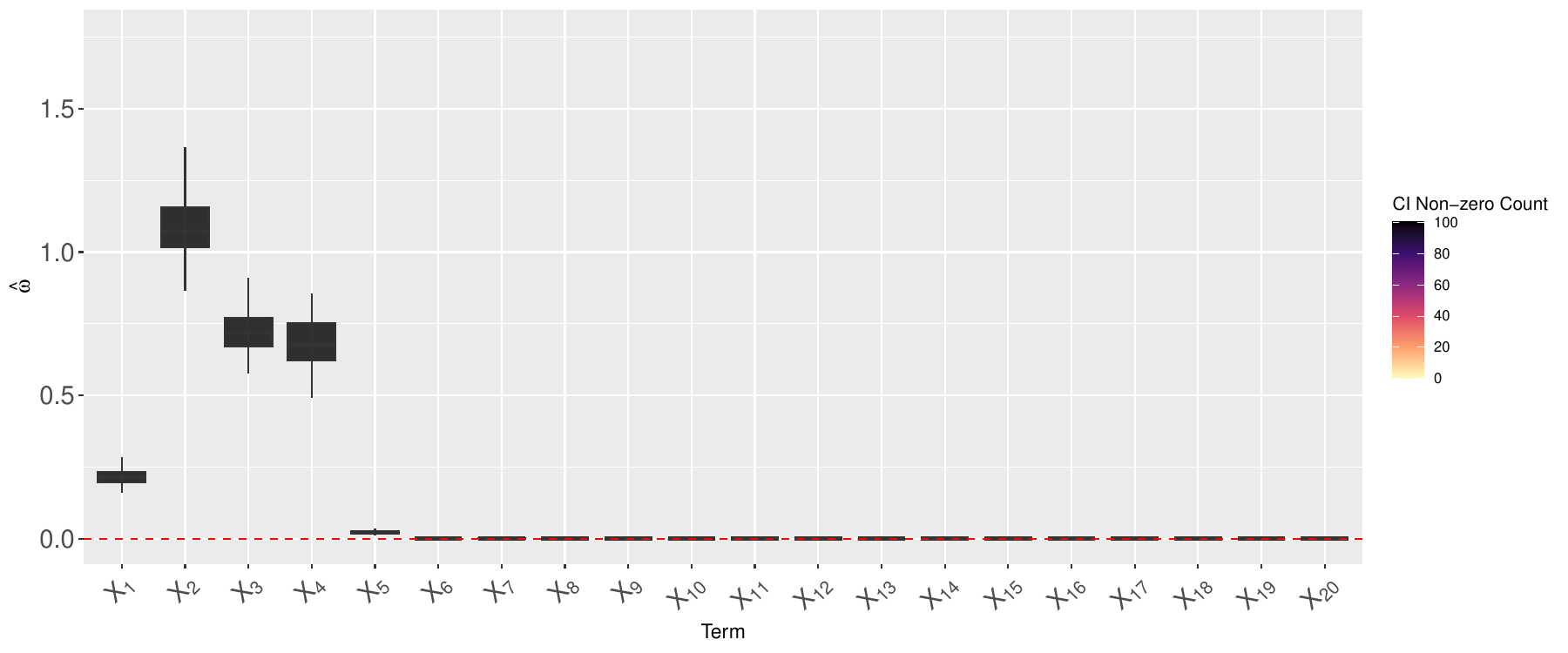}
    \caption{Box plots of point estimates for $\bm \omega$.}
    \end{subfigure}
    \caption{Box plot of point estimates for parameters for the Piston function with $n=200$ and $d=20$.}\label{fig:boxplot_piston_20_para_200}
\end{figure}

\begin{figure}[tbp]
    \centering
    \begin{subfigure}[t]{.9\linewidth}
    \includegraphics[width=\linewidth,height=4cm]{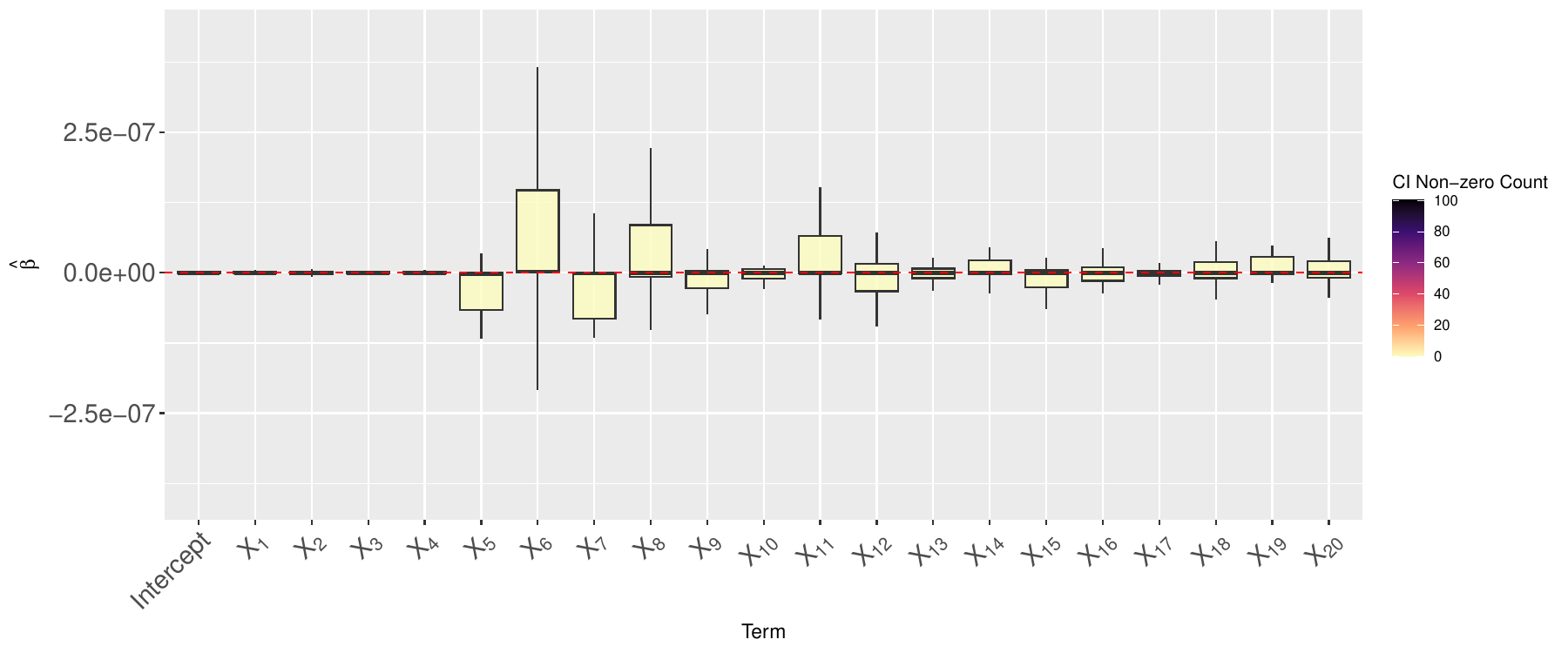}
    \caption{Box plots of point estimates for $\bm \beta$.}
    \end{subfigure}
    \begin{subfigure}[t]{.9\linewidth}
    \includegraphics[width=\linewidth,height=4cm]{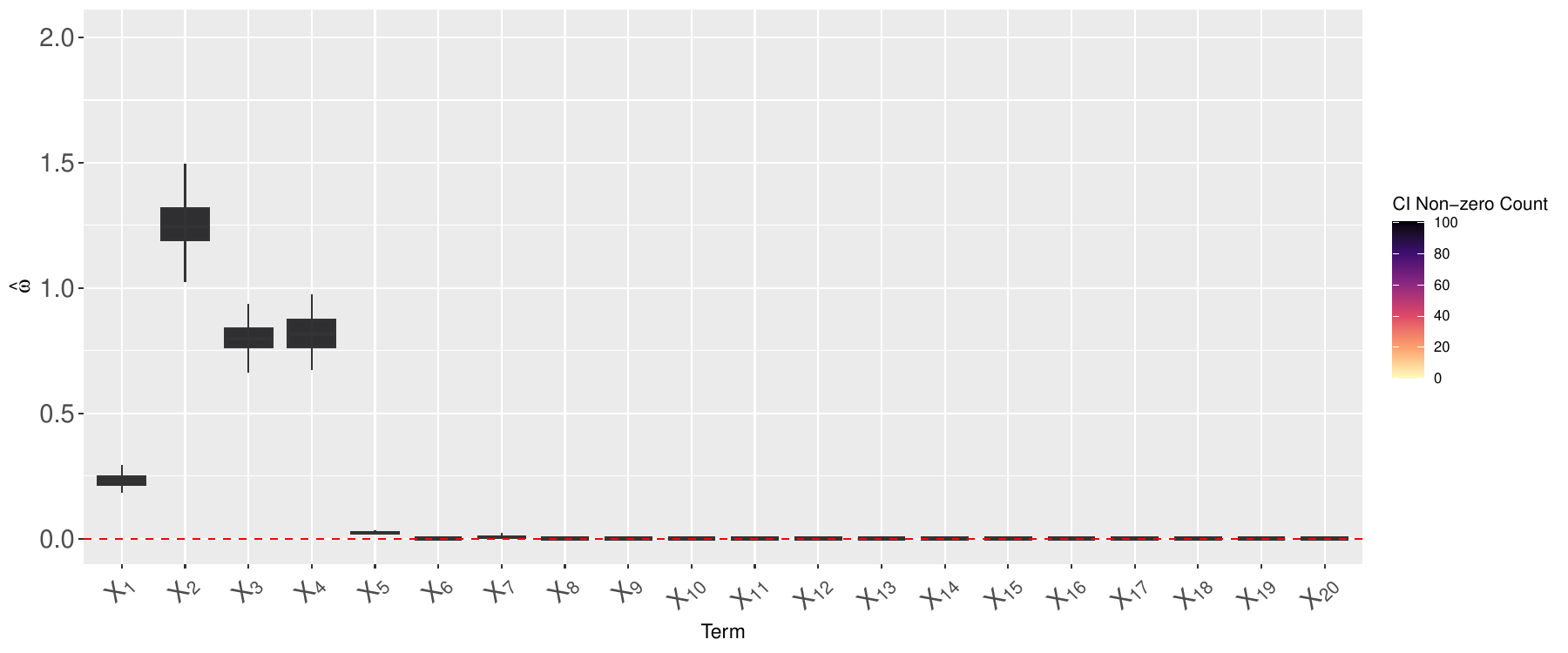}
    \caption{Box plots of point estimates for $\bm \omega$.}
    \end{subfigure}
    \caption{Box plot of point estimates for parameters for the Piston function with $n=500$ and $d=20$.}\label{fig:boxplot_piston_20_para_500}
\end{figure}

\begin{figure}[p]
    \centering
    \begin{subfigure}[t]{.96\linewidth}
    \includegraphics[width=\linewidth]{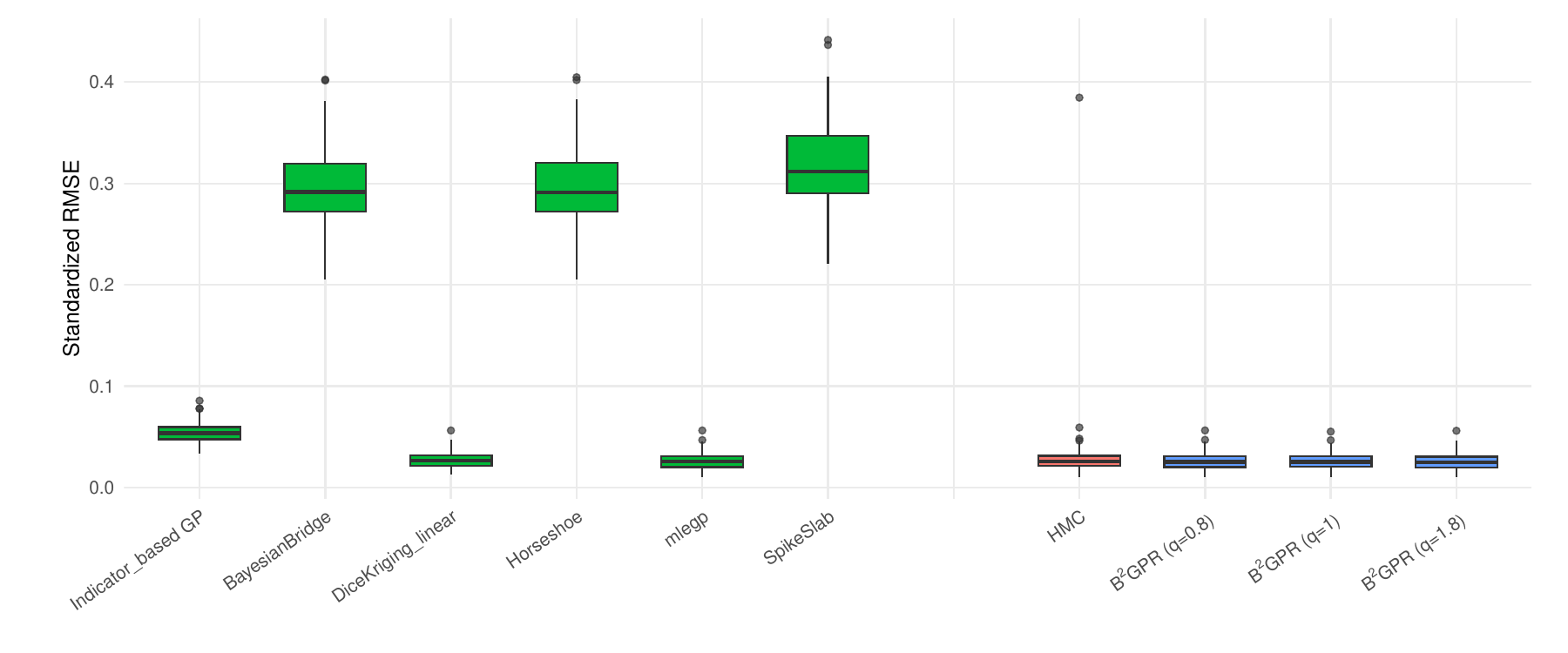}
    \caption{Standardized RMSE.}
    \end{subfigure}
    
    \begin{subfigure}[t]{.96\linewidth}
    \includegraphics[width=\linewidth]{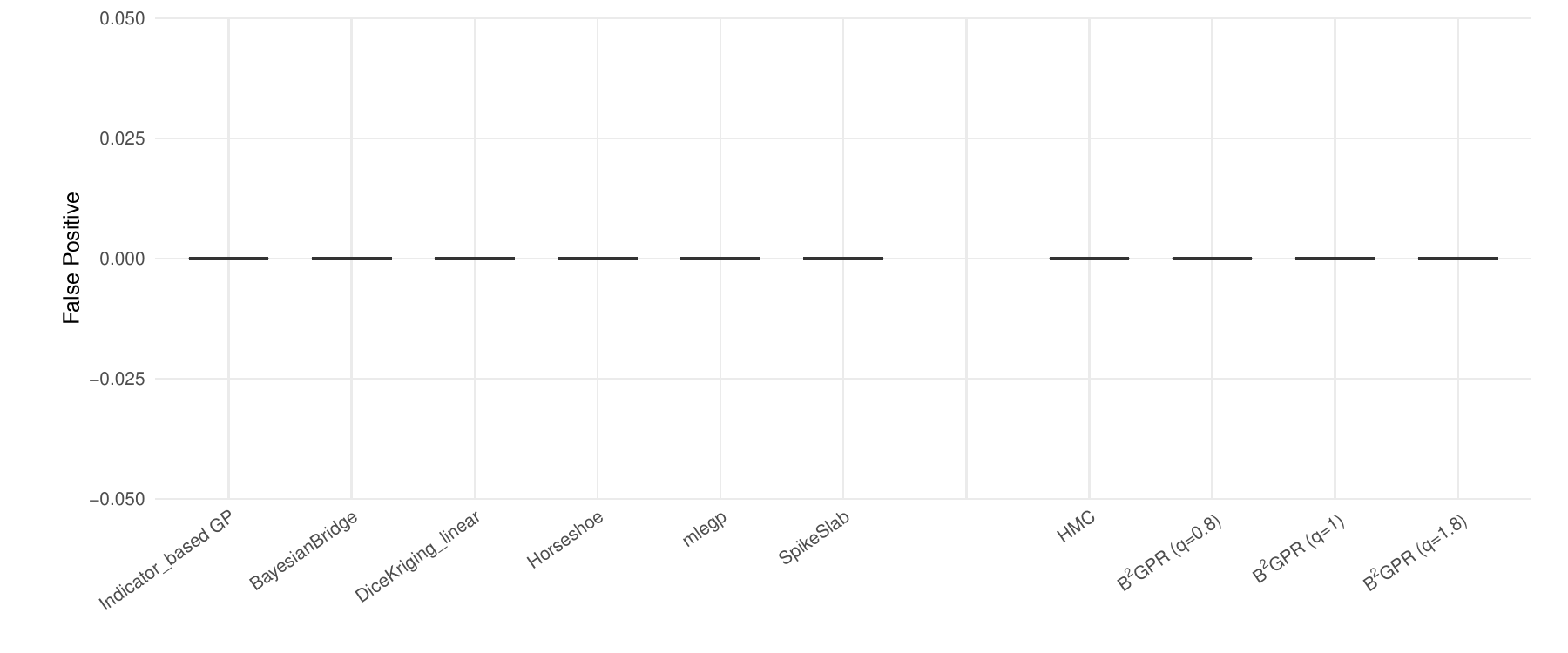}
    \caption{Number of false positives.}
    \end{subfigure}
    
    \begin{subfigure}[t]{.96\linewidth}
    \includegraphics[width=\linewidth]{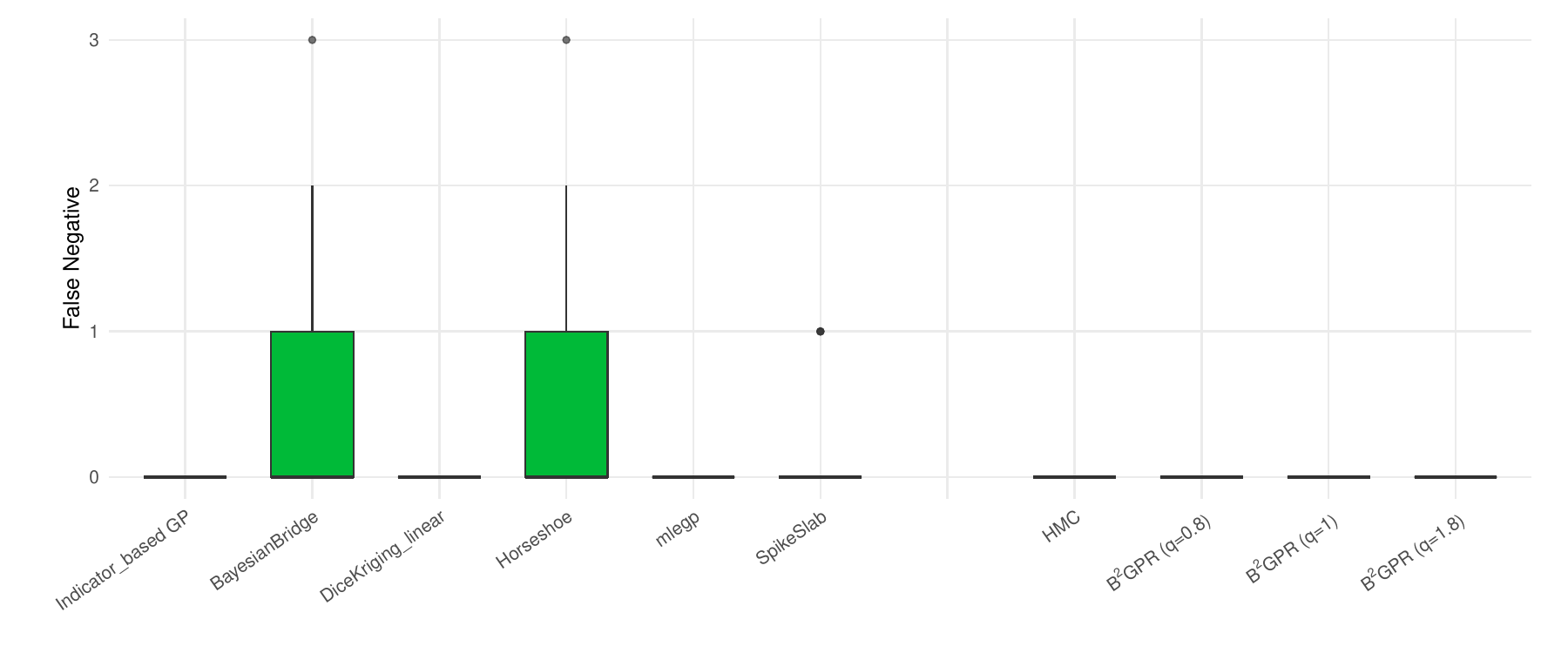}
    \caption{Number of false negatives.}
    \end{subfigure}
    \caption{Box plots of standardized RMSE and variable-selection errors across $100$ replicates for the pre-specified Gaussian process example with $n=200$.}\label{fig:ggplot_simulation2_200}
\end{figure}

\begin{figure}[p]
    \centering
    \begin{subfigure}[t]{.96\linewidth}
    \includegraphics[width=\linewidth]{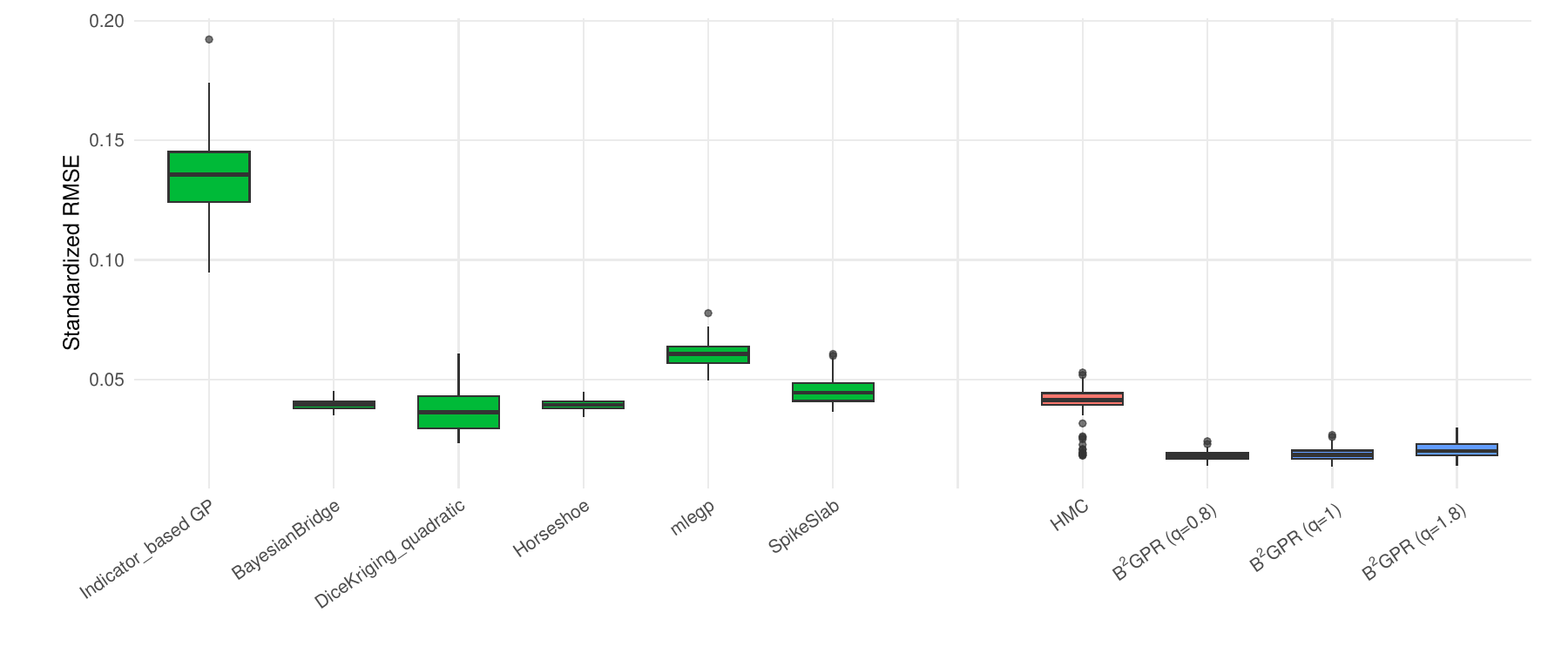}
    \caption{Standardized RMSE.}
    \end{subfigure}
    
    \begin{subfigure}[t]{.96\linewidth}
    \includegraphics[width=\linewidth]{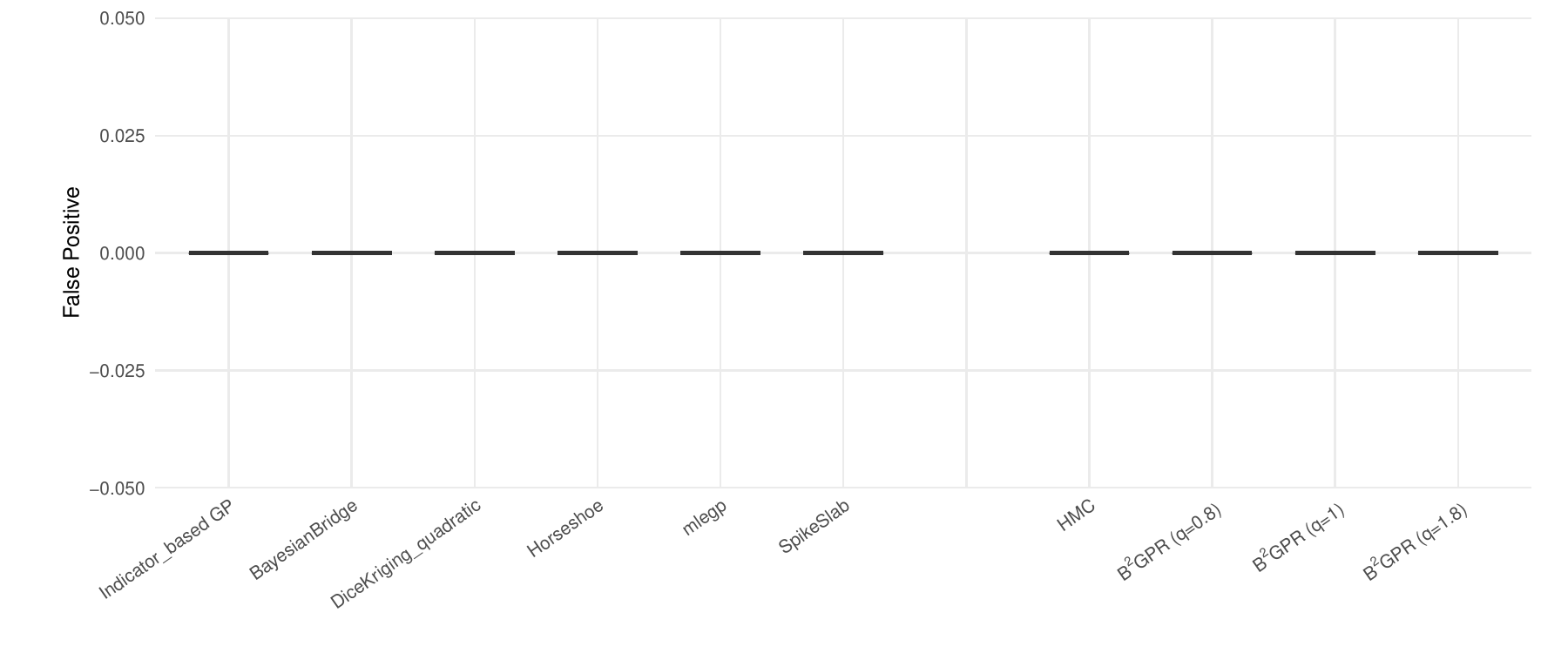}
    \caption{Number of false positives.}
    \end{subfigure}
    
    \begin{subfigure}[t]{.96\linewidth}
    \includegraphics[width=\linewidth]{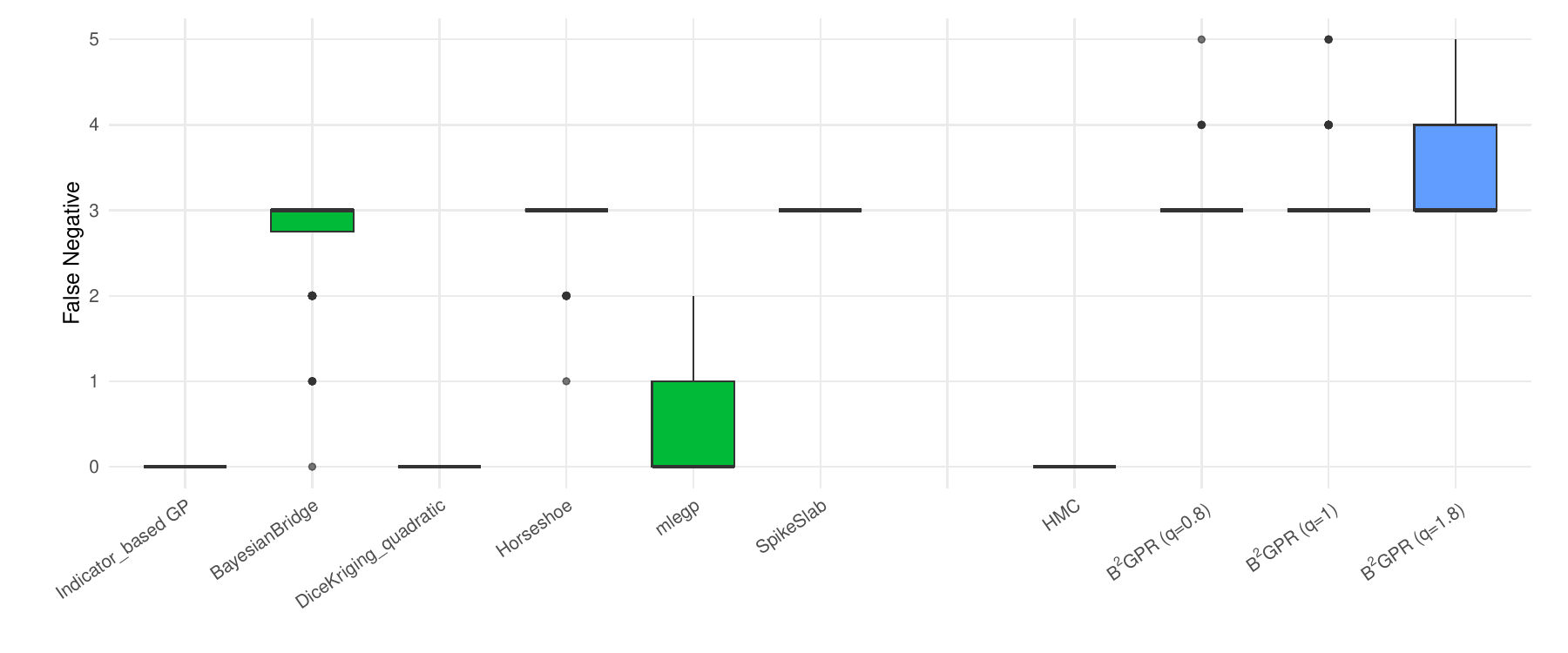}
    \caption{Number of false negatives.}
    \end{subfigure}
    \caption{Box plots of standardized RMSE and variable-selection errors across $100$ replicates for the Borehole function with $n=200$.}\label{fig:ggplot_borehole_200}
\end{figure}

\begin{figure}[p]
    \centering
    \begin{subfigure}[t]{.96\linewidth}
    \includegraphics[width=\linewidth]{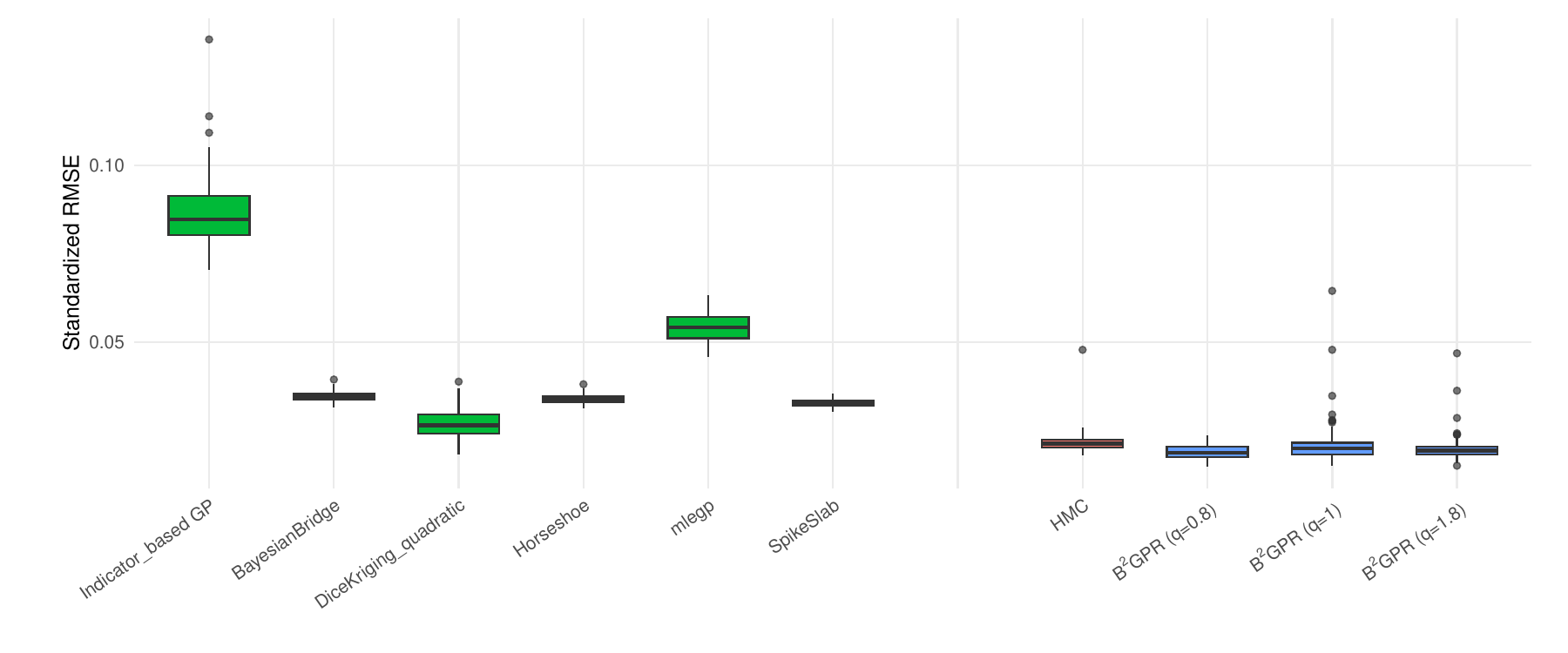}
    \caption{Standardized RMSE.}
    \end{subfigure}
    
    \begin{subfigure}[t]{.96\linewidth}
    \includegraphics[width=\linewidth]{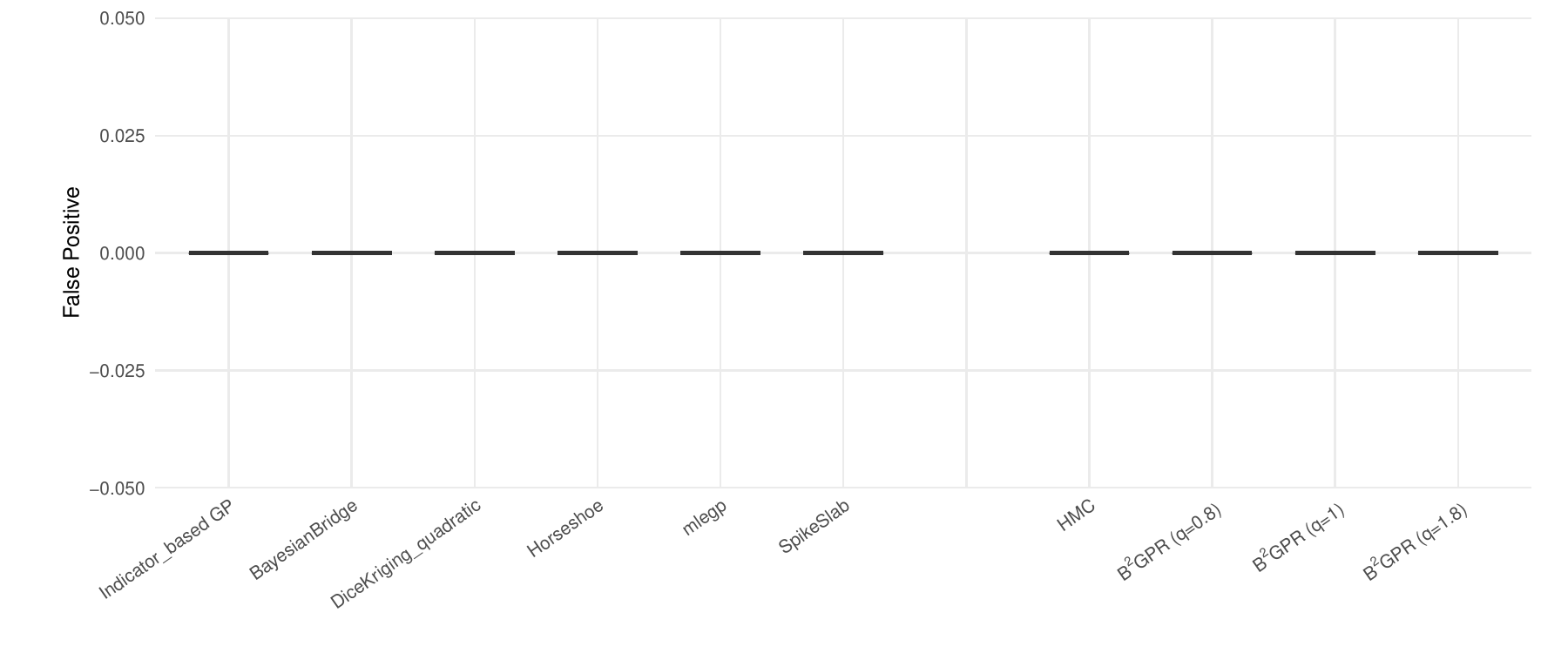}
    \caption{Number of false positives.}
    \end{subfigure}
    
    \begin{subfigure}[t]{.96\linewidth}
    \includegraphics[width=\linewidth]{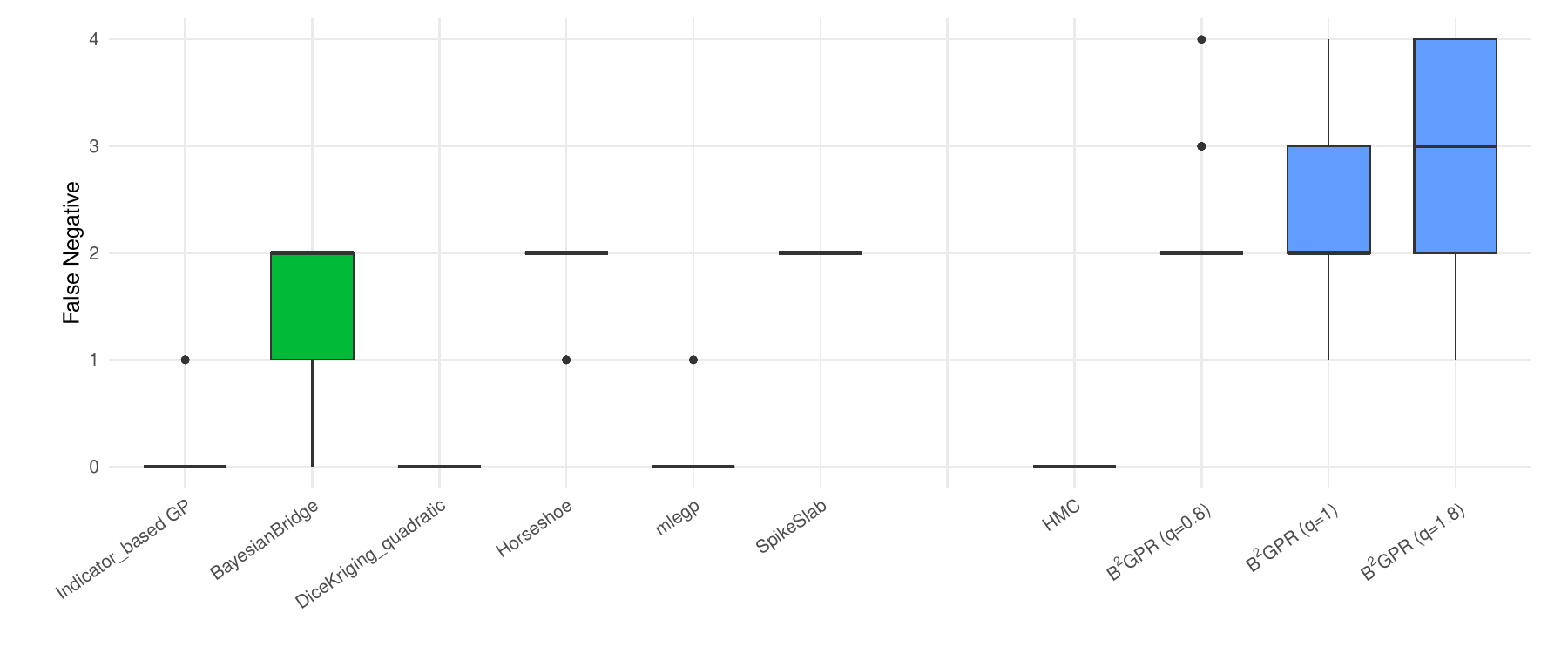}
    \caption{Number of false negatives.}
    \end{subfigure}
    \caption{Box plots of standardized RMSE and variable-selection errors across $100$ replicates for the OTL Circuit function with $n=200$.}\label{fig:ggplot_otlcircuit_200}
\end{figure}

\begin{figure}[p]
    \centering
    \begin{subfigure}[t]{.96\linewidth}
    \includegraphics[width=\linewidth]{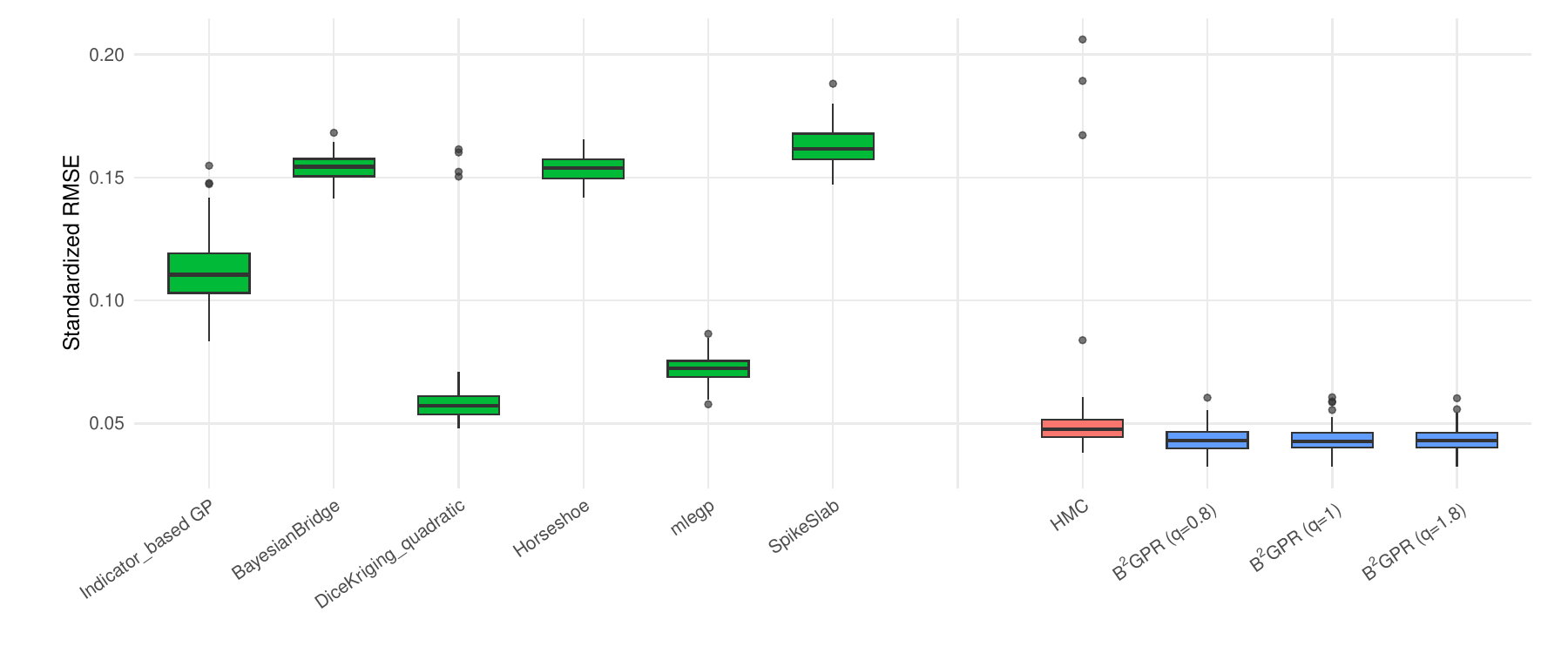}
    \caption{Standardized RMSE.}
    \end{subfigure}
    
    \begin{subfigure}[t]{.96\linewidth}
    \includegraphics[width=\linewidth]{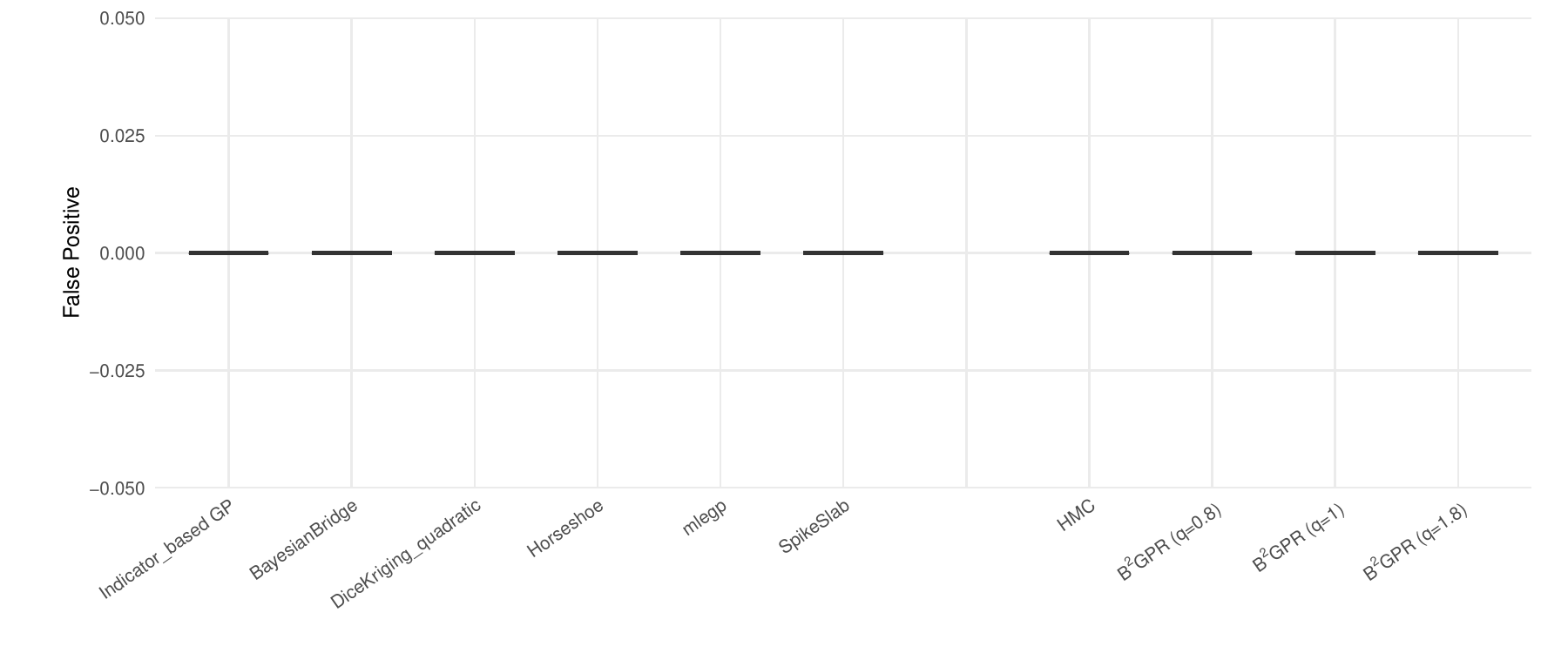}
    \caption{Number of false positives.}
    \end{subfigure}
    
    \begin{subfigure}[t]{.96\linewidth}
    \includegraphics[width=\linewidth]{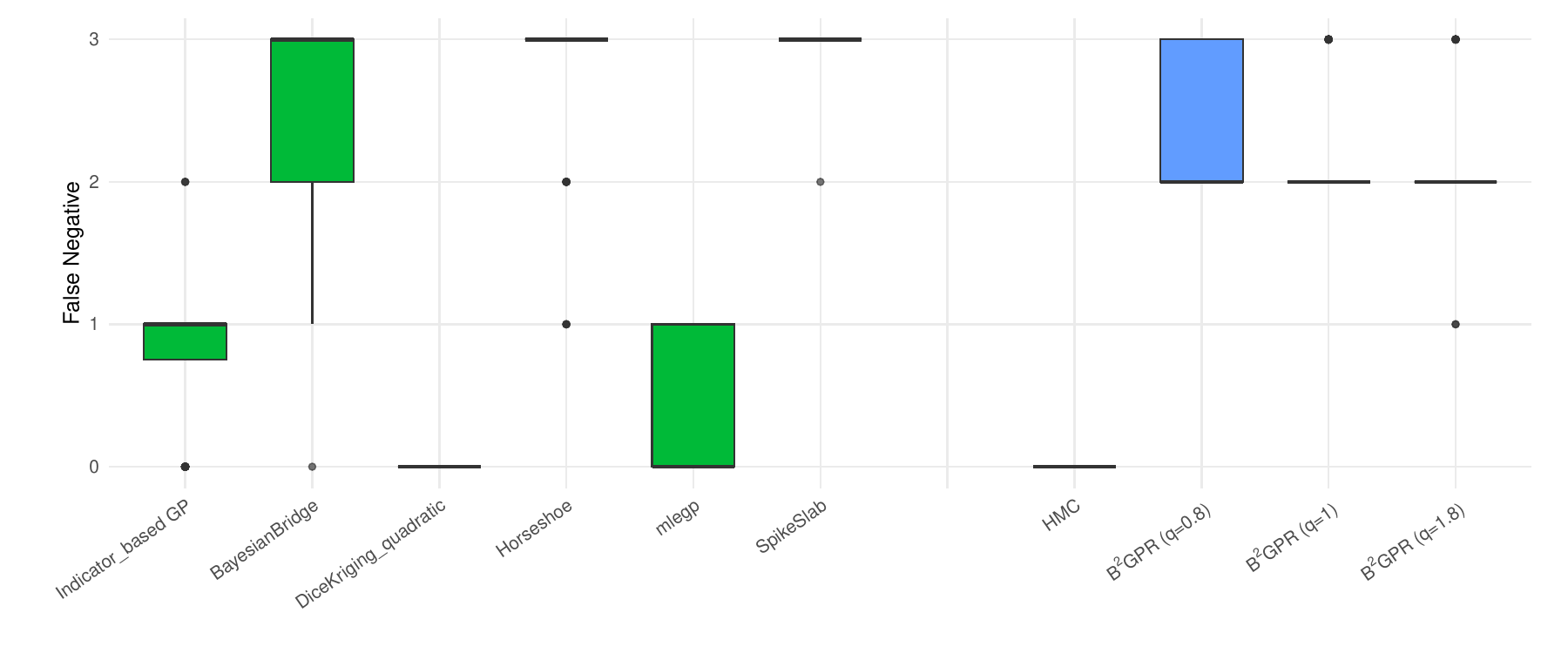}
    \caption{Number of false negatives.}
    \end{subfigure}
    \caption{Box plots of standardized RMSE and variable-selection errors across $100$ replicates for the Piston function with $n=200$.}\label{fig:ggplot_piston_200}
\end{figure}

\begin{figure}[p]
    \centering
    \begin{subfigure}[t]{.96\linewidth}
    \includegraphics[width=\linewidth]{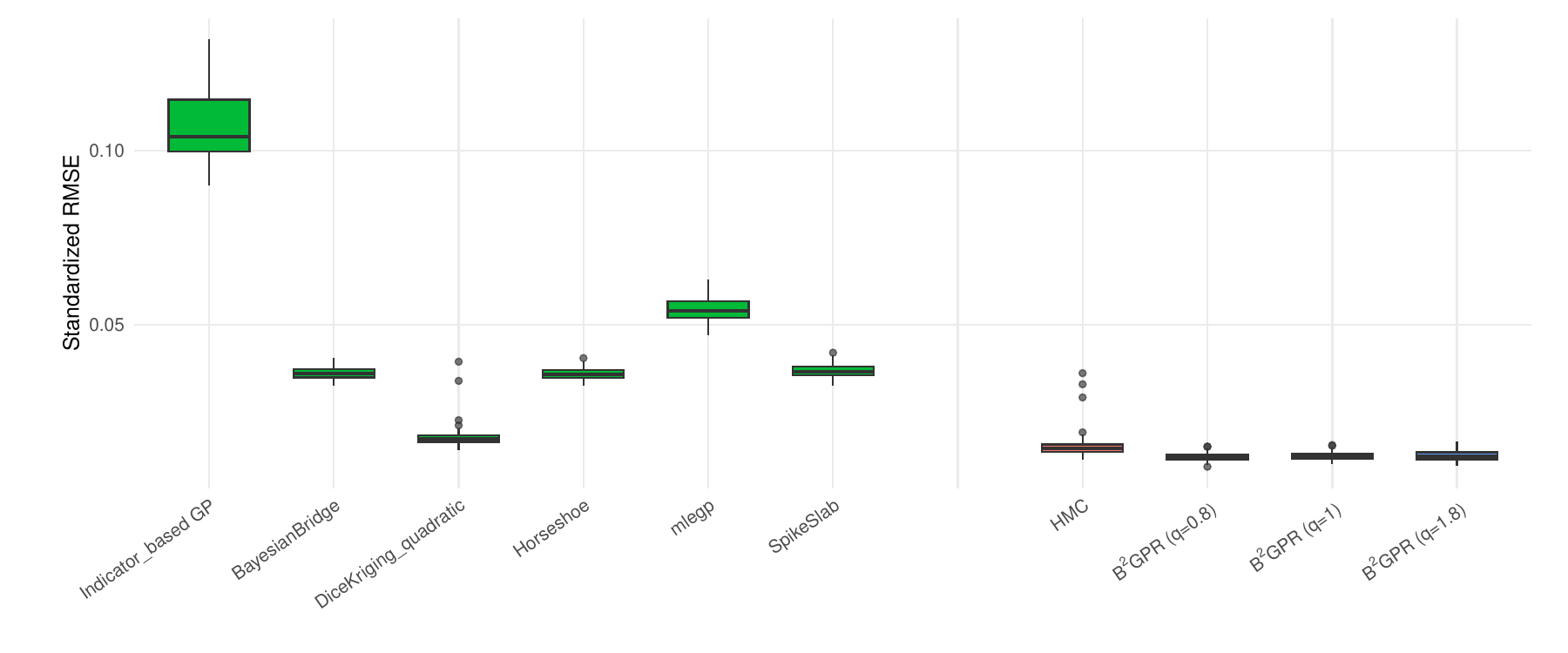}
    \caption{Standardized RMSE.}
    \end{subfigure}
    
    \begin{subfigure}[t]{.96\linewidth}
    \includegraphics[width=\linewidth]{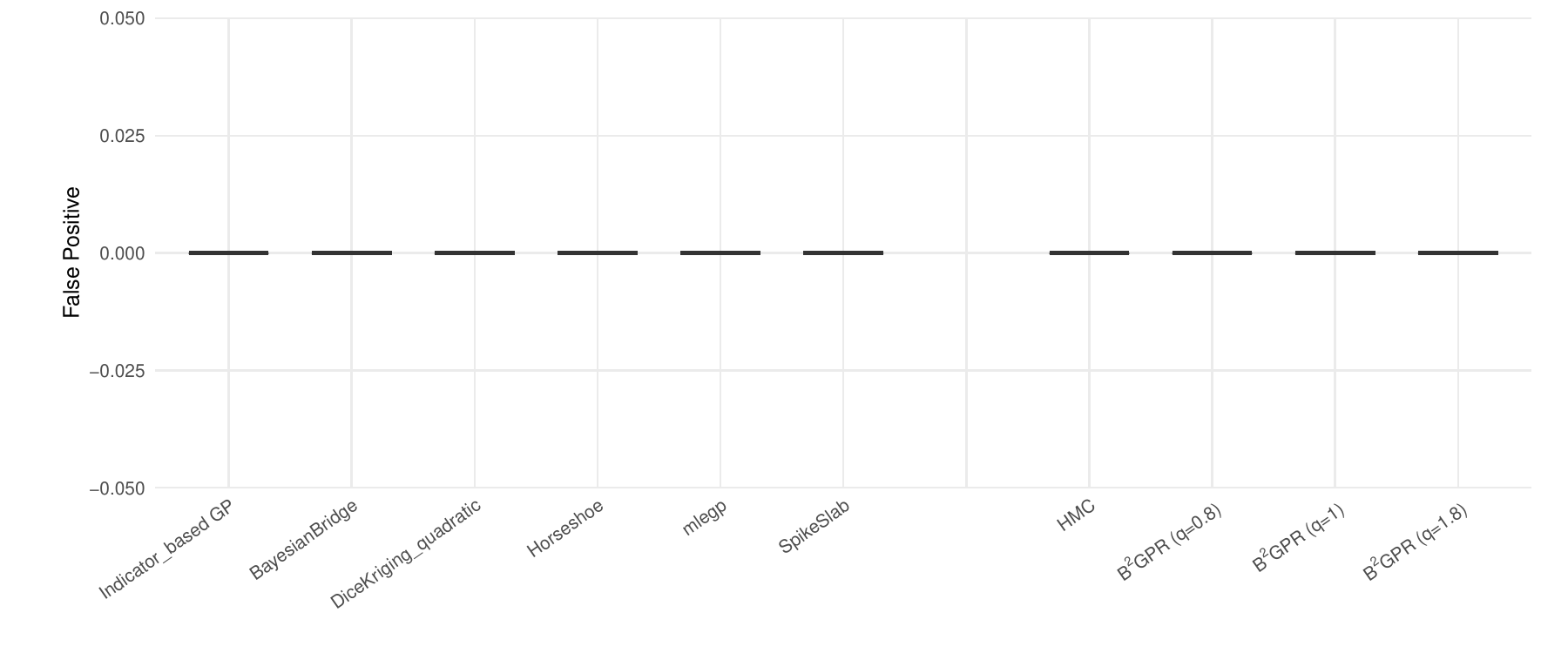}
    \caption{Number of false positives.}
    \end{subfigure}
    
    \begin{subfigure}[t]{.96\linewidth}
    \includegraphics[width=\linewidth]{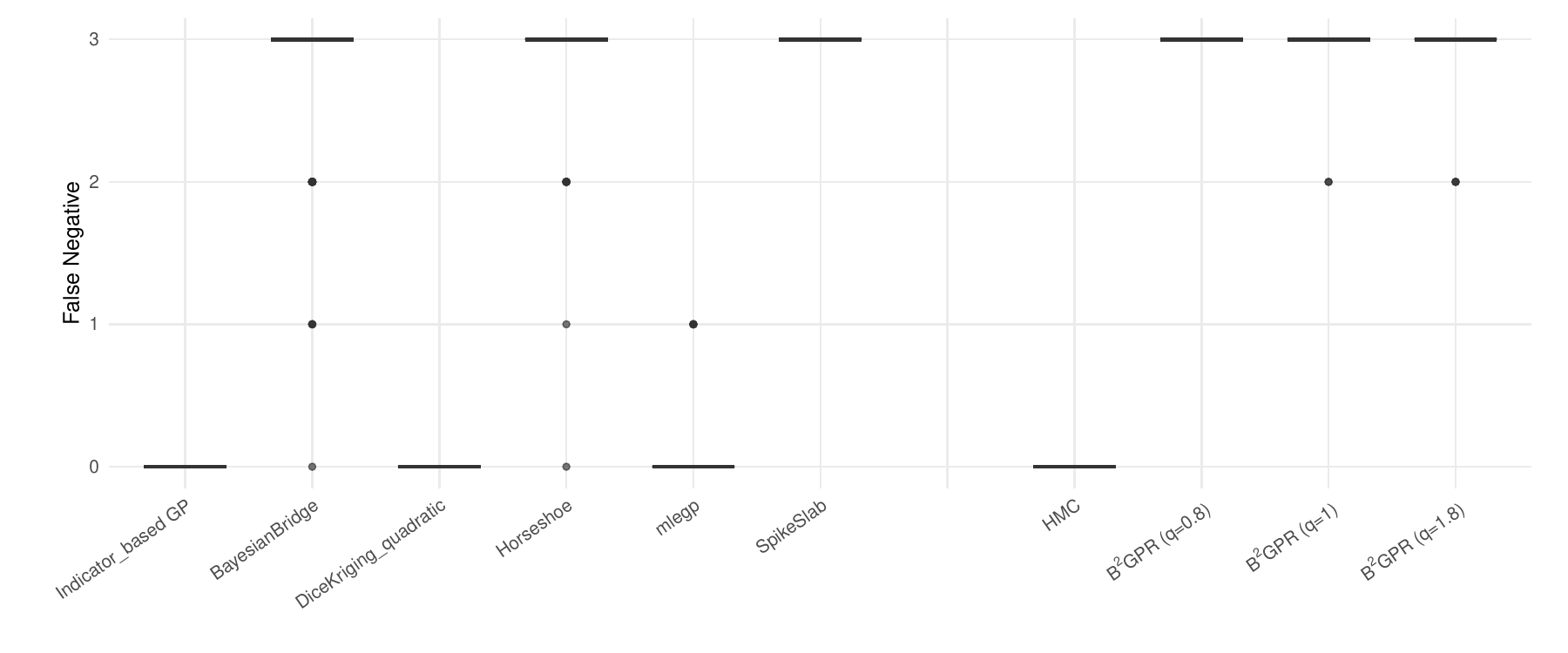}
    \caption{Number of false negatives.}
    \end{subfigure}
    \caption{Box plots of standardized RMSE and variable-selection errors across $100$ replicates for the Borehole function with $n=500$.}\label{fig:ggplot_borehole_500}
\end{figure}

\begin{figure}[p]
    \centering
    \begin{subfigure}[t]{.96\linewidth}
    \includegraphics[width=\linewidth]{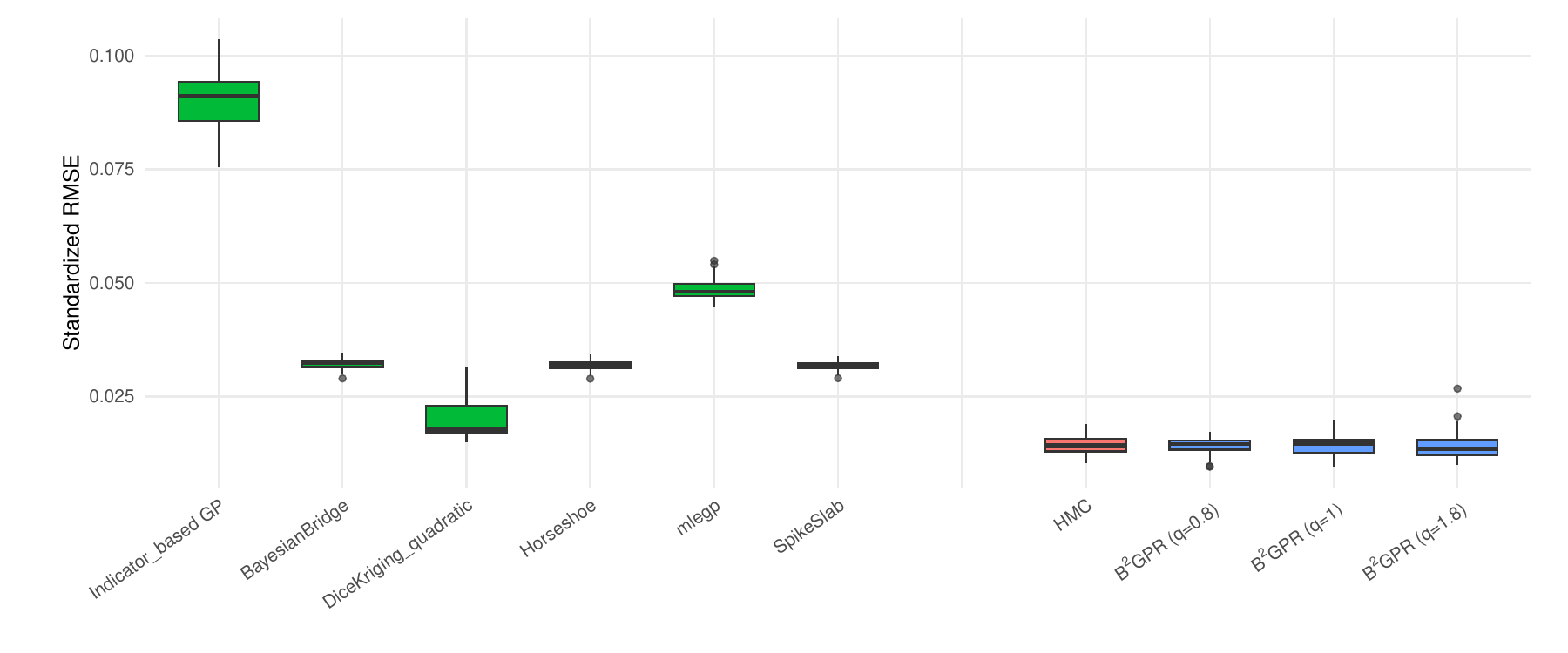}
    \caption{Standardized RMSE.}
    \end{subfigure}
    
    \begin{subfigure}[t]{.96\linewidth}
    \includegraphics[width=\linewidth]{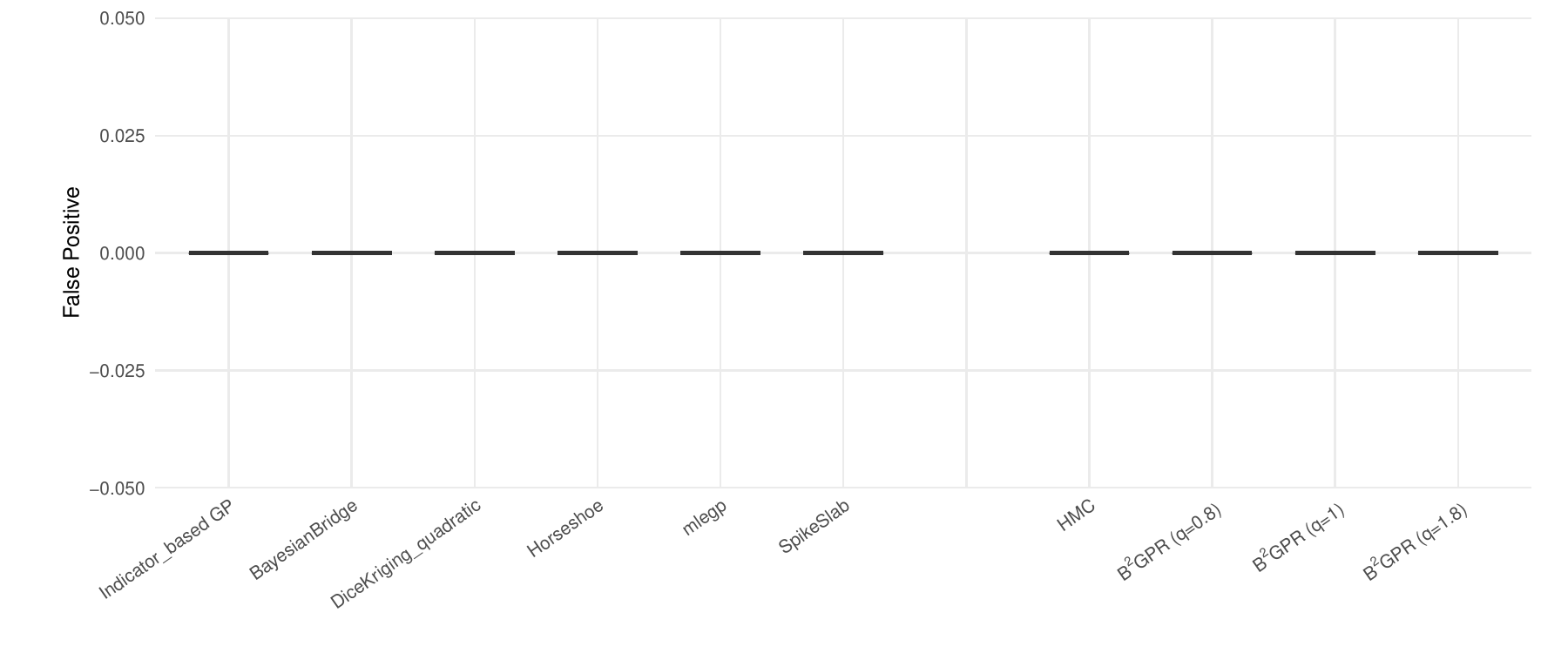}
    \caption{Number of false positives.}
    \end{subfigure}
    
    \begin{subfigure}[t]{.96\linewidth}
    \includegraphics[width=\linewidth]{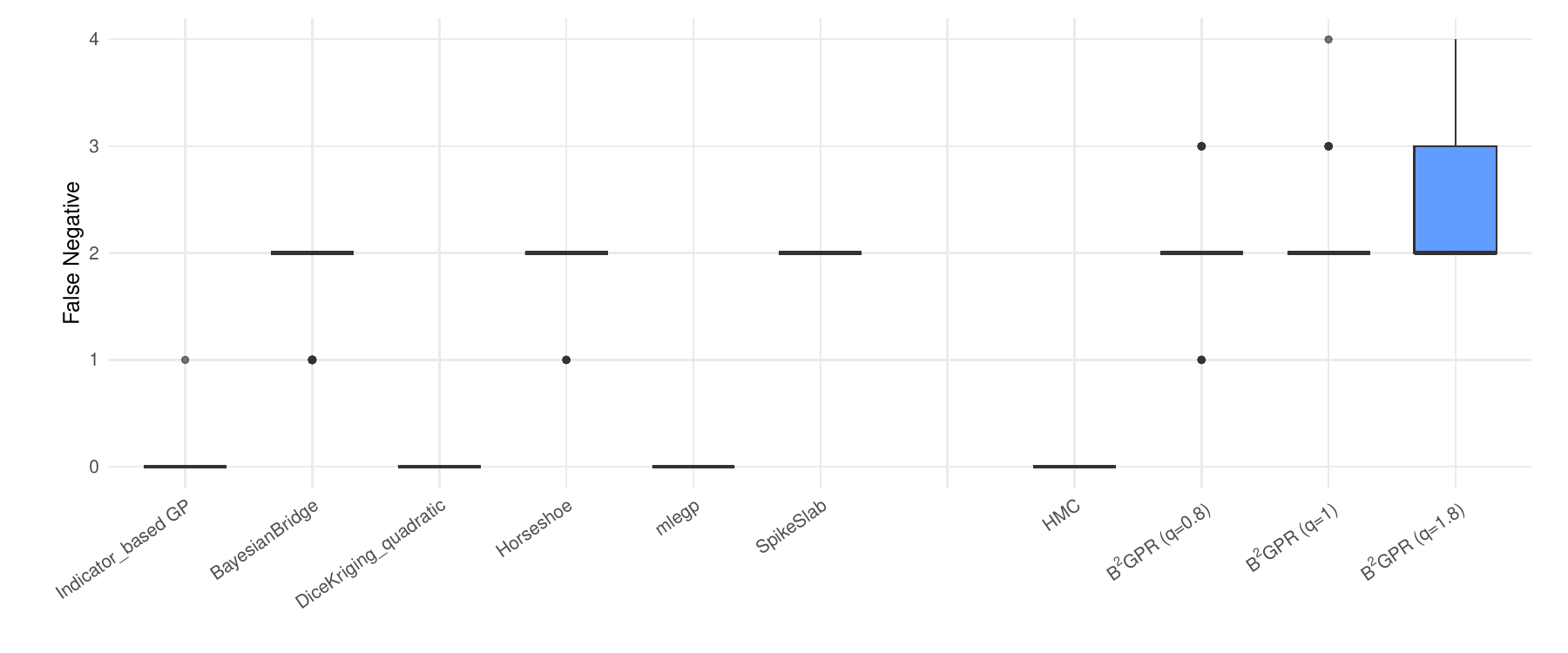}
    \caption{Number of false negatives.}
    \end{subfigure}
    \caption{Box plots of standardized RMSE and variable-selection errors across $100$ replicates for the OTL Circuit function with $n=500$.}\label{fig:ggplot_otlcircuit_500}
\end{figure}

\begin{figure}[p]
    \centering
    \begin{subfigure}[t]{.96\linewidth}
    \includegraphics[width=\linewidth]{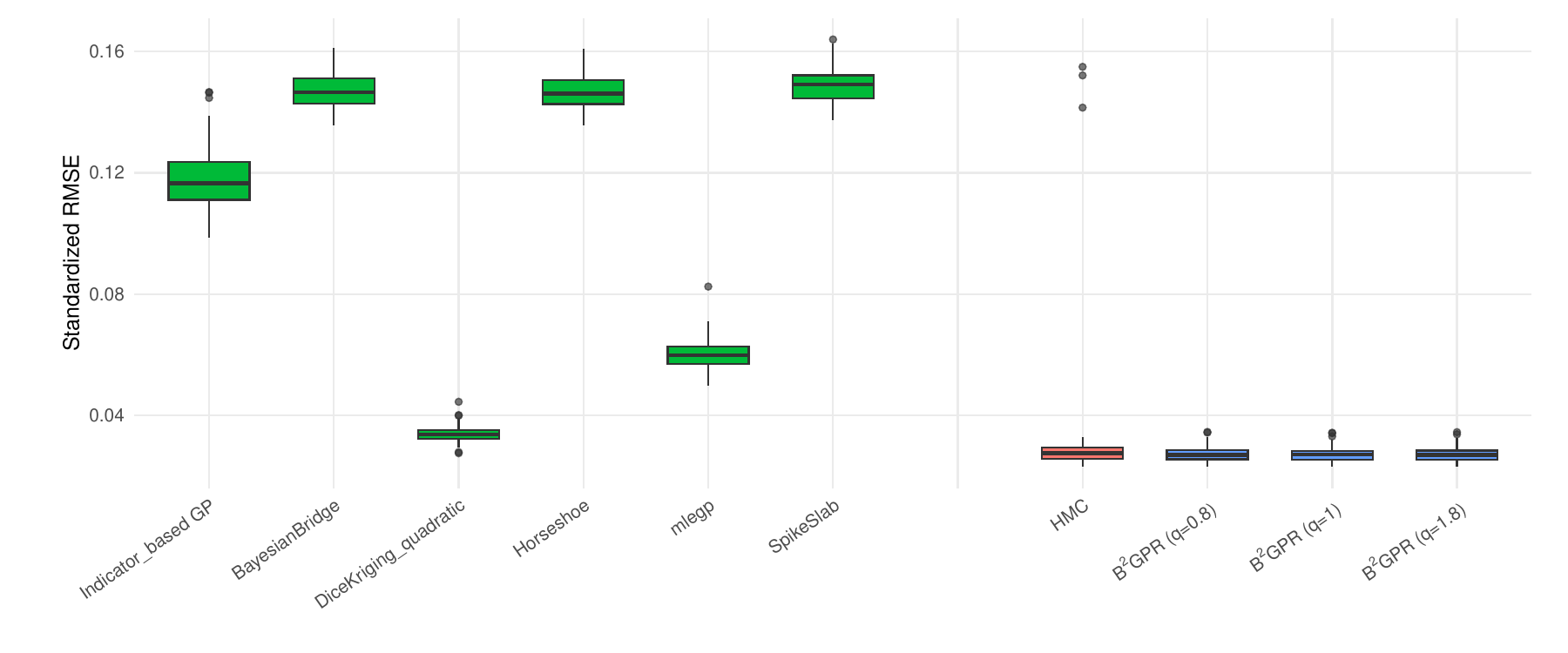}
    \caption{Standardized RMSE.}
    \end{subfigure}
    
    \begin{subfigure}[t]{.96\linewidth}
    \includegraphics[width=\linewidth]{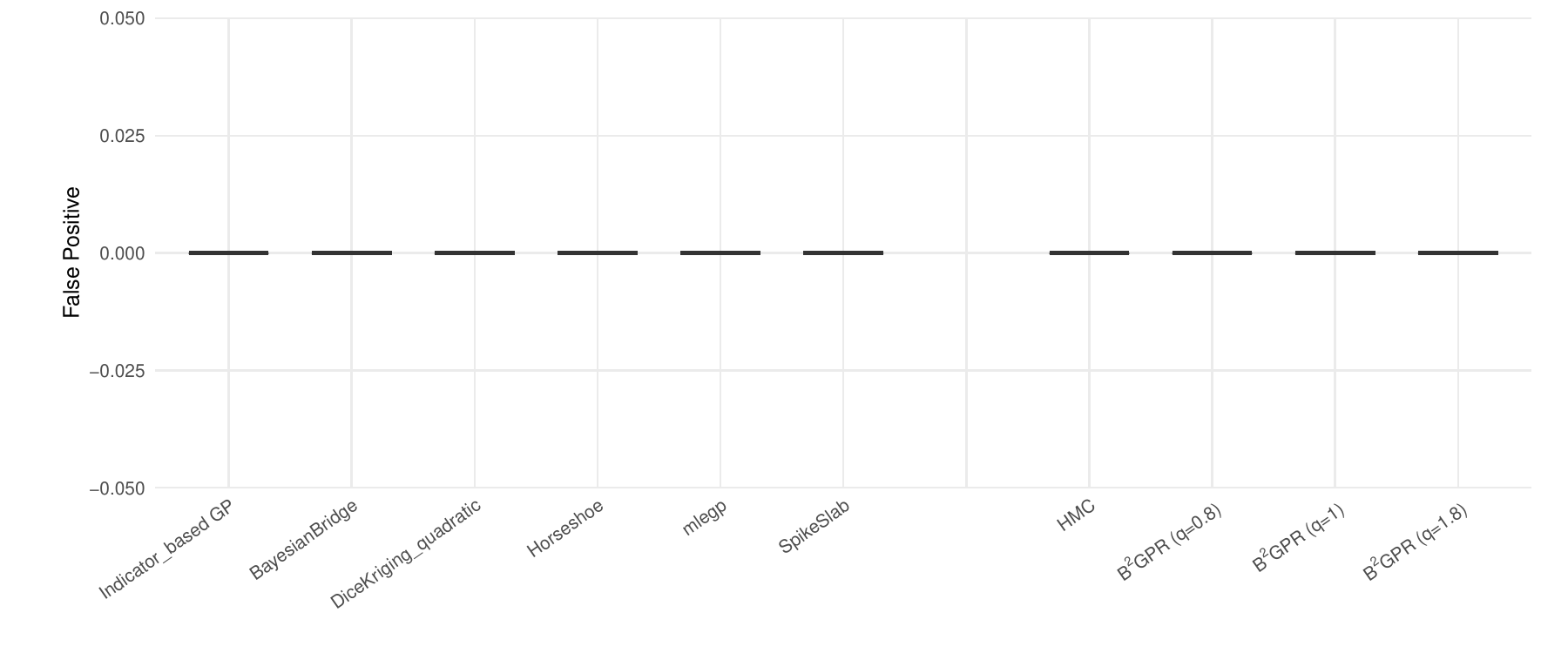}
    \caption{Number of false positives.}
    \end{subfigure}
    
    \begin{subfigure}[t]{.96\linewidth}
    \includegraphics[width=\linewidth]{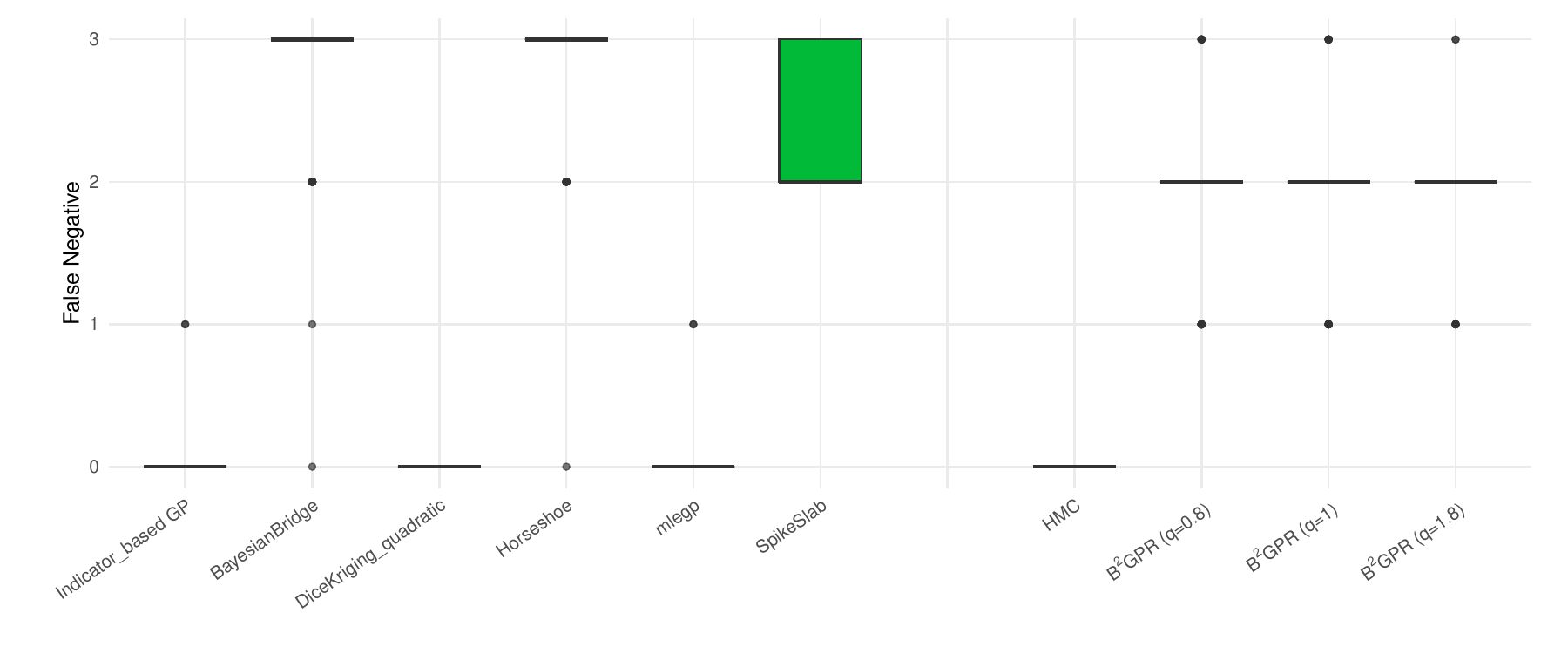}
    \caption{Number of false negatives.}
    \end{subfigure}
    \caption{Box plots of standardized RMSE and variable-selection errors across $100$ replicates for the Piston function with $n=500$.}\label{fig:ggplot_piston_500}
\end{figure}

\begin{figure}[p]
    \centering
    \begin{subfigure}[t]{.96\linewidth}
    \includegraphics[width=\linewidth]{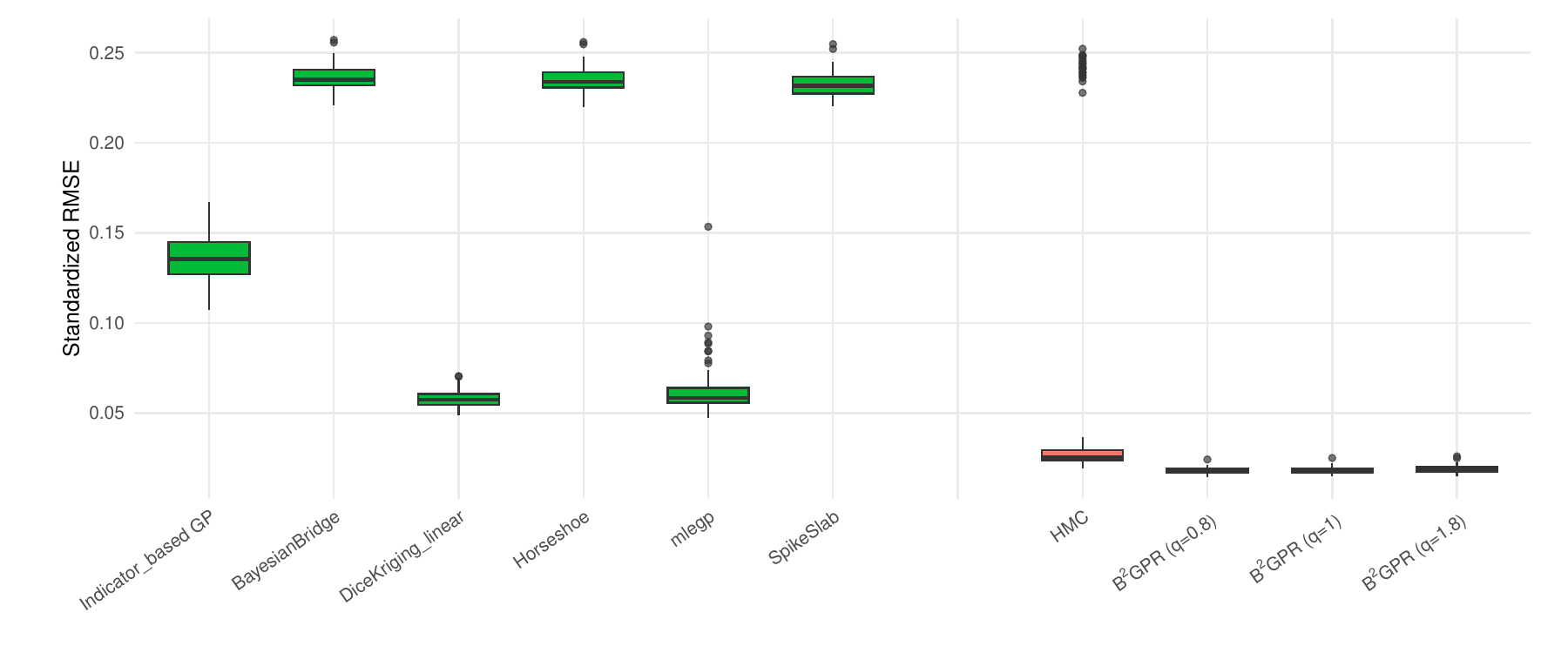}
    \caption{Standardized RMSE.}
    \end{subfigure}
    
    \begin{subfigure}[t]{.96\linewidth}
    \includegraphics[width=\linewidth]{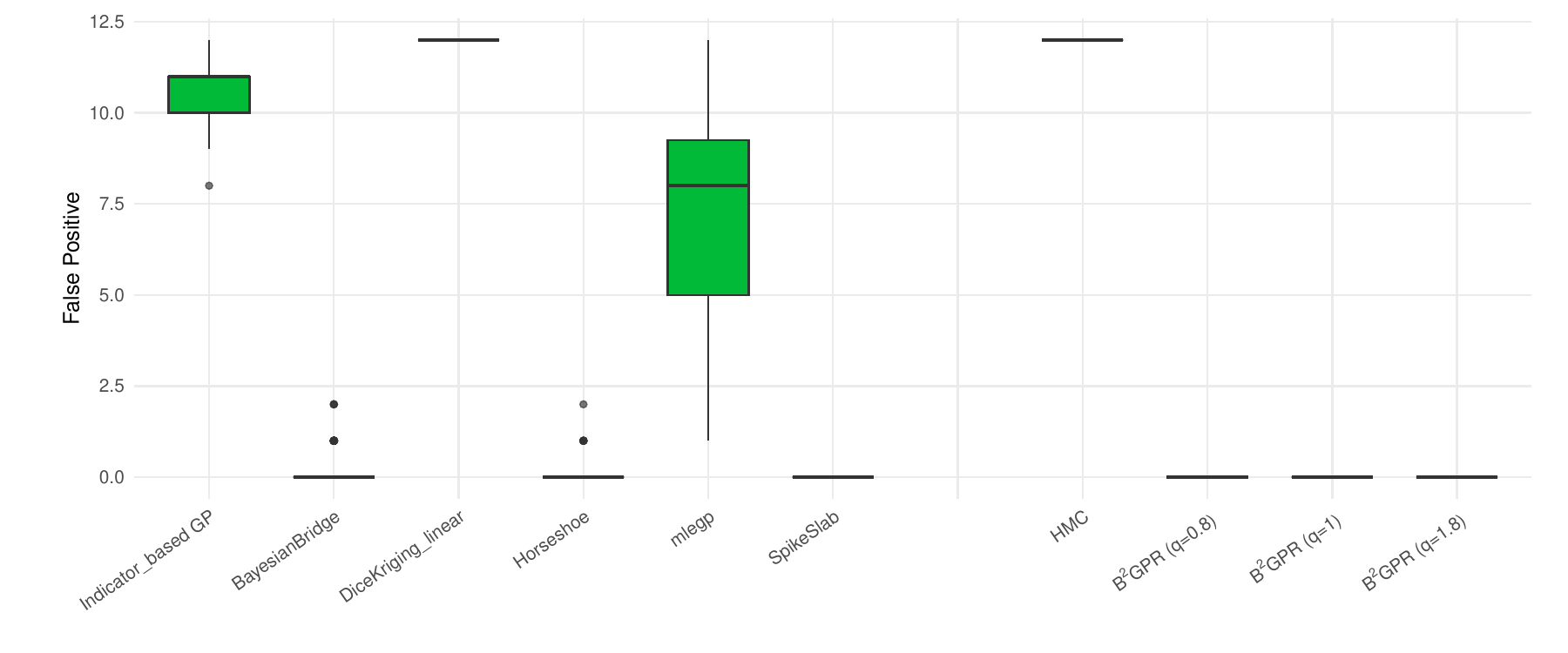}
    \caption{Number of false positives.}
    \end{subfigure}
    
    \begin{subfigure}[t]{.96\linewidth}
    \includegraphics[width=\linewidth]{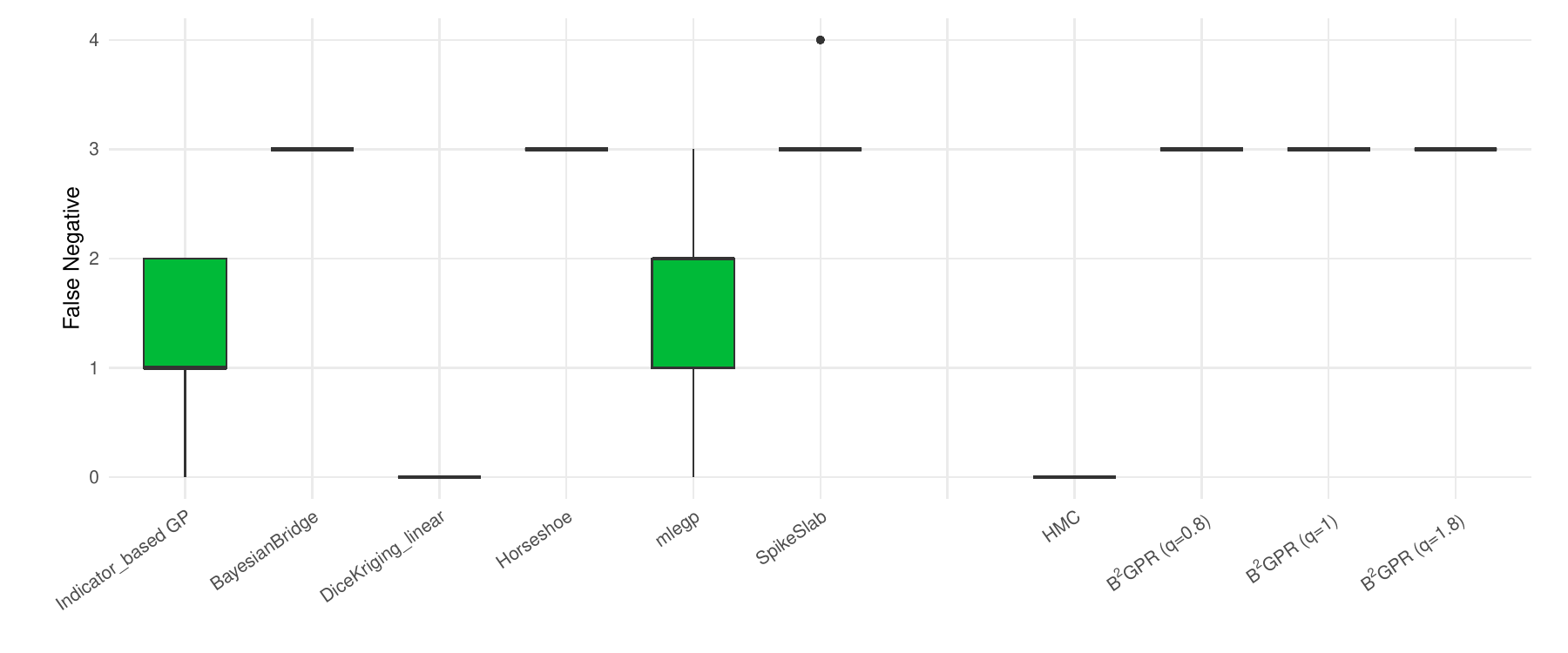}
    \caption{Number of false negatives.}
    \end{subfigure}
    \caption{Box plots of standardized RMSE and variable-selection errors across $100$ replicates for the Borehole function with increased input dimension $d=20$ and $n=200$.}\label{fig:ggplot_borehole_200_20}
\end{figure}

\begin{figure}[p]
    \centering
    \begin{subfigure}[t]{.96\linewidth}
    \includegraphics[width=\linewidth]{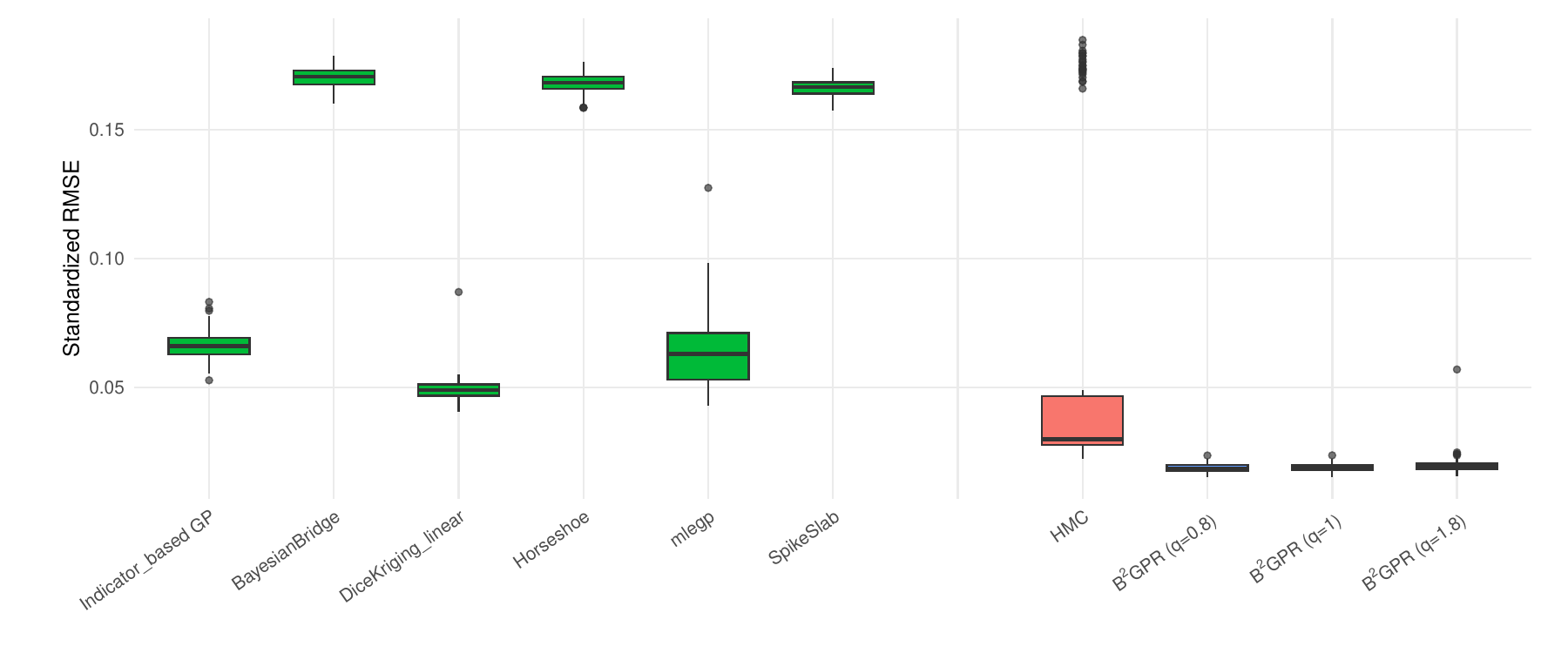}
    \caption{Standardized RMSE.}
    \end{subfigure}
    
    \begin{subfigure}[t]{.96\linewidth}
    \includegraphics[width=\linewidth]{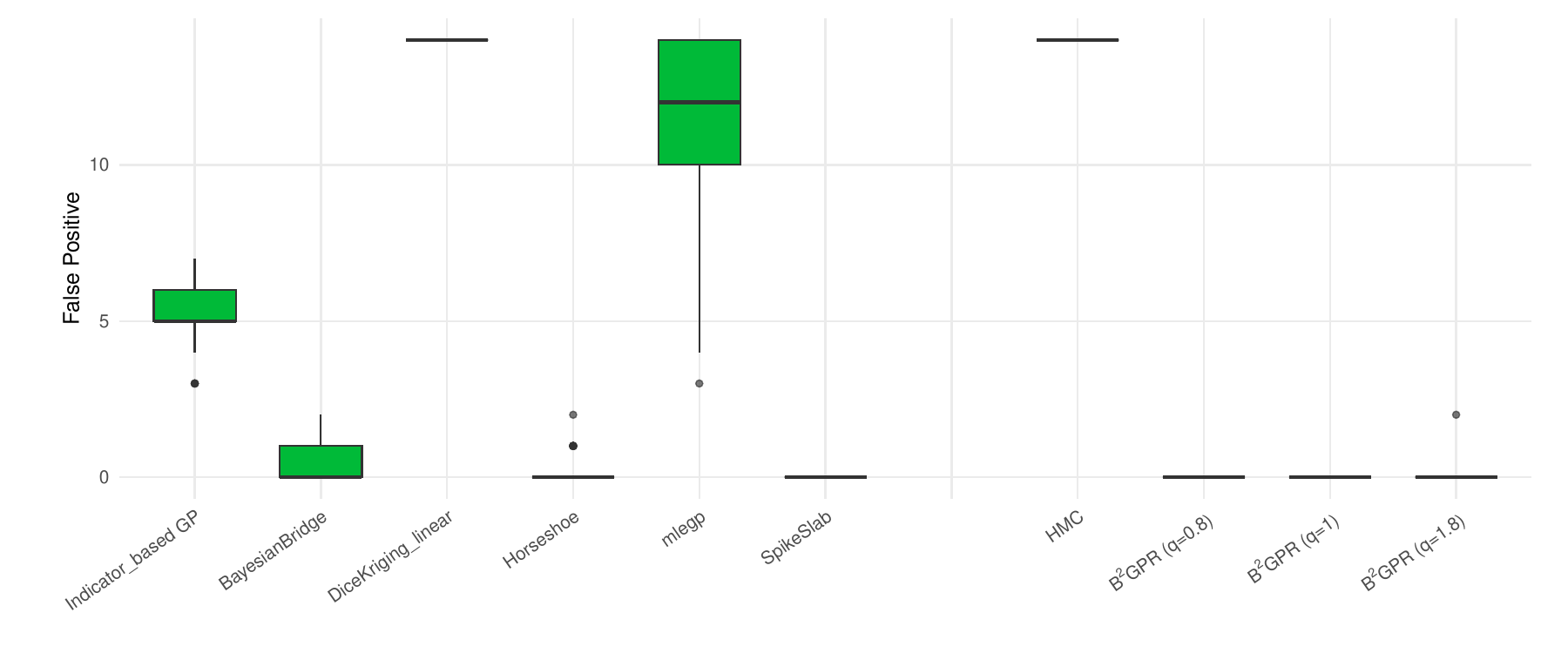}
    \caption{Number of false positives.}
    \end{subfigure}
    
    \begin{subfigure}[t]{.96\linewidth}
    \includegraphics[width=\linewidth]{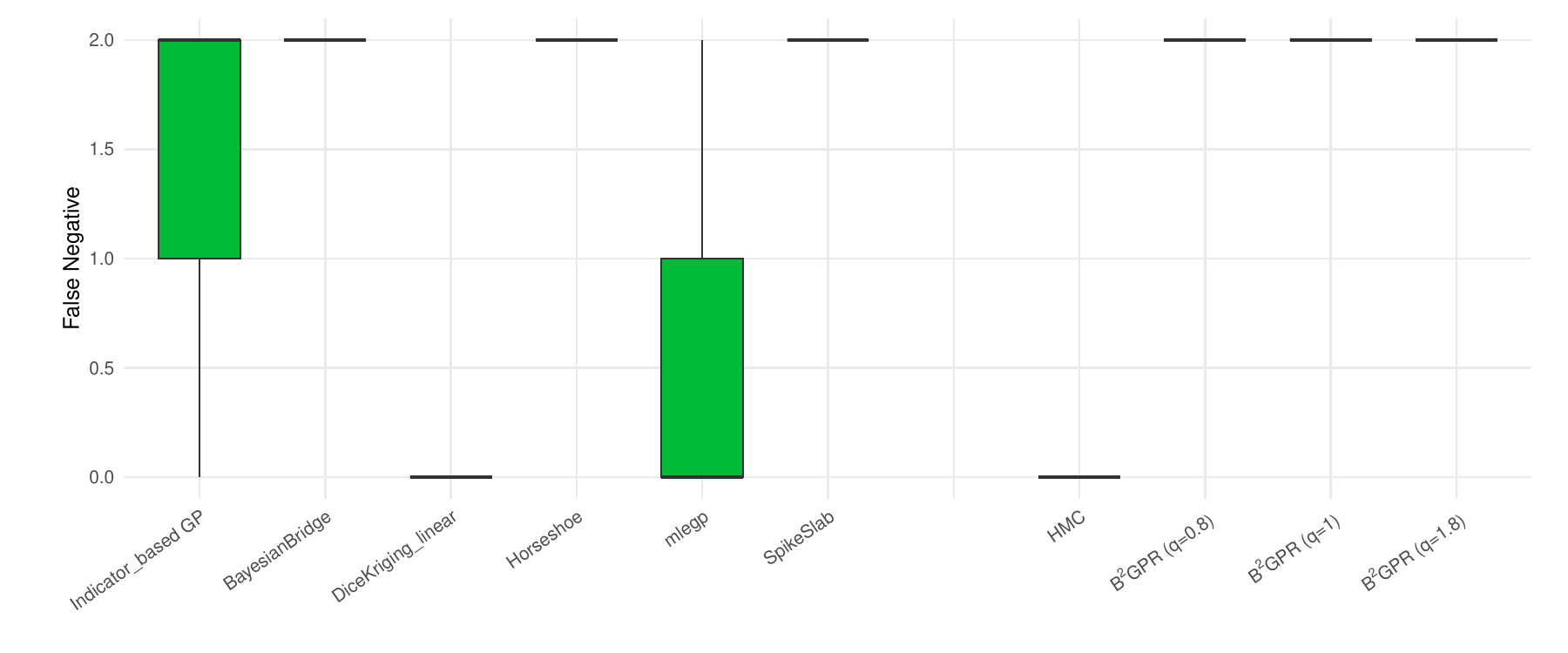}
    \caption{Number of false negatives.}
    \end{subfigure}
    \caption{Box plots of standardized RMSE and variable-selection errors across $100$ replicates for the OTL Circuit function with increased input dimension $d=20$ and $n=200$.}\label{fig:ggplot_otlcircuit_200_20}
\end{figure}

\begin{figure}[p]
    \centering
    \begin{subfigure}[t]{.96\linewidth}
    \includegraphics[width=\linewidth]{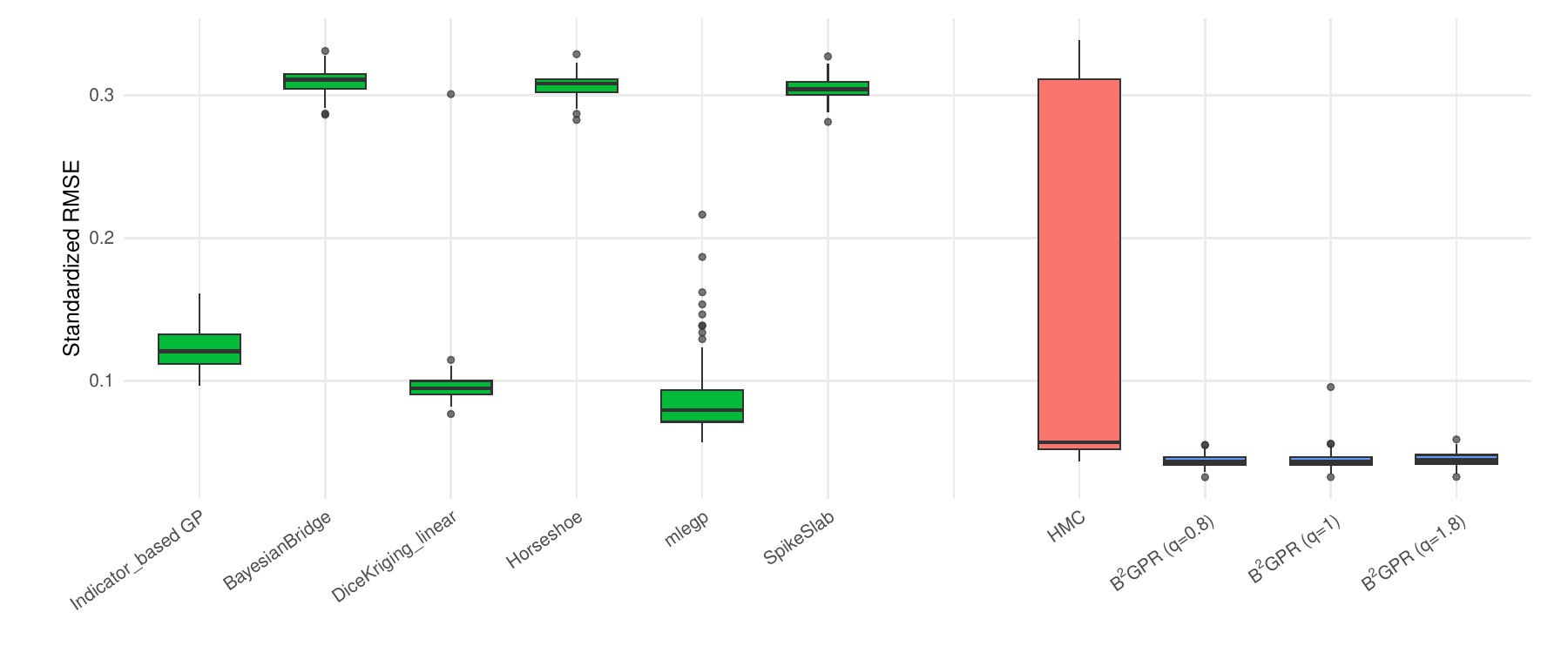}
    \caption{Standardized RMSE.}
    \end{subfigure}
    
    \begin{subfigure}[t]{.96\linewidth}
    \includegraphics[width=\linewidth]{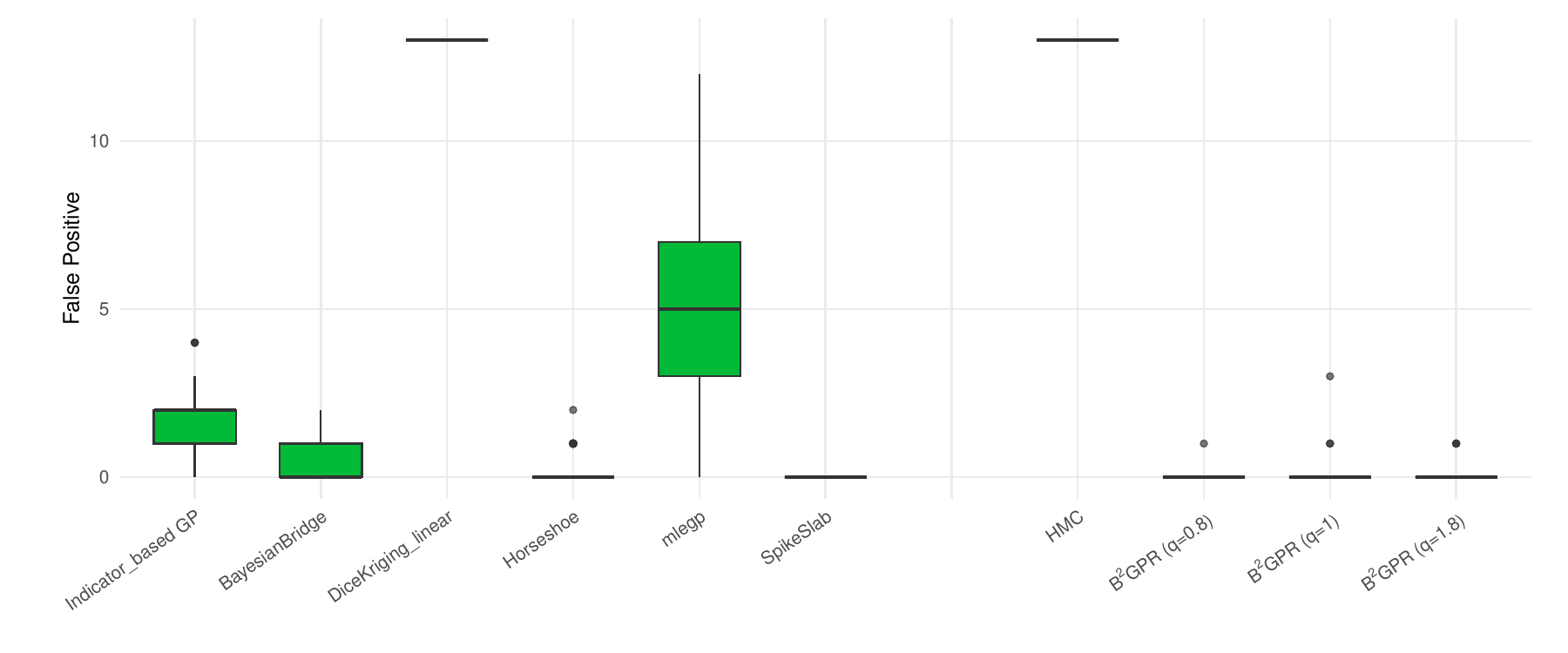}
    \caption{Number of false positives.}
    \end{subfigure}
    
    \begin{subfigure}[t]{.96\linewidth}
    \includegraphics[width=\linewidth]{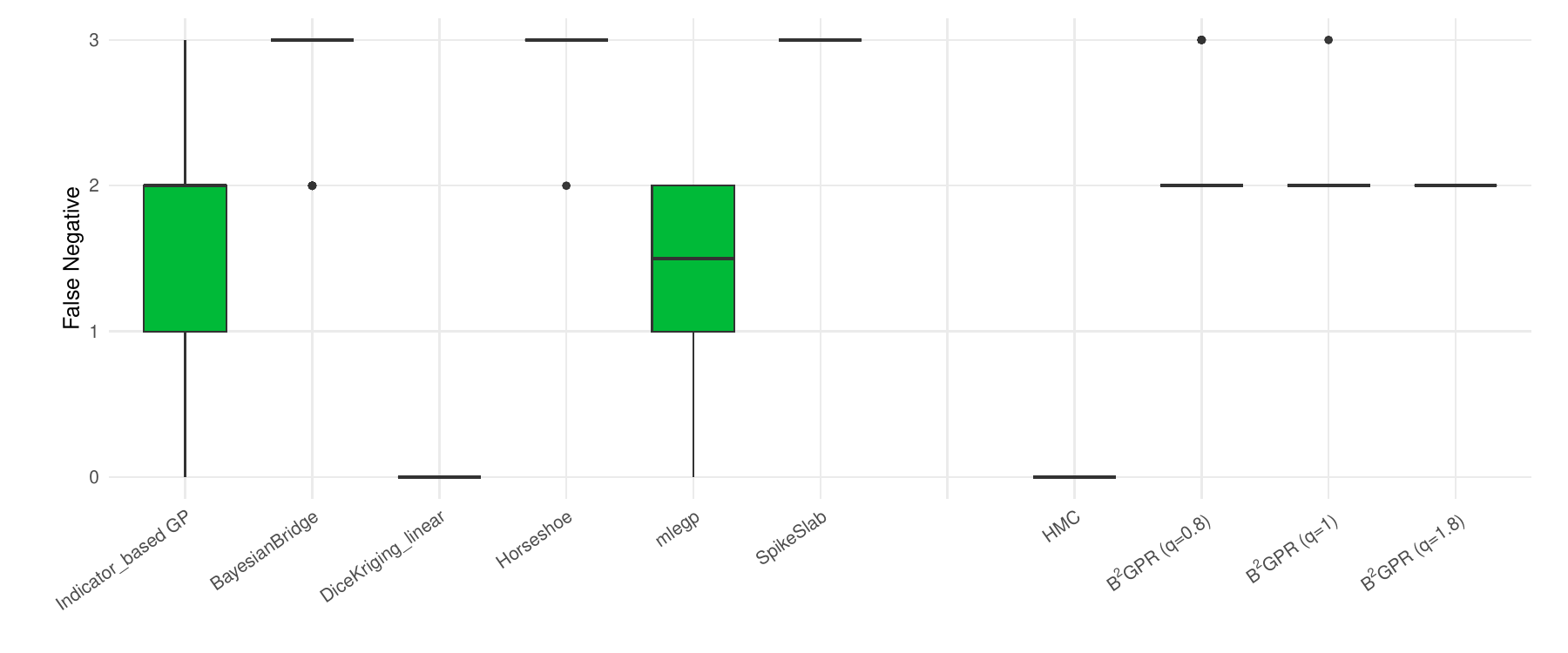}
    \caption{Number of false negatives.}
    \end{subfigure}
    \caption{Box plots of standardized RMSE and variable-selection errors across $100$ replicates for the Piston function with increased input dimension $d=20$ and $n=200$.}\label{fig:ggplot_piston_200_20}
\end{figure}

\begin{figure}[p]
    \centering
    \begin{subfigure}[t]{.96\linewidth}
    \includegraphics[width=\linewidth]{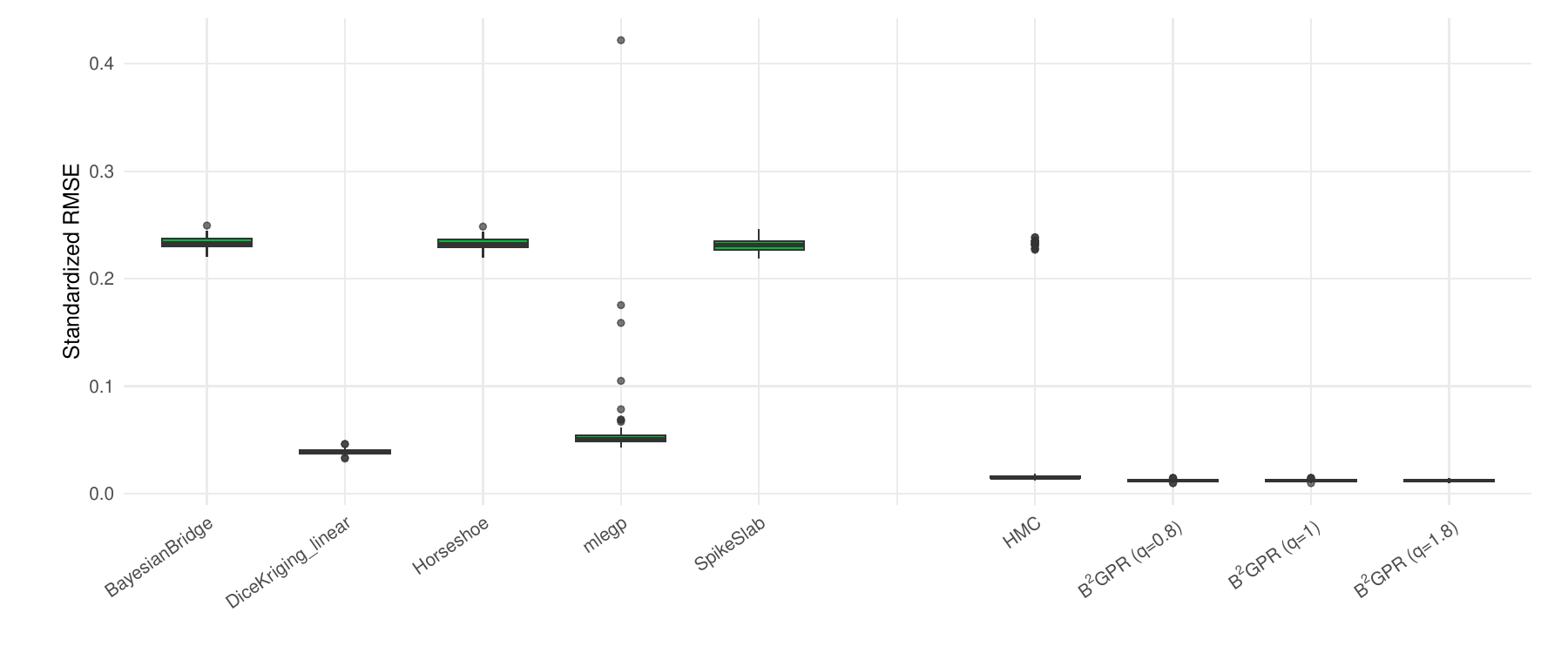}
    \caption{Standardized RMSE.}
    \end{subfigure}
    
    \begin{subfigure}[t]{.96\linewidth}
    \includegraphics[width=\linewidth]{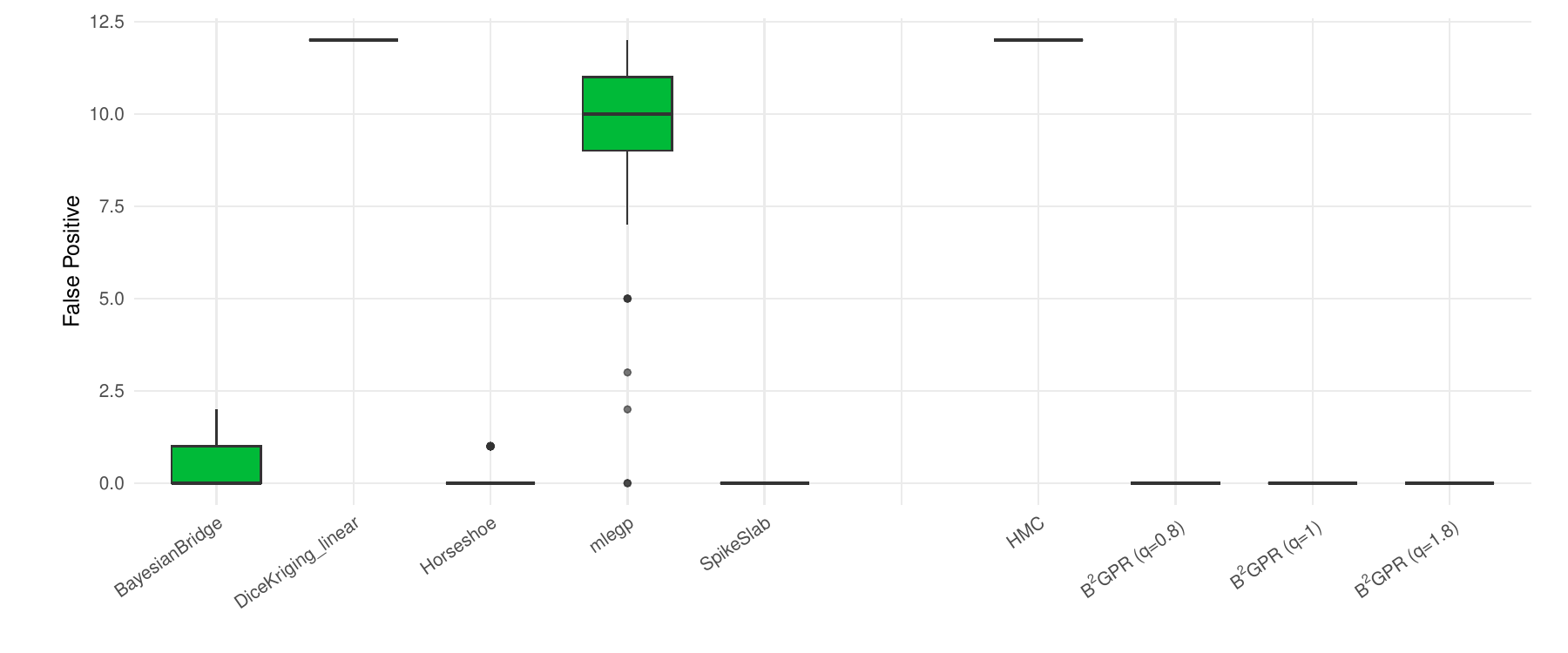}
    \caption{Number of false positives.}
    \end{subfigure}
    
    \begin{subfigure}[t]{.96\linewidth}
    \includegraphics[width=\linewidth]{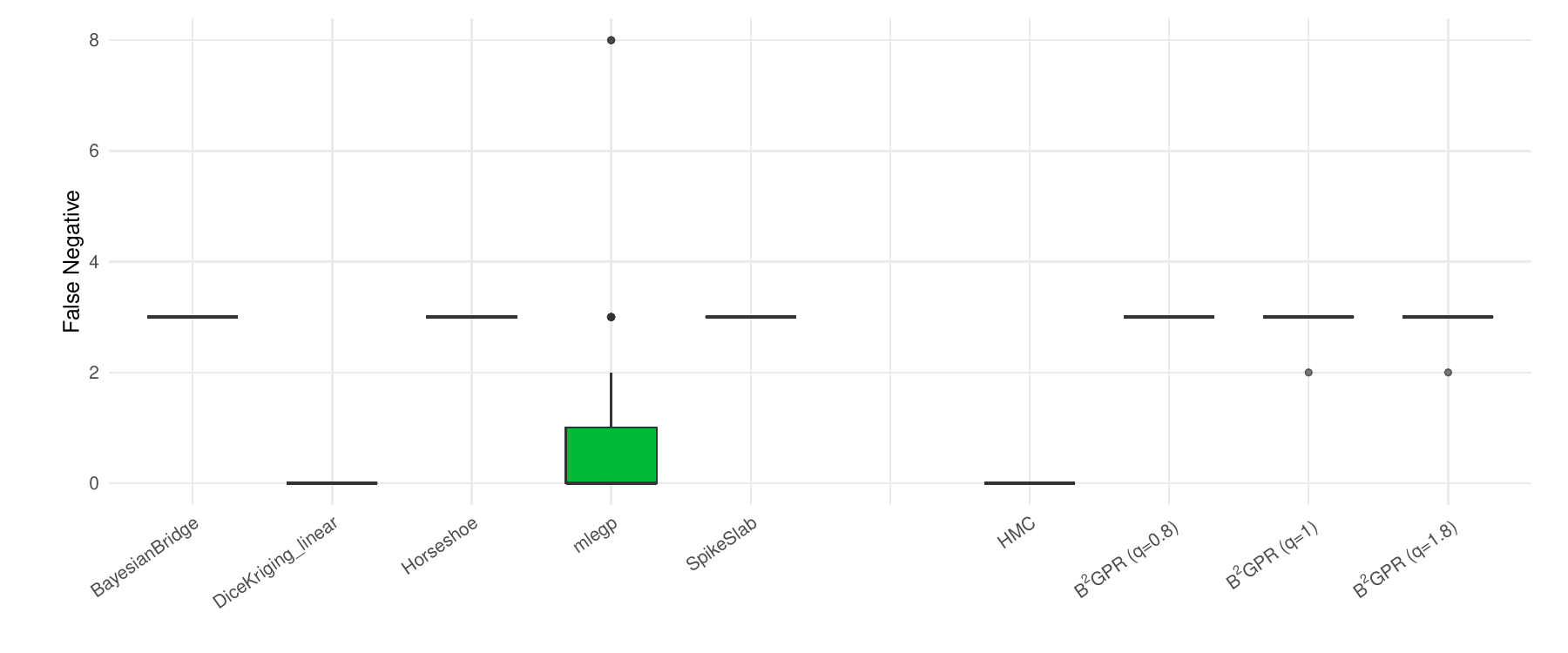}
    \caption{Number of false negatives.}
    \end{subfigure}
    \caption{Box plots of standardized RMSE and variable-selection errors across $100$ replicates for the Borehole function with increased input dimension $d=20$ and $n=500$.}\label{fig:ggplot_borehole_500_20}
\end{figure}

\begin{figure}[p]
    \centering
    \begin{subfigure}[t]{.96\linewidth}
    \includegraphics[width=\linewidth]{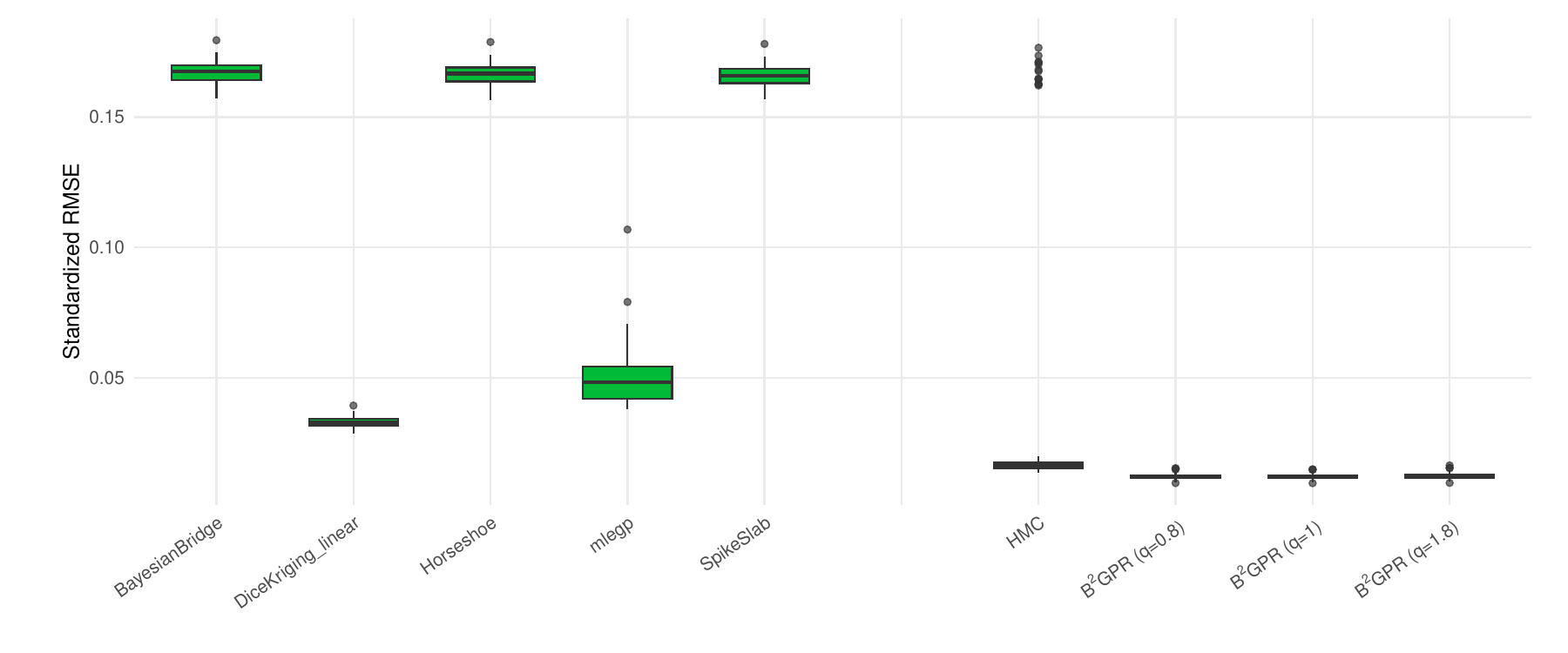}
    \caption{Standardized RMSE.}
    \end{subfigure}
    
    \begin{subfigure}[t]{.96\linewidth}
    \includegraphics[width=\linewidth]{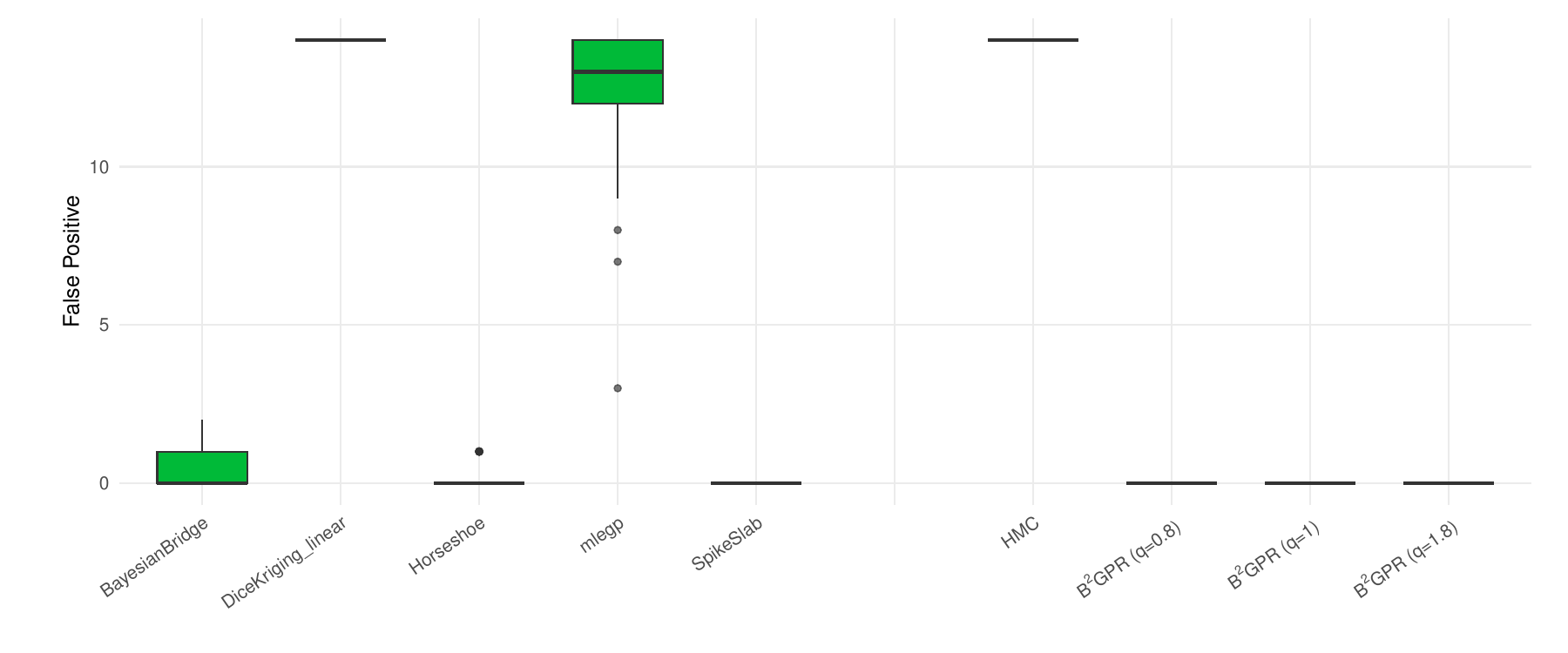}
    \caption{Number of false positives.}
    \end{subfigure}
    
    \begin{subfigure}[t]{.96\linewidth}
    \includegraphics[width=\linewidth]{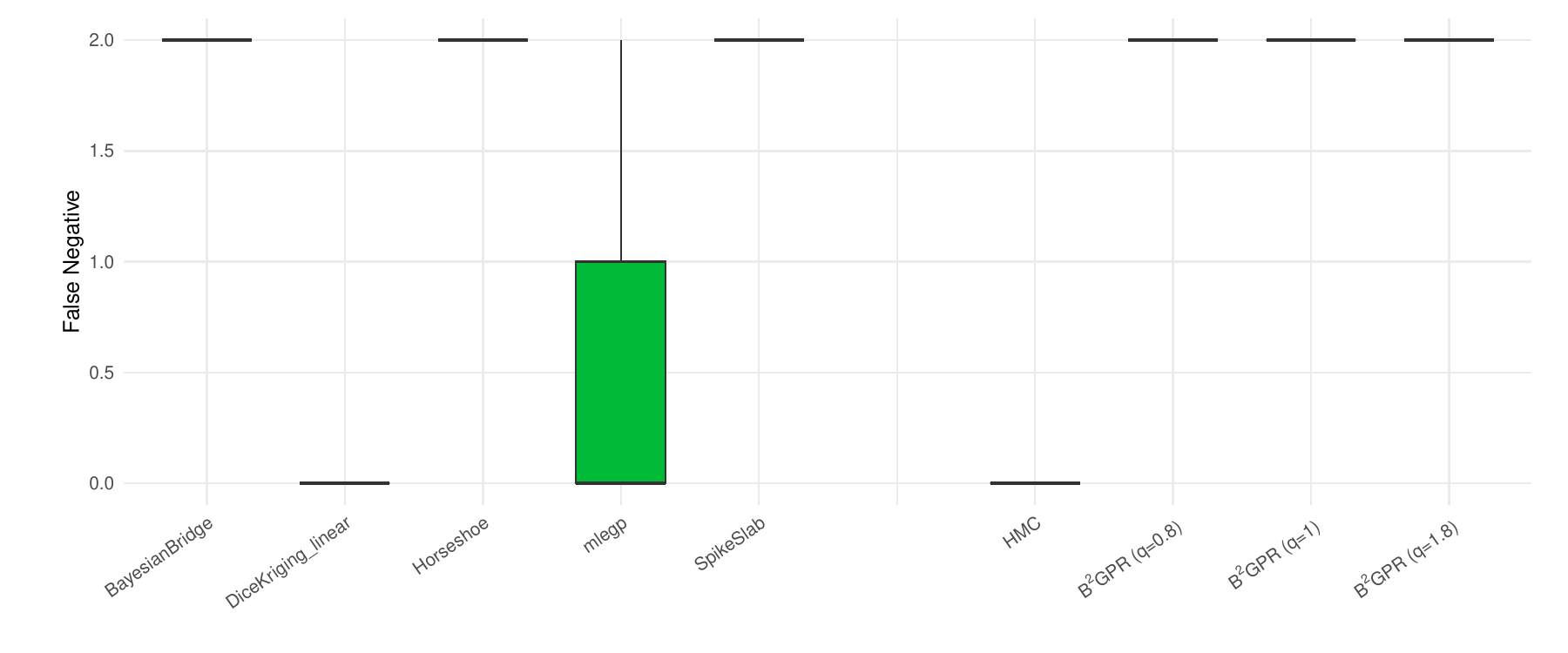}
    \caption{Number of false negatives.}
    \end{subfigure}
    \caption{Box plots of standardized RMSE and variable-selection errors across $100$ replicates for the OTL Circuit function with increased input dimension $d=20$ and $n=500$.}\label{fig:ggplot_otlcircuit_500_20}
\end{figure}

\begin{figure}[p]
    \centering
    \begin{subfigure}[t]{.96\linewidth}
    \includegraphics[width=\linewidth]{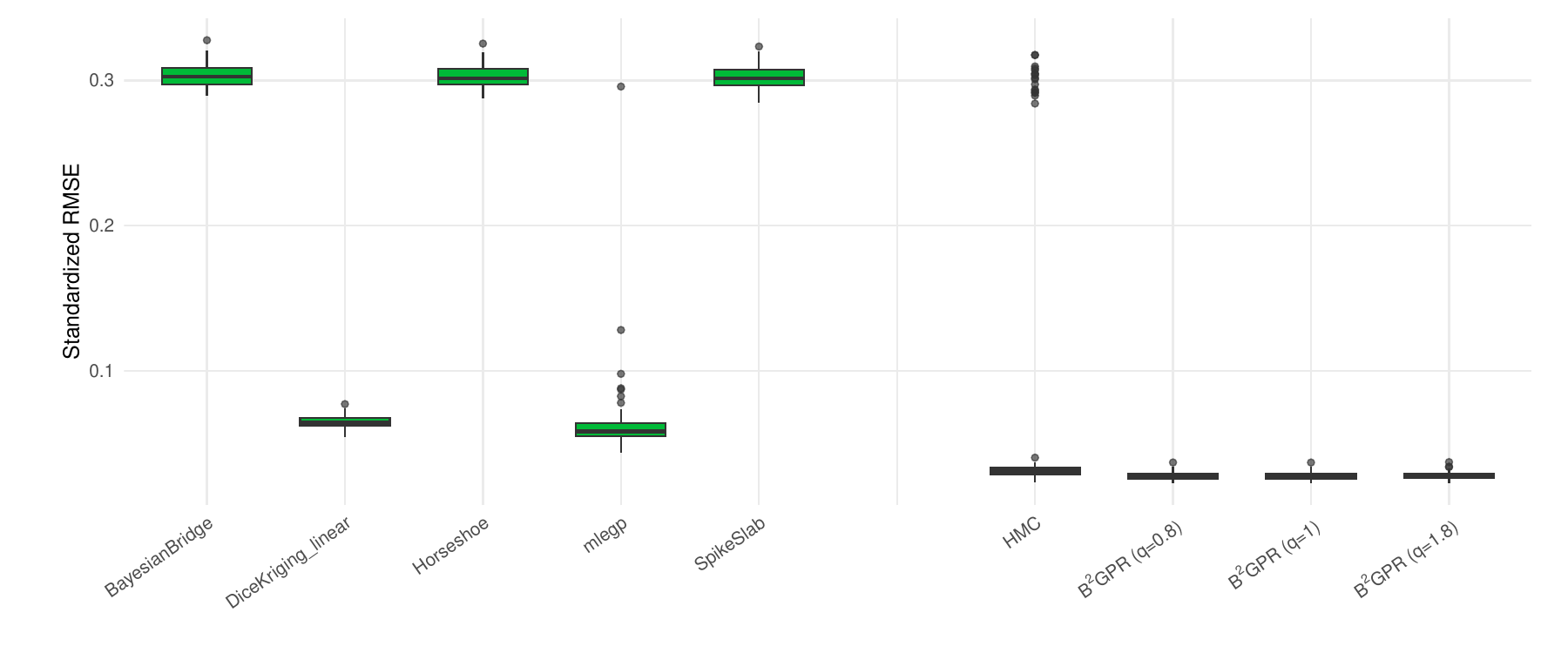}
    \caption{Standardized RMSE.}
    \end{subfigure}
    
    \begin{subfigure}[t]{.96\linewidth}
    \includegraphics[width=\linewidth]{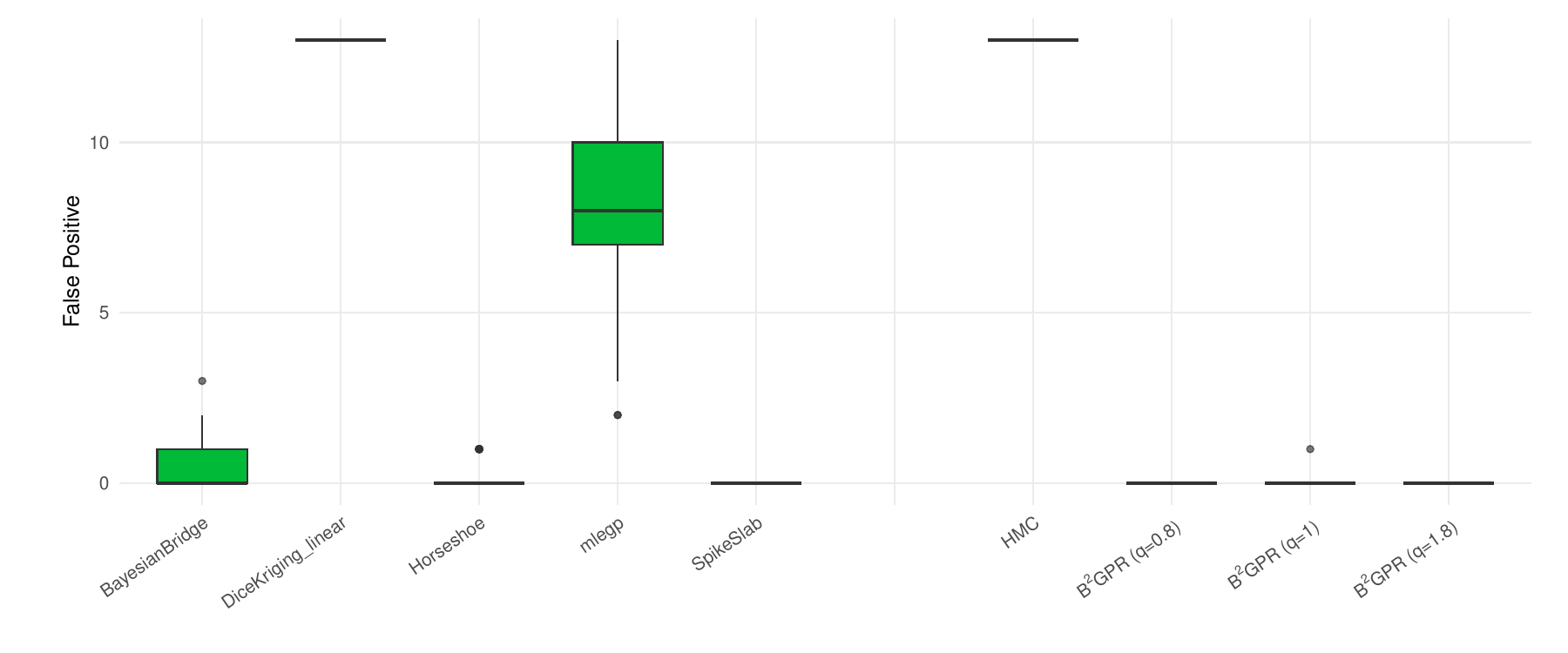}
    \caption{Number of false positives.}
    \end{subfigure}
    
    \begin{subfigure}[t]{.96\linewidth}
    \includegraphics[width=\linewidth]{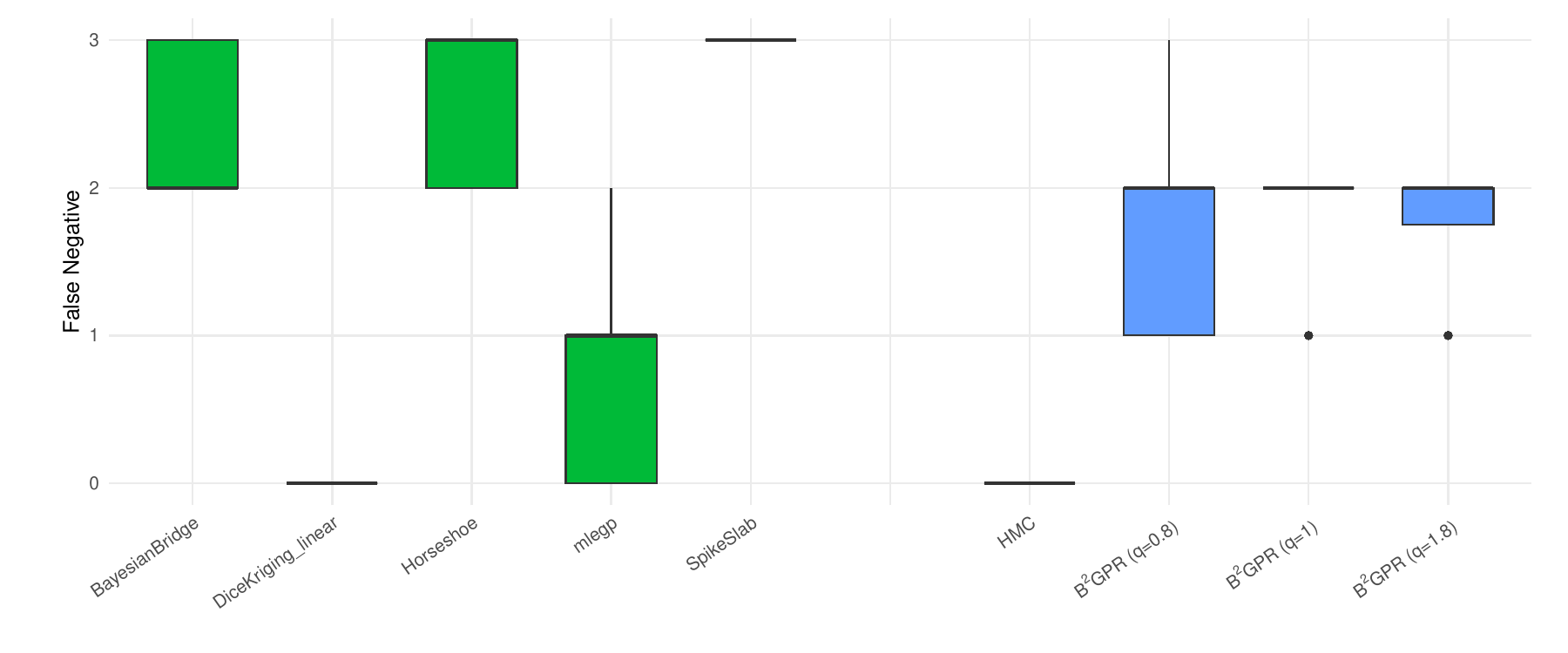}
    \caption{Number of false negatives.}
    \end{subfigure}
    \caption{Box plots of standardized RMSE and variable-selection errors across $100$ replicates for the Piston function with increased input dimension $d=20$ and $n=500$.}\label{fig:ggplot_piston_500_20}
\end{figure}

\clearpage

\section*{Derivation of Proposition~\ref{prop:sph}}
Given the prior as follows:
\begin{equation*} 
\begin{aligned}
& p(\bm \beta|r_{\beta}) \propto \mathbb{I}(\|\bm \beta\|_q \leq r_{\beta}), &&\text{constrained non-informative prior},\\
& p(r_{\beta}) \propto 1, &&\text{non-informative prior},\\
& p(\bm \omega|r_{\omega}) \propto \mathbb{I}(\|\bm \omega\|_q \leq r_{\omega}), &&\text{constrained non-informative prior},\\
& p(r_{\omega}) \propto 1, &&\text{non-informative prior},\\
& \tau^2 \sim \text{Inverse-}\chi^2(df_{\tau^2}),&& \eta \sim \Gamma(a_{\eta}, b_{\eta}).
\end{aligned}
\end{equation*}

\begin{equation*}
    \begin{aligned}
        p(\tau^2) & \propto (\tau^2)^{-(\frac{df_{\tau^2}}{2}+1)} \exp\left(-\frac{1}{2 \tau^2}\right)\\
        p(\eta) & \propto \eta^{a_{\eta}-1} \exp\left(-b_{\eta} \eta\right)\\
    \end{aligned}
\end{equation*}

Given data, the sampling distribution is:
\begin{equation} \label{eq:samp}
    \begin{aligned}
        \bm y_n | \bm\beta, \bm\omega, \tau^2, \eta &\sim \mathcal{MVN}_n\left(\bm G\bm \beta, \tau^2(\bm K_n+\eta \bm I_n)\right)\\
        p(\bm y_n|\bm \beta, \bm \omega, \tau^2, \eta) &\propto (\tau^2)^{-\frac{n}{2}}\det(\bm K_n+\eta\bm I_n)^{-1/2} \\
        &\quad \times \exp\left(-\frac{1}{2\tau^2}(\bm y_n-\bm G\bm \beta)^\top (\bm K_n+\eta \bm I_n)^{-1} (\bm y_n-\bm G\bm \beta)\right)\\
    \end{aligned}
\end{equation}

Through the Bayes rule and \eqref{eq:samp}, we have the following conditional posterior distributions for $\bm \beta$:
\begin{equation*}
    \begin{aligned}
         & p(\bm \beta|\bm y_n, \bm \omega, \tau^2, \eta) \propto p(\bm y_n|\bm \beta, \bm \omega, \tau^2, \eta)p(\bm \beta)\\
        &\propto (\tau^2)^{-\frac{n}{2}}\det(\bm K_n+\eta\bm I_n)^{-1/2}\exp\left(-\frac{1}{2\tau^2}(\bm y_n-\bm G\bm \beta)^\top (\bm K_n+\eta \bm I_n)^{-1} (\bm y_n-\bm G\bm \beta)\right)\\
        & \quad \times \mathbb{I}(\|\bm \beta\|_q \leq r_{\beta})\\
        &\propto \exp\left(-\frac{1}{2\tau^2}(\bm y_n-\bm G\bm \beta)^\top (\bm K_n+\eta \bm I_n)^{-1} (\bm y_n-\bm G\bm \beta)\right) \mathbb{I}(\|\bm \beta\|_q \leq r_{\beta})\\
        &\propto \exp\left(-\frac{1}{2\tau^2}(\bm y_n^\top (\bm K_n+\eta \bm I_n)^{-1} \bm y_n - 2\bm \beta^\top \bm G^\top (\bm K_n+\eta \bm I_n)^{-1} \bm y_n + \bm \beta^\top \bm G^\top (\bm K_n+\eta \bm I_n)^{-1} \bm G\bm \beta)\right)\\
        & \quad \times \mathbb{I}(\|\bm \beta\|_q \leq r_{\beta})\\
        &\propto \exp\left(-\frac{1}{2\tau^2}(\bm \beta^\top \bm G^\top (\bm K_n+\eta \bm I_n)^{-1} \bm G\bm \beta - 2\bm \beta^\top \bm G^\top (\bm K_n+\eta \bm I_n)^{-1} \bm y_n)\right) \mathbb{I}(\|\bm \beta\|_q \leq r_{\beta})\\
    \end{aligned}
\end{equation*}

Following the above posterior density form, we have the following conditional posterior distribution for $\bm \beta$:
\begin{equation*}
    \begin{aligned}
        \bm \beta | \bm y_n,\bm \omega,\tau^2,\eta &\sim \mathcal{MVN}_px\left(\hat{\bm \beta}_n, \bm \Sigma_{\bm \beta|n}\right) \mathbb{I}(\|\bm \beta\|_q \leq r_{\beta}) \\
        \bm \Sigma_{\bm \beta|n} &= \tau^2 (\bm G^\top (\bm K_n+\eta \bm I_n)^{-1} \bm G)^{-1}\\
        \hat{\bm \beta}_n &= \bm \Sigma_{\bm \beta|n}\frac{(\bm G^\top (\bm K_n+\eta \bm I_n)^{-1}\bm y_n)}{\tau^2}\\
        &=(\bm G^\top (\bm K_n+\eta \bm I_n)^{-1} \bm G)^{-1} \bm G^\top (\bm K_n+\eta \bm I_n)^{-1} \bm y_n\\
    \end{aligned}
\end{equation*}

Since we have the mapping $\theta_{\beta_i} \mapsto \beta_i = r_{\beta} \cdot \text{sgn}(\theta_{\beta_i}) |\theta_{\beta_i}|^{2/q}$, the sampling distribution \ref{eq:samp} can be re-expressed as:

\begin{equation} \label{eq:samp2}
    \begin{aligned}
        p(\bm y_n|\bm \theta_{\beta}, r_{\beta}, \bm \omega, \tau^2, \eta) 
        &\propto (\tau^2)^{-\frac{n}{2}}\det(\bm K_n+\eta\bm I_n)^{-1/2} \\
        &\quad \times \exp\Biggl(
            -\frac{1}{2\tau^2}
            (\bm y_n-\bm G(\text{sgn}(\bm \theta_{\beta}) \circ |\bm \theta_{\beta}|^{\frac{2}{q}} r_{\beta}))^\top \\
        &\qquad\qquad\qquad \times
            (\bm K_n+\eta \bm I_n)^{-1}
            (\bm y_n-\bm G(\text{sgn}(\bm \theta_{\beta}) \circ |\bm \theta_{\beta}|^{\frac{2}{q}} r_{\beta}))
        \Biggr)
    \end{aligned}
\end{equation}
where $\circ$ denotes the element-wise Hadamard product.

Through the Bayes rule and \eqref{eq:samp2}, we have the following conditional posterior distributions for $r_{\beta}$:

\begin{equation*}
    \begin{aligned}
        p(r_{\beta} | \bm y_n, \bm \theta_{\beta}, \bm \omega, \tau^2, \eta) & \propto p(\bm y_n|\bm \theta_{\beta}, r_{\beta}, \bm \omega, \tau^2, \eta) p(r_{\beta})\\
        & \propto \exp\Biggl(-\frac{1}{2\tau^2}\Bigl(\bm y_n-\bm G \bigl(r_{\beta} \, \text{sgn}(\bm \theta_{\beta}) \circ |\bm \theta_{\beta}|^{\frac{2}{q}}  \bigr)\Bigr)^\top (\bm K_n+\eta \bm I_n)^{-1} \\
        & \quad \times \Bigl(\bm y_n-\bm G \bigl(r_{\beta} \, \text{sgn}(\bm \theta_{\beta}) \circ |\bm \theta_{\beta}|^{\frac{2}{q}}  \bigr)\Bigr)\Biggr)\\
        -\log f(r_{\beta} | \bm y_n, \bm \theta_{\beta}, \bm \omega, \tau^2, \eta) &= \frac{1}{2\tau^2}\Bigl(\bm y_n-\bm G \bigl(r_{\beta} \, \text{sgn}(\bm \theta_{\beta}) \circ |\bm \theta_{\beta}|^{\frac{2}{q}}  \bigr)\Bigr)^\top (\bm K_n+\eta \bm I_n)^{-1}\\
         & \quad \times \Bigl(\bm y_n-\bm G \bigl(r_{\beta} \, \text{sgn}(\bm \theta_{\beta}) \circ |\bm \theta_{\beta}|^{\frac{2}{q}}  \bigr)\Bigr)\\
    \end{aligned}
\end{equation*}

Through the Bayes rule and \eqref{eq:samp}, we have the following conditional posterior distributions for $\bm \omega$:
\begin{equation*}
    \begin{aligned}
        p(\bm \omega|\bm y_n, \bm \beta, \tau^2, \eta) &\propto p(\bm y_n|\bm \beta, \bm \omega, \tau^2, \eta)p(\bm \omega)\\
        &\propto \det(\bm K_n+\eta\bm I_n)^{-\frac{1}{2}}\exp\left(-\frac{1}{2\tau^2}(\bm y_n-\bm G\bm \beta)^\top (\bm K_n+\eta \bm I_n)^{-1} (\bm y_n-\bm G\bm \beta)\right)\\
        & \quad \times \mathbb{I}(\|\bm \omega\|_q \leq r_{\omega})\\
        \bm K_n(\bm x_i, \bm x_j; \bm \omega)&=\exp\left\{-\sum_{k=1}^d \omega_k^2(x_{ik}-x_{jk})^2\right\}\\
    \end{aligned}
\end{equation*}
Omit the indicator function part for gradient calculation, we have:
\begin{equation*}
    \begin{aligned}
        -\log f(\bm \omega | \bm \beta,\tau^2, \eta, \bm y_n) &= \frac{1}{2} \log \det(\bm K_n+\eta\bm I_n)+\frac{1}{2\tau^2}(\bm y_n-\bm G\bm \beta)^\top (\bm K_n+\eta \bm I_n)^{-1} (\bm y_n-\bm G\bm \beta)\\
        \frac{\partial (-\log f(\bm \omega | \bm \beta,\tau^2, \eta, \bm y_n))}{\partial \omega_k}&= \frac{1}{2} \mathrm{Tr}((\bm K_n+\eta \bm I_n)^{-1} \frac{\partial \bm K_n}{\partial \omega_k})\\
        & \qquad -\frac{1}{2\tau^2}(\bm y_n-\bm G\bm \beta)^\top (\bm K_n+\eta \bm I_n)^{-1} \frac{\partial \bm K_n}{\partial \omega_k} (\bm K_n+\eta \bm I_n)^{-1} (\bm y_n-\bm G\bm \beta)\\
        \frac{\partial \bm K_n(\bm x_i, \bm x_j; \bm \omega)}{\partial \omega_k}&=\bm K_n(\bm x_i, \bm x_j; \bm \omega) (-2\omega_k(x_{ik}-x_{jk})^2)\\
        \nabla_{\bm \omega} (-\log f(\bm \omega | \bm \beta, \tau^2, \eta, \bm y_n)) &=(\frac{\partial}{\partial \omega_1},\dots,\frac{\partial}{\partial \omega_d}) (-\log f(\bm \omega | \bm \beta, \tau^2, \eta, \bm y_n))\\
    \end{aligned}
\end{equation*}

Since we have the mapping $\theta_{\omega_i} \mapsto \omega_i = r_{\omega} \cdot \text{sgn}(\theta_{\omega_i}) |\theta_{\omega_i}|^{2/q}$, the sampling distribution \eqref{eq:samp} can be re-expressed as:
\begin{equation} \label{eq:samp3}
    \begin{aligned}
        p(\bm y_n|\bm \beta, \bm \theta_{\omega}, r_{\omega}, \tau^2, \eta) & \propto (\tau^2)^{-\frac{n}{2}}\det(\bm K_n+\eta\bm I_n)^{-1/2}\exp\left(-\frac{1}{2\tau^2}(\bm y_n-\bm G\bm \beta)^\top (\bm K_n+\eta \bm I_n)^{-1} (\bm y_n-\bm G\bm \beta)\right)\\
        \text{where}\quad \bm K_n(\bm x_i, \bm x_j; \bm \omega)&=\exp\left\{-\sum_{k=1}^d \omega_k^2(x_{ik}-x_{jk})^2\right\}\\
         &=\exp\left\{-\sum_{k=1}^d (\text{sgn}(\theta_{\omega_k}) |\theta_{\omega_k}|^{\frac{2}{q}} r_{\omega})^2(x_{ik}-x_{jk})^2\right\}\\
         &=\exp\left\{-r_{\omega}^2\sum_{k=1}^d |\theta_{\omega_k}|^{\frac{4}{q}} (x_{ik}-x_{jk})^2\right\}\\
    \end{aligned}
\end{equation}

Through the Bayes rule and \eqref{eq:samp3}, we have the following conditional posterior distributions for $r_{\omega}$:

\begin{equation*}
   \begin{aligned}
        p(r_{\omega} | \bm y_n, \bm \beta, \bm \theta_{\omega}, \tau^2, \eta) & \propto p(\bm y_n|\bm \beta, \bm \theta_{\omega}, r_{\omega}, \tau^2, \eta) p(r_{\omega})\\
         & \propto \det(\bm K_n+\eta\bm I_n)^{-\frac{1}{2}}\exp\left(-\frac{1}{2\tau^2}(\bm y_n-\bm G\bm \beta)^\top (\bm K_n+\eta \bm I_n)^{-1} (\bm y_n-\bm G\bm \beta)\right)\\
        -\log f(r_{\omega} | \bm y_n, \bm \beta, \bm \theta_{\omega}, \tau^2, \eta) &=\frac{1}{2} \log \det(\bm K_n+\eta\bm I_n)+\frac{1}{2\tau^2}(\bm y_n-\bm G\bm \beta)^\top (\bm K_n+\eta \bm I_n)^{-1} (\bm y_n-\bm G\bm \beta)\\
        \bm K_n(\bm x_i, \bm x_j; \bm \omega)&=\exp\left\{-r_{\omega}^2\sum_{k=1}^d |\theta_{\omega_k}|^{\frac{4}{q}} (x_{ik}-x_{jk})^2\right\}\\
   \end{aligned} 
\end{equation*}

Through the Bayes rule and \eqref{eq:samp}, we have the following conditional posterior distributions for $\tau^2$:
\begin{equation*}
    \begin{aligned}
        p(\tau^2 | \bm \beta, \bm \omega, \eta, \bm y_n) &\propto p(\tau^2) p(\bm y_n|\bm \beta, \bm \omega, \tau^2, \eta)\\
        &\propto (\tau^2)^{-(\frac{df_{\tau^2}}{2}+1)} \exp(-\frac{1}{2 \tau^2}) (\tau^2)^{-\frac{n}{2}} \\ 
        &\quad \times \exp\left(-\frac{1}{2\tau^2}(\bm y_n-\bm G\bm \beta)^\top (\bm K_n+\eta \bm I_n)^{-1} (\bm y_n-\bm G\bm \beta)\right)\\
        \text{Let} \quad S_n^2 &=(\bm y_n-\bm G\bm \beta)^\top (\bm K_n+\eta \bm I_n)^{-1} (\bm y_n-\bm G\bm \beta)\\
        p(\tau^2 | \bm \beta, \bm \omega, \eta, \bm y_n) &\propto (\tau^2)^{-(\frac{1}{2}(df_{\tau^2}+n)+1)} \exp(-\frac{1+S_n^2}{2 \tau^2})\\
        \text{which implies}\quad &\tau^2 | \bm \beta, \bm \omega, \eta, \bm y_n \sim \text{Scaled Inverse-}\chi^2(df_{\tau^2}+n, \frac{1+S_n^2}{df_{\tau^2}+n})\\
    \end{aligned}
\end{equation*}

Through the Bayes rule and \eqref{eq:samp}, we have the following conditional posterior distributions for $\eta$:
\begin{equation*}
    \begin{aligned}
        p(\eta | \bm \beta, \bm \omega, \tau^2, \bm y_n)&\propto p(\eta) p(\bm y_n|\bm \beta, \bm \omega, \tau^2, \eta)\\
        &\propto \eta^{a_{\eta}-1} e^{-b_{\eta} \eta} \det(\bm K_n+\eta\bm I_n)^{-\frac{1}{2}} \\
        &\quad \times \exp\left(-\frac{1}{2\tau^2}(\bm y_n-\bm G\bm \beta)^\top (\bm K_n+\eta \bm I_n)^{-1} (\bm y_n-\bm G\bm \beta)\right)\\
        -\log f(\eta | \bm \beta, \bm \omega, \tau^2,  \bm y_n) &= (1-a_{\eta}) \log \eta + b_{\eta} \eta + \frac{1}{2} \log \det(\bm K_n+\eta\bm I_n)\\
        & \quad +\frac{1}{2\tau^2}(\bm y_n-\bm G\bm \beta)^\top (\bm K_n+\eta \bm I_n)^{-1} (\bm y_n-\bm G\bm \beta)\\
    \end{aligned}
\end{equation*}

\section*{Derivation of Proposition~\ref{prop:hmc}}
Given the prior as follows:

\begin{equation*}
    \begin{aligned}
        & \bm \beta | \nu_{\beta}^2 \sim \mathcal{MVN}_p({\bf 0}, \nu_{\beta}^2 \bm R) \\
        & \nu_{\beta}^2 \sim \text{Inverse-} \Gamma (a_{\beta},b_{\beta}) \\
        & \omega_i | \nu_{\omega}^2 \overset{\text{i.i.d.}}{\sim} \mathcal{N}(0, \nu_{\omega}^2), \text{ for } i=1,\ldots, d \\
        & \nu_{\omega}^2 \sim \text{Inverse-} \Gamma (a_{\omega},b_{\omega}) \\
        & \tau^2 \sim \text{Inverse-}\chi^2(df_{\tau^2}) \\
        & \eta \sim \Gamma(a_{\eta}, b_{\eta}) \\
    \end{aligned}
\end{equation*}

Given data, the sampling distribution is the same as \eqref{eq:samp}.

Through the Bayes rule and \eqref{eq:samp}, we have the following conditional posterior distributions for $\bm \beta$:

\begin{equation*}
    \begin{aligned}
        p(\bm \beta| \nu_{\beta}^2, \bm \omega, \nu_{\omega}^2,\tau^2, \eta,\bm y_n) &\propto p(\bm \beta | \nu_{\beta}^2) p(\bm y_n|\bm \beta, \nu_{\beta}^2, \bm \omega, \nu_{\omega}^2,\tau^2, \eta) \\
        &\propto \det(\nu_{\beta}^2 \bm R)^{-\frac{1}{2}} \exp(-\frac{1}{2} \bm \beta^\top (\nu_{\beta}^2 \bm R)^{-1} \bm \beta)\\
         &\quad \times \exp\left(-\frac{1}{2\tau^2}(\bm y_n-\bm G\bm \beta)^\top (\bm K_n+\eta \bm I_n)^{-1} (\bm y_n-\bm G\bm \beta)\right)\\
        &\propto \exp(-\frac{1}{2} \bm \beta^\top (\nu_{\beta}^2 \bm R)^{-1} \bm \beta) \exp\left(-\frac{1}{2\tau^2}(\bm y_n-\bm G\bm \beta)^\top (\bm K_n+\eta \bm I_n)^{-1} (\bm y_n-\bm G\bm \beta)\right)\\
    \end{aligned}
\end{equation*}

Due to the conditional conjugacy or from the formula above (we can combine the terms in two exponential functions), we can find that the conditional posterior distribution of $\bm \beta$ is still a multivariate normal distribution.

\begin{equation*}
    \begin{aligned}
        \bm \beta | \bm y_n, \nu_{\beta}^2, \bm \omega, \nu_{\omega}^2,\tau^2,\eta &\sim \mathcal{MVN}_p\left(\hat{\bm \beta}_n, \bm \Sigma_{\bm \beta|n}\right)\\
        \bm \Sigma_{\bm \beta|n} &=\left[\frac{1}{\tau^2}\bm G^\top (\bm K_n+\eta \bm I_n)^{-1}\bm G+\frac{1}{\nu_{\beta}^2} \bm R^{-1}\right]^{-1}\\
        \hat{\bm \beta}_n &=\bm \Sigma_{\bm \beta|n}\frac{(\bm G^\top (\bm K_n+\eta \bm I_n)^{-1}\bm y_n)}{\tau^2}\\
    \end{aligned}
\end{equation*}

\begin{equation*}
    \begin{aligned}
        p(\bm \beta | \bm y_n, \nu_{\beta}^2, \bm \omega, \nu_{\omega}^2,\tau^2,\eta) &\propto \exp(-\frac{1}{2} (\bm \beta-\hat{\bm \beta}_n)^\top (\bm \Sigma_{\bm \beta|n})^{-1} (\bm \beta-\hat{\bm \beta}_n))\\
        -\log f(\bm \beta | \bm y_n, \nu_{\beta}^2, \bm \omega, \nu_{\omega}^2,\tau^2,\eta) &= \frac{1}{2} (\bm \beta-\hat{\bm \beta}_n)^\top (\bm \Sigma_{\bm \beta|n})^{-1} (\bm \beta-\hat{\bm \beta}_n)\\
        \nabla_{\bm \beta} -\log f(\bm \beta | \bm y_n, \nu_{\beta}^2, \bm \omega, \nu_{\omega}^2,\tau^2,\eta) &= (\bm \Sigma_{\bm \beta|n})^{-1} (\bm \beta-\hat{\bm \beta}_n)\\
    \end{aligned}
\end{equation*}

Similarly, we can get the conditional posterior distribution of $\nu_{\beta}^2$ as follows:

\begin{equation*}
    \begin{aligned}
        p(\nu_{\beta}^2 | \bm \beta, \bm \omega, \nu_{\omega}^2,\tau^2, \eta, \bm y_n) &= \frac{p(\nu_{\beta}^2 , \bm \beta, \bm \omega, \nu_{\omega}^2,\tau^2, \eta, \bm y_n)}{p(\bm \beta, \bm \omega, \nu_{\omega}^2,\tau^2, \eta, \bm y_n)}\\
        &\propto p(\bm \beta | \nu_{\beta}^2) p(\nu_{\beta}^2) p(\bm y_n|\bm \beta, \nu_{\beta}^2, \bm \omega, \nu_{\omega}^2,\tau^2, \eta) \\
        &\propto \det(\nu_{\beta}^2 \bm R)^{-\frac{1}{2}} \exp(-\frac{1}{2} \bm \beta^\top (\nu_{\beta}^2 \bm R)^{-1} \bm \beta) (\nu_{\beta}^2)^{-(a_{\beta}+1)} \exp(-\frac{b_{\beta}}{\nu_{\beta}^2}) \\
        & \qquad \times \exp\left(-\frac{1}{2\tau^2}(\bm y_n-\bm G\bm \beta)^\top (\bm K_n+\eta \bm I_n)^{-1} (\bm y_n-\bm G\bm \beta)\right)\\
        &\propto \det(\bm R)^{-\frac{1}{2}}(\nu_{\beta}^2)^{-(a_{\beta}+\frac{p}{2}+1)} \exp\left(-\frac{1}{\nu_{\beta}^2}\left(\frac{1}{2} \bm \beta^\top \bm R^{-1} \bm \beta+b_{\beta}\right)\right)\\
        \nu_{\beta}^2 | \bm \beta, \bm \omega,\tau^2, \eta, \bm y_n &\sim \text{Inverse-} \Gamma \left(a_{\beta}+\frac{p}{2},\frac{1}{2} \bm \beta^\top \bm R^{-1} \bm \beta+b_{\beta}\right)\\
    \end{aligned}
\end{equation*}

If $R$ is identity matrix, then

\begin{equation*}
    \begin{aligned}
        p(\nu_{\beta}^2 | \bm \beta, \bm \omega, \nu_{\omega}^2,\tau^2, \eta, \bm y_n) &\propto (\nu_{\beta}^2)^{-(a_{\beta}+\frac{p}{2}+1)} \exp\left(-\frac{1}{\nu_{\beta}^2}\left(\frac{1}{2} \bm \beta^\top \bm \beta+b_{\beta}\right)\right)\\
        \nu_{\beta}^2 | \bm \beta, \bm \omega, \nu_{\omega}^2,\tau^2, \eta, \bm y_n &\sim \text{Inverse-} \Gamma \left(a_{\beta}+\frac{p}{2},\frac{1}{2} \bm \beta^\top \bm \beta+b_{\beta}\right)\\
    \end{aligned}
\end{equation*}

We can get the conditional posterior distribution of $\bm \omega$ as follows:

\begin{equation*}
    \begin{aligned}
        p(\bm \omega | \bm \beta, \nu_{\beta}^2, \nu_{\omega}^2,\tau^2, \eta, \bm y_n) &\propto \left(\prod_{i=1}^d p(\omega_i | \nu_{\omega}^2)\right) p(\bm y_n|\bm \beta, \nu_{\beta}^2, \bm \omega, \nu_{\omega}^2,\tau^2, \eta)\\
        p(\bm \omega | \bm \beta, \nu_{\beta}^2, \nu_{\omega}^2,\tau^2, \eta, \bm y_n) &\propto \left(\prod_{i=1}^{d} (\nu_{\omega}^2)^{-\frac{1}{2}} \exp(-\frac{\omega_i^2}{2\nu_{\omega}^2})\right) \det(\bm K_n+\eta\bm I_n)^{-\frac{1}{2}}\\
        & \qquad \times \exp\left(-\frac{1}{2\tau^2}(\bm y_n-\bm G\bm \beta)^\top (\bm K_n+\eta \bm I_n)^{-1} (\bm y_n-\bm G\bm \beta)\right)\\
        &\propto \left(\prod_{i=1}^{d} \exp(-\frac{\omega_i^2}{2\nu_{\omega}^2})\right) \det(\bm K_n+\eta\bm I_n)^{-\frac{1}{2}}\\
        & \qquad \times \exp\left(-\frac{1}{2\tau^2}(\bm y_n-\bm G\bm \beta)^\top (\bm K_n+\eta \bm I_n)^{-1} (\bm y_n-\bm G\bm \beta)\right)\\
        -\log f(\bm \omega | \bm \beta, \nu_{\beta}^2, \nu_{\omega}^2,\tau^2, \eta, \bm y_n) &= \sum_{i=1}^{d} \left(\frac{\omega_i^2}{2\nu_{\omega}^2}\right) + \frac{1}{2} \log \det(\bm K_n+\eta\bm I_n)\\
        & \quad +\frac{1}{2\tau^2}(\bm y_n-\bm G\bm \beta)^\top (\bm K_n+\eta \bm I_n)^{-1} (\bm y_n-\bm G\bm \beta)\\
        \frac{\partial (-\log f(\bm \omega | \bm \beta, \nu_{\beta}^2, \nu_{\omega}^2,\tau^2, \eta, \bm y_n))}{\partial \omega_k}&= \frac{\omega_k}{\nu_{\omega}^2}+ \frac{1}{2} \text{Tr}((\bm K_n+\eta \bm I_n)^{-1} \frac{\partial \bm K_n}{\partial \omega_k})\\
        & \quad -\frac{1}{2\tau^2}(\bm y_n-\bm G\bm \beta)^\top (\bm K_n+\eta \bm I_n)^{-1} \frac{\partial \bm K_n}{\partial \omega_k} (\bm K_n+\eta \bm I_n)^{-1} (\bm y_n-\bm G\bm \beta)\\
        \nabla_{\bm \omega} (-\log f(\bm \omega | \bm \beta, \nu_{\beta}^2, \nu_{\omega}^2,\tau^2, \eta, \bm y_n)) &=(\frac{\partial}{\partial \omega_1},\dots,\frac{\partial}{\partial \omega_d}) (-\log f(\bm \omega | \bm \beta, \nu_{\beta}^2, \nu_{\omega}^2,\tau^2, \eta, \bm y_n))\\
    \end{aligned}
\end{equation*}
Considering the specific element of the kernel matrix $\bm K_n$, we have

\begin{equation*}
    \begin{aligned}
        \bm K_n(\bm x_i, \bm x_j; \bm \omega)&=\exp\left\{-\sum_{k=1}^d \omega_k^2(x_{ik}-x_{jk})^2\right\}\\
        \frac{\partial \bm K_n(\bm x_i, \bm x_j; \bm \omega)}{\partial \omega_k}&=\bm K_n(\bm x_i, \bm x_j; \bm \omega) (-2\omega_k(x_{ik}-x_{jk})^2)\\
    \end{aligned}
\end{equation*}

Then we can get the conditional posterior distribution of $\nu_{\omega}^2$ as follows:

\begin{equation*}
    \begin{aligned}
        p(\nu_{\omega}^2 | \bm \beta, \nu_{\beta}^2, \bm \omega, \tau^2, \eta, \bm y_n) &= \frac{p(\nu_{\omega}^2 , \bm \beta, \nu_{\beta}^2, \bm \omega, \tau^2, \eta, \bm y_n)}{p(\bm \beta, \nu_{\beta}^2, \bm \omega, \tau^2, \eta, \bm y_n)}\\
        &\propto p(\bm \omega | \nu_{\omega}^2) p(\nu_{\omega}^2) p(\bm y_n|\bm \beta, \nu_{\beta}^2, \bm \omega, \nu_{\omega}^2,\tau^2, \eta) \\
        &\propto \left(\prod_{i=1}^{d} (\nu_{\omega}^2)^{-\frac{1}{2}} \exp(-\frac{\omega_i^2}{2\nu_{\omega}^2})\right) (\nu_{\omega}^2)^{-(a_{\omega}+1)} \exp(-\frac{b_{\omega}}{\nu_{\omega}^2})\\
        & \qquad \det(\bm K_n+\eta\bm I_n)^{-\frac{1}{2}} \exp\left(-\frac{1}{2\tau^2}(\bm y_n-\bm G\bm \beta)^\top (\bm K_n+\eta \bm I_n)^{-1} (\bm y_n-\bm G\bm \beta)\right)\\
        &\propto (\nu_{\omega}^2)^{-(a_{\omega}+\frac{d}{2}+1)} \exp\left(-\frac{1}{\nu_{\omega}^2}\left(\frac{1}{2} \sum_{i=1}^{d} \omega_i^2+b_{\omega}\right)\right)\\
    \end{aligned}
\end{equation*}

\begin{equation*}
    \begin{aligned}
        \nu_{\omega}^2 | \bm \beta, \nu_{\beta}^2, \bm \omega, \tau^2, \eta, \bm y_n &\sim \text{Inverse-} \Gamma \left(a_{\omega}+\frac{d}{2},\frac{1}{2} \sum_{i=1}^{d} \omega_i^2+b_{\omega}\right)\\
    \end{aligned}
\end{equation*}

Similarly, we can get the conditional posterior distribution of $\tau^2$ as follows:

\begin{equation*}
    \begin{aligned}
        p(\tau^2 | \bm \beta, \nu_{\beta}^2,\bm \omega, \nu_{\omega}^2,\eta, \bm y_n) &\propto p(\tau^2) p(\bm y_n|\bm \beta, \nu_{\beta}^2,\bm \omega, \nu_{\omega}^2,\tau^2, \eta)\\
        &\propto (\tau^2)^{-(\frac{df_{\tau^2}}{2}+1)} \exp(-\frac{1}{2 \tau^2}) (\tau^2)^{-\frac{n}{2}} \\
        &\quad \times \exp\left(-\frac{1}{2\tau^2}(\bm y_n-\bm G\bm \beta)^\top (\bm K_n+\eta \bm I_n)^{-1} (\bm y_n-\bm G\bm \beta)\right)\\
    \end{aligned}
\end{equation*}
Let $S_n^2 =(\bm y_n-\bm G\bm \beta)^\top (\bm K_n+\eta \bm I_n)^{-1} (\bm y_n-\bm G\bm \beta)$, we have
\begin{equation*}
    \begin{aligned}
        p(\tau^2 | \bm \beta, \nu_{\beta}^2,\bm \omega, \nu_{\omega}^2,\eta, \bm y_n) &\propto (\tau^2)^{-(\frac{1}{2}(df_{\tau^2}+n)+1)} \exp(-\frac{1+S_n^2}{2 \tau^2})\\
        \text{which implies} \quad \tau^2 | \bm \beta, \nu_{\beta}^2, \bm \omega, \nu_{\omega}^2,\eta, \bm y_n  &\sim \text{Scaled Inverse-}\chi^2(df_{\tau^2}+n, \frac{1+S_n^2}{df_{\tau^2}+n})
    \end{aligned}
\end{equation*}

Then we can get the conditional posterior distribution of $\eta$ as follows:
\begin{equation*}
    \begin{aligned}
        p(\eta | \bm \beta, \nu_{\beta}^2, \bm \omega, \nu_{\omega}^2,\tau^2, \bm y_n)&\propto p(\eta) p(\bm y_n|\bm \beta, \nu_{\beta}^2, \bm \omega, \nu_{\omega}^2,\tau^2, \eta)\\
        &\propto \eta^{a_{\eta}-1} e^{-b_{\eta} \eta} \det(\bm K_n+\eta\bm I_n)^{-\frac{1}{2}} \\
        &\quad \times \exp\left(-\frac{1}{2\tau^2}(\bm y_n-\bm G\bm \beta)^\top (\bm K_n+\eta \bm I_n)^{-1} (\bm y_n-\bm G\bm \beta)\right)\\
        -\log f(\eta | \bm \beta, \nu_{\beta}^2, \bm \omega, \nu_{\omega}^2,\tau^2,  \bm y_n) &= (1-a_{\eta}) \log \eta + b_{\eta} \eta + \frac{1}{2} \log \det(\bm K_n+\eta\bm I_n)\\
        & \quad +\frac{1}{2\tau^2}(\bm y_n-\bm G\bm \beta)^\top (\bm K_n+\eta \bm I_n)^{-1} (\bm y_n-\bm G\bm \beta)\\
    \end{aligned}
\end{equation*}

\end{document}